\documentclass[acmtog,nonacm]{acmart}
\usepackage{booktabs} 
\pdfoutput=1
\usepackage[ruled]{algorithm2e} 

\SetAlFnt{\small}
\SetAlCapFnt{\small}
\SetAlCapNameFnt{\small}
\SetAlCapHSkip{0pt}

\setcopyright{none}	

\usepackage{amsmath}
\usepackage{amsfonts}
\usepackage{enumitem}
\usepackage[ruled]{algorithm2e}
\usepackage[normalem]{ulem}
\usepackage{microtype}
\usepackage{units}
\usepackage{gensymb}
\usepackage{threeparttable}
\usepackage{array}
\usepackage{subfig}

\usepackage{dsfont}
\usepackage{wrapfig}

\usepackage{mdframed}
\mdfsetup{skipabove=\topskip,skipbelow=\topskip}
\newmdenv[	linewidth=0.75pt,
			leftline=false,
			rightline=false,
   			innerleftmargin=0pt,
			innerrightmargin=0pt]{topbot}

\newcommand{\figref}[1]{Fig.\ \ref{#1}}
\newcommand{\secref}[1]{Sec.\ \ref{#1}}
\newcommand{\equref}[1]{Eq.\ \eqref{#1}}

\DeclareMathOperator*{\argmin}{arg\,min}

\newcommand{\set}[1]{\mathcal{#1}}

\newcommand{\tr}{{\mathsf{T}}}

\newcommand{\R}{\mathds{R}} 
\newcommand{\Sdev}{\mathcal{S}} 

\newcommand{\rperp}{r^\perp}

\begin{document}
	\title{Dev2PQ: Planar Quadrilateral Strip Remeshing of Developable Surfaces}

	\author{Floor Verhoeven}
	\affiliation{%
		\institution{ETH Zurich}
	}
	\email{vfloor@inf.ethz.ch}
	\author{Amir Vaxman}
	\affiliation{%
		\institution{Utrecht University}
	}
	\email{a.vaxman@uu.nl}
	\author{Tim Hoffmann}
	\affiliation{%
		\institution{TU Munich}
	}
	\email{tim.hoffmann@ma.tum.de}
	\author{Olga Sorkine-Hornung}
	\affiliation{%
		\institution{ETH Zurich}
	}
	\email{sorkine@inf.ethz.ch}

\begin{teaserfigure}
  \centering
  \setlength{\tabcolsep}{2pt}
  \begin{tabular}{ccccc}
\includegraphics[height=0.12\linewidth]{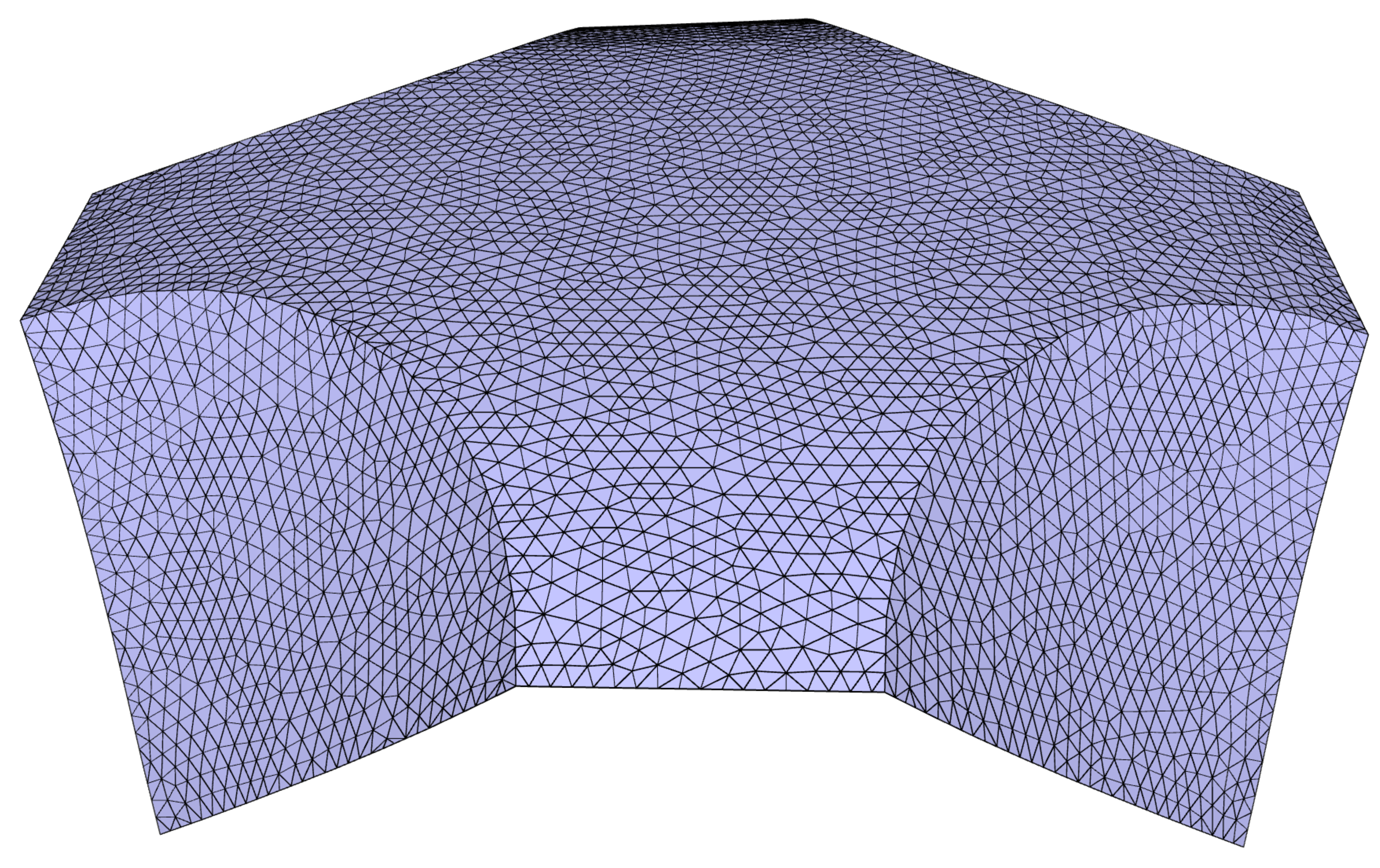} &
 \includegraphics[height=0.12\linewidth]{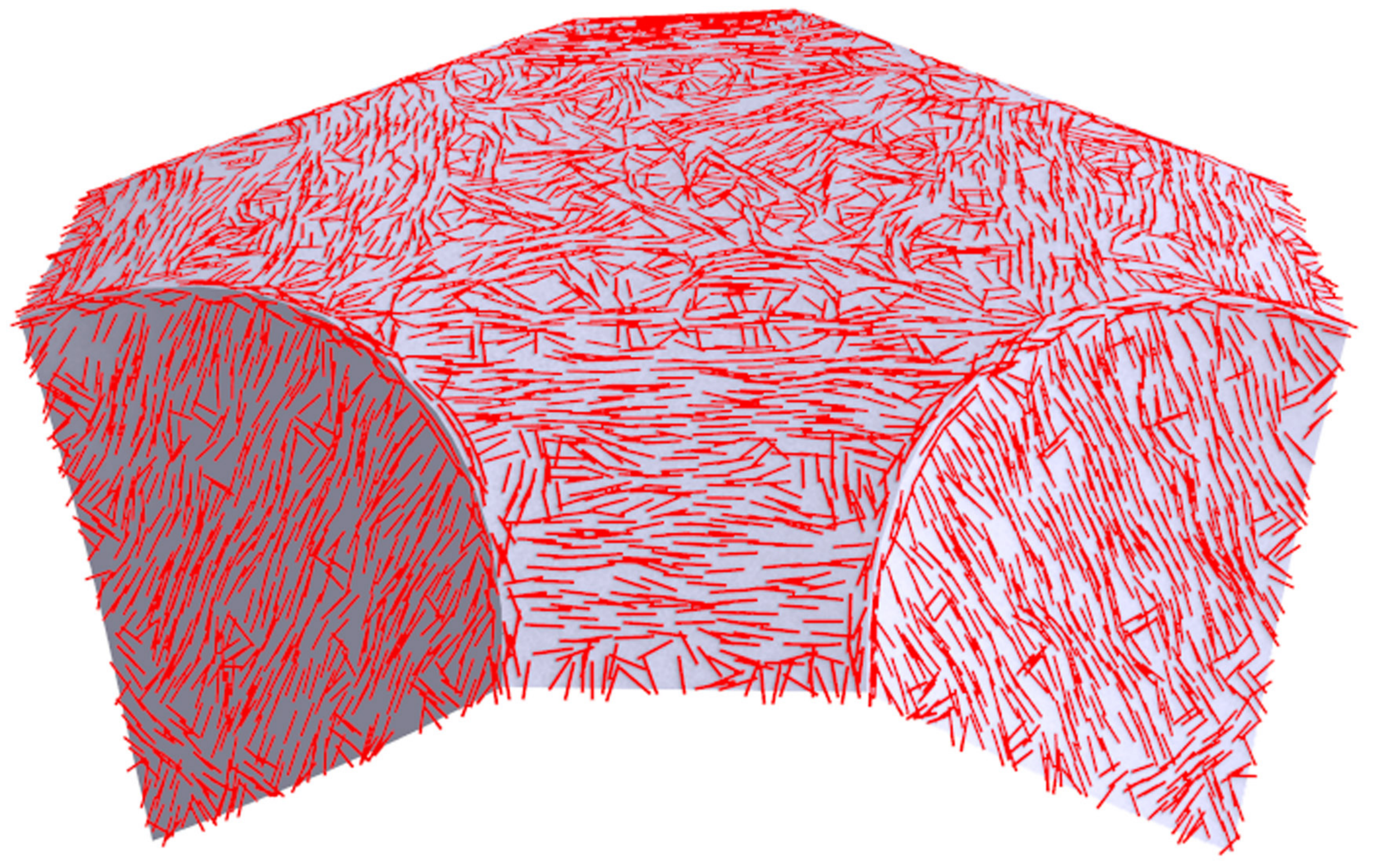} &
 \includegraphics[height=0.12\linewidth]{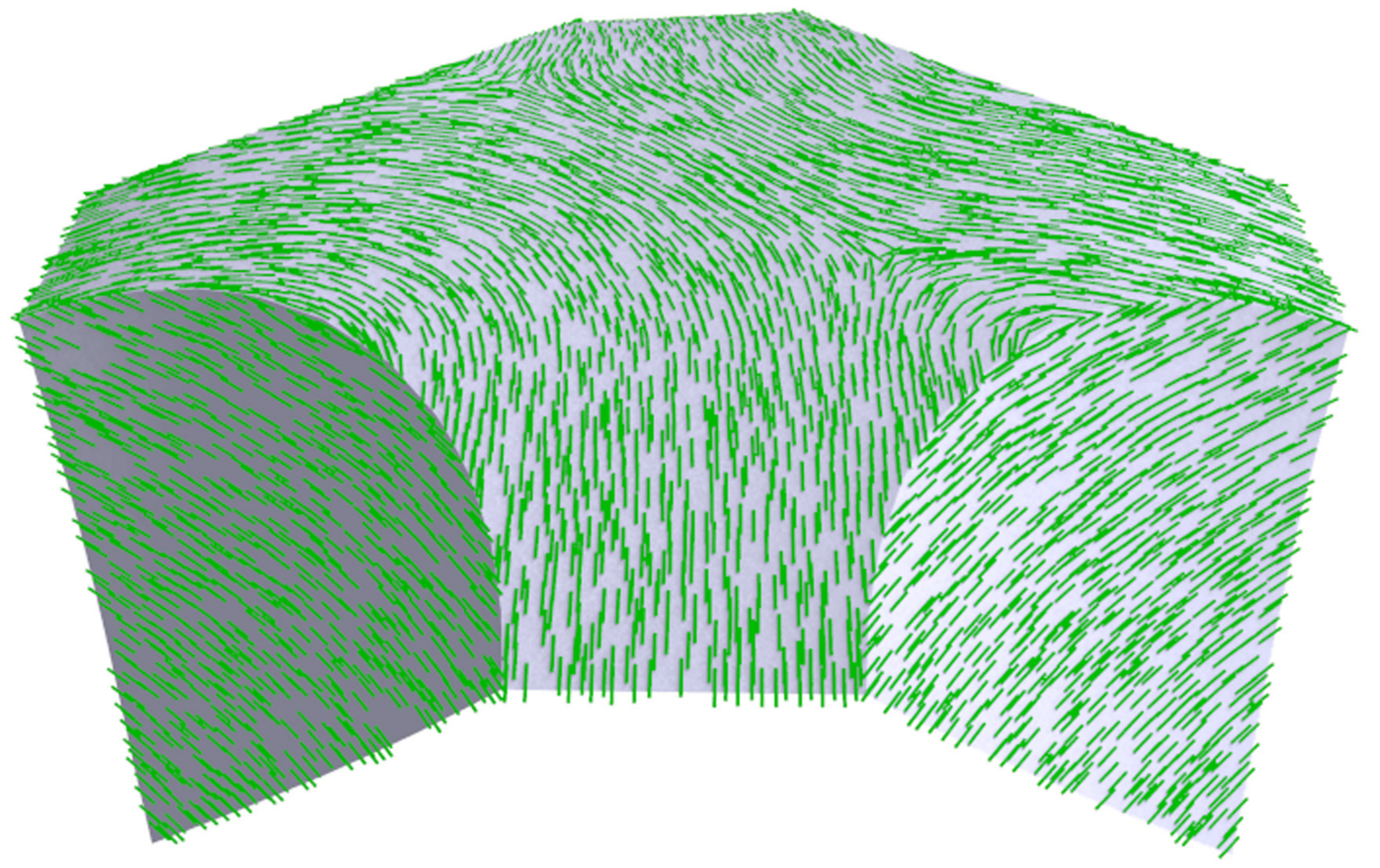} &
 \includegraphics[height=0.12\linewidth]{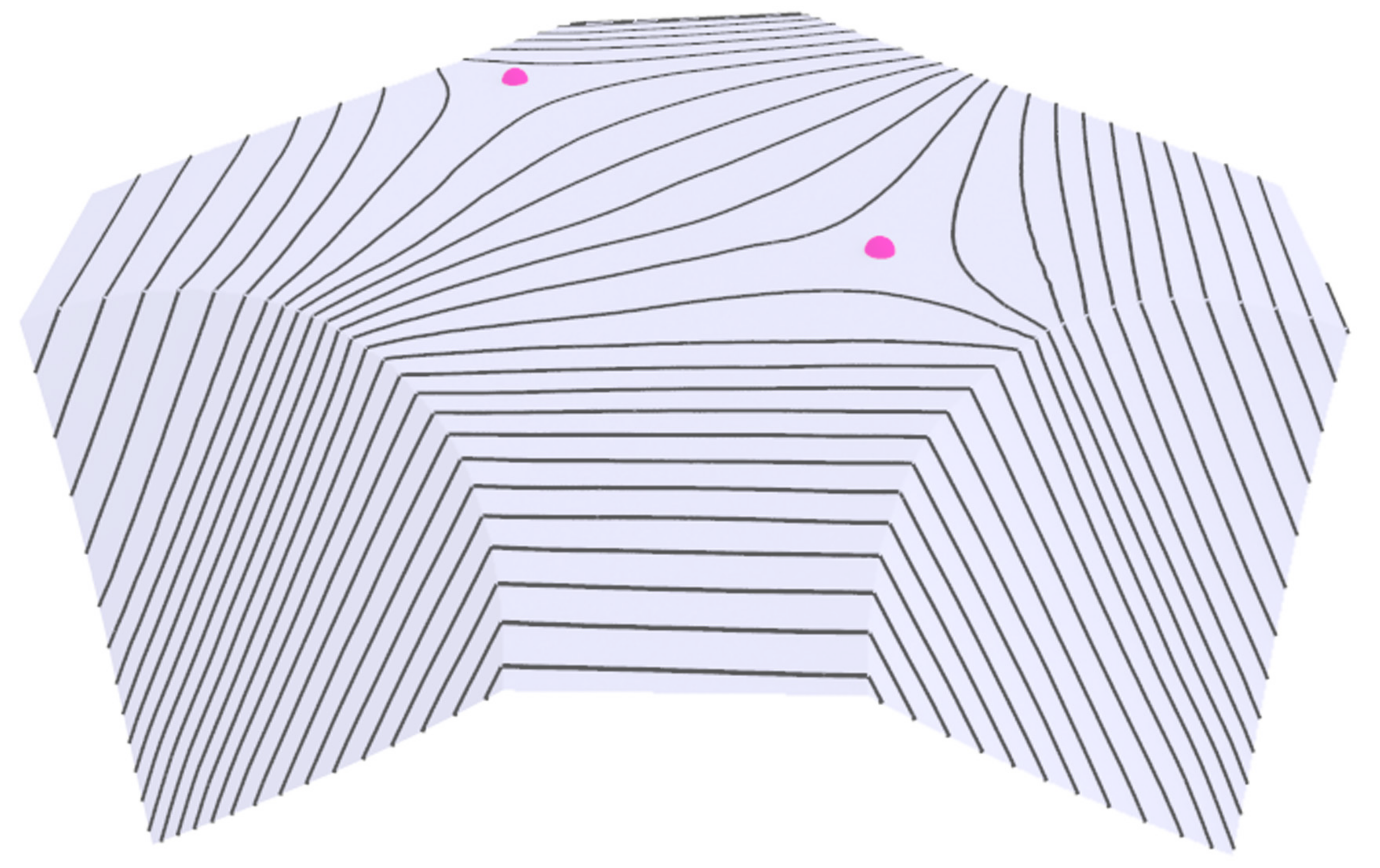} &
 \includegraphics[height=0.12\linewidth]{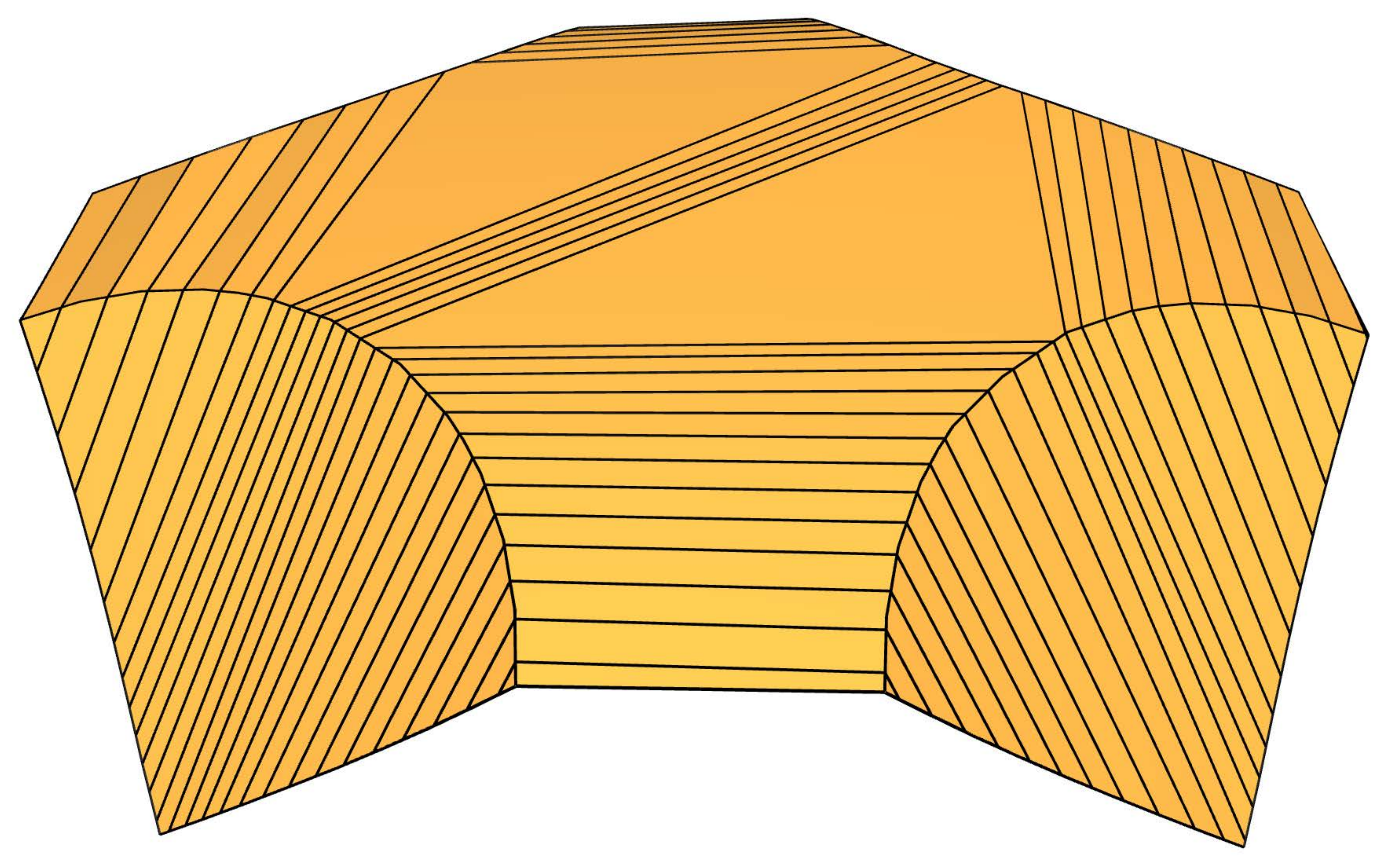}\\
       \small (a) input mesh &
       \small (b) locally estimated rulings &
       \small (c) gradient direction field &
       \small (d) optimized scalar field &
       \small (e) remeshing result\\
  \end{tabular}
  \caption{
  Developable shapes can be digitally acquired by 3D scanning or freeform modeling (a). In such scenarios, the meshing is typically not aligned to principal curvature directions, which hampers practical applications, such as fabrication with flat polygonal panels (\figref{fig:teaser2}). Our method remeshes an input mesh of a (piecewise) developable surface into a curvature aligned, planar polygonal mesh (e) by computing a vector field (c), from which we integrate a function whose level sets (d) align as well as possible to the locally estimated rulings (b). Our vector field contains automatically placed singularities 
  in the planar region (d), which result in naturally placed triangular patches.}
  \label{fig:teaser}
\end{teaserfigure}

\begin{abstract}

We introduce an algorithm to remesh triangle meshes representing developable surfaces to planar quad dominant meshes. The output of our algorithm consists of planar quadrilateral (PQ) strips that are aligned to principal curvature directions and closely approximate the curved parts of the input developable, and planar polygons representing the flat parts of the input.
Developable PQ-strip meshes are useful in many areas of shape modeling, thanks to the simplicity of fabrication from flat sheet material. Unfortunately, they are difficult to model due to their restrictive combinatorics and locking issues. Other representations of developable surfaces, such as arbitrary triangle or quad meshes, are more suitable for interactive freeform modeling, but generally have non-planar faces or are not aligned to principal curvatures. Our method leverages the modeling flexibility of non-ruling based representations of developable surfaces, while still obtaining developable, curvature aligned PQ-strip meshes. 
Our algorithm optimizes for a scalar function on the input mesh, such that its level sets are extrinsically straight and align well to the locally estimated ruling directions. The condition that guarantees straight level sets is nonlinear of high order and numerically difficult to enforce in a straightforward manner. We devise an alternating optimization method that makes our problem tractable and practical to compute. Our method works automatically on any developable input, including multiple patches and curved folds, without explicit domain decomposition.
We demonstrate the effectiveness of our approach on a variety of developable surfaces and show how our remeshing can be used alongside handle based interactive freeform modeling of developable shapes.

\end{abstract}

%
%

\ccsdesc[500]{Computing methodologies~Mesh models}
\ccsdesc[500]{Computing methodologies~Mesh geometry models}

%
%

\keywords{Developable surfaces, planar quadrilateral meshes, curvature line nets, remeshing}

\maketitle


\begin{figure}[b]
\includegraphics[width=\linewidth]{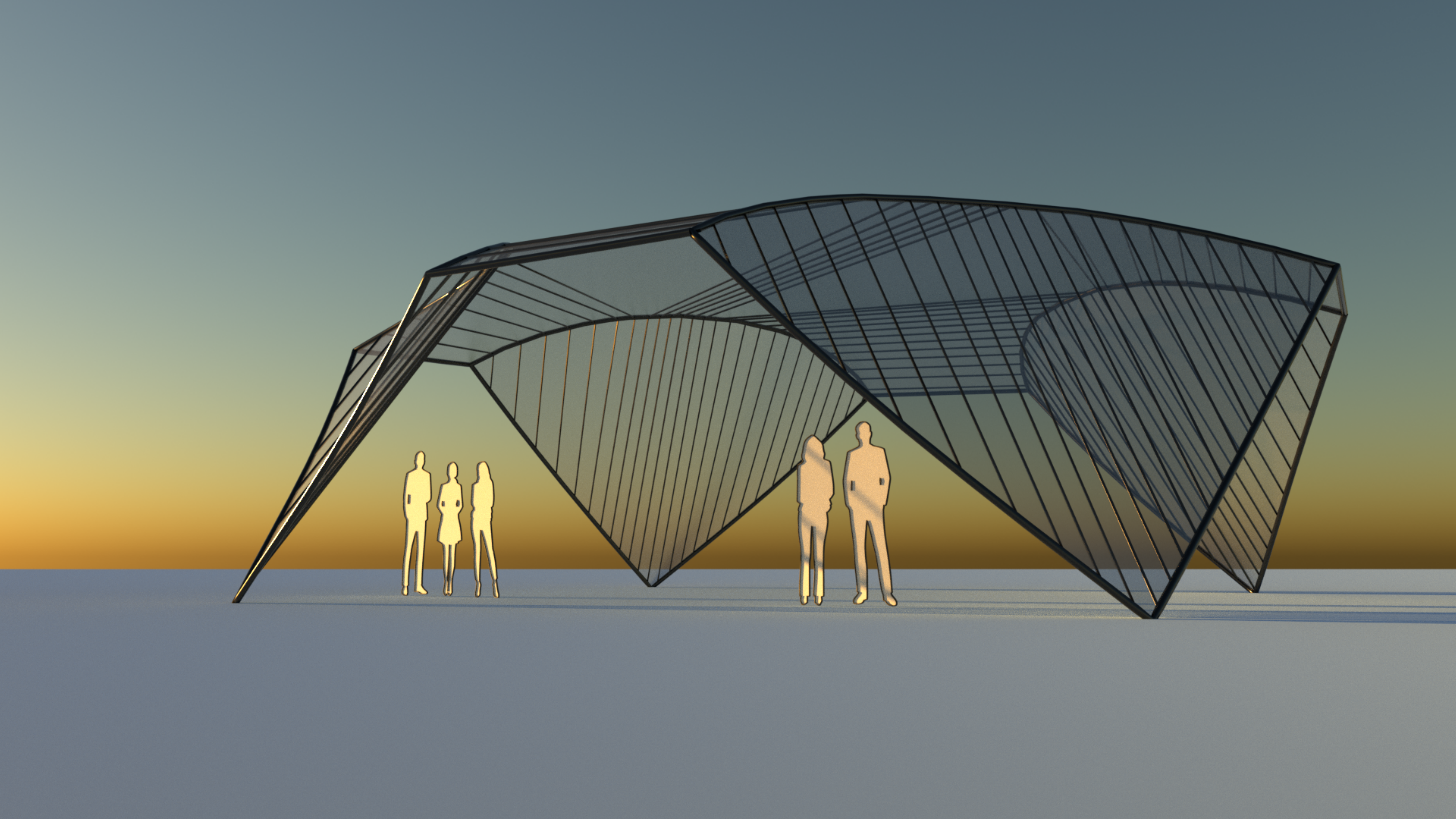} 
\caption{\label{fig:teaser2}
An architectural illustration of our result from \figref{fig:teaser}, fitting with flat glass-like panels.}
\end{figure}

\section{Introduction}

Developable surfaces are commonly used in architecture and product design due to the simplicity of their fabrication. Such surfaces are locally isometric to a planar domain, which means they can be manufactured by mere bending of sheet material, such as metal. 
Freeform developable surfaces form a rich and interesting shape space, but they are notoriously difficult to design due to their highly constrained nature. Therefore in most cases, only simple forms are used, such as cylinders and cones. 

The majority of methods for developable surface modeling use rulings based representations (see, e.g., \cite{pottmann_new,solomon}), or isometry optimization (see, e.g., \cite{shells,froh_botsch}). Using such representations in the common handle-based editing paradigm is hindered by locking issues that make the exploration of the developable shape space difficult, as described in \cite{locking1,locking2,pottmann_new,rabi18}. 
The recently proposed discrete orthogonal geodesic nets \cite{rabi18} and the checkerboard pattern isometries \cite{CBP_Pottmann_Isometry:2020}, lift this limitation by representing developable surfaces without explicitly accounting for principal curvature directions. These methods enable intuitive surface modeling via freeform deformation, as if bending and manipulating a piece of paper. 

While these discrete representations of developable surfaces are excellently suited for creative exploration and design of freeform developable \emph{shapes}, they fall short of providing a suitable final representation for manufacturing. For that purpose, it is especially important to have planar mesh faces aligned to principal curvature directions \cite{AnisotropicPolygonalRemeshing:Alliez:2003,liu:conicalmeshes:2006,pottmann_new}. A curvature-line representation of a developable surface possesses the desired properties for fabrication once the shape is fixed, since the minimal curvature lines on a developable surface are rulings with a constant normal along them, such that they can be easily tessellated into planar polygons that approximate the surface shape well. In fact, meshes comprised of planar quadrilateral strips (with no interior vertices) constitute a well known model for discrete developable surfaces, whose refinement and convergence properties have been studied \cite{liu:conicalmeshes:2006}.

In this paper, we develop a method to convert a triangle mesh representation of a developable surface into discrete curvature-line representation in order to reap the benefits of both worlds: the support for unhindered developable shape creation provided by a representation of choice, and the desirable properties for fabrication offered by the curvature line representation. 
Our method produces strips of planar quadrilaterals (PQ) aligned to principal curvature directions that closely approximate the curved parts of a given input mesh, along with planar polygons representing the flat parts of the input (see \figref{fig:teaser}).
In particular, our method produces precisely straight rulings, modeled as individual edges in the output mesh.

The past decade has seen a highly active stream of fruitful research on field aligned quad meshing, where principal curvature fields have naturally received special attention \cite{QUADSTAR2012,vaxman2016directional,FernandoVF}. However, to the best of our knowledge, no existing general remeshing method is guaranteed to perfectly align to principal directions and produce edge flows that are entirely consistent with curvature lines, which are often difficult to obtain fully and faithfully for discrete meshes.
In this work we exploit the specific constrained setting and the geometric structure of developable shapes to reproduce straight minimum-curvature lines, as well as automatically segment the input into curved and planar parts in a robust manner that is consistent with the structure dictated by developability.

Our method is based on fitting a scalar function on the input mesh, such that its level sets are straight and align as well as possible to the \emph{locally} estimated rulings on non-planar regions (see \figref{fig:teaser}). The condition that guarantees straight level sets on developable surfaces is simple to formulate: the normalized gradient of the scalar field needs to be divergence free. This nonlinear and high order condition is numerically difficult to enforce in a straightforward manner. We therefore devise a dedicated optimization scheme that factors the problem into a divergence free and integrable directional field optimization that is subsequently integrated into a scalar function. This makes our problem tractable and practical to optimize. We extract the level sets of the obtained scalar field at the desired resolution, and remesh the input into strips of planar quads whose chordal edges are the level sets, i.e., the rulings. We supplement the mesh by planar polygonal faces that represent the planar patches of the input surface. The flexibility of the field-to-function design allows for the automatic inclusion of singularities, flat regions and curved folds without explicitly segmenting different curvature regions on the mesh.

We demonstrate the effectiveness of our approach on a variety of input developable shapes represented by general triangle meshes (and triangulated quad meshes) and show how our remeshing can be used side-by-side with freeform modeling of developables. 

\section{Related Work}
\label{sec:previous_work}

Remeshing general meshes into (planar) quad meshes is an active area of research. A comprehensive review is beyond the scope of this paper, but we highlight the main features of existing approaches most closely related to our work. 

\begin{figure}
    \centering
    \includegraphics[width=\linewidth]{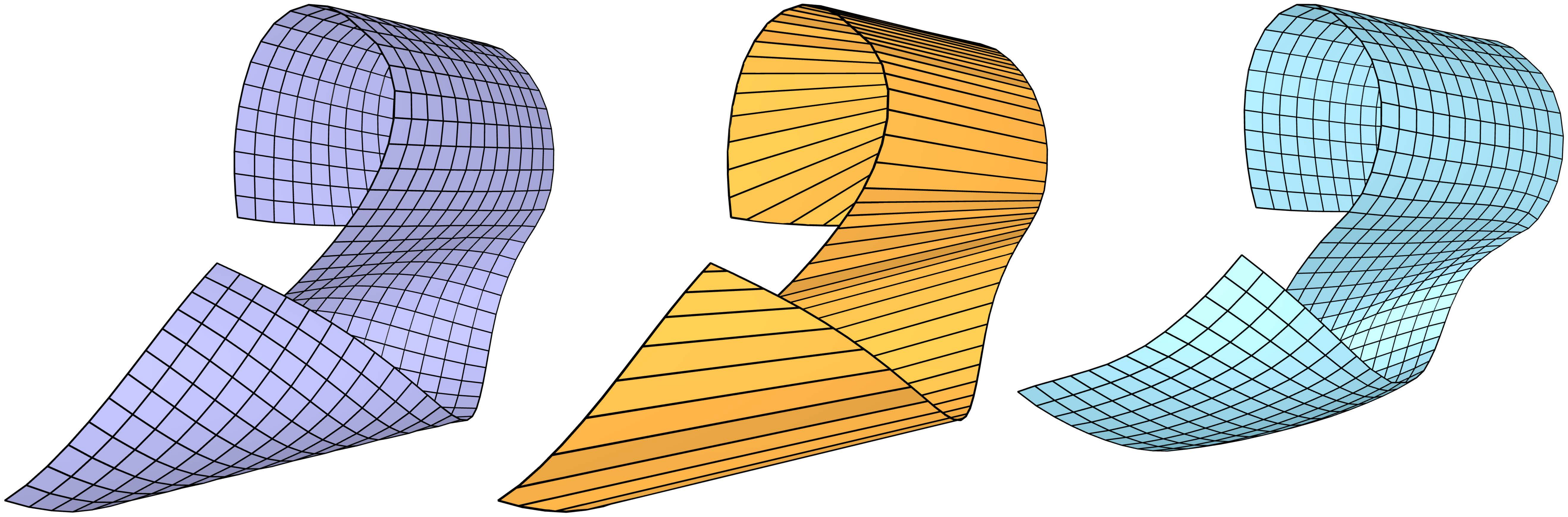}
    \setlength{\tabcolsep}{0pt}
    \begin{tabular}{>{\centering\arraybackslash}m{0.155\textwidth}>{\centering\arraybackslash}m{0.155\textwidth}>{\centering\arraybackslash}m{0.155\textwidth}}
%
\small input & 
\small ours &
\small ShapeUp \\
& 
\small $h = 0.67\%$ &
\small $h = 9.74\%$ \\
&
\small $p = 1.80\%$ &
\small $p = 0.50\%$ \\

\end{tabular}
    \caption{Attempting to convert a quad mesh of a developable shape to a PQ mesh using a general-purpose planarization technique (ShapeUp~\cite{shape_up}) significantly alters the shape and makes it non-developable. This happens because the edges of the input mesh are generally not aligned to principal curvature directions. Our method is applied to a trivial triangulation of the input mesh. We report Hausdorff distance $h$ with respect to bounding box diagonal and the maximal planarity error $p$.
    }
    \label{fig:general_planarization_fail}
\end{figure}

As stated in the introduction, the quadrilaterals in freeform models of developable surfaces are usually non-planar, and typically neither triangle nor quad meshes are curvature aligned. Our goal is to obtain a curvature-aligned remeshing with planar faces. Planarization of general polygonal meshes has been explored in several works~\cite{alexa_planarization,shape_up,tang-2014-ff,Diamanti2014,Poranne2013}. These methods take arbitrary shapes as input and are not specifically targeted at developable surfaces. Typically, applying a general planarization method to developable surfaces leads to poor results in terms of curvature alignment and shape approximation (see \figref{fig:general_planarization_fail}).

\begin{figure*}
    \centering
   \setlength{\tabcolsep}{1mm}
    \begin{tabular}{cccc}
		\small input &
    	\small our result &
    	\small Diamanti et al.~\shortcite{Diamanti2014} &
    	\small Instant Meshes \cite{Jakob2015Instant} \\
        \includegraphics[width=0.19\linewidth]{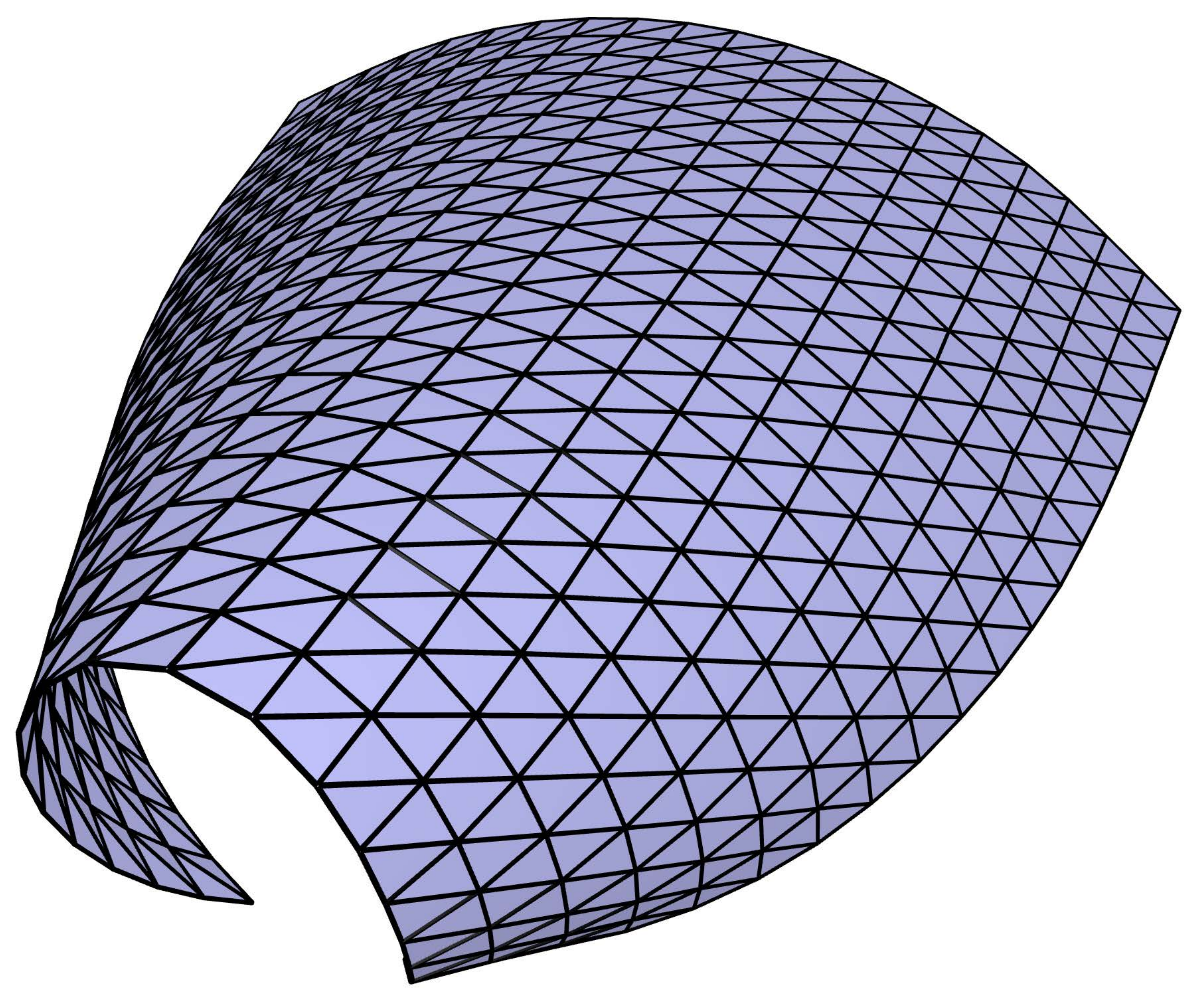}   & 
        \includegraphics[width=0.19\linewidth]{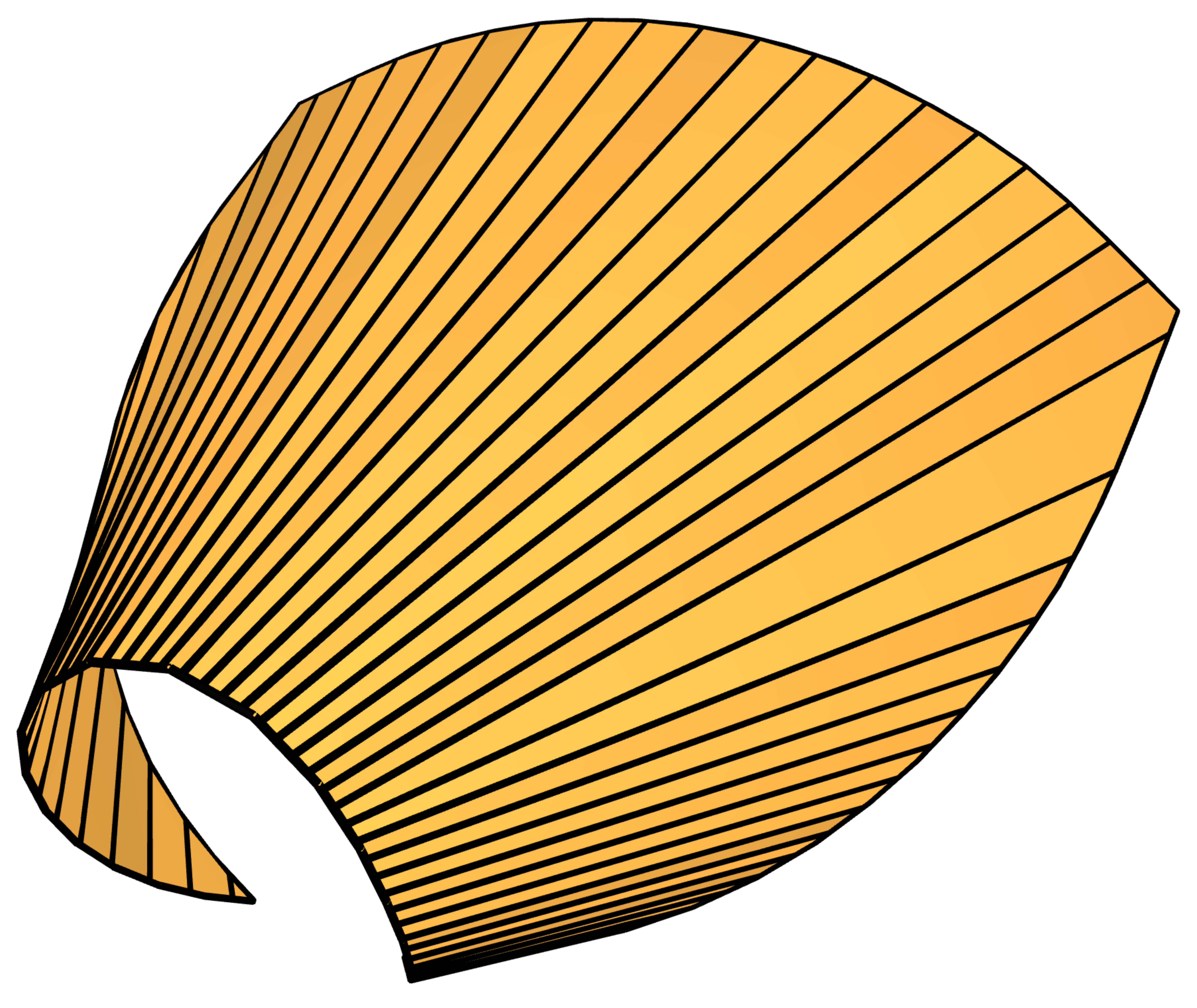} &
        \includegraphics[width=0.19\linewidth]{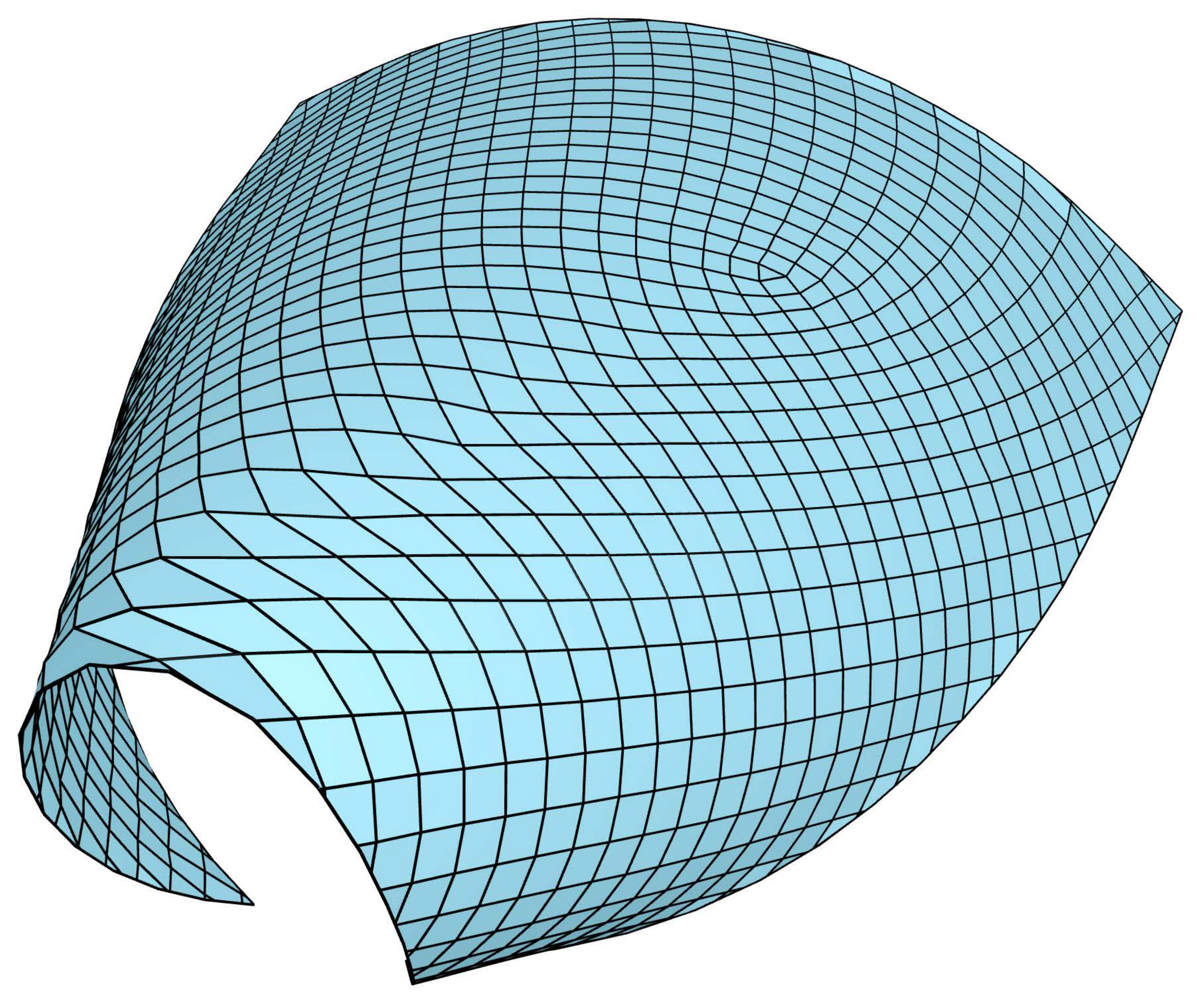}  & 
        \includegraphics[width=0.19\linewidth]{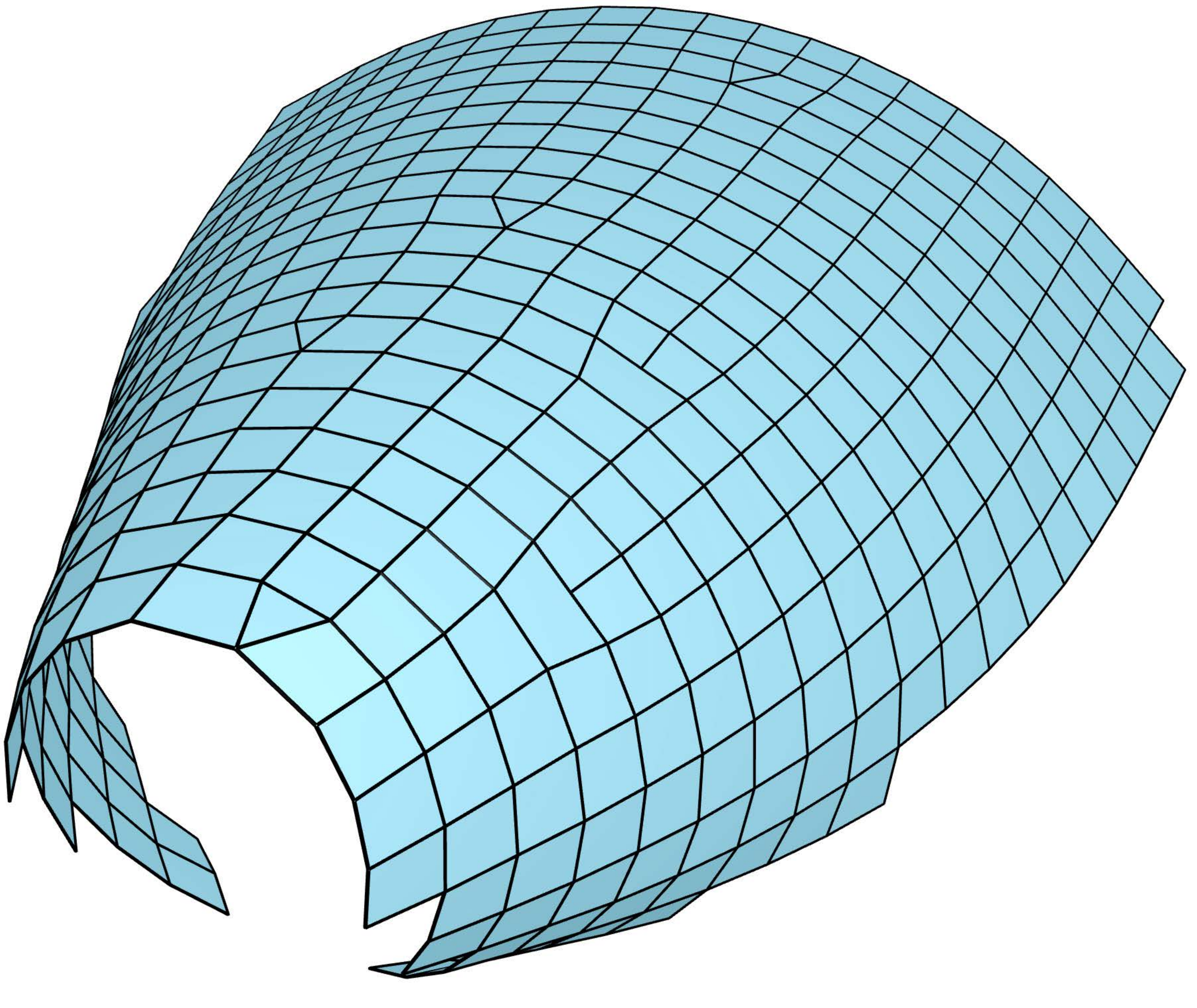} \\
        \includegraphics[width=0.23\linewidth]{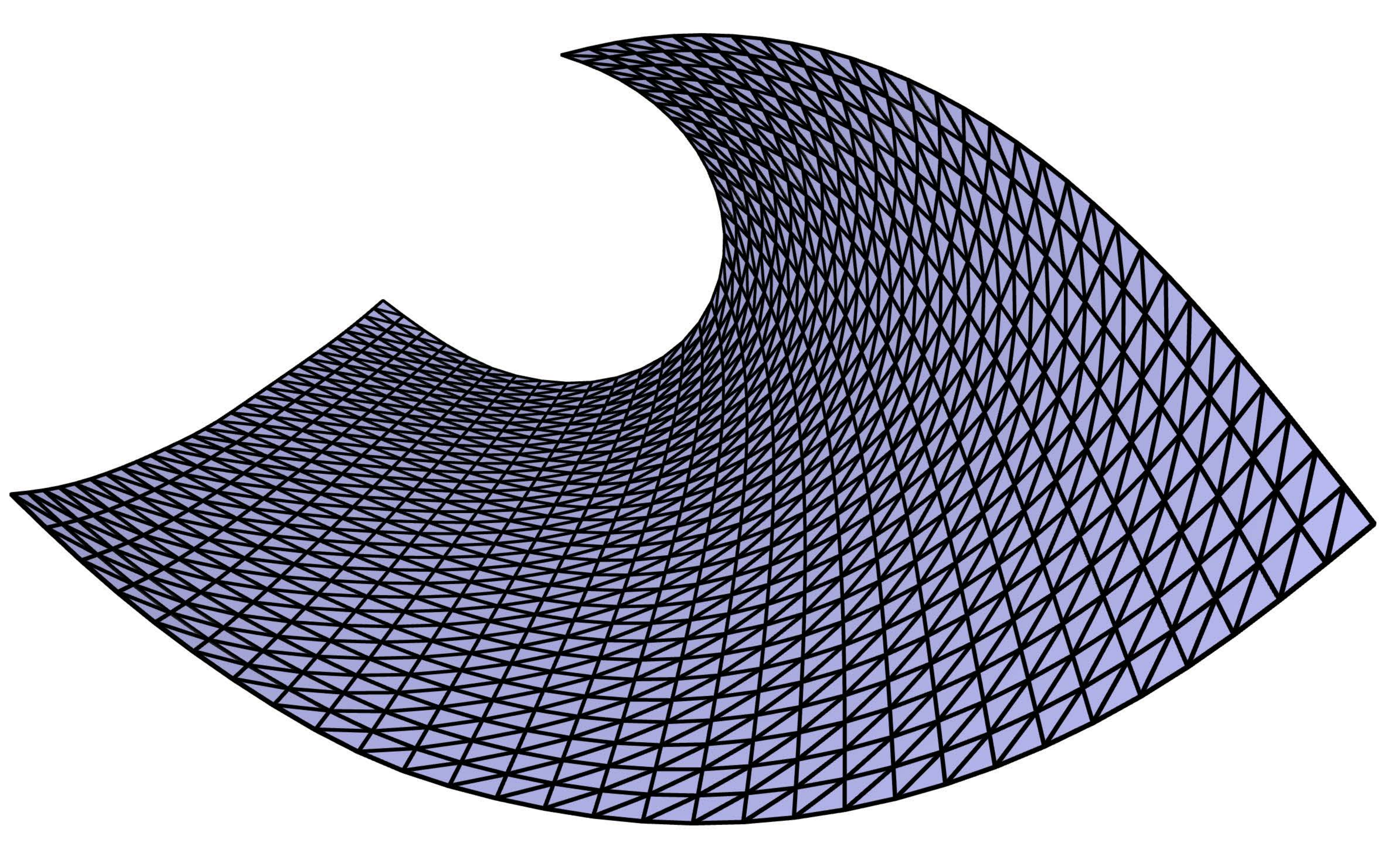}   & 
        \includegraphics[width=0.23\linewidth]{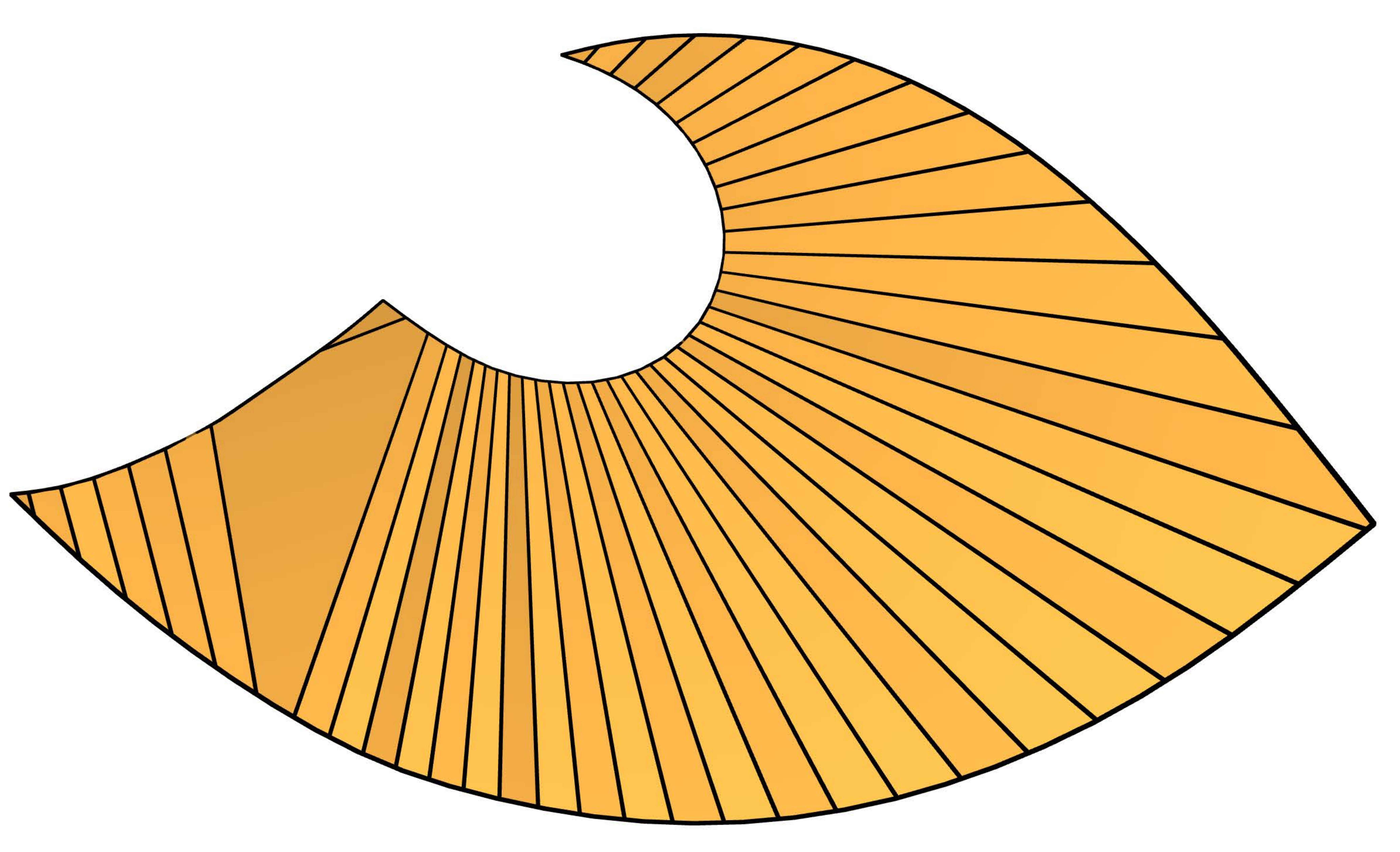} &
        \includegraphics[width=0.23\linewidth]{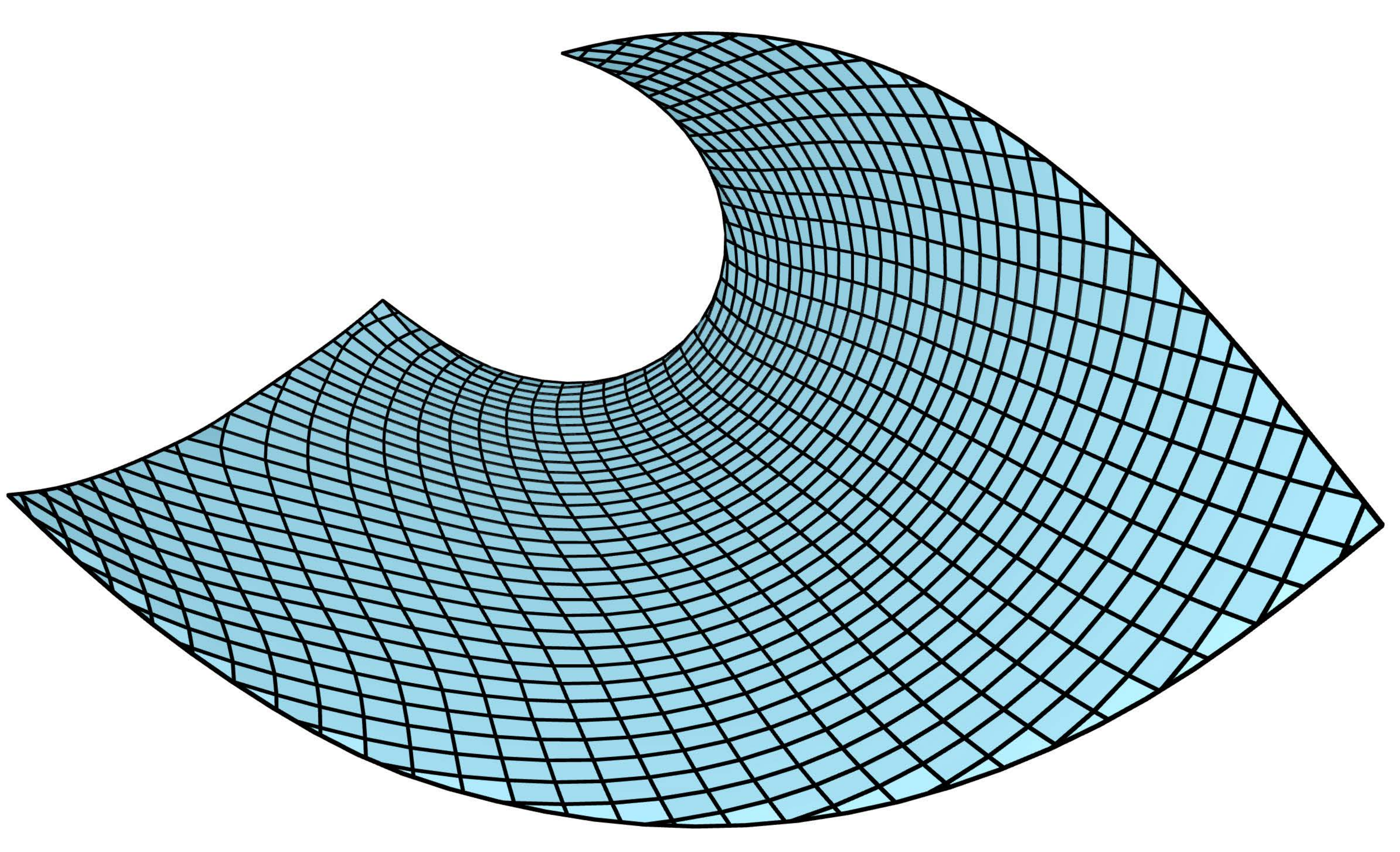}  & 
        \includegraphics[width=0.23\linewidth]{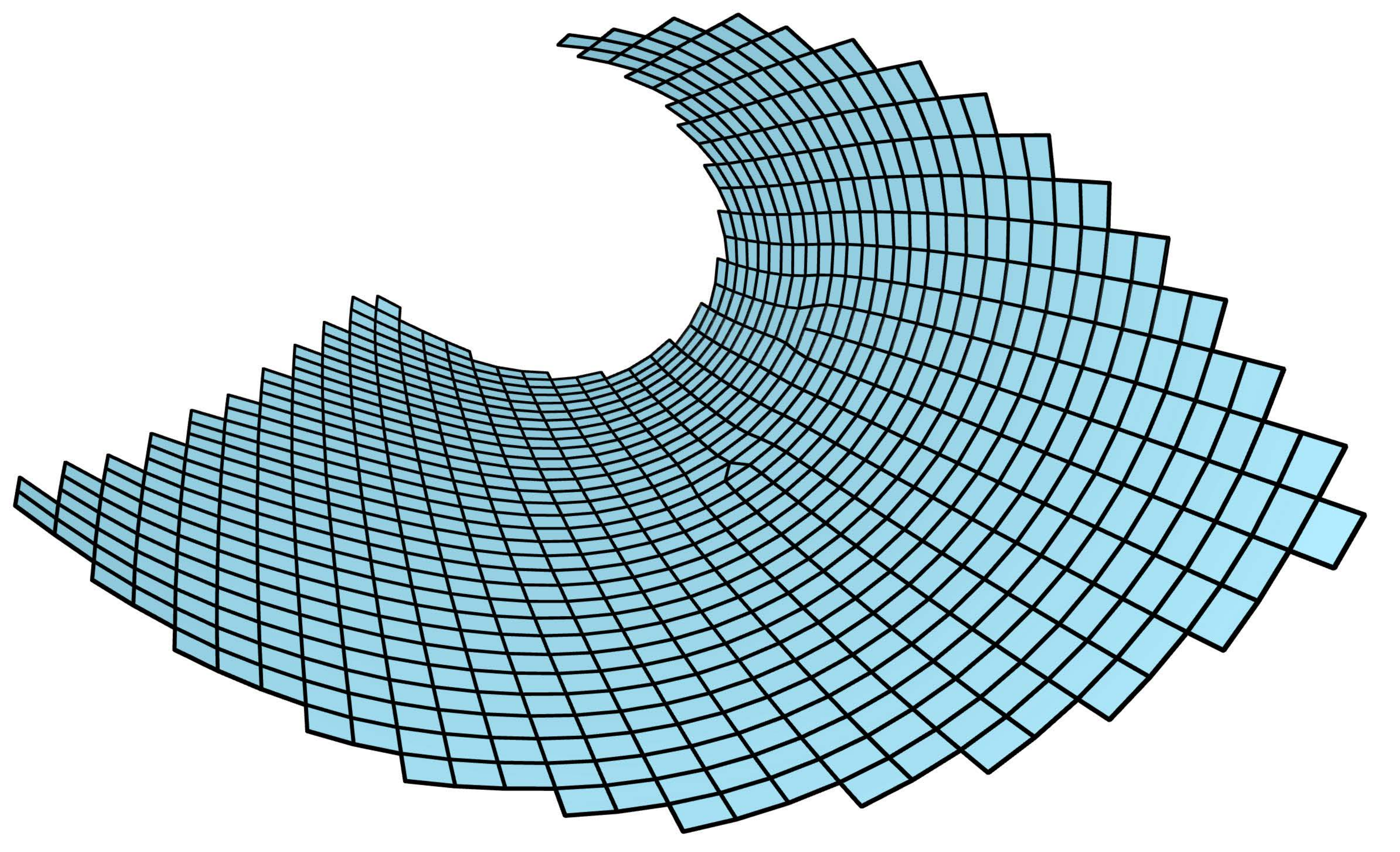} \\
        \includegraphics[width=0.20\linewidth]{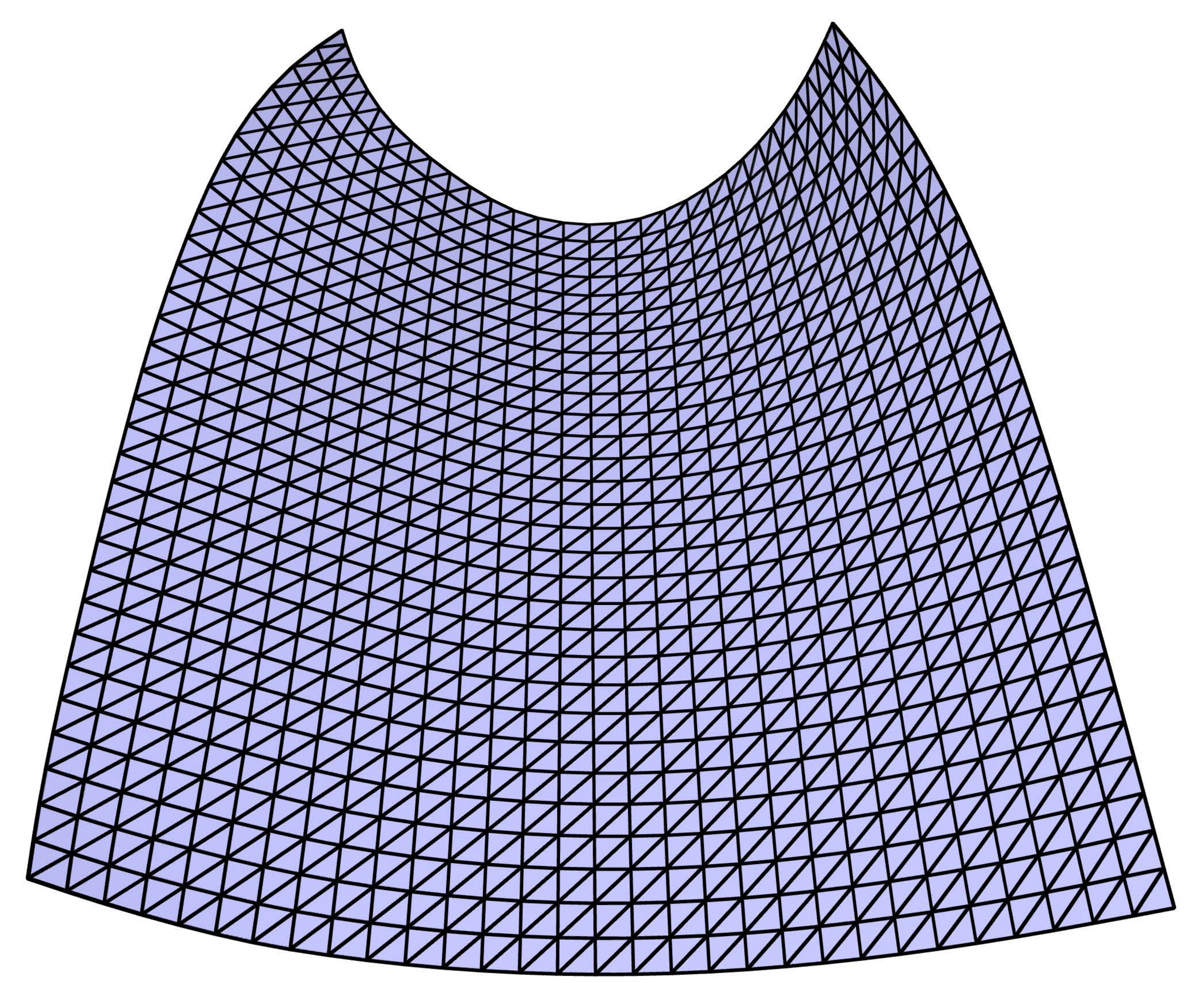}   & 
		\includegraphics[width=0.20\linewidth]{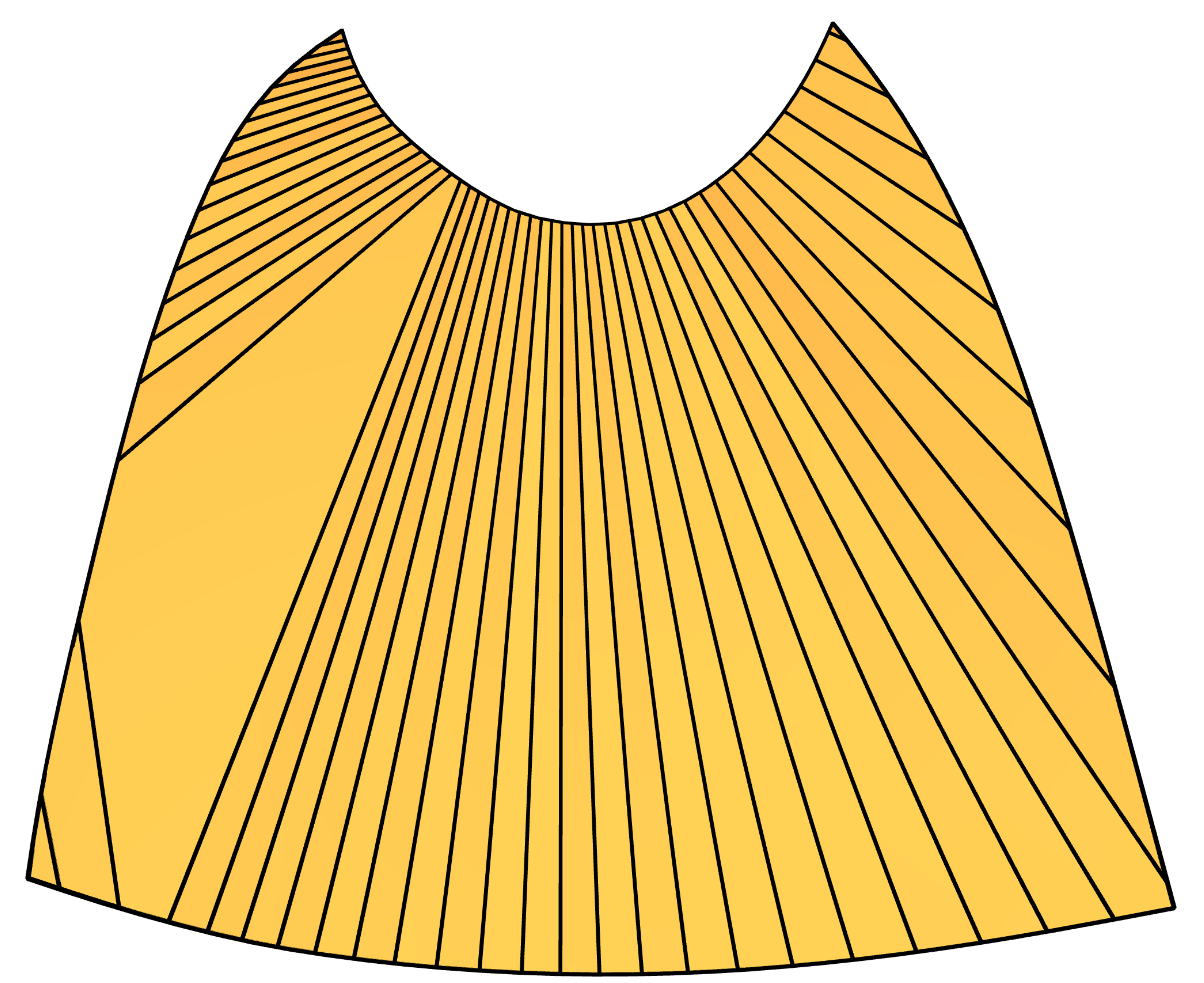} &
		\includegraphics[width=0.20\linewidth]{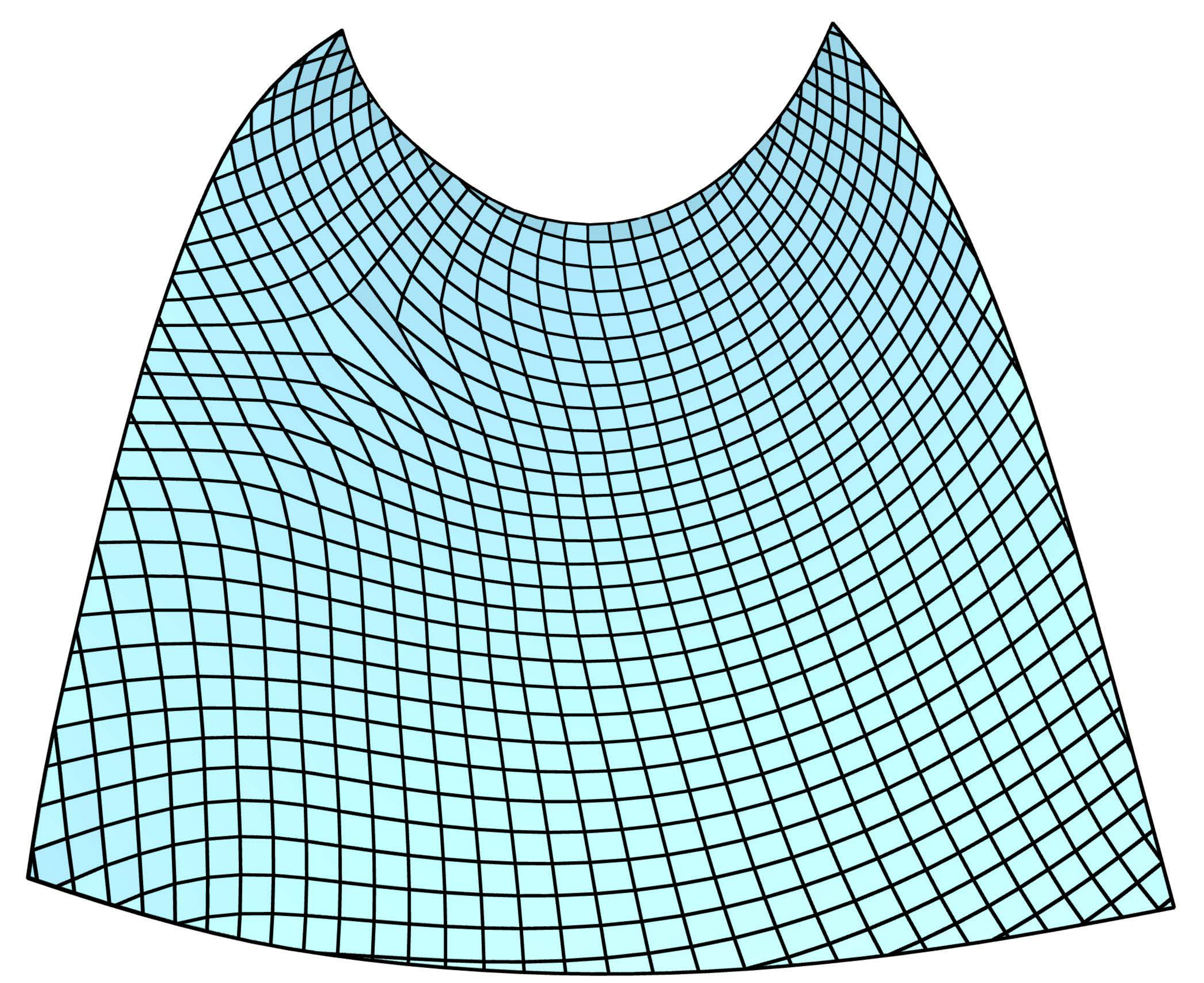}  & 
		\includegraphics[width=0.20\linewidth]{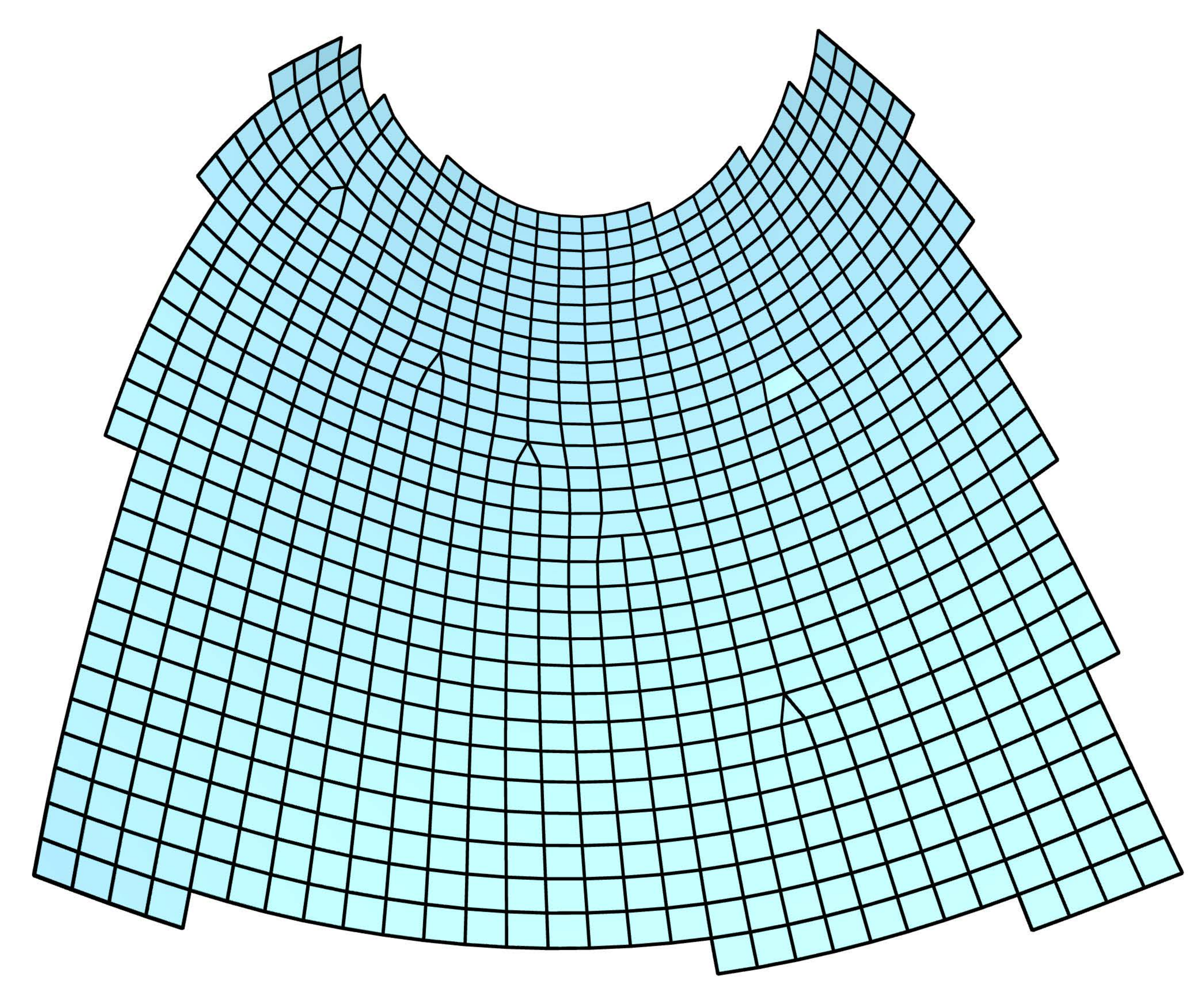} \\        
		\includegraphics[width=0.23\linewidth]{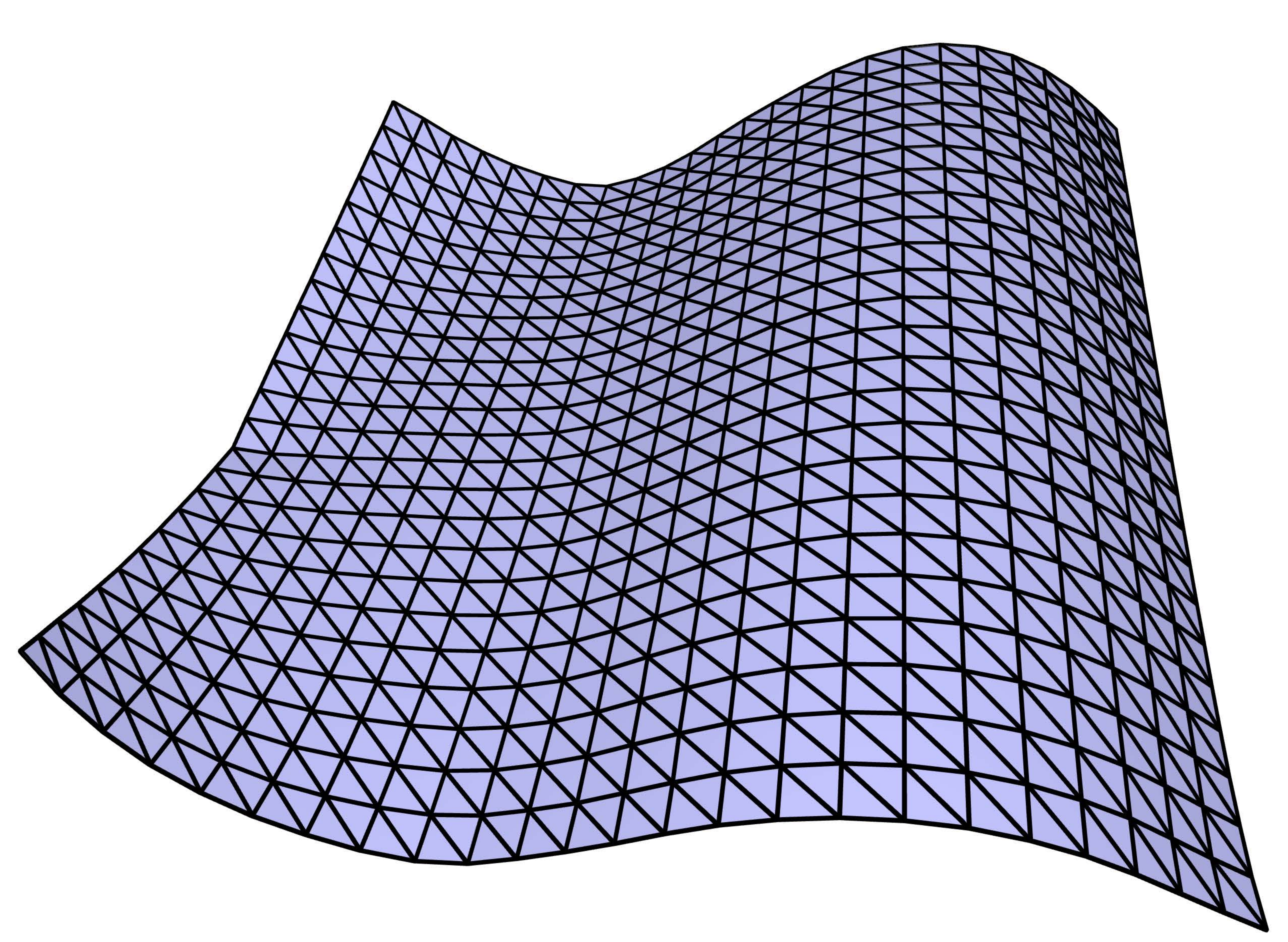}   & 
		\includegraphics[width=0.23\linewidth]{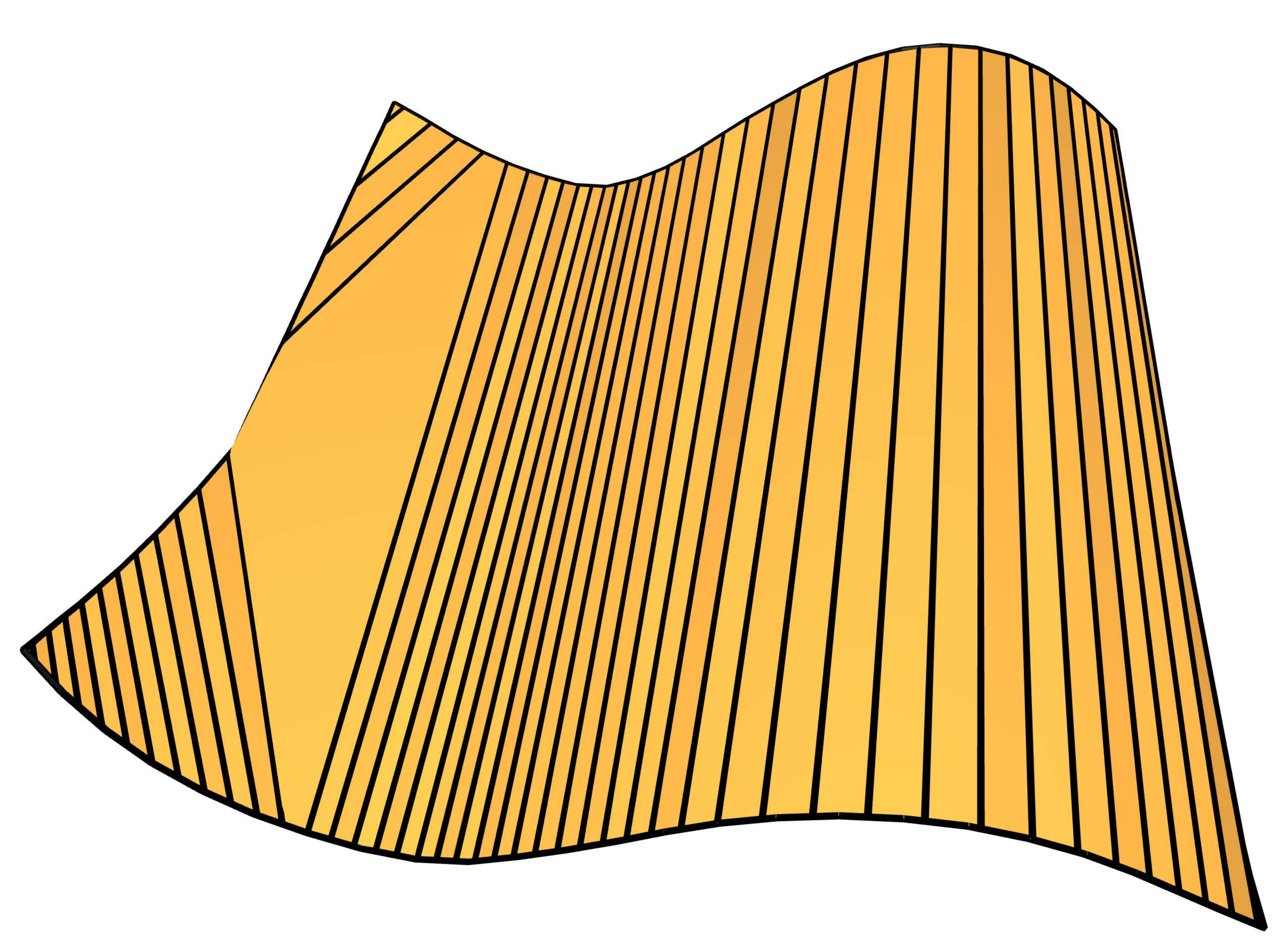} &
		\includegraphics[width=0.23\linewidth]{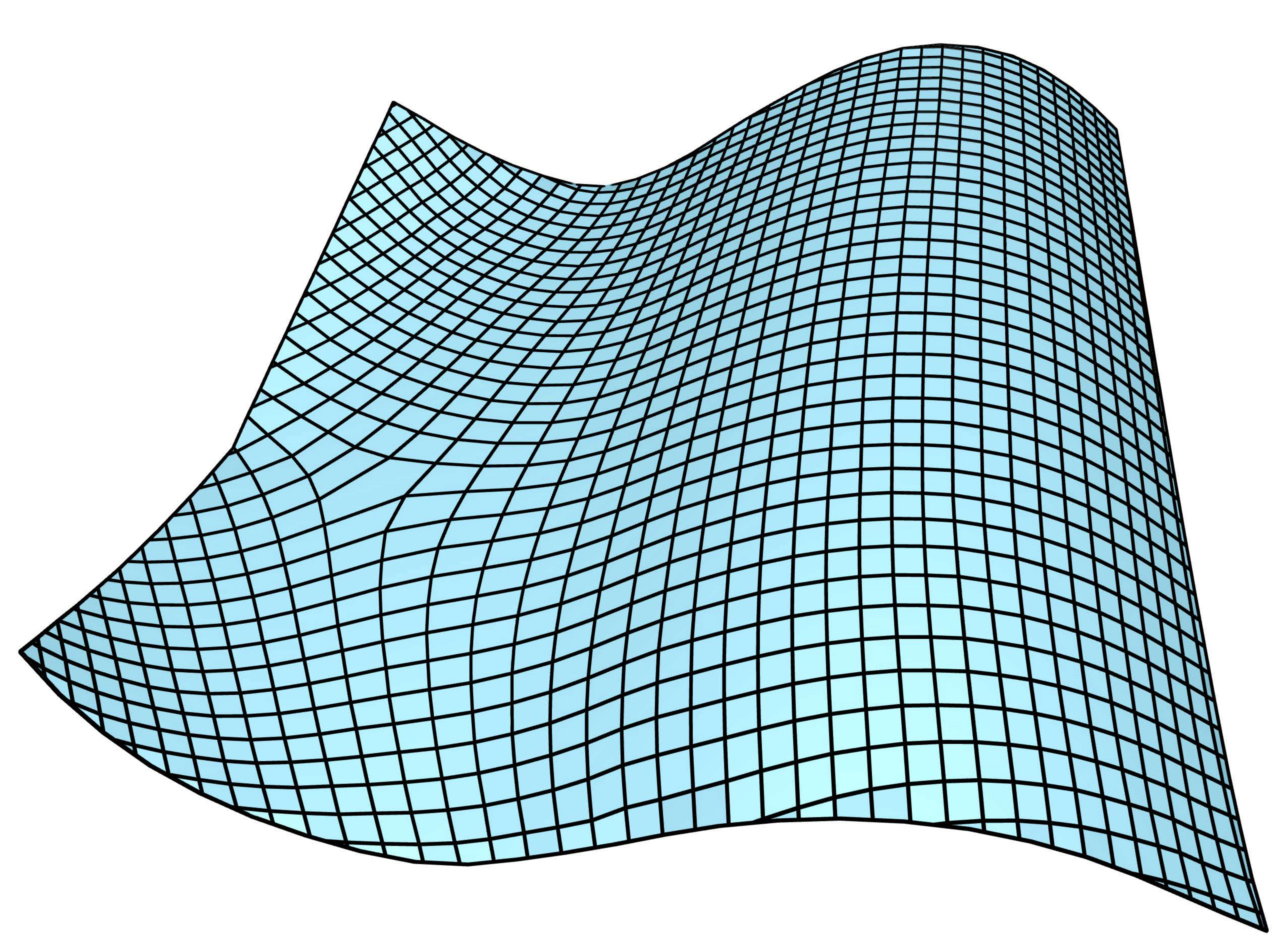}  & 
		\includegraphics[width=0.23\linewidth]{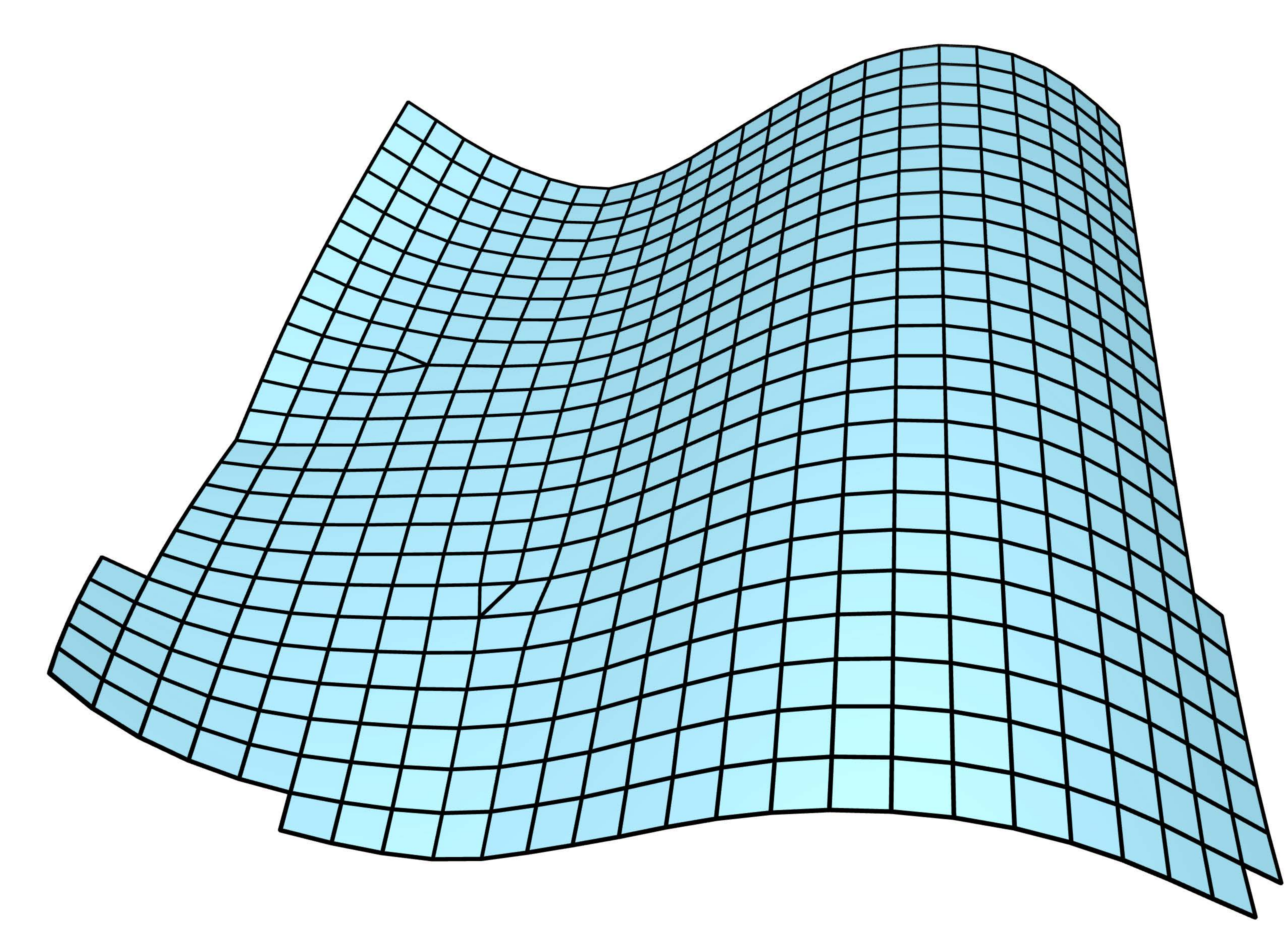} \\
    \end{tabular}
    \caption{Remeshing an input developable surface using the directional field design of \cite{Diamanti2014} does not result in globally straight edge sequences. Instant Meshes, the curvature-aligned quad dominant remeshing technique of \cite{Jakob2015Instant}, introduces superfluous singularities and does not always succeed in finding the exact rulings. For Instant Meshes we use the following settings: 4-RoSy extraction, quad-dominant mesh extraction, no boundary alignment (to ensure better curvature alignment; trimming can be done in a post-processing step).}
    \label{fig:quad_remesher_fail}
\end{figure*}

A different approach to obtaining PQ meshes from general developable input meshes is to utilize the fact that PQ meshes are a discrete model for conjugate nets and seek a remeshing that is aligned to ruling directions. Many curvature-aligned or just conjugate quad remeshing techniques for general shapes exist, see e.g.~\cite{QUADSTAR2012,Jakob2015Instant,Diamanti2014,Liu2011conjugate,Zadravec2010conjugate}. Similar to our method, these techniques rely on numerical estimation of the principal curvature directions, but they do not guarantee exact alignment or straight edge sequences and may introduce unnecessary singularities on developable shapes. Their optimization process might fail to create precise, straight rulings on developable surfaces, unlike the algorithm we propose in this work (see \figref{fig:quad_remesher_fail}).

A more promising approach to PQ meshing of developable shapes is a dedicated technique that utilizes their specific properties. Peternell~\shortcite{peternell2004developable} converts a scan of a single torsal developable patch into a PQ mesh by thinning its tangent space representation into a one-dimensional object (a simple curve). This approach is not immediately applicable to composite and possibly piecewise developable surfaces that consist of multiple torsal patches and planar regions. Kilian and colleagues~\shortcite{curved_folding_kilian} compute a torsal patch decomposition for 3D scans of physical developable surfaces by estimating flat regions and ruling directions. 
This approach may struggle with developable meshes that are coarse in comparison to scans due to insufficient data density for reliable {ruling} fitting. {Their method relies on the ruling estimates to initialize a planar mesh development, which is used in the subsequent optimization. The connectivity of this initial mesh cannot be changed during the optimization, and thus determines the approximation quality that can be obtained.}
Locally estimated rulings on developable meshes can be quite noisy and inaccurate, as we discuss in \secref{sec:discretization}. We avoid a direct domain decomposition based on rulings employed in~\cite{curved_folding_kilian} and instead devise a global constrained optimization approach. As a result, our method is successful on coarse and noisy inputs.

Wang and colleagues~\shortcite{Wang2019geodesic} use discrete parallel geodesic nets as a discrete model for developable surfaces. They require the geodesic strips to be of constant width for a surface to be developable, but do not impose any requirements on the directions of these strips and as such do not have a ruling-aligned representation for developables. They also use parallel geodesic nets to approximate surfaces by piecewise developable strips. The individual geodesic strips of a parallel geodesic net are approximated by piecewise planar faces in a postprocessing step, but this does not guarantee that compatibility between neighboring strips is preserved. In contrast, our method produces a complete, connected remeshing of the input developable surface with strips aligned to the rulings, rather than in the orthogonal direction.

While targeting geodesic fields, rather than planar quad remeshing, the works by Vekhter et al.~\shortcite{Vekhter2019weaving} and Pottmann et al.~\shortcite{pottmannGeodesicPatterns} show parallels to our proposed method. They compute a unit curl free field, while we compute a divergence free field---these two kinds of fields are in fact duals. Nevertheless, our field has further constraints in terms of ruling alignment, which we take into account. In addition, our optimization strategy is different, interlacing integrability optimization with divergence reduction. We discuss this in further detail in \secref{sec:continuous-and-discrete-developable-surfaces}.

\begin{figure}[b]
	\centering
	\includegraphics[width=0.9\linewidth]{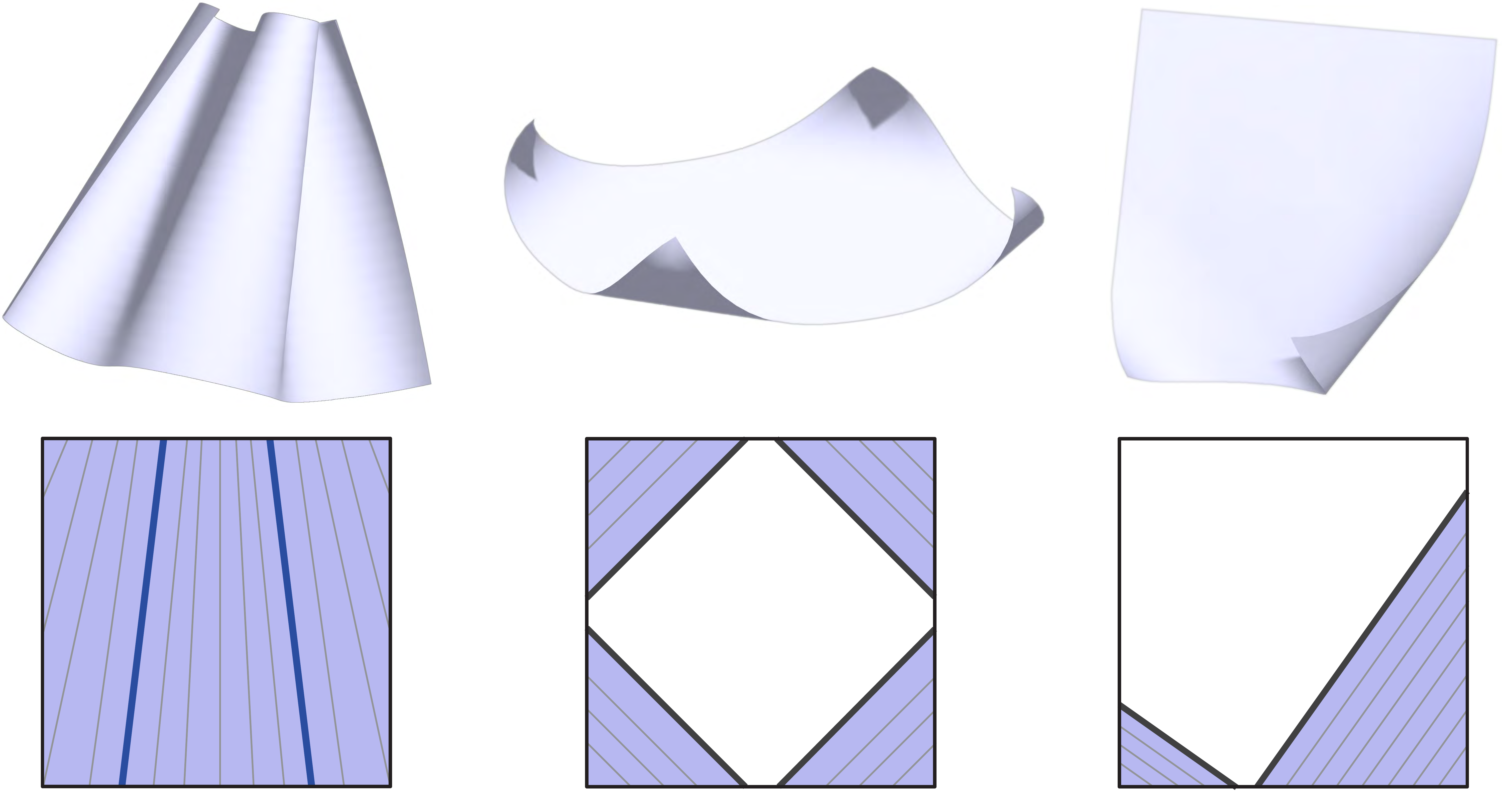}
	\caption{Developable surfaces (top row) and their decompositions into planar and curved (torsal) patches, shown on the 2D development (bottom row). We display the planar patches in white and the curved patches in purple. The rulings are illustrated as thin grey lines, with the borders between curved and flat patches in thick black and inflection lines in blue.}
	\label{fig:general-developables}
\end{figure}

\section{Continuous and Discrete Developable Surfaces}
\label{sec:continuous-and-discrete-developable-surfaces}
	
In this section, we summarize relevant facts about developable surfaces that inform our algorithm and offer a directional field based definition of developable surfaces. We provide a discrete setup for these fields in \secref{sec:discretization}, and an optimization scheme in \secref{sec:method}.

\subsection{Developable surface parameterization}
\label{sec:developable_background}
A $C^2$-continuous surface $\Sdev$ that has vanishing Gauss curvature everywhere is a smooth developable surface. A general developable comprises multiple developable patches $\left\{\Sdev_i\right\},\ \bigcup \Sdev_i = \Sdev$, where each such patch is either a \emph{torsal} patch (a \emph{curved} ruled surface with constant normal along each ruling) or a \emph{planar} patch. The rulings are completely contained in each $\Sdev_i$, i.e., they extend up to the boundary $\partial \Sdev_i$ \cite{massey1962surfaces}. The \emph{planar patches} are regions with vanishing mean curvature $H = \kappa_2 / 2$, where $\kappa_2$ is the max curvature. They are bounded by rulings of torsal patches and the boundary of the surface, as shown in \figref{fig:general-developables}.

\paragraph{Non-smooth developables} We also consider more general, piecewise developable surfaces. One type is \emph{creased} shapes, where several smooth developable surfaces are joined along curves with only $C^0$-continuity \cite{huffman}. These curves are termed \emph{curved folds} when the surface is globally isometric to a planar domain (as in \figref{fig:curvedfolds}), and \emph{creases} when this is not the case (e.g., \figref{fig:sphericonsdforms}). We treat curved folds and creases identically in the rest of this paper and refer to them as creases from now on. Another type is surfaces that contain \emph{point singularities}, such as cone apexes (see \figref{fig:glued}). These surfaces are locally non-developable at the singularities; they can be constructed by gluing parts of the boundary of a developable surface together while allowing isometric deformation. The cone apexes are easily identified, and to run our method we remove the apex vertices that don't coincide with a crease together with their incident faces. If desired, they can be added back in post-processing. We note that our method requires that the individual pieces are sufficiently smooth, and allow a definition of rulings whose endpoints are always on boundaries or creases (as is the case for the surface types described above), therefore explicitly excludes surfaces that look like crumpled paper.

\paragraph{Conjugate nets} Consider a single torsal patch $\Sdev_i$, where we parameterize the patch with coordinates $\Sdev_i(u,w)$ as follows:
\begin{equation}
    \Sdev_i(u,w) = p(u) + w\, r(u),
\label{eq:developable-parameterization}
\end{equation}
where $p(u):\R \rightarrow \R^3$ is a \emph{generating curve,} and every $u$-level-set is a straight line with direction $r(u):\R \rightarrow \mathbb{S}^2$, i.e., a ruling. The Gauss map $n(u,w)$ must be constant on the $u$-level-sets in order for $\Sdev_i$ to be developable: $n(u,w) = n(u)$. This means that the $u$-level-sets are \emph{extrinsically flat}; they constitute lines in $\R^3$.

The rulings are the minimum curvature lines of $\Sdev_i$. The $uw$-parameterization constitutes a \emph{conjugate net}~\cite{liu:conicalmeshes:2006}. In particular, choosing $p(u)$ to be a max curvature line (i.e., having the $p(u)$ curve intersect all rulings at right angles) makes $\Sdev_i(u,w)$ a principal curvature line parameterization. 

\paragraph{Seamless parameterization}
When $\set{S}$ is  $C^2$-continuous, every torsal patch either borders a planar patch, or the outer boundary of $\set{S}$.
\setlength{\intextsep}{0pt}%
\setlength{\columnsep}{5pt}%
\begin{wrapfigure}{r}{0.15\textwidth} 
	\centering
	\includegraphics[width=0.15\textwidth]{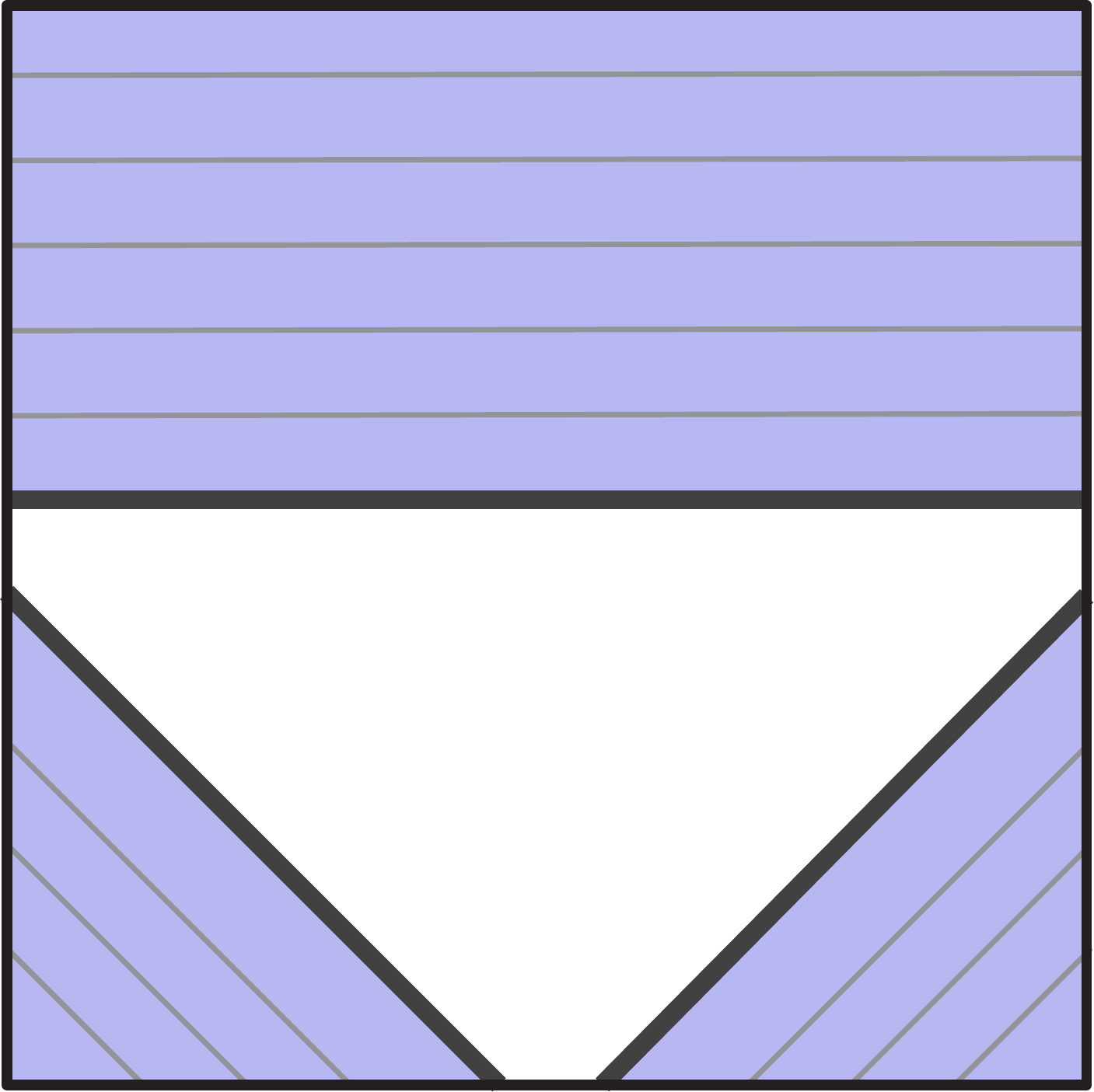}
\end{wrapfigure}
Planar patches allow to connect several different ruling strips, with multiple orientations, by introducing \emph{singularities} within the planar patch {(see inset)}.
Since the rulings are by definition $2$-symmetric, being lines (i.e., they are invariant to the sign of $r(u)$), the singularities can be of indices $\pm \frac{1}{2}$, and the function $u$ is only \emph{seamless} rather than continuous on $\set{S}$, as in the stripe patterns of~\cite{Knoppel2015}. Note however that $u$ is locally fully continuous (it can be ``combed'') on each torsal patch $\Sdev_i$. Across creases, rulings are only $C^0$-continuous, and thus we constrain $u$ to be $C^0$ there as well.

We further note that works such as~\cite{Diamanti2014,Liu2011conjugate, Jakob2015Instant} intermix between the $u$ and the $w$ coordinates, and enable full quad-mesh $\pm \frac{1}{4}$ singularities (see Fig.~\ref{fig:quad_remesher_fail}). While this provides more meshing flexibility for general curved surfaces, it in fact hinders the ability to correctly capture the pure foliation topology of the ruling stripes comprising the developable surfaces.

\subsection{Ruling fields}
\label{subsec:ruling-fields}
Our work focuses on designing directional fields that generate the rulings of a developable surface from other representations, and integrate them to compute $u$. Consider the gradient vector field $\nabla u$, which is by definition orthogonal to the level sets of $u$. The geodesic curvature of level sets is defined as:
$\kappa_g(u)= \nabla \cdot \frac{\nabla u}{\|\nabla u\|}$~\cite{Sethian:1999}. Since the $u$ level-sets of the $uw$-parameterization of a developable surface following \equref{eq:developable-parameterization} are extrinsically flat, we have $\forall u,\ \kappa_g(u)=0$. As such, ruling fields are {both} geodesic and principal.

Denote by $\rperp$ a unit-length vector field orthogonal to the ruling directions $r$ in the tangent bundle of $\Sdev$, such that $\frac{\nabla u}{\|\nabla{u}\|} = \rperp$. 
Next, consider a unit length $2$-directional field $\gamma$ on a developable surface $\Sdev$, which is the assignment of a tangent vector $\gamma$ to every point $p \in \Sdev$, and which is defined up to sign. If we align $\gamma$ with $\rperp$, we have by definition
\begin{equation}
	\nabla u \parallel \gamma.
	\label{eq:grad_u_perp_r}
\end{equation}

For simplicity we first consider the case where $\gamma$ does not have singularities, and the surface $\Sdev$ does not contain creases. We then get:
\begin{equation}
	\nabla \cdot \gamma = \nabla \cdot \frac{\nabla u}{\|\nabla u\|} = 0.
	\label{eq:div_free_norm_grad}
\end{equation}
This means that $\gamma$ is a divergence free unit vector field. 
Our objective is to design $\gamma$ and integrate $u$ from it, which leads to the question for which divergence free unit fields such a $u$ exists. The field must be \emph{integrable} up to a scalar. That is, there must exist a positive scalar function $s$, $s(p)>0,\ \forall p \in \Sdev$, for which:
\begin{equation}
\label{eq:curl_free}
    \nabla \times (s\gamma) = 0.
\end{equation}
The geometric meaning of the scalar function $s$ is the \emph{density} of the level sets of $u$ at point $p$. A non-constant $s$ comes up naturally when the level sets have a fan-like structure (for instance, the rulings of a cone). 

\paragraph{Singularities and combing} We design $\gamma$ as a $2$-directional field, where it is only defined up to sign. Therefore, the divergence and curl operators do not automatically apply. Rather, in every local surface patch that does not contain singularities, $\gamma$ can be \emph{combed} by consistently choosing one of the directions to obtain a smooth single-vector field on which we adhere to conditions~\eqref{eq:div_free_norm_grad} and~\eqref{eq:curl_free}. 

Since our field $\gamma$ is $2$-symmetric, singularities have indices that are integer multiples of $\pm \frac{1}{2}$. Therefore, the field $\gamma$ is not defined there, and neither is $u$. As a consequence, it is not divergence free in any neighborhood that contains the singularity, and the level sets of $u$ are not straight there, see \figref{fig:teaser}. Note that singularities either arise on planar patches, or on cone apexes, and therefore do not compromise the properties of the field on torsal patches. Following the common paradigm of seamless parameterization (e.g.,~\cite{MIQBommes,Diamanti2015}), this is the reason why we design the field $sY$ as curl free, rather than as the \emph{conservative} $\nabla u$, which is only locally defined in simply-connected non-singular patches.
%

\paragraph{Relation to geodesic fields} 
Vekhter et al.~\shortcite{Vekhter2019weaving} and Pottmann et al.~\shortcite{pottmannGeodesicPatterns} both apply the unit-length divergence property to design geodesic fields; more precisely, Vekhter et al.~\shortcite{Vekhter2019weaving} work with the dual curl-free vector field $\gamma^{\perp}$ and define a similar integrability measure. Nevertheless, our work handles further challenges, as it is not enough to target geodesic fields to guarantee that they follow rulings, even though rulings are geodesics. It is in fact theoretically impossible to characterize rulings of a developable merely as geodesics, since they depend on the shape operator and are thus extrinsic. Therefore, $\gamma$ has to be designed such that $\gamma^{\perp}$ aligns to prescribed rulings. As we see in \secref{sec:discretization}, estimating and aligning to reliable rulings is a challenging task that must include completion in unreliable regions. 

\paragraph{Ruling field at creases}
Rulings on two developable patches adjacent to a crease typically do not form a single, intrinsically straight line, but rather meet at an angle (see e.g.\ \figref{fig:piecewisedev}, \ref{fig:curvedfolds}). We therefore do not require $\gamma$ to be divergence-free near creases, effectively allowing the vector field to break across them. 

\subsection{Discrete ruling-aligned developable meshes}
\label{subsec:discrete-ruling-aligned-developable-surface}
A discrete sampling of the $u$ level sets of a principally-aligned creates a quadrilateral mesh whose faces are planar up to second order \cite{liu:conicalmeshes:2006}. Anisotropic quadrilateral meshing aligned to principal directions is known to have optimality properties in terms of approximation quality (see e.g.~\cite{AnisotropicPolygonalRemeshing:Alliez:2003}). These facts motivate curvature-aligned polygonal remeshing, in particular for fabrication purposes.

Since we only design and discretize $u$, leaving the coordinate $w$ as a degree of freedom, our discretization for a torsal patch is that of a mesh comprising long planar polygons. These polygons are for the most part quadrilaterals whose edges are two boundary curves and two straight rulings; thus, a torsal patch is represented as a PQ-strip model. Planar patches are represented as big flat polygons, where the non-straight level sets are fully contained in the plane, and we are therefore allowed to straighten them out. The planar polygons are in general non-quad, since they may contain singularities; nevertheless, their planarity makes them easy to refine if required.

\section{Discretization}
\label{sec:discretization}

The input to our algorithm is a triangle mesh $\set{M} = \{\set{V},\set{E},\set{F}\}$ representing the (piecewise) developable surface, where $\set{V}$ denotes the set of vertices, $\set{E}$ the set of edges and $\set{F}$ the faces. To regularize the scale of surface curvature between different surfaces, and our optimization parameters, we scale the input $\set{M}$ to have unit-length bounding box diagonal. We define $u$ as a piecewise-linear vertex-based function $u(v), \ v \in \set{V}$, and consequently represent $r$, $r^\perp$, and $\nabla u$ as face-based piecewise-constant tangent fields; we denote this space as $\set{X}$. We use the conforming discrete gradient $G: (\set{V}\rightarrow \R)\rightarrow\set{X}$ and divergence $D:\set{X}\rightarrow (\set{V}\rightarrow \R)$ operators, and the non-conforming discrete curl operator $C:\set{X}\rightarrow (\set{E}\rightarrow \R)$. Their explicit expressions can be found in, e.g.,~\cite{Brandt2017spectral}.

\paragraph{Estimating rulings}
We compute a ruling direction $r(f),\ \forall f \in \set{F}$, as the eigenvector corresponding to the minimal eigenvalue of the face-based shape operator ${S}(f)$, as defined in~\cite{FernandoPoly}. Since we know the ruling only up to sign, we represent it unambiguously using a \emph{power representation}~\cite{azencot2017consistent, Knoppel2013}: we first represent $r(f)$ as a complex number in a local coordinate system and then square this complex number to have a representation that is invariant to the sign of the direction, i.e., we store $R(f) =r^2(f)$. We also define $R^{\perp}(f) = (r^{\perp}(f))^2$, the power representation of the ruling locally rotated by 90 degrees.

\begin{figure}[t]
	\centering
	\setlength{\tabcolsep}{2pt}
	\begin{tabular}{ccc}
	\includegraphics[width=0.3\linewidth]{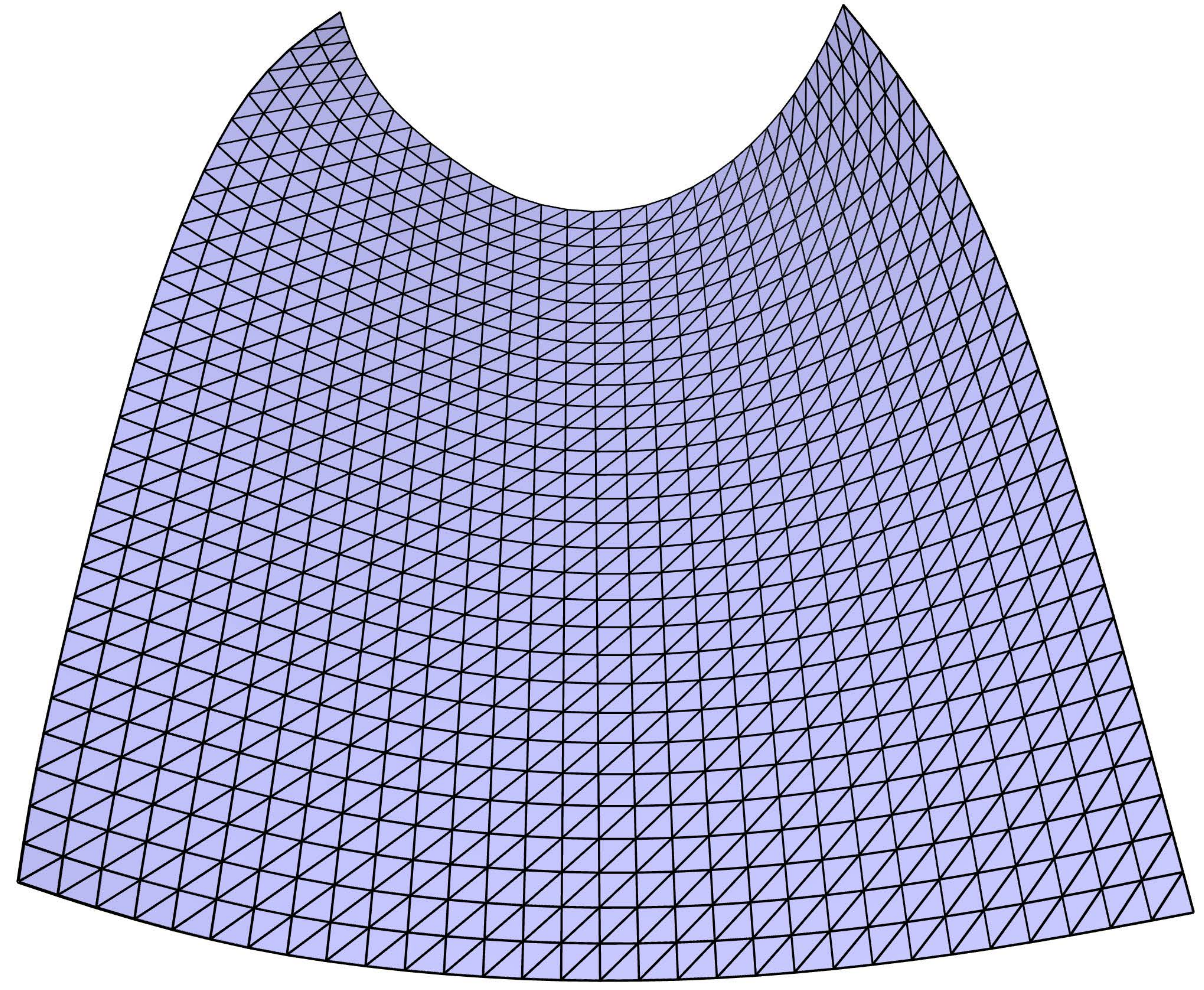} &
	\includegraphics[width=0.3\linewidth]{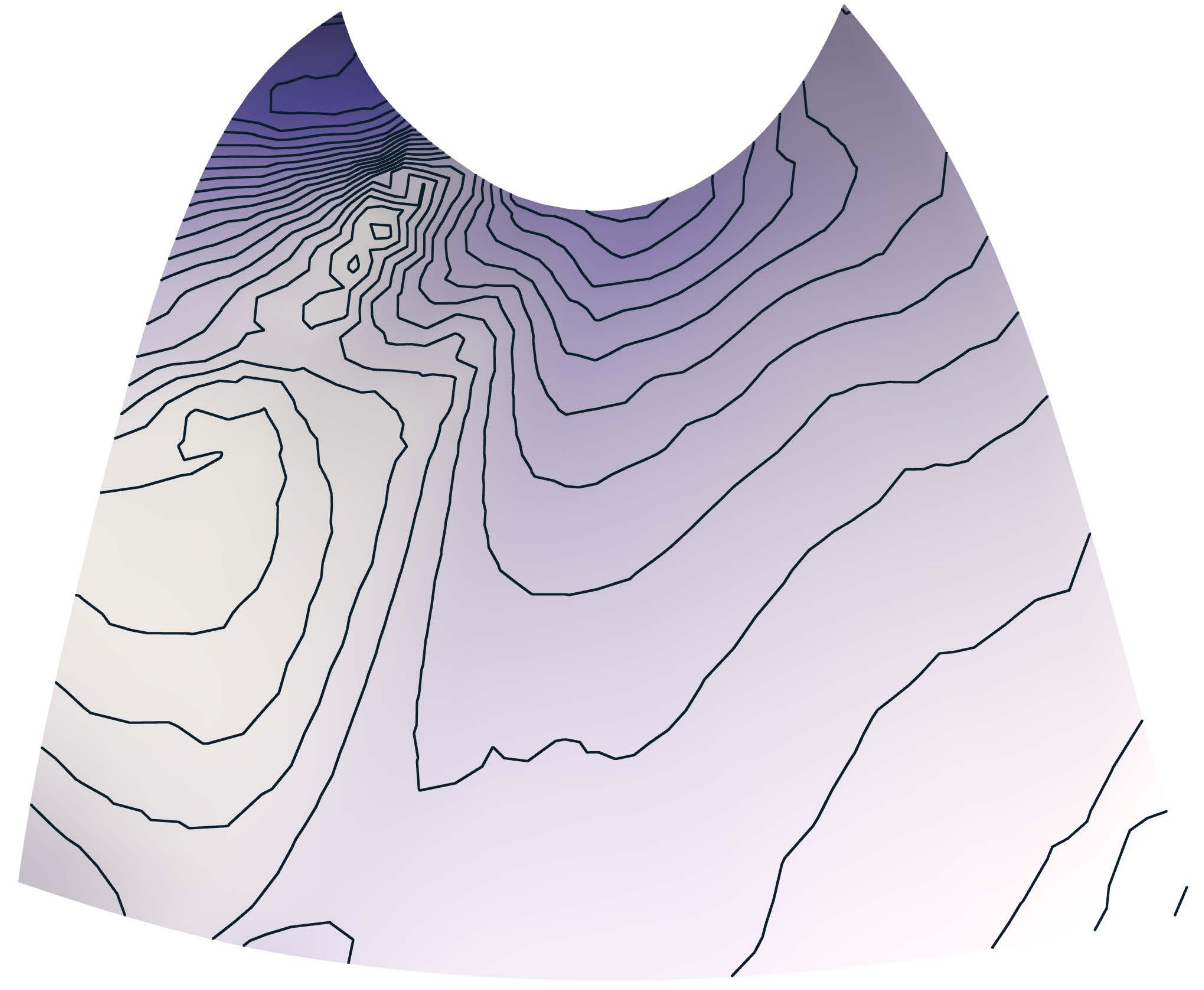} &
	\includegraphics[width=0.3\linewidth]{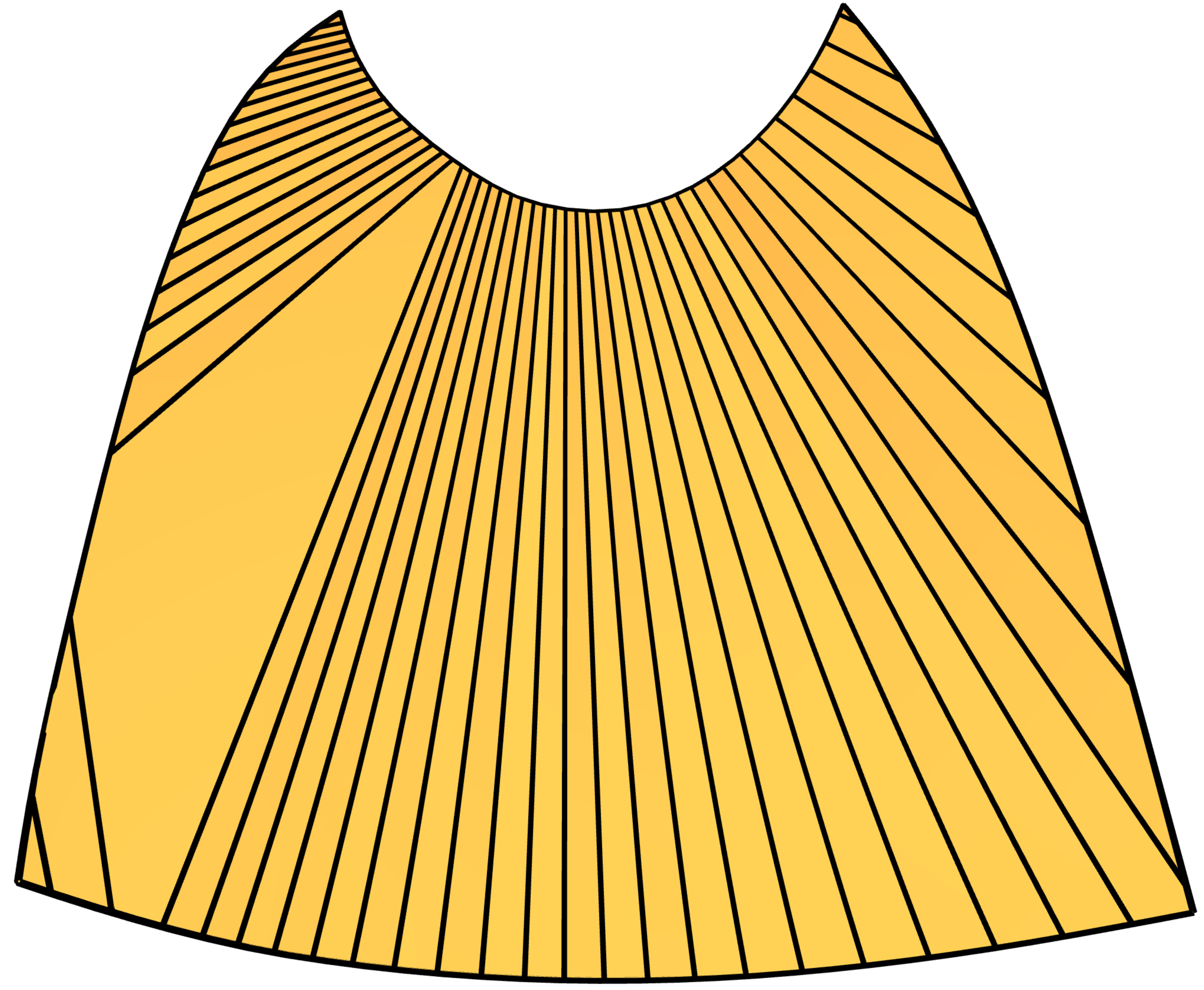}\\
	\end{tabular}
	\caption{The level sets of the curvature measure $|\kappa_2-\kappa_1|$ (middle) do not provide a clean delineation between torsal and planar parts. However, our method automatically places a planar polygon in the appropriate region, without being provided with an explicit decomposition (right).}
	\label{fig:curvedness}
\end{figure}

\begin{figure*}[t]
	\centering
	\setlength{\tabcolsep}{1pt}
	\begin{tabular}{ccccc}
		\includegraphics[width=0.19\linewidth]{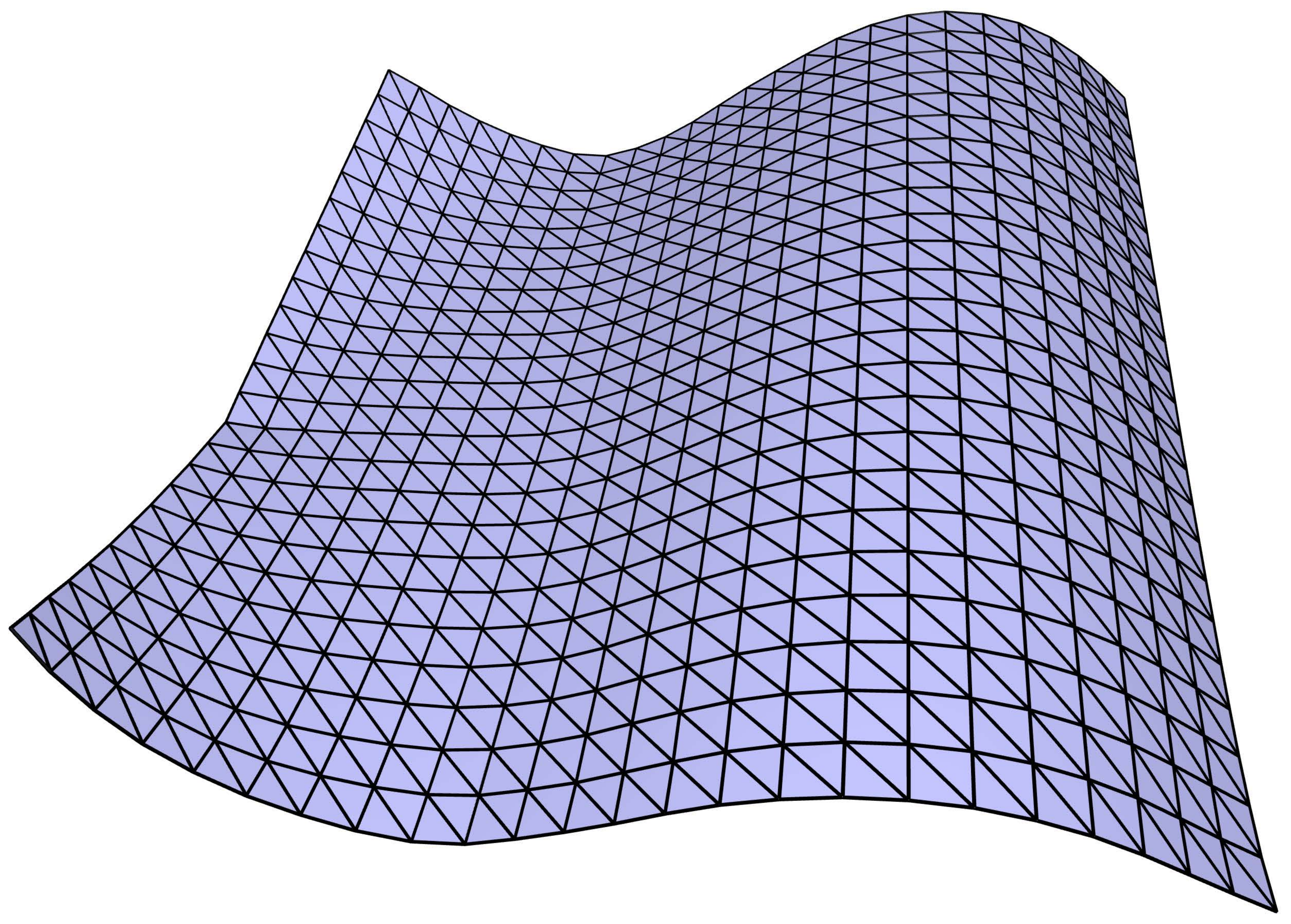}&
		\includegraphics[width=0.2\linewidth]{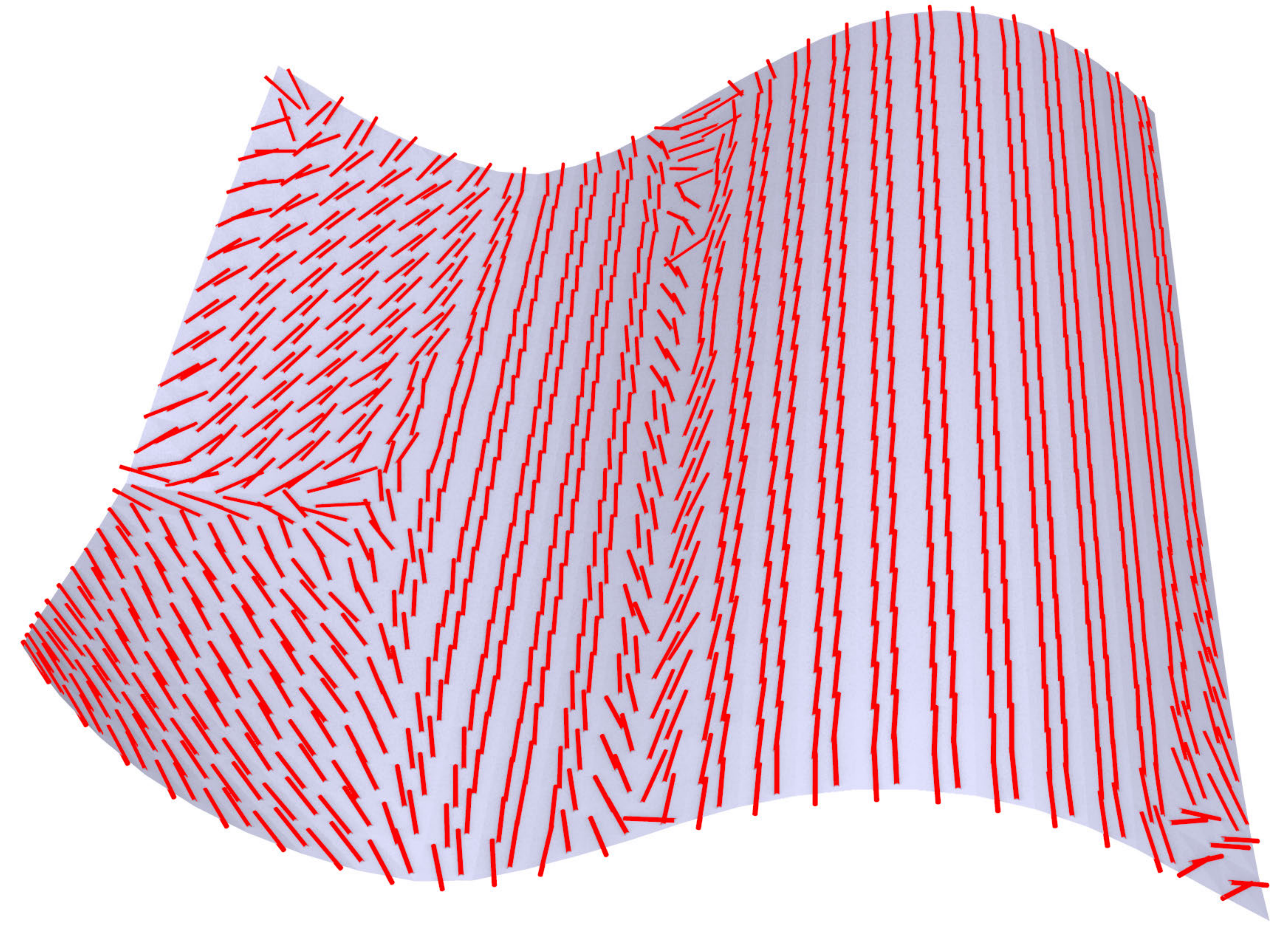}&
		\includegraphics[width=0.2\linewidth]{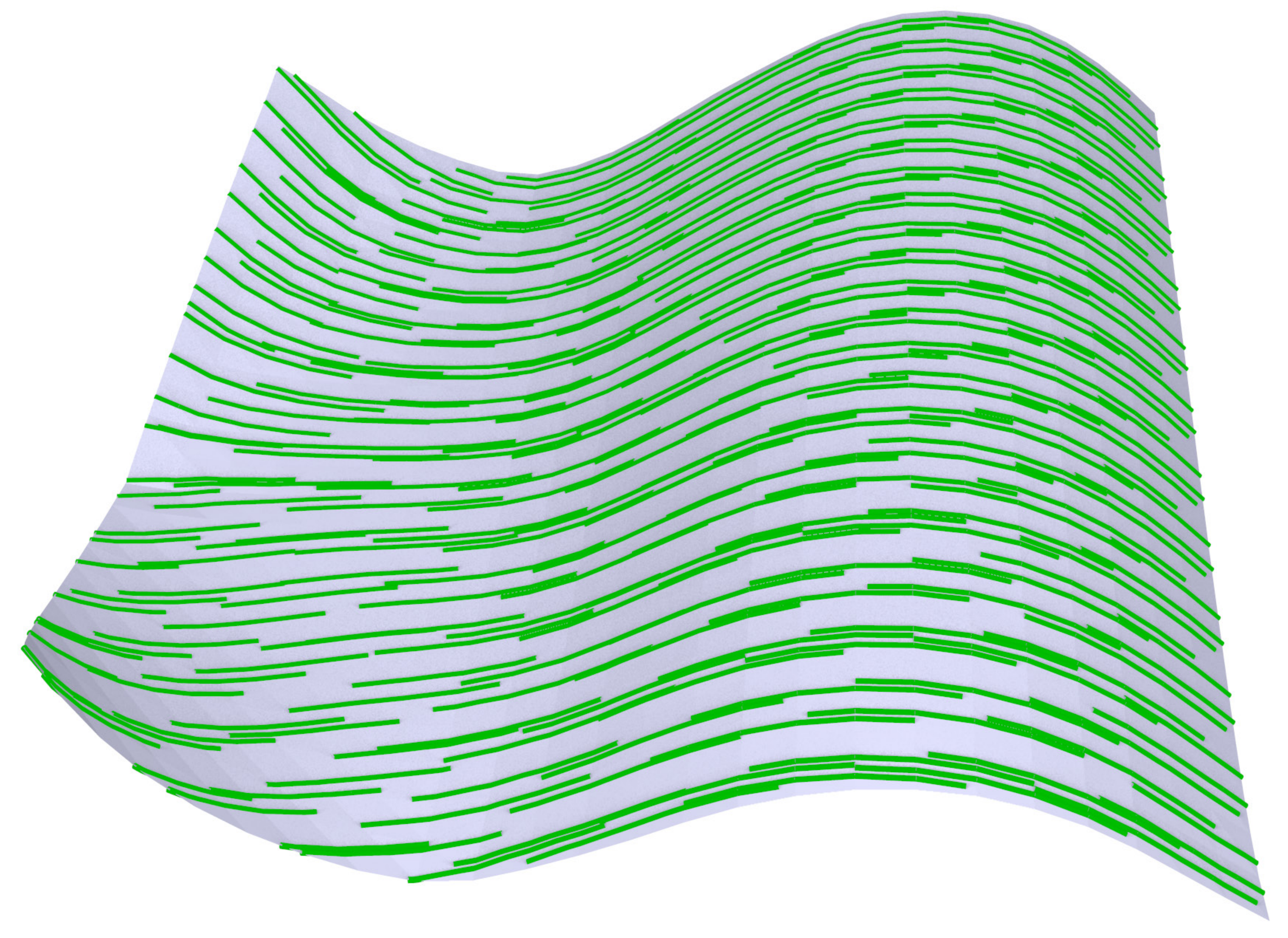}&
		\includegraphics[width=0.2\linewidth]{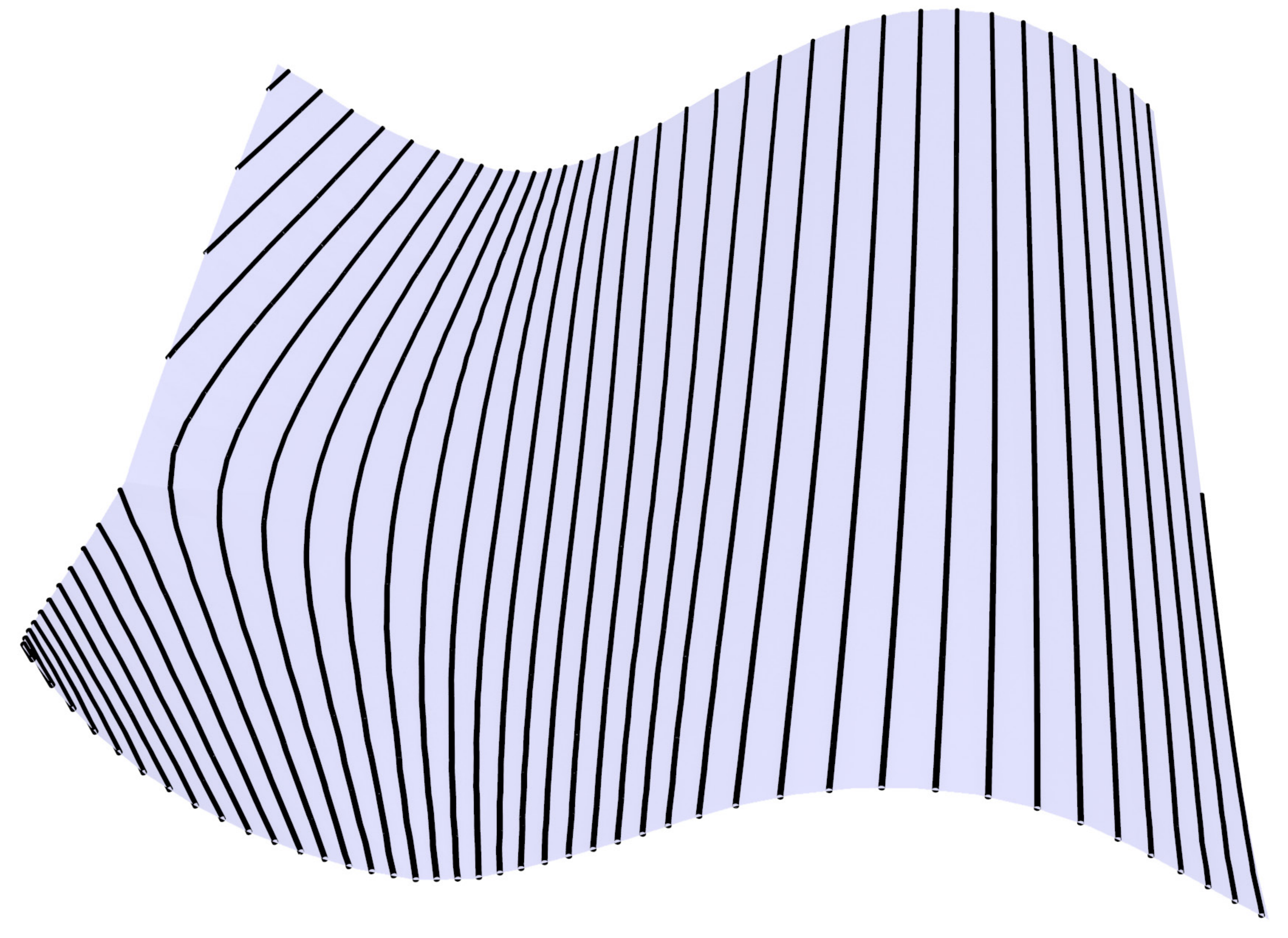}&
		\includegraphics[width=0.19\linewidth]{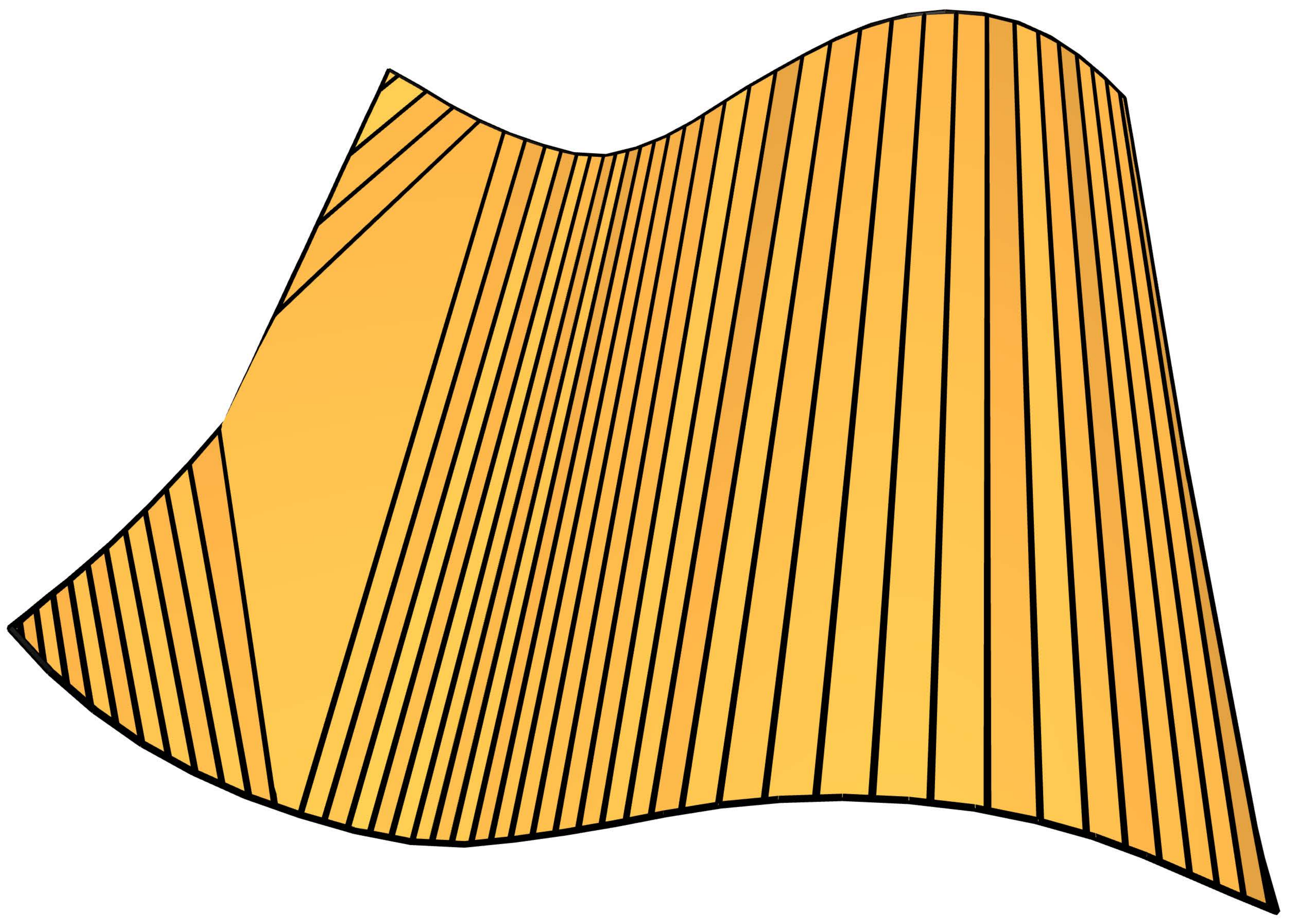}\\
		\\
		\small (a) input mesh $\set{M}$ &
		\small (b) input rulings $r$&
		\small (c) streamlines of $\gamma$&
		\small (d) level sets of $u$ &
		\small (e) our result $\set{M}'$\\
	\end{tabular}
	\caption{Our remeshing pipeline: 
		(a) The original input $\set{M}$; 
		(b) the noisy input rulings $r$; 
		(c) our computed $\gamma$ field, visualized with streamlines;
		(d) the level sets of the optimized function $u$;
		(e) the final remeshing result $\set{M}'$. Note how the level sets in (d) bend inside the planar region, which gets meshed as one large polygon, but are straight in the torsal regions, which result in PQ-strips.}
	\label{fig:pipeline}
\end{figure*}

\paragraph{Confidence weights}
A clean domain decomposition into planar and torsal regions would significantly simplify the fitting of individual developable patches. Unfortunately, we cannot obtain such a clean segmentation directly, because the curvature measure (like the ruling estimates) is noisy and does not delineate planar and torsal patches nicely (\figref{fig:curvedness}). Therefore, we model on the assumption that the rulings are least reliable in planar or near-planar regions, and mostly consistent in strongly curved areas (see Figs. \ref{fig:teaser} and \ref{fig:pipeline}). We thus attach a \emph{relative confidence} weight $w(f)$ to each face $f\in\set{F}$, as a function of the discrete absolute max and min curvatures $\kappa_1(f)$ and $\kappa_2(f)$: 

\begin{equation}
   {w}(f) = \theta_1\left(1 - e^{\theta_2(\kappa_1(f)-\kappa_2(f))^2}\right).
    \label{eq:confweights}
\end{equation}
\setlength{\intextsep}{0pt}%
\setlength{\columnsep}{5pt}%
\begin{wrapfigure}{r}{0.2\textwidth}
	\centering
	\includegraphics[trim={1cm 8cm 2cm 8cm}, width=0.2\textwidth]{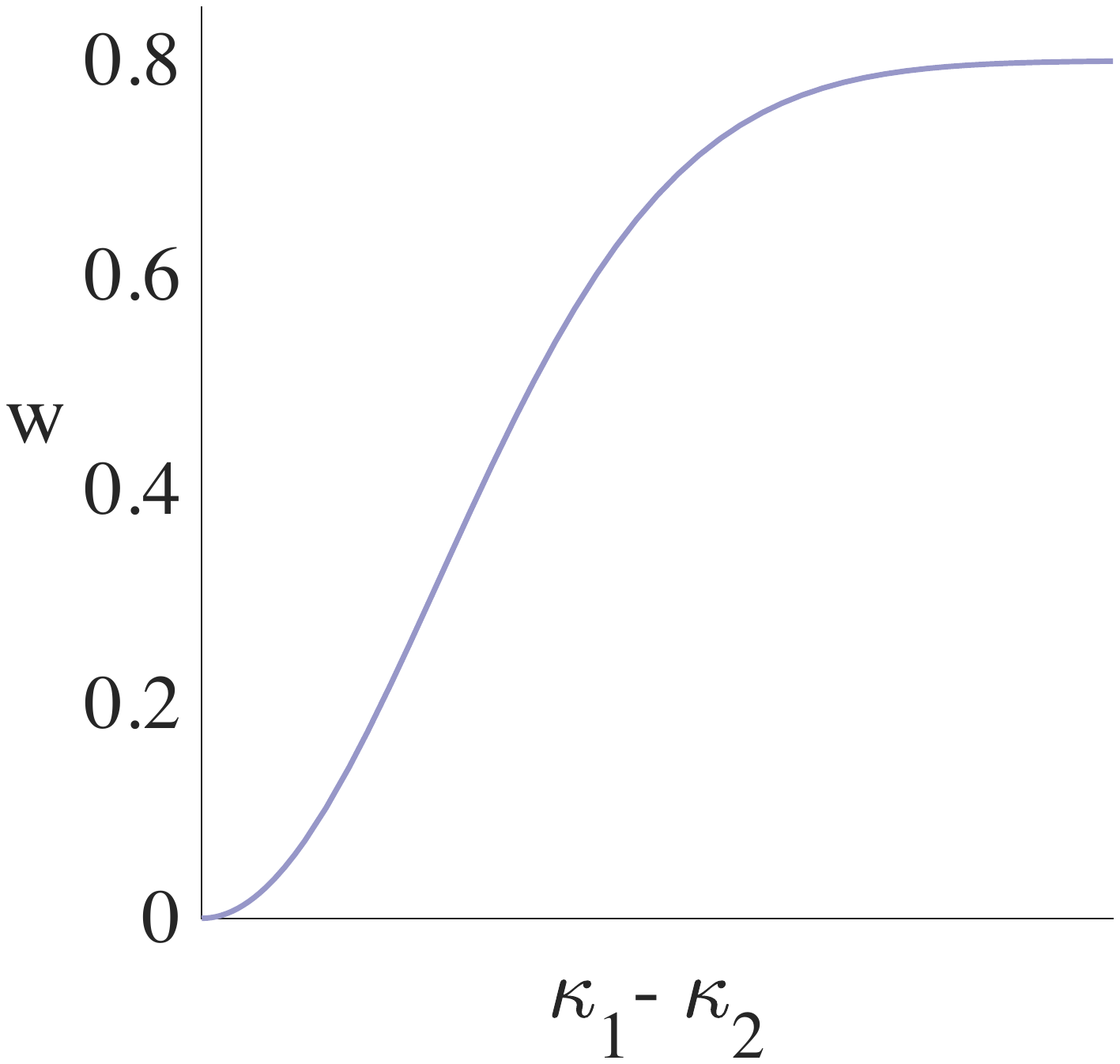}
\end{wrapfigure}
For $\kappa_1(f)$ and $\kappa_2(f)$ we use the absolute largest and smallest eigenvalues of the shape operator $\set{S}(f)$, and set $\theta_1=0.8$, $\theta_2=-0.014$. The confidence function is a logistic curve (see inset), facilitating a stronger distinction between confidence in planar and near-planar regions (albeit still small compared to stronger curved regions). The value of $w(f)$ is capped at $0.8$ by design, ensuring that we never fully rely on a ruling.
We define $\set{V}_b$ to be all vertices on the boundary of $\set{M}$ and $\set{F}_b$ all faces that contain a vertex in $\set{V}_b$, and we set $w(f)=0$ for these faces.

\paragraph{Creases}
Our method requires as input the explicit identification of the set of crease edges $\set{E}_c$ that define curved-fold creases and boundaries to developable pieces.
We define $\set{V}_c$ as all vertices that are incident on an edge in $\set{E}_c$, and from this we define the set of faces adjacent to them: $\set{F}_c$ is the set of all faces that have one or more vertices in $\set{V}_c$. We update the confidence weights by setting $w(f)=0$ for all faces in $\set{F}_c$. 
We make an initial guess for the crease edges based on the dihedral angle of adjacent faces and manually add edges belonging to softer creases. The final set of crease edges should divide the surface into smooth developable surfaces without creases. The required seams for this segmentation are typically easily visually distinguishable (for example using reflection lines). Figures \ref{fig:sphericonsdforms}, \ref{fig:Keenan}, \ref{fig:piecewisedev}, \ref{fig:curvedfolds} and \ref{fig:curvedfoldingKilian} show examples of developable surfaces with creases and \figref{fig:Keenan} explicitly shows them on a complex model. When creases end in the interior of a developable piece rather than on another crease or boundary, so called \emph{open creases}, we duplicate their interior vertices and define them as mesh boundaries. Examples of such open creases can be seen in Figures \ref{fig:glued}, \ref{fig:Keenan} and \ref{fig:curvedfoldingKilian} (the chair model).

\section{Method}
\label{sec:method}

We describe our approach to remeshing a (near-)developable input triangle mesh to a curvature-aligned, planar polygonal mesh consisting primarily of PQ strips. 

\subsection{Optimization problem}
\paragraph{Setup}
We compute the face-based shape operator and the ruling related quantities $r(f)$ and $\rperp(f)$. Our computed field $Y$, and the ruling fields $r(f)$ and $\rperp(f)$ are represented as vectors in $\mathbb{R}^{2|\set{F}|}$, where each two consecutive elements are the components of the vector in a local basis defined on each $f \in \set{F}$. We further compute the power quantities $R(f)=r^2(f),\ R^{\perp}(f)=(r^{\perp})^2(f)$, $\forall f \in \set{F}$, as well as the confidence weights $w(f)$, as detailed in \secref{sec:discretization}. The power operation is understood to act on $r(f)$ (for instance), in an equivalent local complex representation.

We consider the diagonal face-based mass matrix either for vectors $M_{\set{X}}:2\left|\set{F}\right| \times 2\left|\set{F}\right|$, or for scalar or complex quantities $\set{M}_\set{F}: \left|\set{F}\right| \times \left|\set{F}\right|$, holding the face areas $m(f)$. We further consider the edge-based mass matrix $M_{\set{E}}$ holding edge masses 
$$m(e)= \frac{\|e\|}{\|e_\text{dual}\|}\left(m(f)+m(g)\right)/2,$$ where $\|e_\text{dual}\|$ is defined as the summed length of the two dual edges from the midpoint of $e$ to the barycenters of the adjacent faces $f$ and $g$.
Finally, for a vector field $\gamma \in \set{X}$ we use the \emph{integrated} discrete divergence $D\gamma = G^\tr M_{\set{X}}\gamma  \ (\in \R^{|\set{V}	|})$, where $G$ is the discrete gradient operator, which for a triangular face $f$ consisting of vertices $i,j,k$ and scalar function $u$ is defined as $Gu_{ijk}(f) = \frac{1}{2m(f)}(e_{jk}^\perp u_i + e_{ki}^\perp u_j + e_{ij}^\perp u_k)$.

 \paragraph{Objective} We optimize for a gradient ruling field $Y(f)$, according to the requirements of~\secref{subsec:ruling-fields}. That is, $Y(f)$ should have unit norm, it should align to the estimated rulings $r^{\perp}(f)$ up to sign and according to confidence, it should be divergence free away from creases, boundaries, and singularities, and it should be curl free up to scaling. We then use as variables $Y(f)$ itself, its power representation $\Gamma(f)=\gamma^2(f)$, where $\Gamma(f)$ should align to the perpendicular power ruling field $R^{\perp}(f)$ according to the confidence $w(f)$, and where $\gamma$ is divergence-free away from singularities. Furthermore, we optimize for a scalar field $s(f)$, such that $s(f)\cdot\gamma(f)$ is curl-free. Our objective breaks down to the following terms:

\paragraph{Alignment objective} Our alignment term is 
\begin{equation}
    E_a(\Gamma) = \sum_{f \in \set{F}} m(f) w(f)\, \|\Gamma(f) - R^{\perp}(f) \|^2,
    \label{eq:Gamma_R_alignment}
\end{equation}
where $m(f)$ is the face area of $f$ and $w(f)$ is the confidence weight as defined in \equref{eq:confweights}. This can be formulated in matrix form as 
\begin{equation}
\label{eq:E_a_Gamma}
E_a(\Gamma) = \left(\Gamma - R^{\perp}\right)^H M_{\set{F}}W_{\set{F}}\left(\Gamma - R^{\perp}\right),
\end{equation}
where $W_{\set{F}}$ is the diagonal matrix of per-face confidences for complex numbers or scalars, and $\Gamma$ and $R^{\perp}$ are arranged as $\left|\set{F}\right| \times 1$ complex vectors. Note the \emph{conjugate} transpose $\left(\Gamma - R^{\perp}\right)^H$.

\paragraph{Unit-norm divergence-free objective} We ideally want the field to be perfectly divergence-free and have unit norm everywhere. However, this is impossible at singularities (\secref{subsec:ruling-fields}) and in general would only be important on torsal patches. We follow~\cite{viertel2019GL} and~\cite{Sageman2019chebyshev} by using a \emph{Ginzburg-Landau} approach, introducing the following objective term:
\begin{equation}
E_d(\gamma) = \sum_{v\in\set{V}}{\frac{1}{m(v)}\left|D\gamma(v)\right|^2} + \frac{1}{\epsilon^2}\sum_{f\in\set{F}}m(f){\left(\|\gamma(f)\|^2-1\right)},
\label{eq:Gl_div_unit_norm}
\end{equation}
where $m(v)$ is the barycentric Voronoi area of vertex $v$; note that its reciprocal is used since $Dy$ is integrated.
When $\epsilon \rightarrow 0$, this is analogous to minimizing the divergence of a unit-norm field after removing a ball of radius $\epsilon$ around singularities. Since the unit-norm divergence-free condition is satisfiable on torsal patches in a direction that matches with the alignment terms, singularities (if any) will naturally be located inside planar regions.

\paragraph{Smoothness regularizer} To encourage the field to smoothly transition from curved to planar parts, and in general to regularize low-confidence regions, we add a small smoothness term that encodes the smoothness of the power vector field across edges. For each interior edge $e$ adjacent to faces $f$ and $g$, the power smoothness~\cite{Knoppel2013} is measured as:
\begin{equation}
\label{eq:power_smoothness}
	\|\Gamma(f)\, \bar{e}^2_f - \Gamma(g) \, \bar{e}^2_g\|^2.
\end{equation}
Here, $\bar{e}_f$ is the conjugate of $e_f$, which is the complex representation of the normalized edge vector $e$ in the basis of $f$, and similarly for $g$.
Our smoothness regularizer then becomes:
\begin{equation}
\label{eq:smoothness_term}
    E_s(\Gamma) = 
    \sum_{e \in \set{E}} m(e)\left(1-w(e)\right)\|\Gamma(f)\, \bar{e}^2_f - \Gamma(g)\, \bar{e}^2_g\|^2,
\end{equation}
where $w(e) = (w(f)+w(g))/2$. In matrix form, we write this energy as $E_s(\Gamma) = \Gamma^H L_2 \Gamma$, where
\begin{equation}
\label{eq:E_s_Gamma_matrix}
L_2 = G^{H}_{\set{E}}M_{\set{E}}\left(I-W_{\set{E}}\right)G_{\set{E}},
\end{equation}
where $G_{\set{E}}$ stacks the differences $\Gamma(f)\, \bar{e}^2_f - \Gamma(g) \, \bar{e}^2_g$ from \equref{eq:power_smoothness}.

\paragraph{Integrability} We use the discrete curl operator $C$ to measure integrability of the (per-face) scaled field $s\gamma $:
\begin{equation}
C s\gamma(e)  = \left\langle s(f)\gamma(f) - s(g)\gamma(g), \ e \right\rangle.
\end{equation}
We constrain 
\begin{equation}
Cs\gamma = 0.
\end{equation}
To constrain $s$ to be positive and prevent large density variations, we further bound
\begin{equation}
\forall f \in \set{F},\  s_\text{low} < s(f) < s_\text{high}\,.
\end{equation}
We provide the values used for $s_\text{low}$ and $s_\text{high}$ in \secref{subsec:optimproblem}.

\paragraph{Branching and singularities} Generating $\Gamma$ from $\gamma$ is well-defined. However, the inverse has a sign degree of freedom. We follow common practice by arbitrarily choosing a sign in each face, and relating $\gamma$ values across faces by using \emph{principal matching}~\cite{Diamanti2014}; in our context, this means we match vectors according to the smallest rotation angle. The curl and divergence operators are always understood to be defined with relation to the matching at every edge and vertex, with the exception of singularities, creases, and boundary vertices (where we do not optimize for divergence). 

\paragraph{Full optimization problem} Our optimization problem can then be finally formulated as follows:
\begin{align}
	\label{eq:optim_full_begin}
&    \left(\Gamma, \gamma,s\right)  = \argmin \ \omega_a E_a(\Gamma) + \omega_d E_d(\gamma)+\omega_sE_s(\Gamma),\ \ s.t.\\
&    \phantom{\left(\Gamma, \gamma,s\right) =\ } 
     \forall f \in \set{F},\ \Gamma(f) = \gamma^2(f), \\
    \label{eq:optim_full_curl}
&   \phantom{\left(\Gamma, \gamma,s\right) = \ }
     Cs\gamma = 0,\ \ \\
    \label{eq:optim_full_end}
&    \phantom{\left(\Gamma, \gamma,s\right) = \ }
\forall f \in \set{F},\  s_\text{low} < s(f) < s_\text{high}\,.
\end{align}
Here, $\omega_a, \omega_d, \omega_s$ are scalar weights. Similar to~\cite{Sageman2019chebyshev}, we seek solutions where 
$\frac{\omega_s}{\omega_d}\rightarrow 0$ and $\frac{\omega_s}{\omega_a}\rightarrow 0$ to allow the solution to converge to a divergence-free unit-norm field aligned to rulings away from planar regions and singularities.

\subsection{Optimization algorithm}
\label{subsec:optimproblem}
Our energy is nonlinear and its constraints use discrete variable quantities such as the matching. 
As the optimization problem is separable in the $\Gamma$, $\gamma$ and $s$ variables, we optimize for them in an alternating fashion, following the spirit of~\cite{Sageman2019chebyshev}. Our method proceeds as described in Algorithm~\ref{alg:Optimize_gamma}.

\begin{figure*}[t]
\centering
\includegraphics[width=\linewidth]{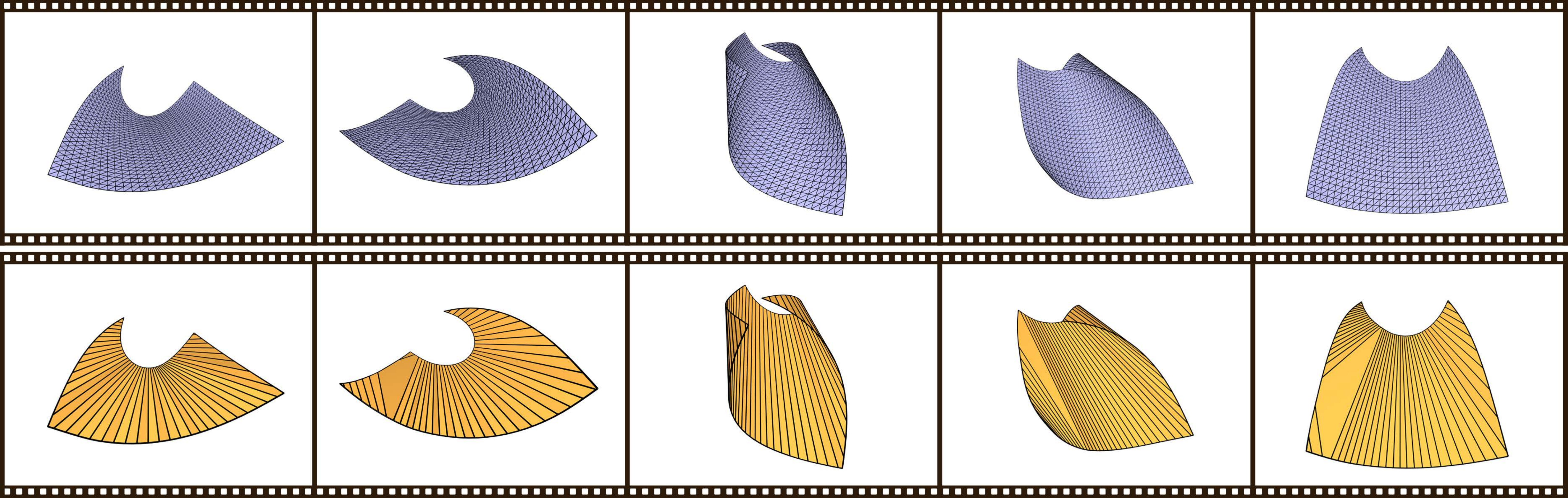}    
    \caption{Several snapshots from an interactive editing session. The user deforms the DOG model by interacting with point handles at some selected vertices. At any time during the interactive session, the user may invoke our remeshing algorithm and view the curvature-aligned remesh nearly instantaneously. Note how the combinatorial structure of the ruled remeshing automatically changes to accommodate the changes in the surface geometry, without forcing the user to specify the patch decomposition manually.}
    \label{fig:editing-sequence}
\end{figure*}

\begin{algorithm}
\SetAlgoLined
 Initialize $\Gamma^0 = R^{\perp},\ k=0, \set{V}^* = \set{V} \setminus (\set{V}_b\cup\set{V}_c)$\\
 \Repeat{$ \max_{f}\|\Gamma^k(f)-\Gamma^{k-1}(f)\| < 10^{-3}$}{
 	$k \leftarrow k+1$\\
    $\Gamma_a^{k} \leftarrow \texttt{ImplicitAlign}(\Gamma^{k-1})$\\
   $\Gamma_s^{k} \leftarrow \texttt{ImplicitSmooth}(\Gamma_a^{k})$\\
  $\forall f\in\set{F},\ \ \Gamma_u^{k}(f) \leftarrow \frac{\Gamma_s^{k}(f)}{\left\|\Gamma_s^{k}(f)\right\|}$\\
    $\left(\gamma_u^{k},C^{k},D^{k},\set{V}^{\ast}\right) \leftarrow \texttt{LocalRawRepresentation}(\Gamma_u^{k})$\\
  $\gamma_d^{k} \leftarrow \texttt{ProjectDivFree}(\gamma_u^{k},D^k,\set{V}^{\ast})$\\
  $\gamma_c^{k} \leftarrow \texttt{ProjectCurlFree}(\gamma_d^{k},C^k)$\\
  $\Gamma^{k} \leftarrow \texttt{PowerRepresentation}(\gamma_c^{k})$\\
 }
 \caption{Optimize for ruling field}
 \label{alg:Optimize_gamma}
\end{algorithm}

The function \texttt{ImplicitAlign($\Gamma^{k-1}$)} reduces the alignment energy $E_a$ by a single implicit Euler step, by solving the following linear system:
\begin{equation}
\left(I + \frac{\omega_a}{\mu_a}W_\set{X}	\right) \Gamma_a^{k}  = \Gamma^{k-1} + \frac{\omega_a}{\mu_a}W_{\set{F}}R^{\perp}.
\end{equation}
%
The implicit step size $w_a$ is resized by $\mu_a$, which is the lowest nonzero eigenvalue of $W_\set{F}$. Note that the mass matrix $M_\set{F}$ is cancelled out in the gradient and eigenvalue.
Similarly, \texttt{ImplicitSmooth($\Gamma_a^k$)} solves the following linear system:
\begin{equation}
(M_\set{F} + \frac{\omega_s}{\mu_s}L_2) \Gamma_s^{k^\prime}  = M_\set{F} \,\Gamma_a^{k}\,.
\end{equation}
with the lowest nonzero eigenvalue $\mu_s$ so that $\exists x \neq 0,\ L_2x = \mu_s M_{\set{F}}x$. The step size $\omega_a$ is fixed to $0.1$, and the step size $\omega_s$ starts as $0.005$ and is halved every 30 iterations, to ensure that the alternation with the renormalization of $\Gamma$ converges. 

After normalizing the current vector field, it is transformed into the ``raw'' representation $Y$, where the principal matching (and consequently, the singularities) are computed, and from them the curl matrix $C$ and divergence matrix $D$ are updated. Furthermore, this function updates $\set{V}^*$ according to the current singularities. Note that the sets of boundary vertices $\set{V}_b$ and crease vertices $\set{V}_c$ (\secref{sec:discretization}) are always mutually exclusive with $\set{V}^*$.

Next, \texttt{ProjectDivFree($\gamma_u^k$)} finds the closest divergence-free solution to $Y^k_u$ by solving the following linear system: 
\begin{equation}
\argmin_{\gamma_d^{k}} \|\gamma_d^{k} - \gamma_u^{k}\|^2 \ \text{ s.t. } \ D\gamma_d^{k}\left(\set{V}^*\right) = 0.
\end{equation}
Specifically we do this by solving 
$$\argmin_{x^*}\|x^*\| \ \ \text{s.t.} \  \ Dx^*\left(\set{V}^*\right) = -D\gamma_u^{k}\left(\set{V}^*\right),$$ 
where $x^*=\gamma_d^{k} - \gamma_u^{k}$. For $x^*$ to be a minimum-norm solution adhering to the constraints, it should be expressible as $x^* = D^\tr w$ for some $w$. So we can solve $DD^\tr w \left(\set{V}^*\right) = -D\gamma_u^{k}\left(\set{V}^*\right)$, set $x^*=D^\tr w$ and finally obtain the divergence-free solution as $x^* + \gamma_u^{k}$.

Finally, the function \texttt{ProjectCurlFree($\gamma_d^k$)} finds the closest scaled curl-free solution by solving the following convex system:
\begin{align}
\label{eq:curlproj1}
&\argmin_{\gamma_c^{k},\ s} \|\gamma_c^{k} - s\gamma_d^{k}\|^2, \\
\label{eq:curlproj2}
&s.t. \ C\gamma_c^{k} = 0, \\
&\phantom{s.t.} \ 0.4 \leq s \leq 1.6.
\label{eq:curlproj}
\end{align}
Although we have no guarantee of convergence (as discussed in \secref{sec:results}), in our experiments it typically takes {50} iterations of our optimization algorithm or less for $\Gamma$ to converge.

\subsection{Vector field integration and meshing}
Having an integrable $\gamma$, we use a mixed-integer integration scheme \cite{MIQBommes} to obtain a seamless globally smooth parameterization which produces $u$. The input triangle mesh is cut into a topological disc, where the singularities are on the boundary, and then a corner-based $u$ function is extracted, which is seamless across the cuts, using integer translations. We configure the integrator to produce $u \in \mathds{Z}+\frac{1}{2}$ values at singular vertices, since then the integer level sets avoid meeting at these singularities, and we obtain a single polygon around the singularity.

To create the final PQ-strip mesh, we trace the integer level sets of $u$ (at a user-specified global resolution) and then collapse all valence-$2$ vertices that are not on the boundary. This effectively straightens the polylines of the level sets, which has little effect in torsal regions, since the level sets are already almost straight by the optimization (see \figref{fig:teaser} and \figref{fig:pipeline}). However, the level sets in planar regions, which might be more curved if a singularity causes the divergence-free constraint to be excluded, become chords between boundary vertices. We note that since we have a full parameterization of the surface, control over individual panel width is also a possibility. 
 We illustrate our remeshing pipeline in \figref{fig:pipeline}.

\begin{figure}[t]
   \setlength{\tabcolsep}{0pt}   
    \begin{tabular}{ccc}
         \includegraphics[width=0.3\columnwidth]{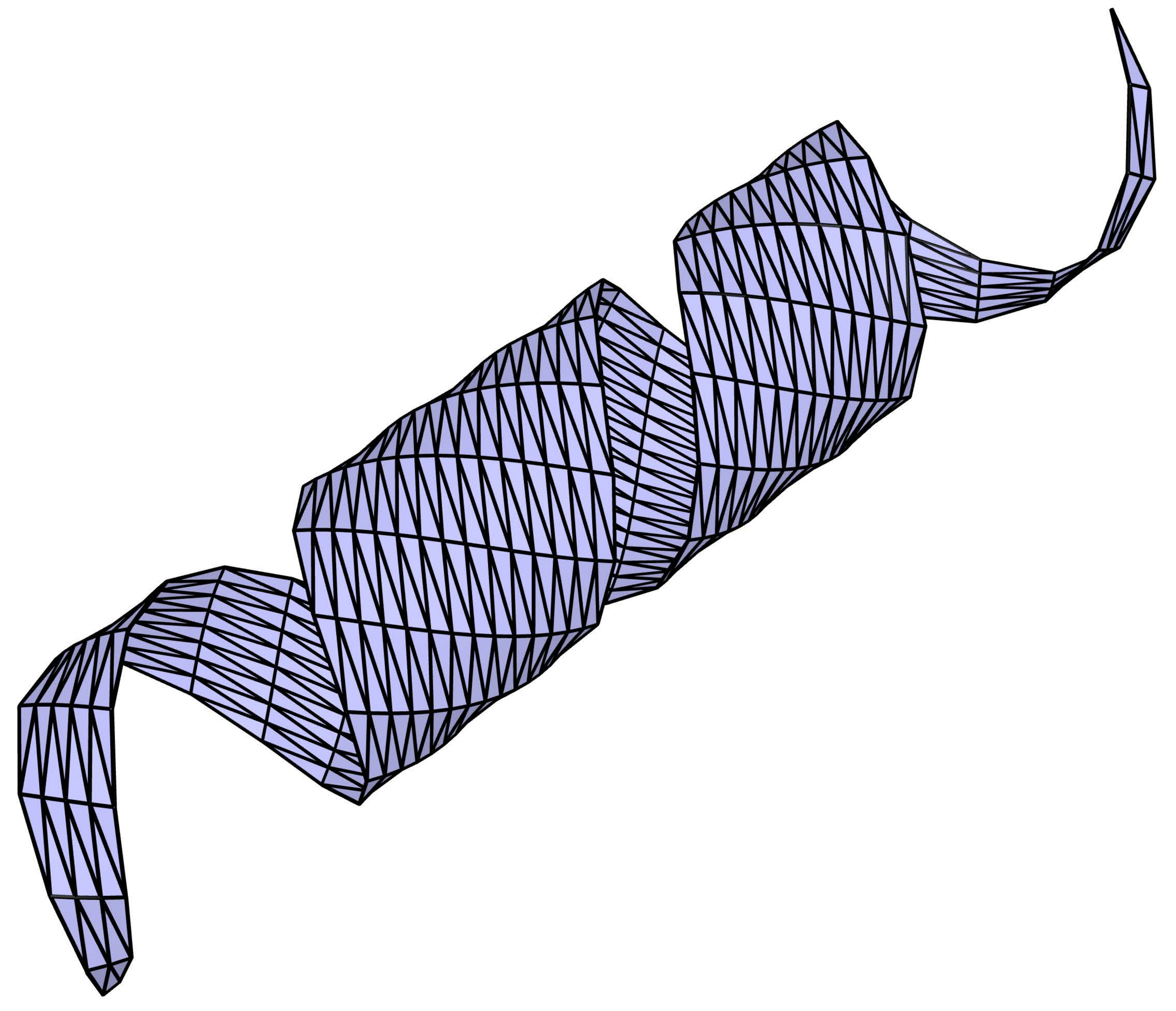} &
         \includegraphics[width=0.3\columnwidth]{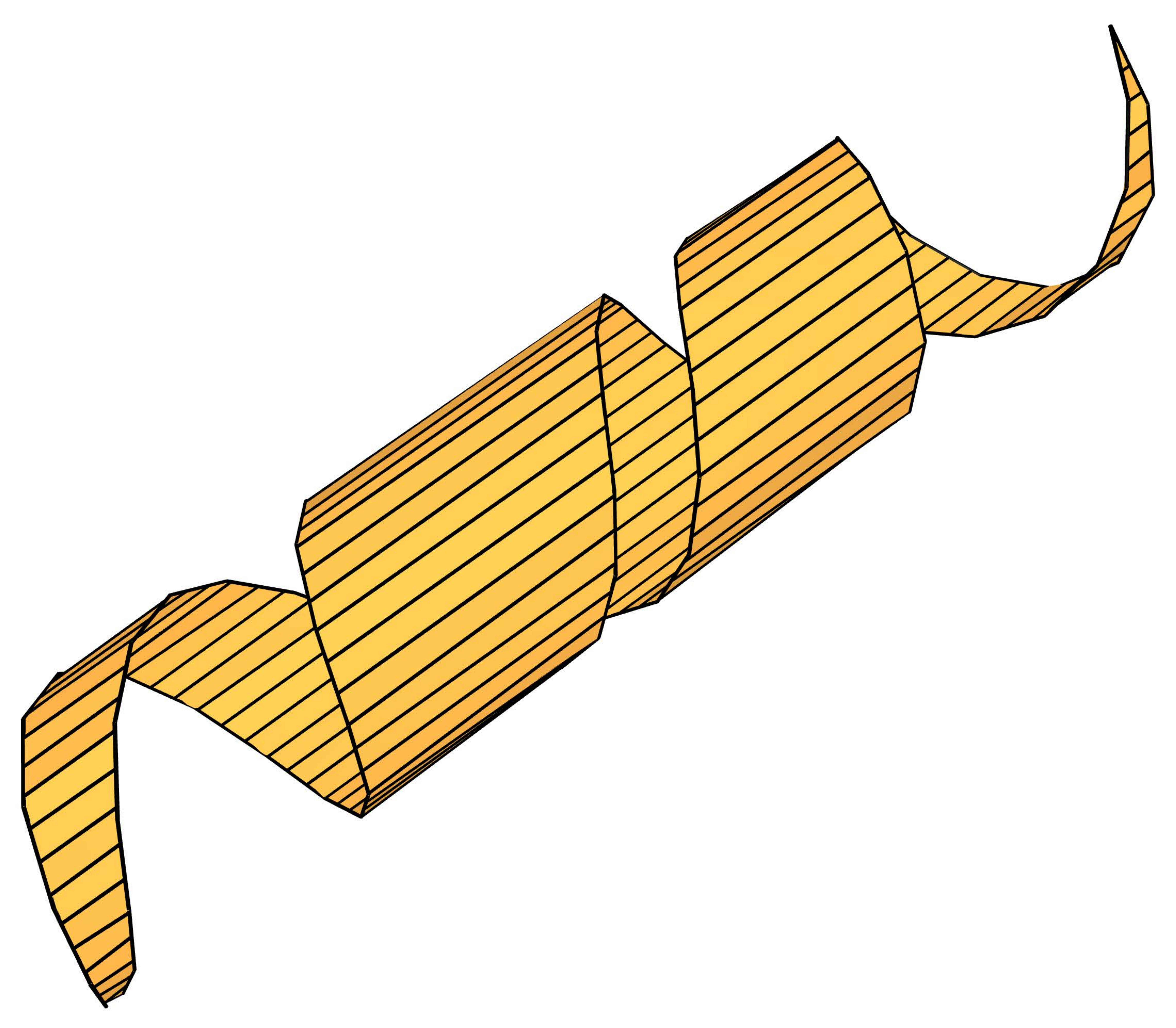} &
         \includegraphics[width=0.3\columnwidth]{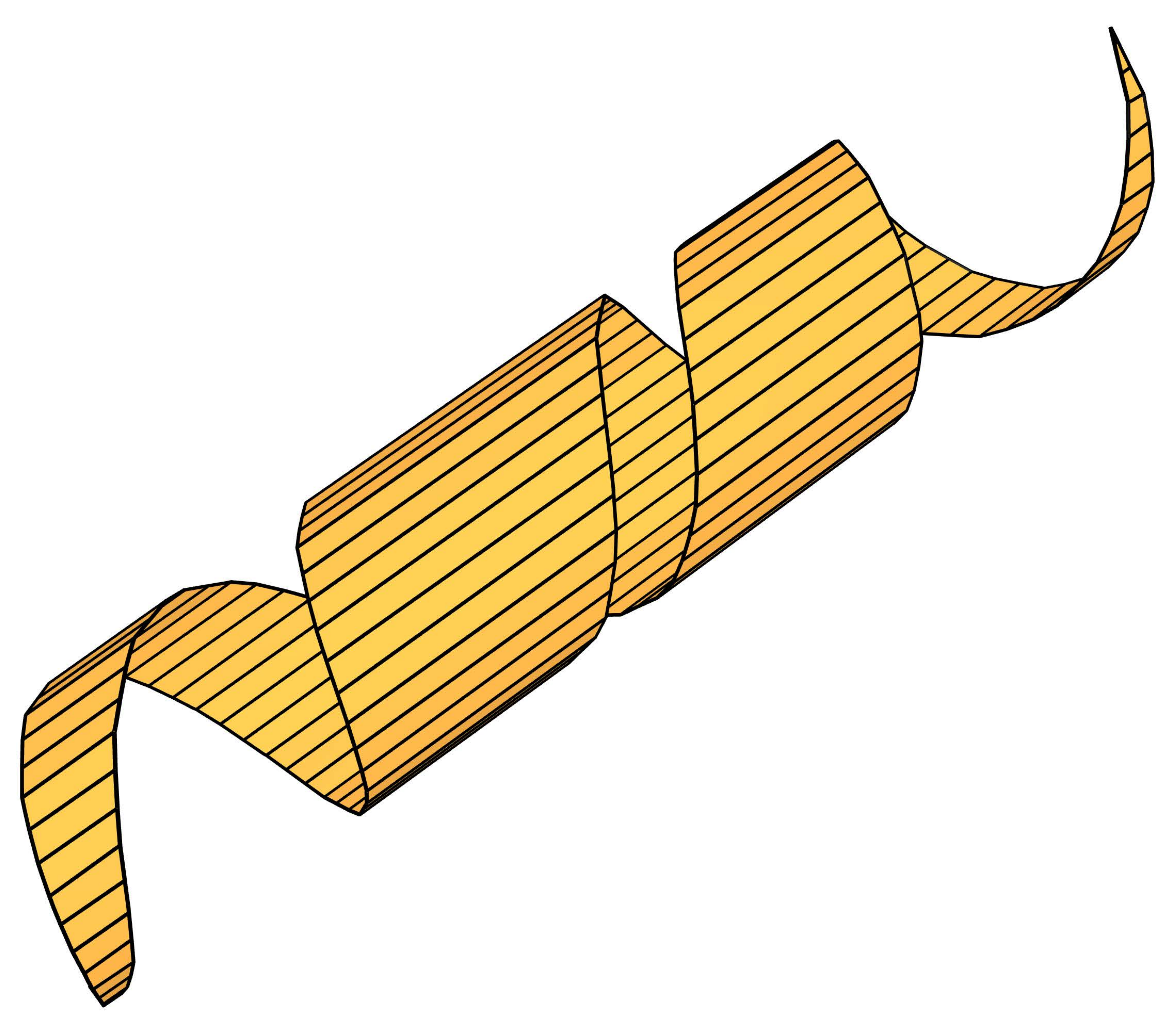}\\
         \small input &
         \small our result&
         \small after planarization \\
         &
         \small $p=11.84\%$ &
         \small $p = 0.0034\%$
    \end{tabular}
    \caption{Our result from \figref{fig:galleryNewwide} is planarized using ShapeUp~\cite{shape_up}, achieving maximal face planarity error of $p = 0.0034\%$, compared with $p=11.84\%$ in our initial result. The visual difference between the results is negligible. The Hausdorff distances are reported in Table~\ref{tab:planarity}.}
      \label{fig:planarization}
\end{figure}

\begin{figure}[t]
	\setlength{\tabcolsep}{0pt}   
	\begin{tabular}{cc}
		\includegraphics[width=0.5\columnwidth]{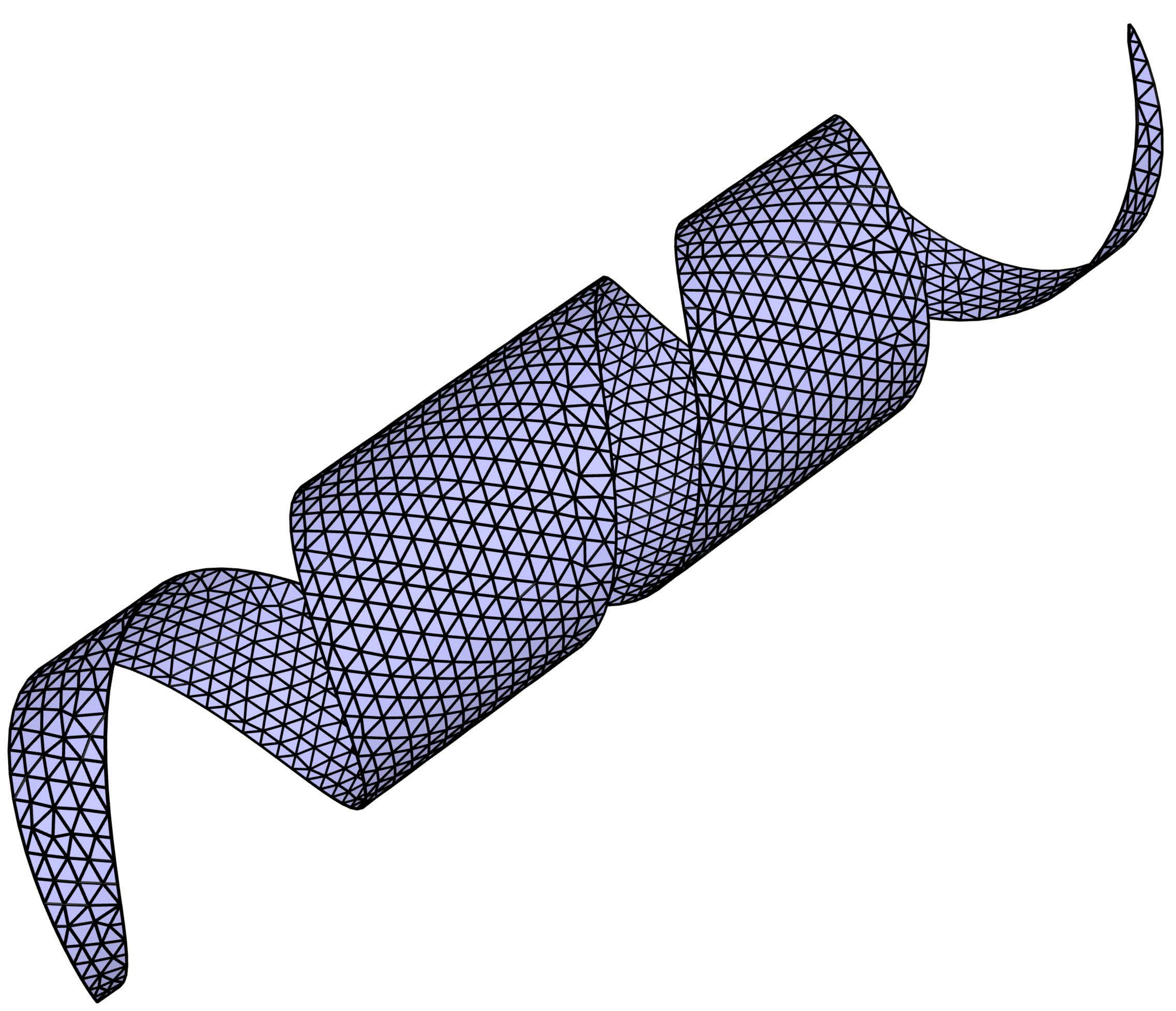}&
		\includegraphics[width=0.5\columnwidth]{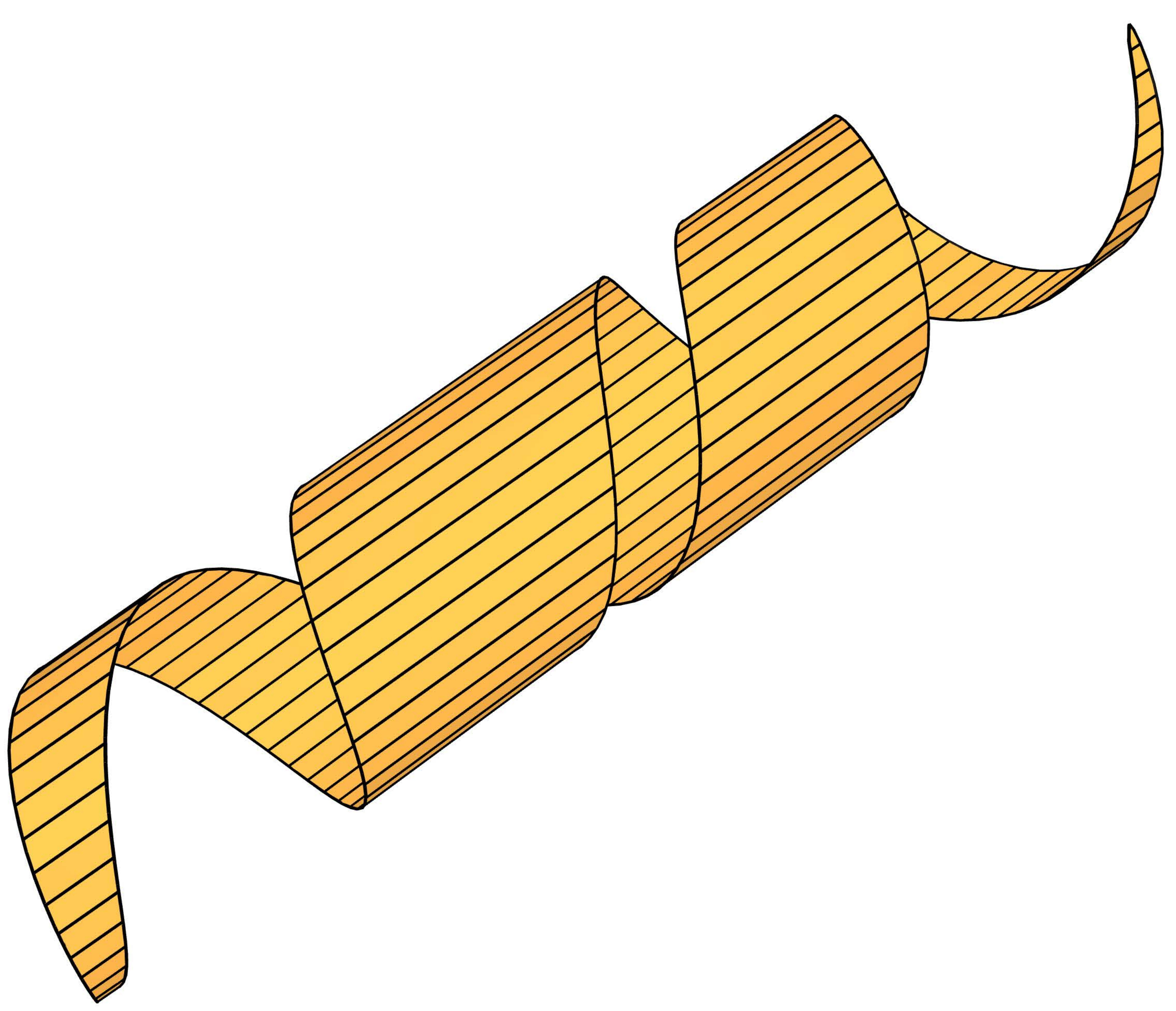}\\
		\small uniformly tessellated input &
		\small output, $p=2.58\%$
	\end{tabular}
	\caption{Our result on a better tessellation of the input from \figref{fig:planarization} has maximal planarity error of $p = 2.58\%$, compared with $p=11.84\%$ initially.}
	\label{fig:schwartztri}
\end{figure}

\section{Results and discussion}
\label{sec:results}

We implemented our algorithm using \textsf{libigl}~\cite{libigl} and \textsf{Directional}~\cite{directional} on a machine with i7-8569U CPU and \unit[16]{GB} RAM. Our typical input mesh resolution is 1800 faces, and for this approximate input size the vector field design part of our method takes 4--5 seconds, of which the majority of the time is spent in the \texttt{ProjectCurlFree} step, i.e., solving the convex optimization \equref{eq:curlproj1}-\eqref{eq:curlproj}. We currently use \textsf{CVX} \cite{cvx}, but this part can be optimized for better speed.
Although we do not have a formal convergence guarantee for our alternating algorithm, we observe that it typically converges to our specified tolerance level within 10--20 iterations for vector fields without singularities and 40--50 iterations for shapes with planar parts that introduce singularities in the vector field. We also test our method on inputs of up to 160k faces, which does not cause problems for convergence. The parameterization part of our method takes approximately 10--15 seconds.

\begin{figure*}[t]
	\centering
	\includegraphics[height=0.18\textheight]{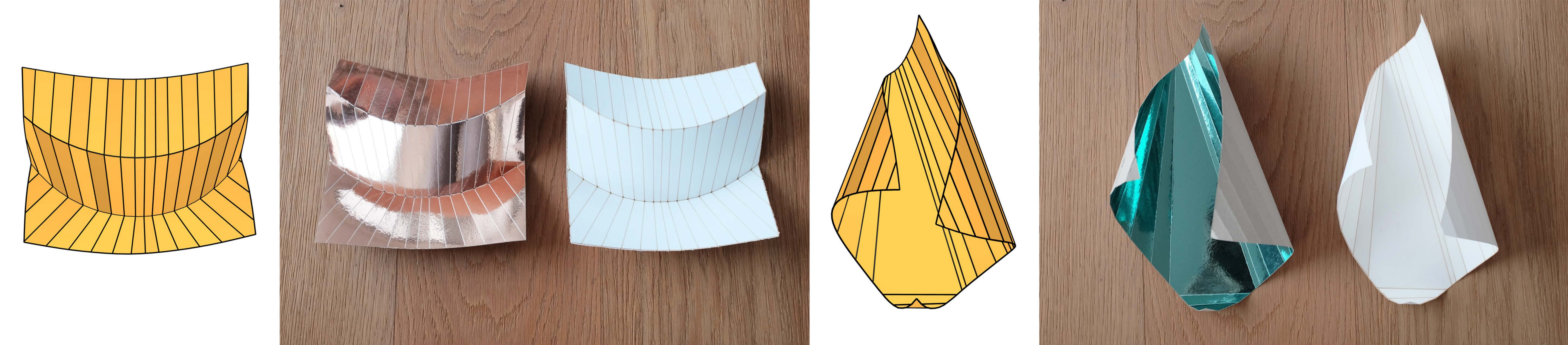}\\
	\caption{Our output meshes can be physically fabricated from planar sheets of stiff material. For this experiment, we parameterize our output mesh to the plane and etch the flattened mesh edges into cardboard using a laser cutter. Appropriately bending the sheet of cardboard along the edges then gives a shape that matches our output.}
	\label{fig:fabrication}
\end{figure*}

A variety of our results can be seen in \figref{fig:galleryNewwide}. 
Note that our method preserves the input boundary vertices, and therefore our output faces are quadrilateral-like higher degree polygons, rather than actual quadrilaterals. Examples of our results with various boundary shapes and non-disk topologies are included in \figref{fig:galleryNewwide}. 
Our method is applicable to developables with curved folds, as seen in \figref{fig:teaser}, \ref{fig:teaser2} and \ref{fig:curvedfolds} (input models from \cite{Rabinovich:CurvedFolds:2019}) as well as in \figref{fig:curvedfoldingKilian}. It can handle piecewise developable shapes, such as D-forms (shapes obtained by gluing together two planar domains with the same perimeter) from \cite{CBP_Pottmann_Isometry:2020} and sphericons from \cite{pottmann_new}, see \figref{fig:sphericonsdforms}, as well as other shapes with creases from \cite{pottmann_new}, see \figref{fig:piecewisedev}. 
Works by Stein et al.~\shortcite{stein_dev}, Sell\'{a}n et al.~\shortcite{heightfield_dev}, Ion et al.~\shortcite{Ion:ApproximatingDOGs:2020} and others produce piecewise developable approximations, but they are not PQ meshes. Our work can be used to remesh those surfaces, see \figref{fig:Keenan} for an example.
We successfully apply our method to glued constructions from \cite{CBP_Pottmann_Isometry:2020}, including point singularities, see \figref{fig:glued}. We have physically fabricated some of our results, shown in \figref{fig:fabrication}.
Table~\ref{tab:planarity} lists the most important statistics about our results.

\paragraph{Developable surface editing with dynamic connectivity} To demonstrate the utility of our approach, we use the point handle-based editing system of \cite{rabi18} to interactively deform an input discrete orthogonal geodesic net (DOG) and create a sequence of a few developable surfaces, on which we run our algorithm after trivial triangulation. See \figref{fig:editing-sequence} and the accompanying video for some examples of such editing sessions. Note the natural change in the combinatorics that our algorithm induces to model exact developability, which can change considerably even for small deformations in the input.

\paragraph{Planarity evaluation.}
Since our output meshes have no interior vertices inside the developable patches, the ultimate accuracy measure for the developability of our results is the planarity of the mesh faces. We measure planarity of each quadrilateral face by the ratio of the distance between the diagonals to their average length, in percent~\cite{liu:conicalmeshes:2006}. For higher-degree polygons, we compute the root-mean-square (RMS) error of all quads constructed from every 4 consecutive vertices in the polygon. An acceptable stringent tolerance for the planarity error is $\leq 1\%$ \cite{libhedra}. It is generally not expected for parameterization based methods to achieve planarity to more than first order, so that usually further planarization post-processing is needed. We show the raw maximum and mean planarity error values of our results without any post-processing in Table~\ref{tab:planarity}. Even though our output meshes are quite coarse, our planarity errors are typically very low without such planarity optimization, and very close to the tolerance. Our worst maximum planarity error is obtained on a mesh with very thin features (\figref{fig:galleryNewwide}, $5^\text{th}$ row, middle column), and the output quad with this maximal planarity error is towards the end of the spiral. As can be seen in \figref{fig:galleryNewwide} and \figref{fig:schwartztri}, the input triangulation is very coarse there. We planarize this example, which has the worst maximum planarity error ($p=11.84\%$) to zero planarity ($p=0.0034\%$) using ShapeUp~\cite{shape_up}, and reach a visually highly similar result, see \figref{fig:planarization}. This demonstrates the capability of our algorithm to utilize the information in the original mesh effectively. Interestingly, the triangulation of the input mesh is similar to a Schwartz lantern, and remeshing this input to a more uniform triangulation already drastically brings down the maximal planarity error of our algorithm to $p=2.58\%$, see \figref{fig:schwartztri}.

\begin{figure}[t]
	\setlength{\tabcolsep}{0pt}
	\begin{tabular}{p{0.25\linewidth}p{0.25\linewidth}p{0.25\linewidth}p{0.25\linewidth}}
		\includegraphics[width=0.8\linewidth]{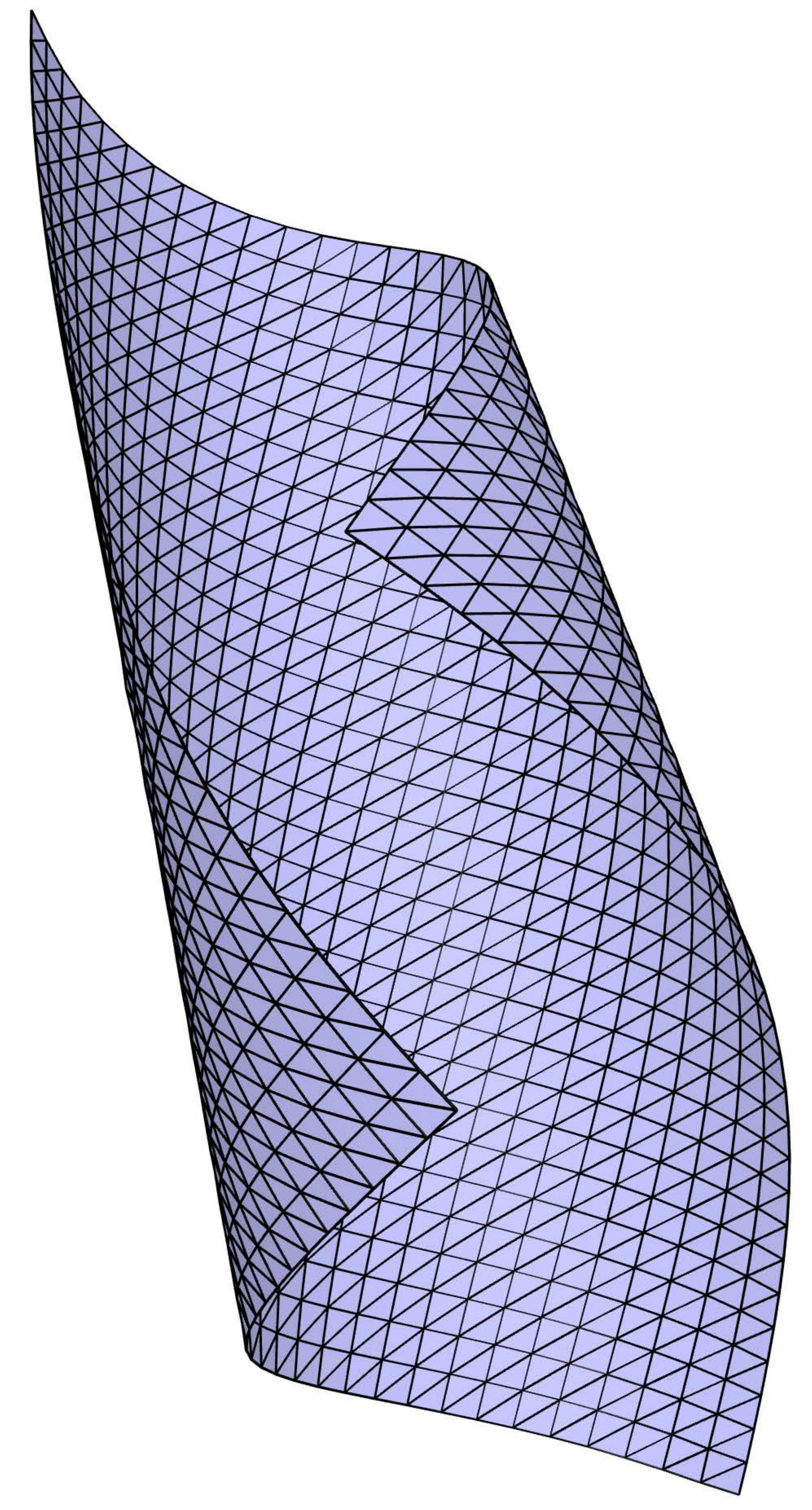}&
		\includegraphics[width=0.8\linewidth]{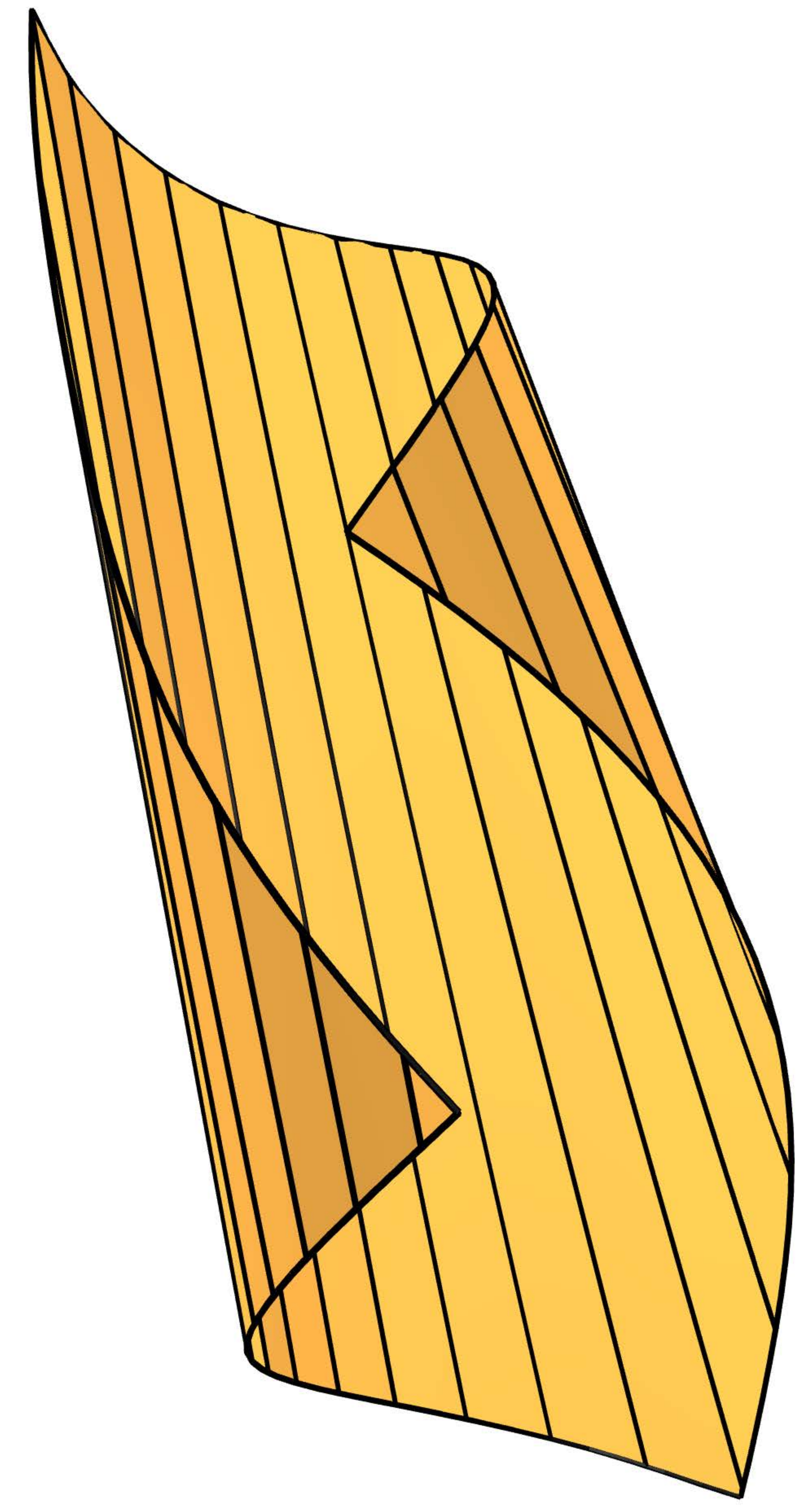}&
		\includegraphics[width=0.8\linewidth]{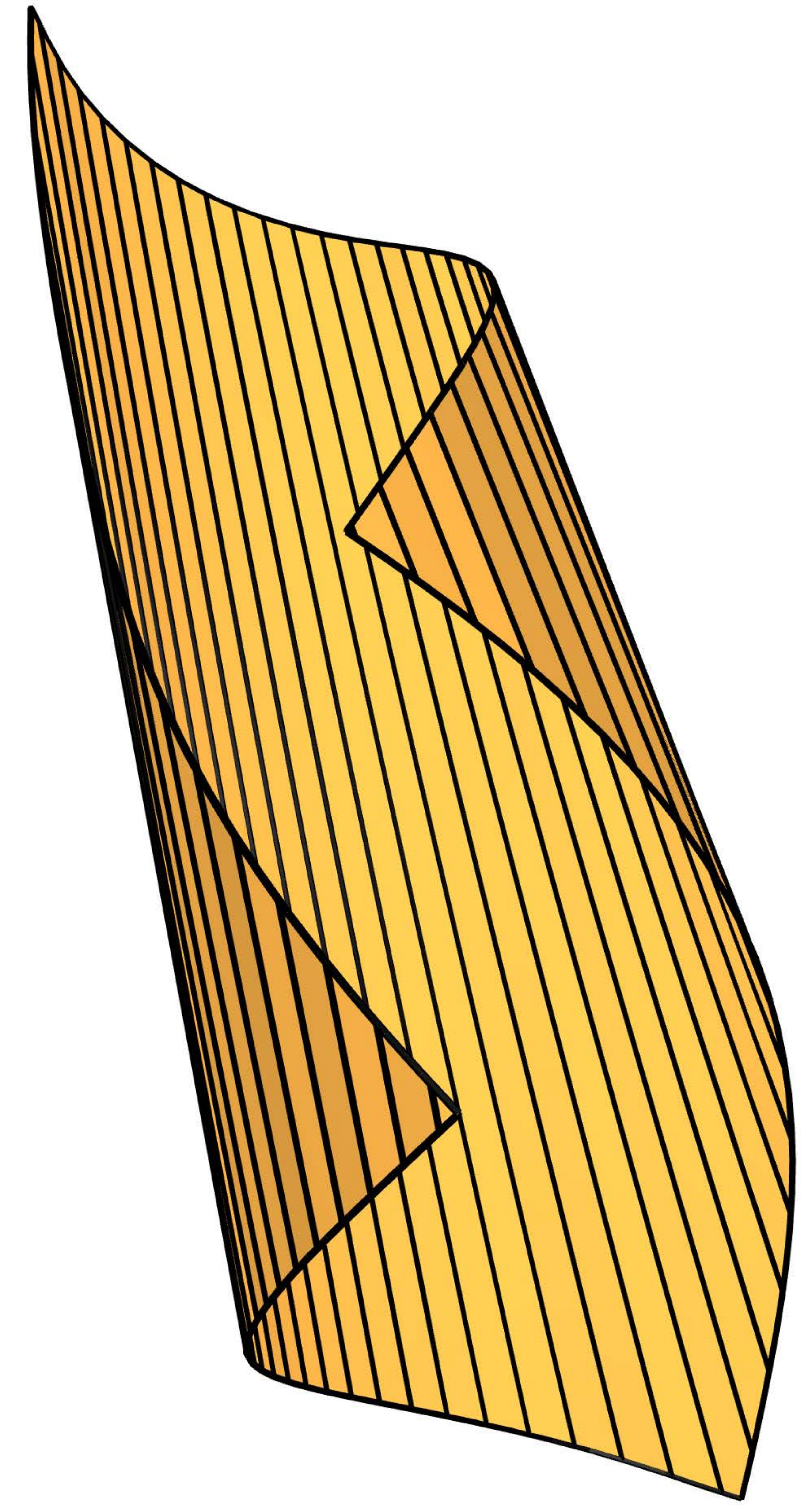}& 
		\includegraphics[width=0.8\linewidth]{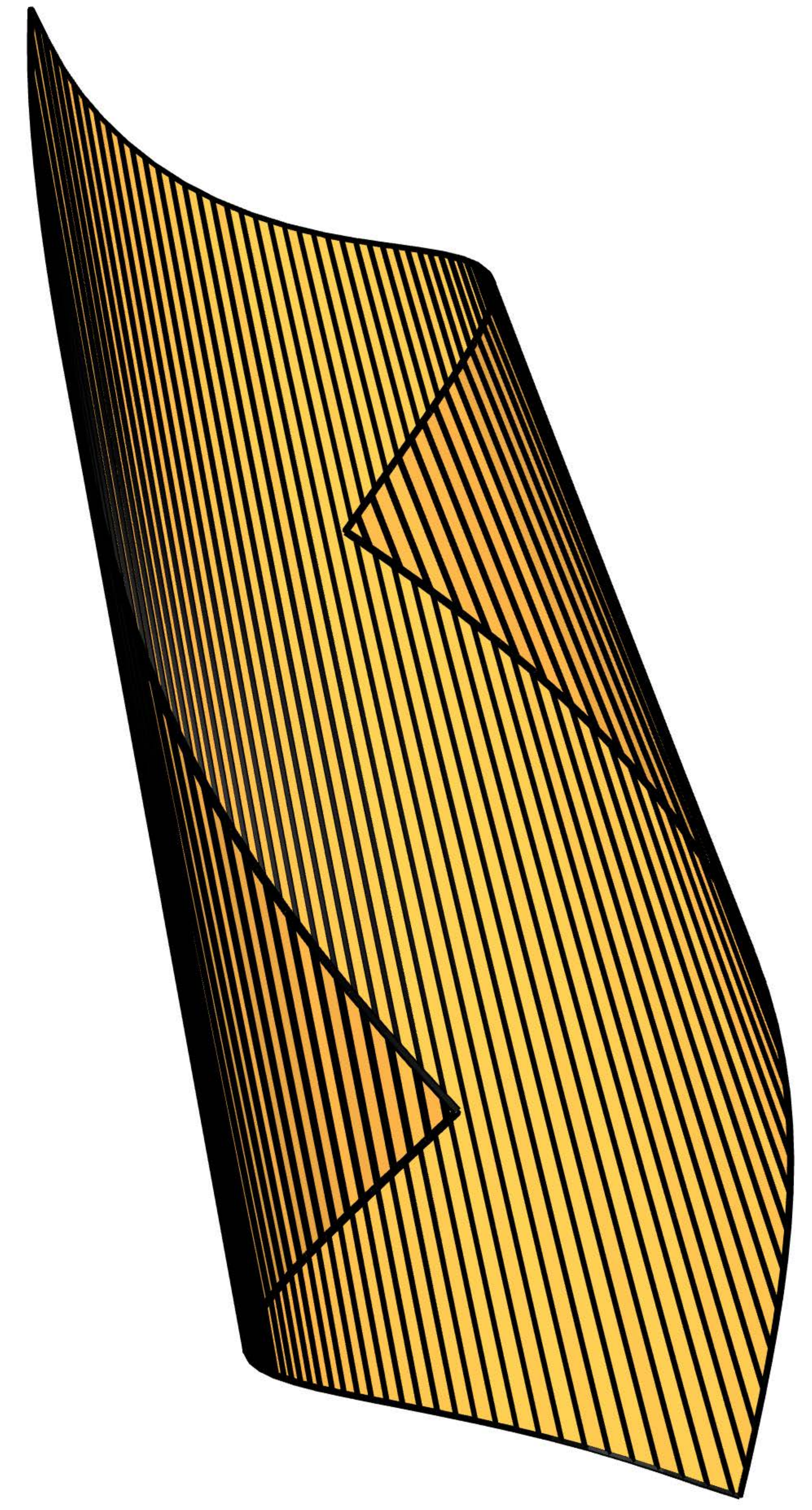}
		\\
		\multicolumn{1}{c}{\small input $\set{M}$} &
		\small $|\set{F}'|=23$ &
		\small $|\set{F}'|=46$ &
		\small $|\set{F}'|=90$ \\
		&
		\small $h = 0.47\%$ &
		\small $h = 0.41\%$ &
		\small $h = 0.41\%$ \\
		&
		\small $p_\text{max} = 0.52\%$ &
		\small $p_\text{max} = 0.54\%$ &
		\small $p_\text{max} = 0.32\%$ \\
		&
		\small $p_\text{mean} = 0.24\%$ &
		\small $p_\text{mean} = 0.16\%$ &
		\small $p_\text{mean} = 0.10\%$ 
	\end{tabular}
	\caption{Sampling the level sets of our optimized function $u$ with increasing density leads to finer remeshing of the input mesh, where the Hausdorff distance to the input $h$, as well as the maximal and mean polygon planarity error, $p_\text{max}$ and $p_\text{mean}$, decrease. The output resolution is denoted by the number of faces $|\set{F}'|$. The Hausdorff distance is reported relative to the bounding box diagonal.}
	\label{fig:different-resolutions}
\end{figure}

\paragraph{Effect of output resolution} 
We vary the number of isovalues and extract varying amounts of level sets of $u$ to create output meshes of different resolutions; we then measure their planarity and approximation quality w.r.t.\ the input mesh in terms of Hausdorff distance, see \figref{fig:different-resolutions}. We note that the approximation quality and the planarity improve with higher resolution, although even for the coarsest resolution these metrics are already below tolerance.

\begin{figure}[t]
	\setlength{\tabcolsep}{2pt}
	\begin{tabular}{cc}
		\includegraphics[width=0.48\columnwidth]{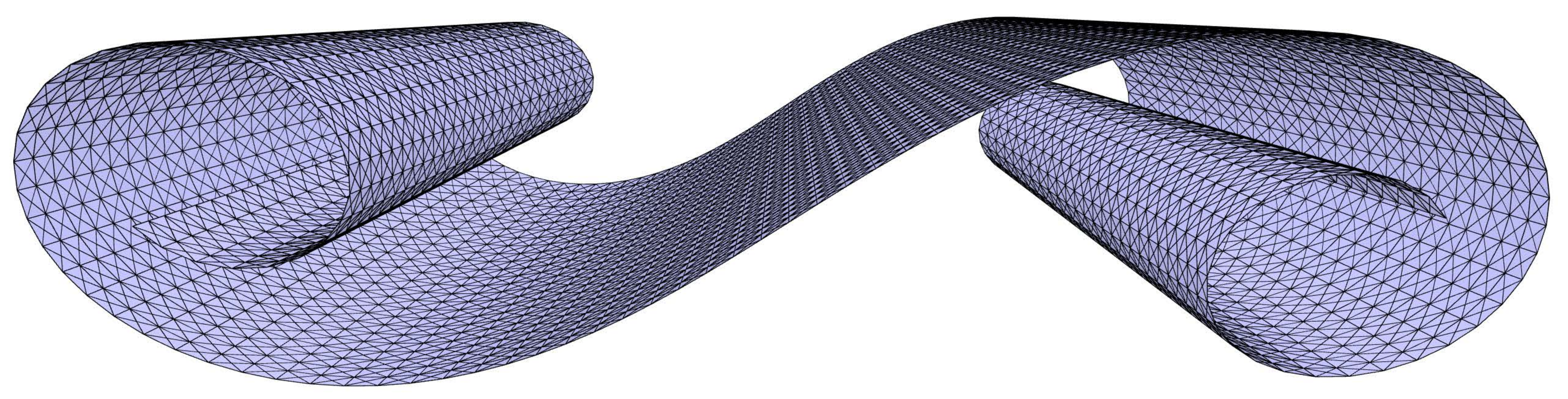}&
		\includegraphics[width=0.48\columnwidth]{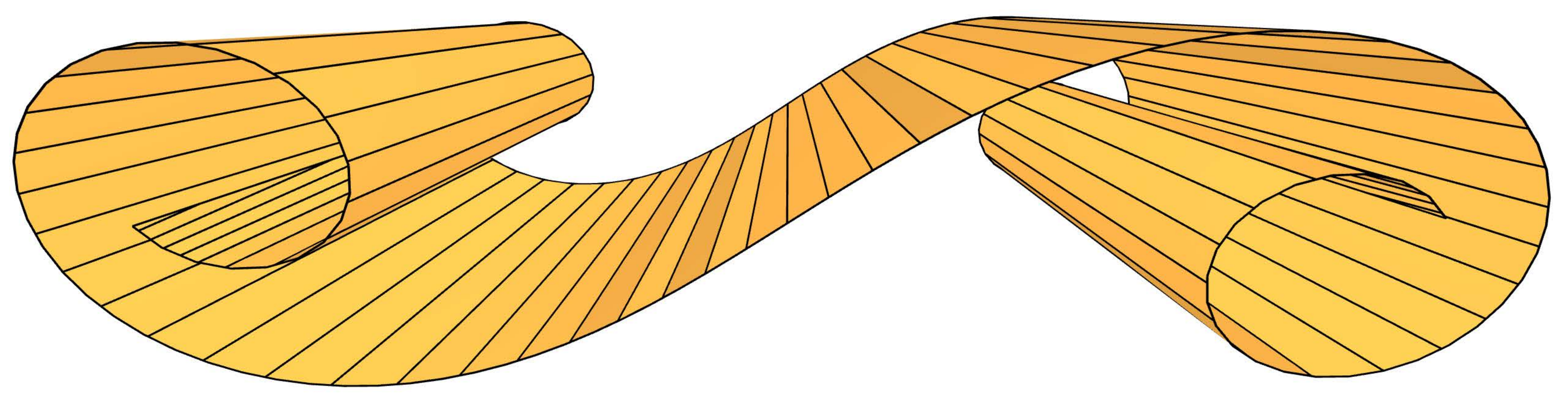}\\  
		\small $|\set{F}| = 10\text{k}$&
		\small $|\set{F}'| = 68$, $h= 0.31\%$\\
		&
		\small $p_\text{max}=1.08\%$, \small $p_\text{mean}=0.34\%$\\                
		\includegraphics[width=0.48\columnwidth]{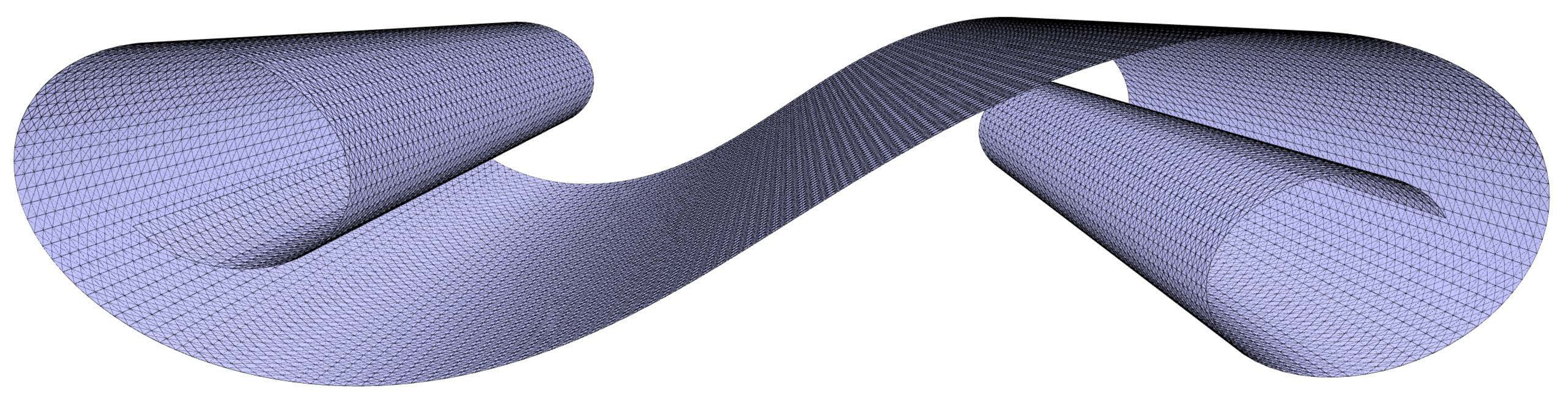}&
		\includegraphics[width=0.48\columnwidth]{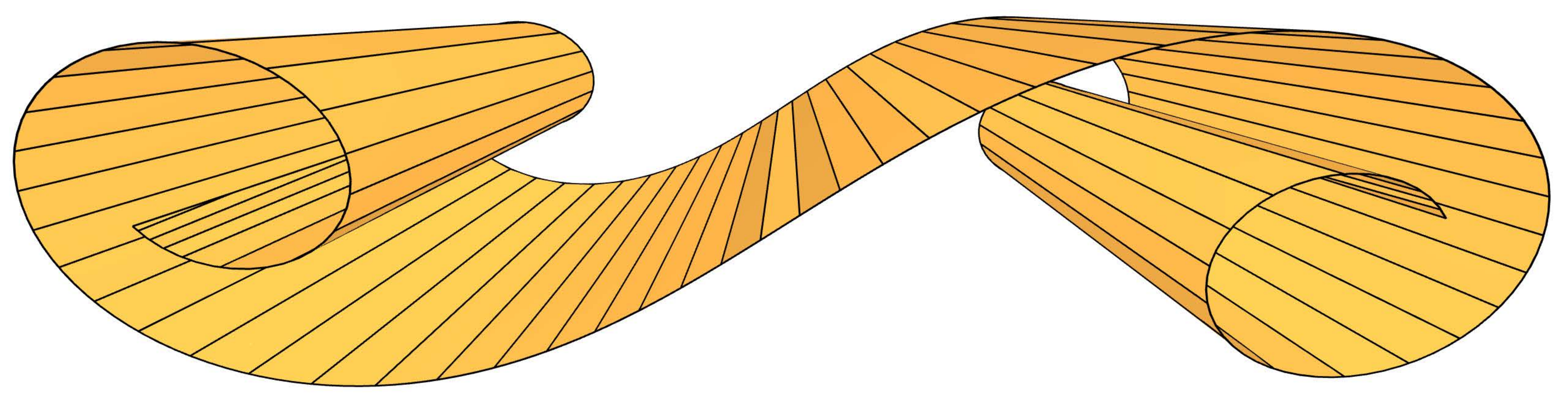}\\    
		\small $|\set{F}| = 40\text{k}$&
		\small $|\set{F}'| = 67$, $h= 0.26\%$\\  
		&
		\small $p_\text{max}=1.22\%$, $p_\text{mean}=0.10\%$\\                    
		\includegraphics[width=0.48\columnwidth]{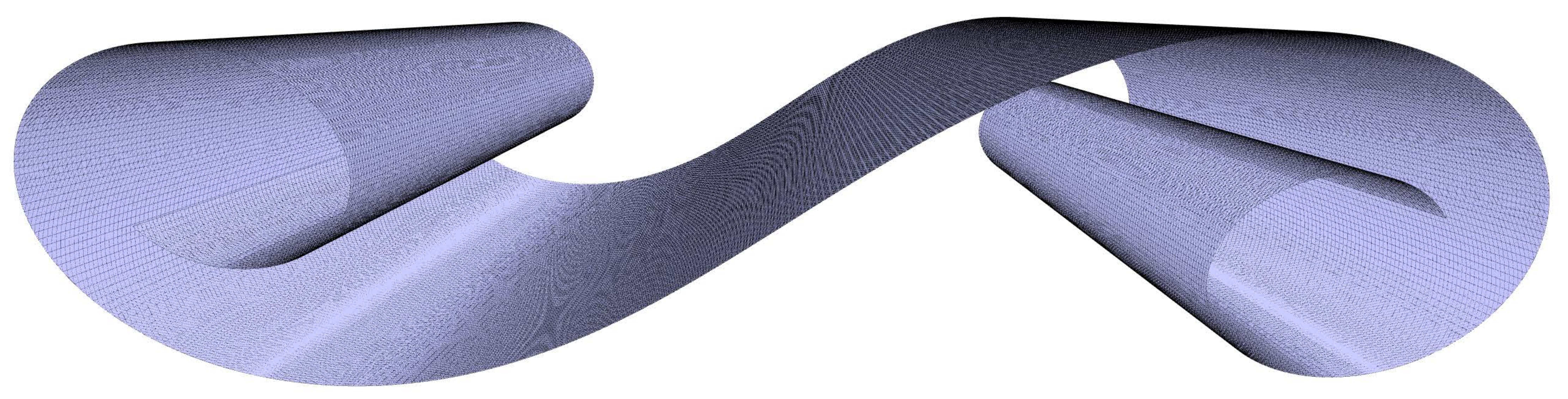}&
		\includegraphics[width=0.48\columnwidth]{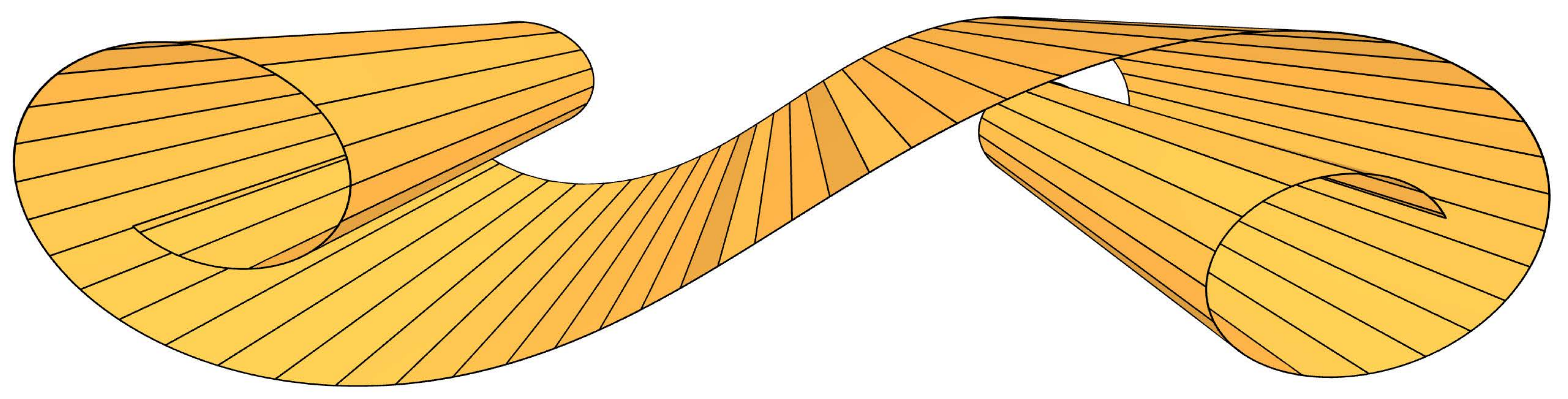}\\     
		\small $|\set{F}| = 160\text{k}$&
		\small $|\set{F}'| = 66$, $h= 0.24\%$\\
		&
		\small $p_\text{max}=0.25\%$, $p_\text{mean}=0.02\%$                 
	\end{tabular}
	\caption{As the input resolution $|\set{F}|$ of a sampled analytical developable surface increases, the approximation accuracy and the planarity of our remeshed result increase. The meshing direction also aligns better with the mesh boundaries that coincide with analytical ruling directions in this case as the resolution increases.}
	\label{fig:analytical-comparison}
\end{figure}

\paragraph{Comparison with analytical principal-curvature directions}
We test our method on an input mesh sampled from an analytical clothoid surface with varying resolution and compare the obtained vector field $\gamma$ with the analytical max curvature directions, see \figref{fig:analytical-comparison} and Table \ref{tab:anglediffanalytic}. Note that the input to our method are \emph{numerically estimated} ruling directions $r$, not their analytical values. As the data shows, upon refinement of the input mesh, our output field converges towards the analytical solution. 

\begin{table}[t]
	\caption{{Difference between our optimized vector field and the analytical principal curvature directions on the clothoid mesh shown in \figref{fig:analytical-comparison} (angular difference reported in degrees).}}
	\label{tab:anglediffanalytic}
	\centering
	\setlength{\tabcolsep}{4.6pt}
	\begin{tabular}{lll}
		\toprule
		\small $|\set{F}|$ \phantom{AAAA} & 
		\small \phantom{AA} max $\degree$ &
		\small mean  $\degree$\\
		\midrule
		\small 10k & 
		\small \phantom{AA} 9.49 &
		\small 2.24
		\\
		\small 40k & 
		\small \phantom{AA} 4.80 &
		\small 1.19
		\\
		\small 160k & 
		\small \phantom{AA} 2.31 &
		\small 0.52
		\\  
		\bottomrule
	\end{tabular}
\end{table}

\begin{figure}[t]
	\setlength{\tabcolsep}{2pt}
	\begin{tabular}{cc}
		\includegraphics[width=0.48\columnwidth]{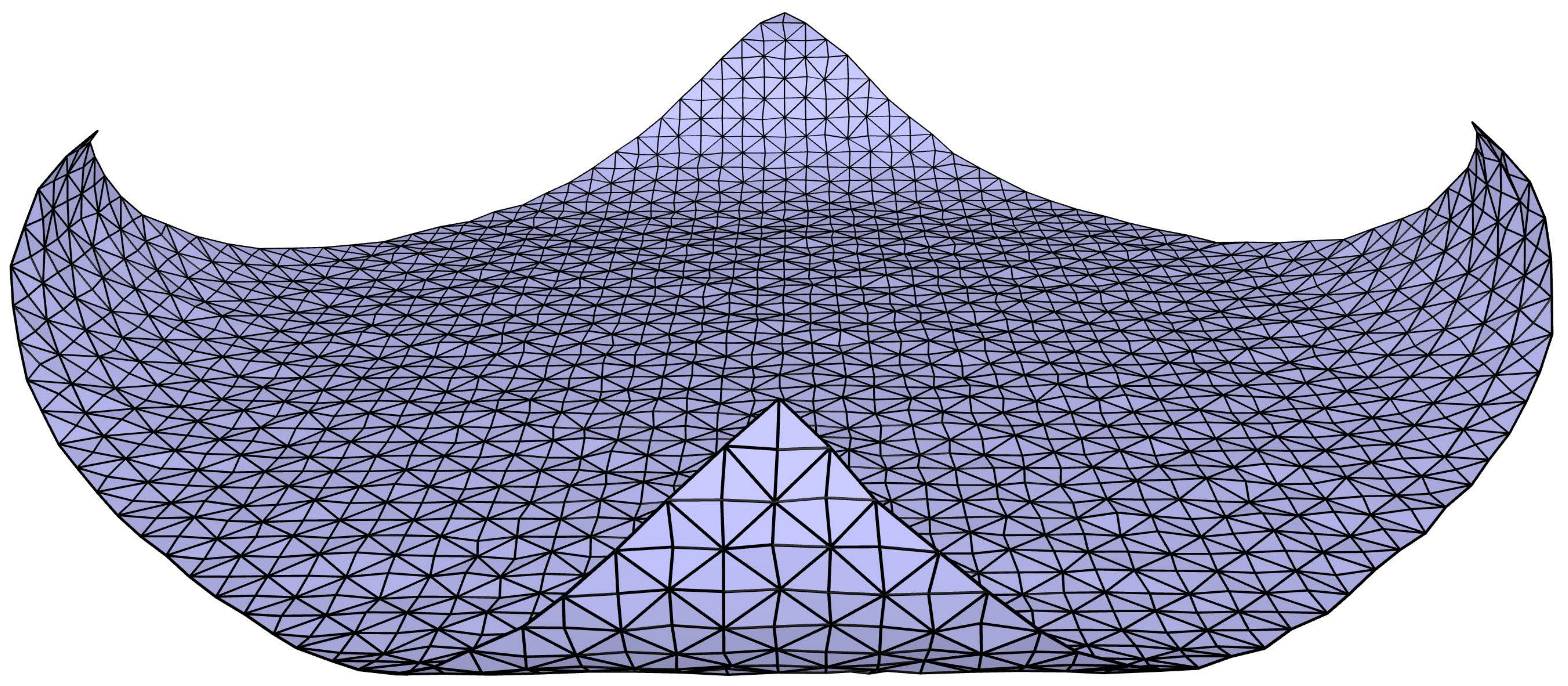}&
		\includegraphics[width=0.48\columnwidth]{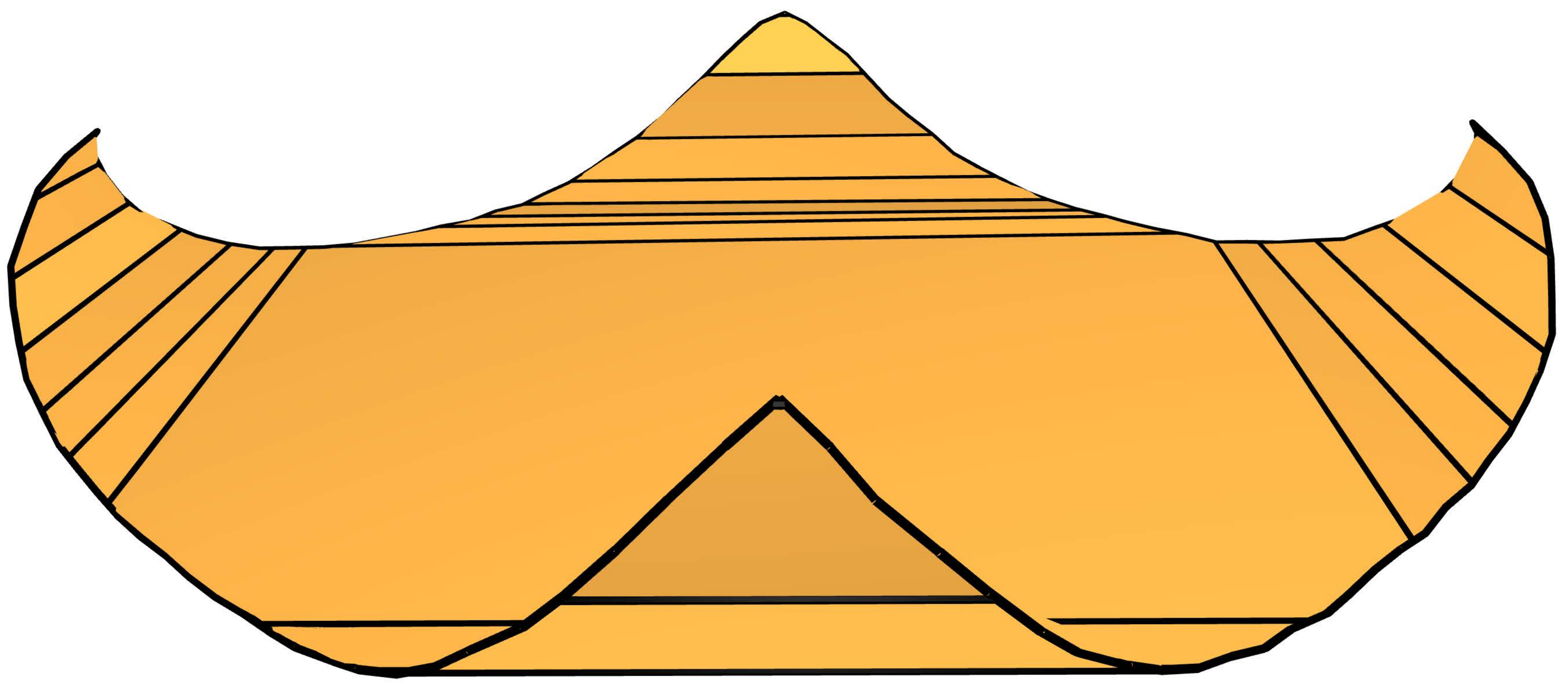}\\                  
		\includegraphics[width=0.48\columnwidth]{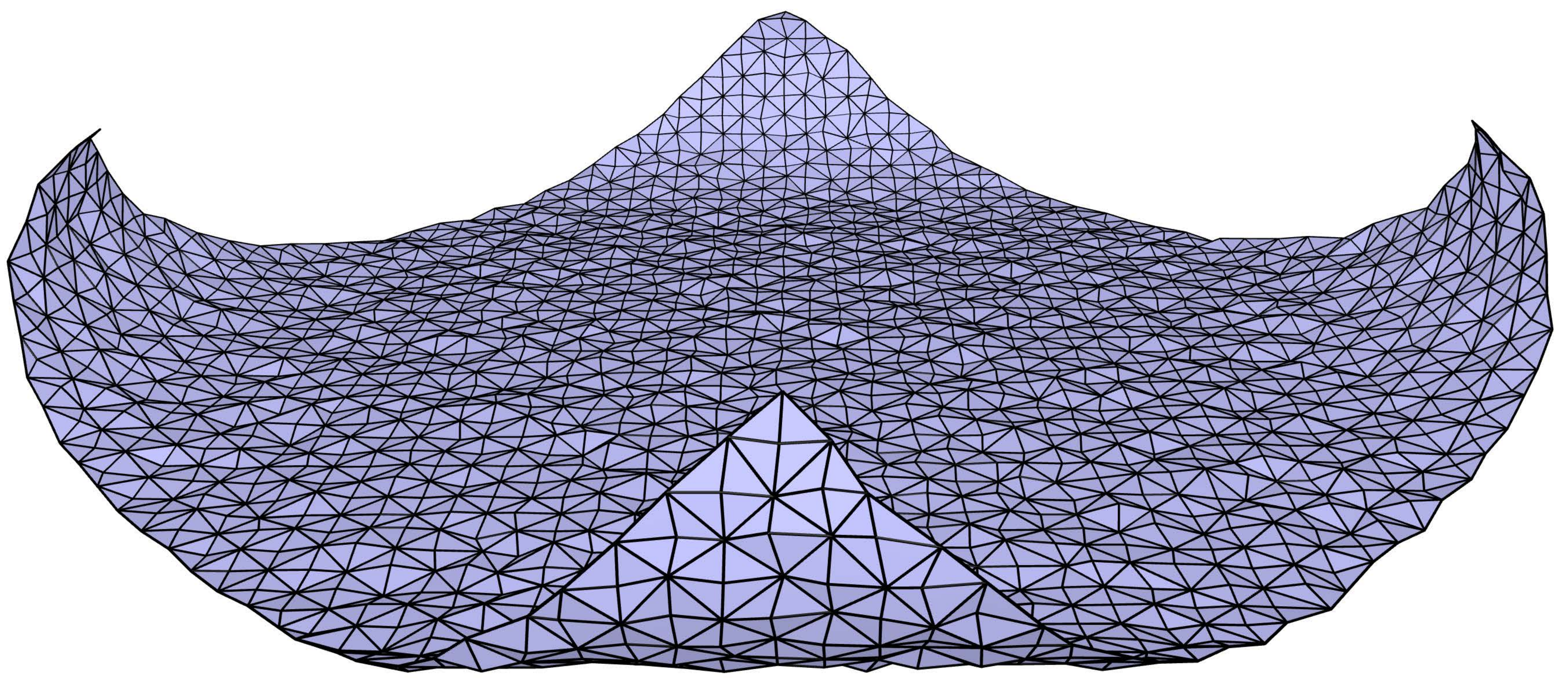}&
		\includegraphics[width=0.48\columnwidth]{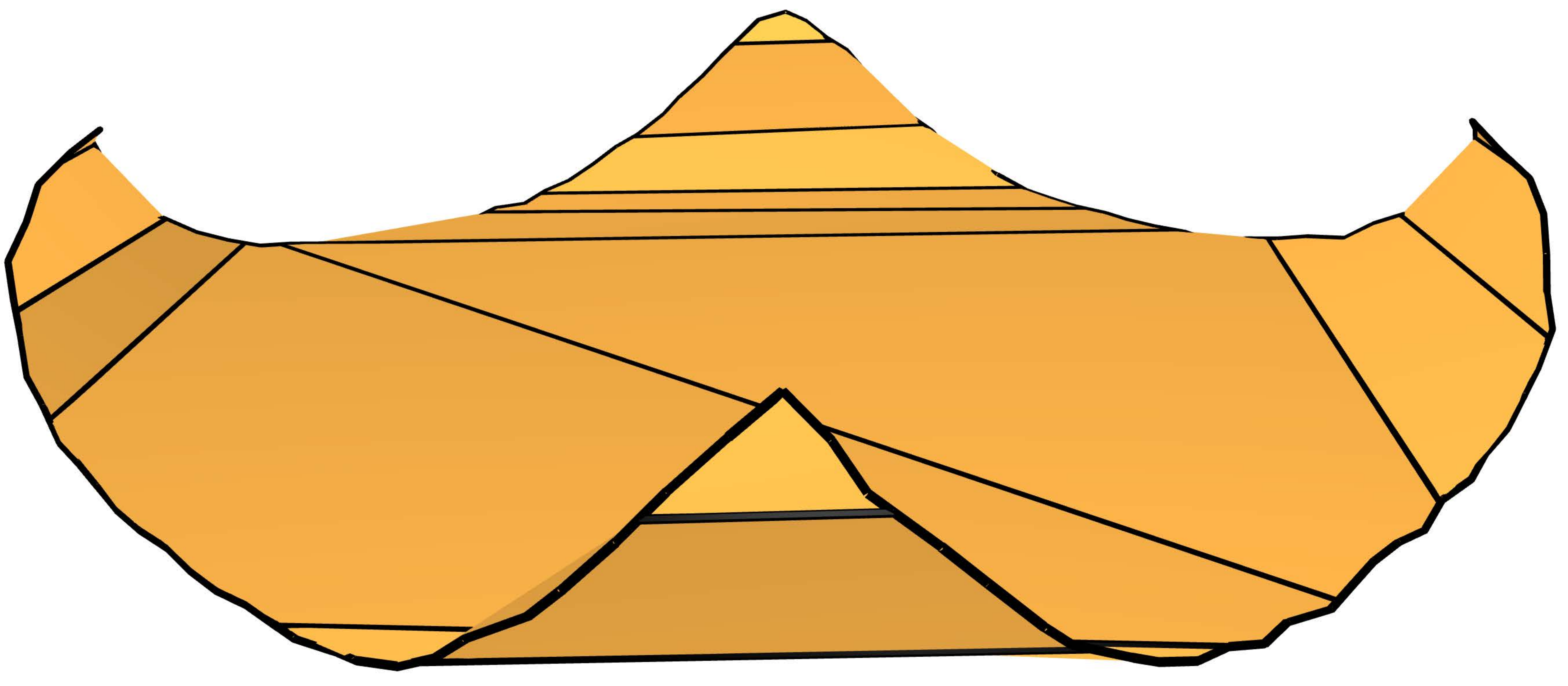}\\                  
	\end{tabular}
	\caption{Our method is robust to noise on developable inputs. These examples show our third example from \figref{fig:triangulation-direction}, but with random vertex displacements applied. Left: a displacement of maximally 12.5\% of the average edge length is applied, right:  maximally 25\% of the average edge length. Our method still recovers a meshing that is compatible with the original principal directions.}
	\label{fig:noisyinput}
\end{figure}

\paragraph{Robustness}
We show that our method is robust with respect to the parameters $\omega_a$ and $\omega_s$. There is a range of values for these parameters that leads to visually very similar results. As the relative weight of $\omega_s$ with respect to $\omega_a$ increases, the vector field turns into a more constant field, reducing alignment quality of the final output mesh. For noisy inputs, as in \figref{fig:noisyinput}, our method does not converge with our standard parameter settings, or it converges but generates a vector field with a large amount of singularities. For these cases, simply increasing $\omega_s$ ensures that the optimization converges, although some small and noisy details may be lost (in \figref{fig:noisyinput} (right) we use $\omega_s = 0.15$). This shows that our method with the help of the smoothness term manages to recover a principally-aligned vector field even if the information from the input ruling directions is very weak.
For optimal alignment the value of $\omega_s$ should be chosen as small as possible; e.g., for the cone in the second to last row of \figref{fig:galleryNewwide} we use $\omega_s = 0.00005$ to emphasize better alignment near the boundary. 

\begin{figure}[t]
	\setlength{\tabcolsep}{2pt}
	\begin{tabular}{cc}
		\includegraphics[width=0.45\columnwidth]{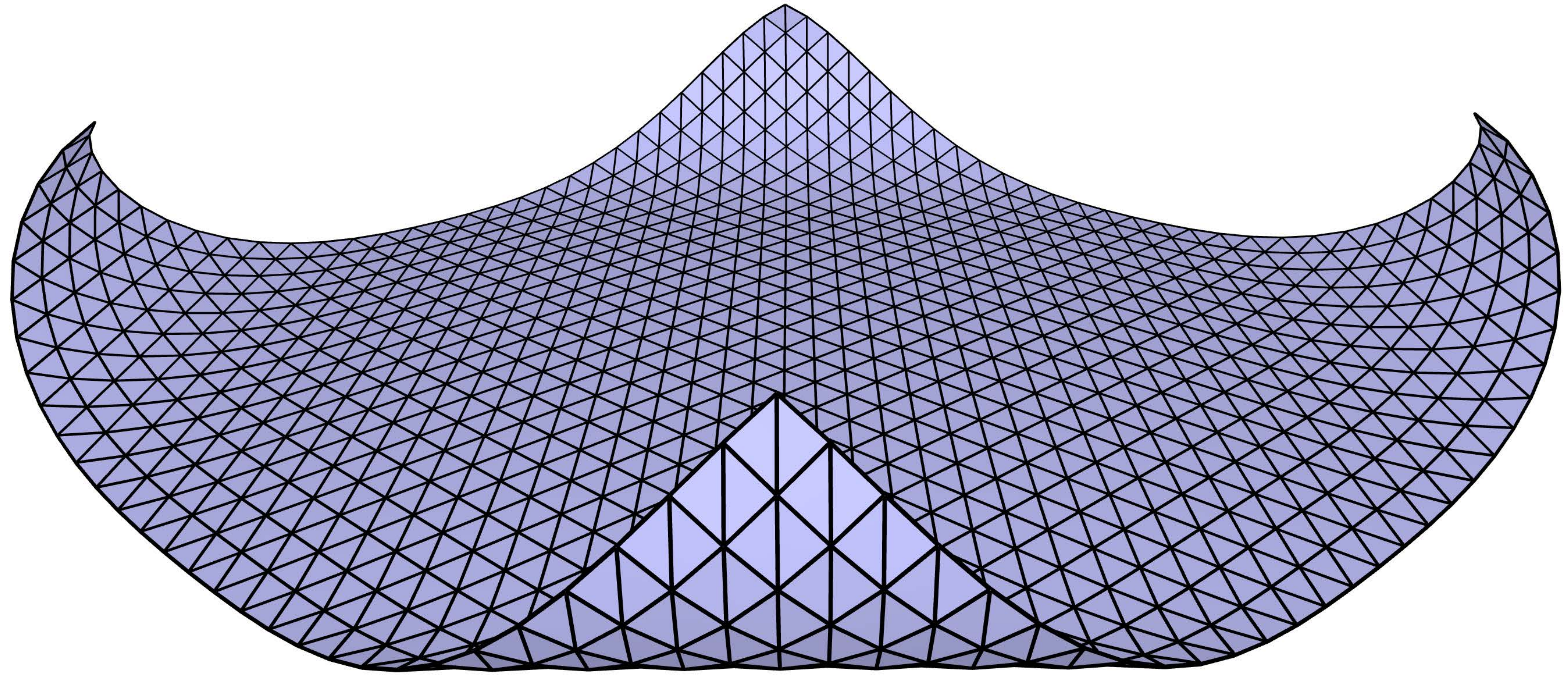}&
		\includegraphics[width=0.45\columnwidth]{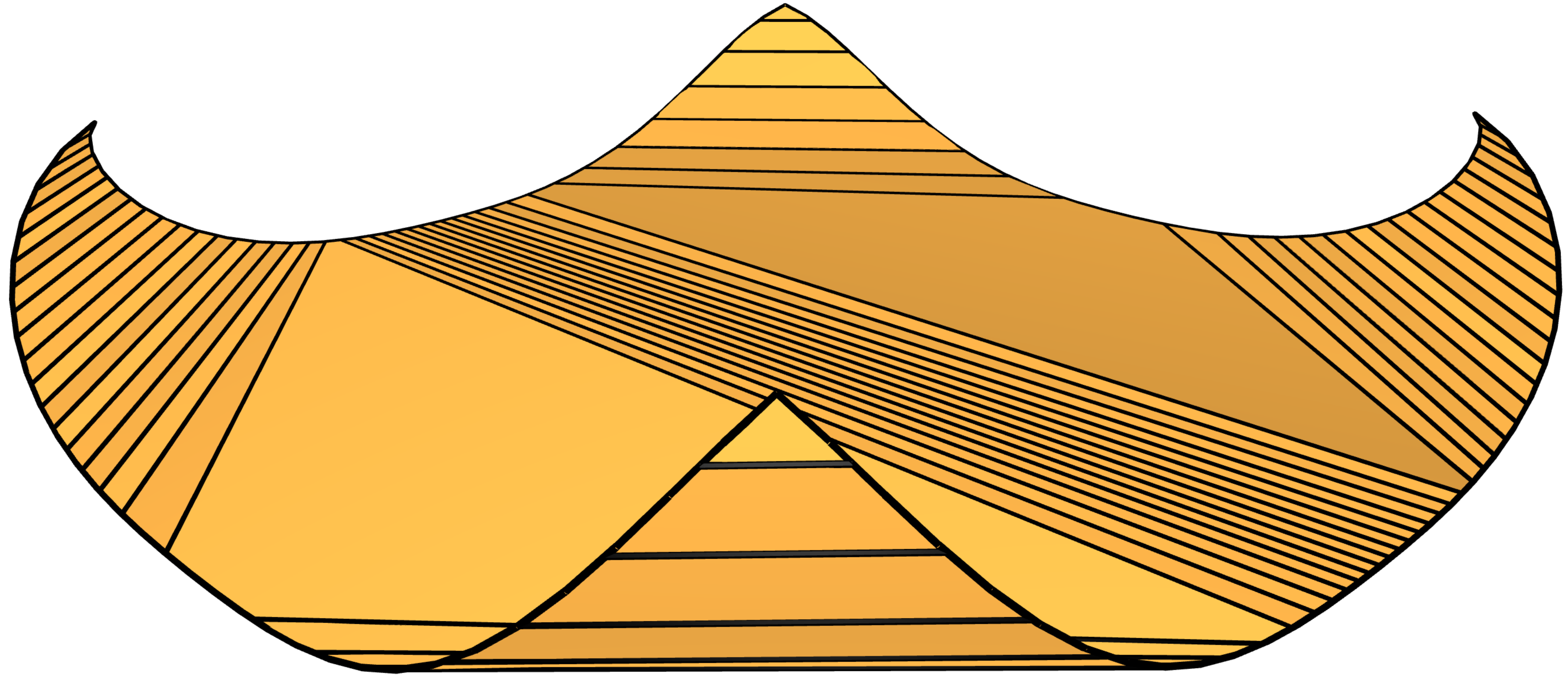}\\                  
		\includegraphics[width=0.45\columnwidth]{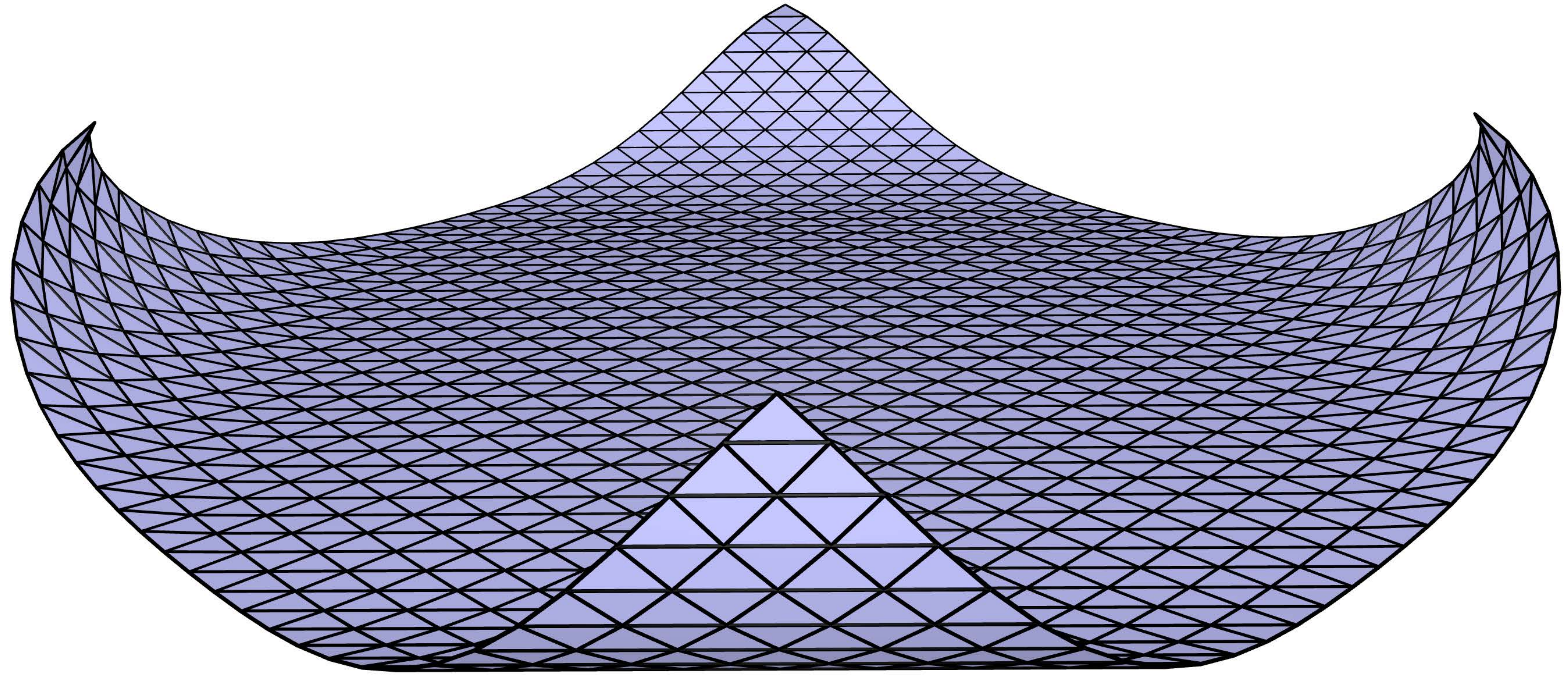}&
		\includegraphics[width=0.45\columnwidth]{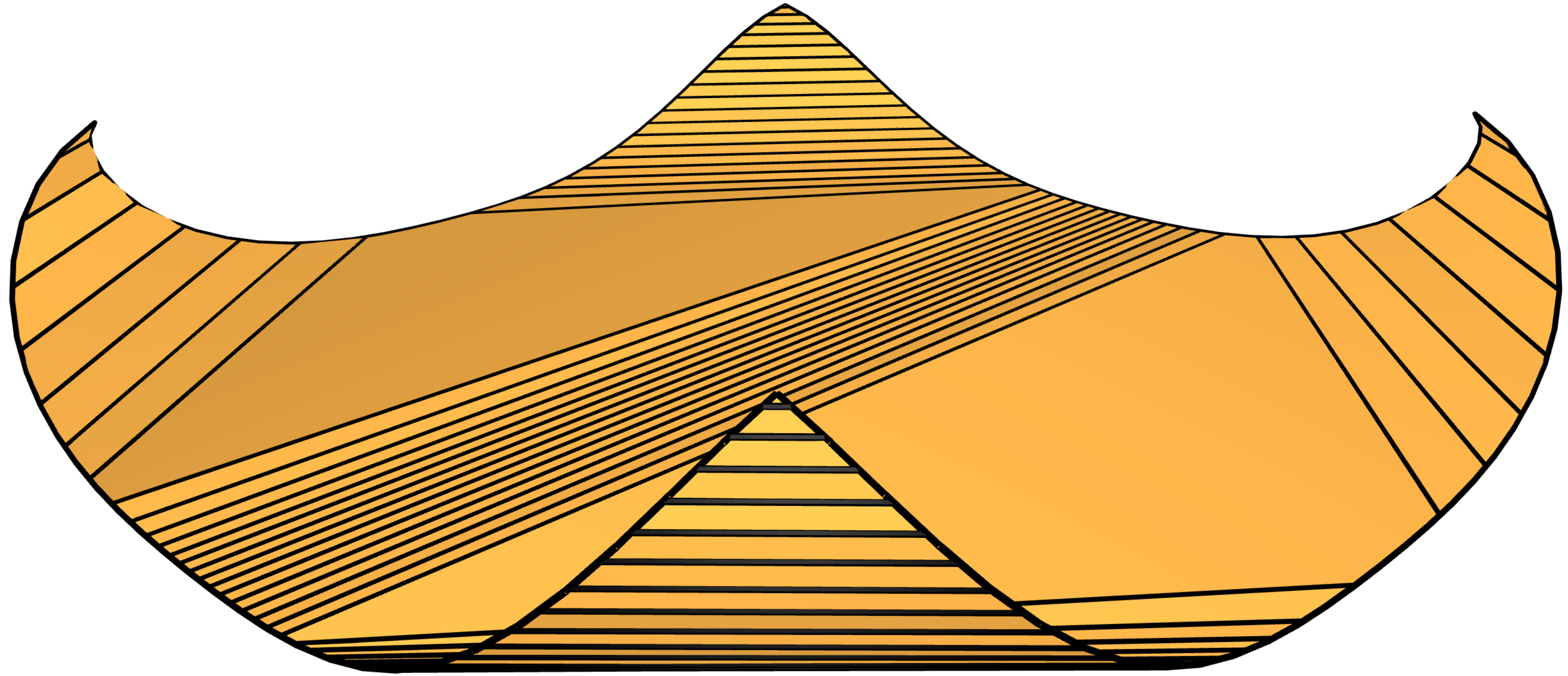}\\                   
		\includegraphics[width=0.45\columnwidth]{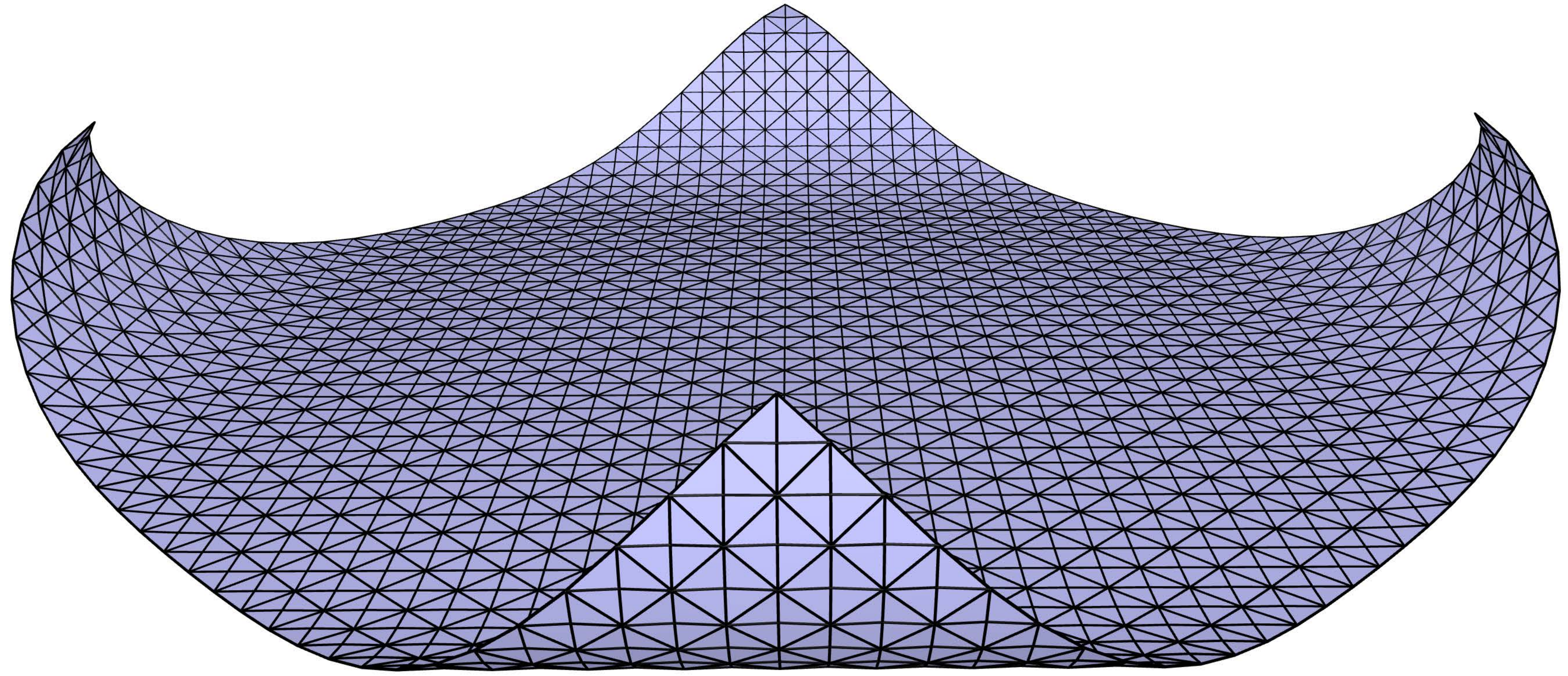}&
		\includegraphics[width=0.45\columnwidth]{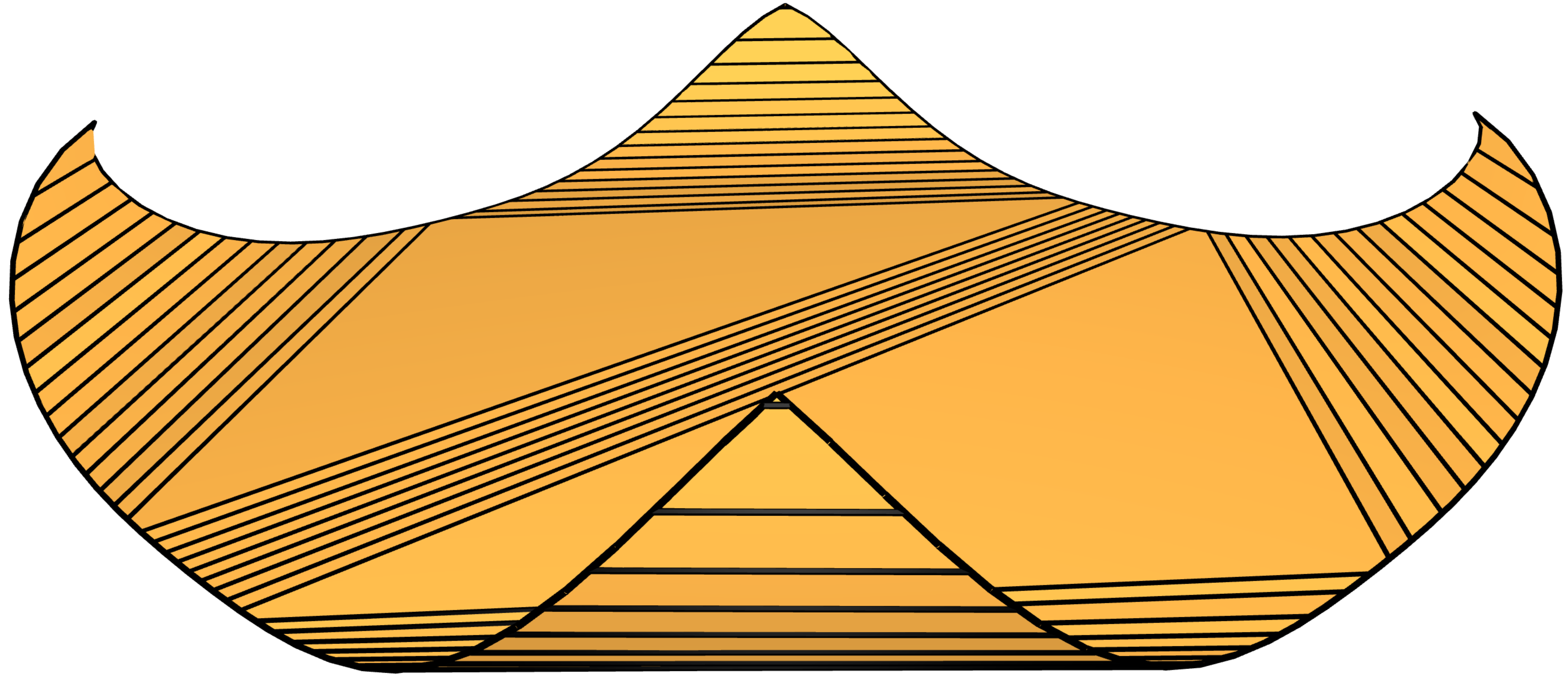}                  
	\end{tabular}
	\caption{Different triangulations of a quad mesh lead to different remeshing results, mainly in the near-planar regions. Nevertheless, all of the resulting meshing directions on the planar region are valid.}
	\label{fig:triangulation-direction}
\end{figure}

\begin{figure}[b]
	\setlength{\tabcolsep}{4pt}
	\begin{tabular}{cc}
		\includegraphics[width=0.46\columnwidth]{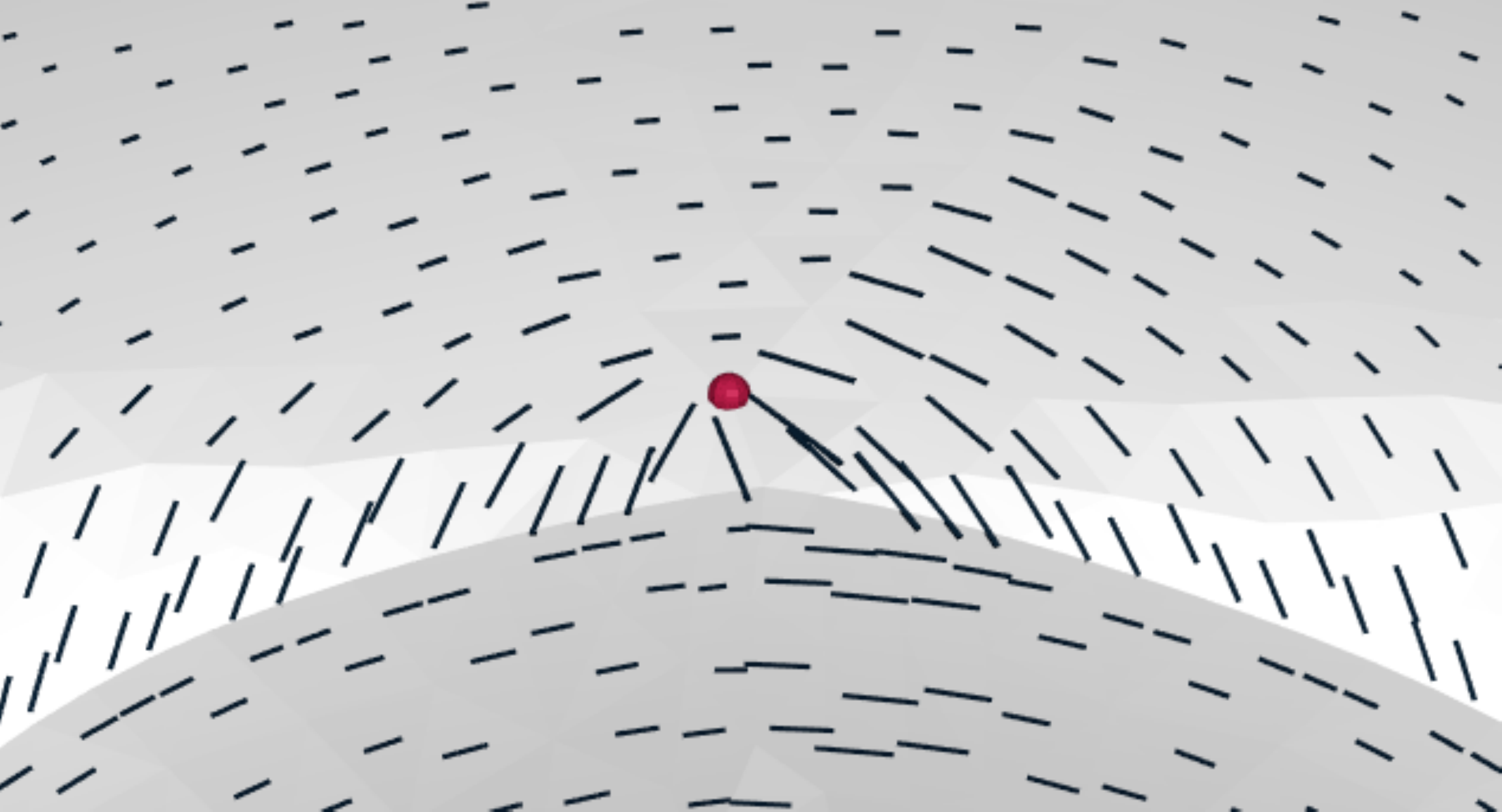}&
		\includegraphics[width=0.46\columnwidth]{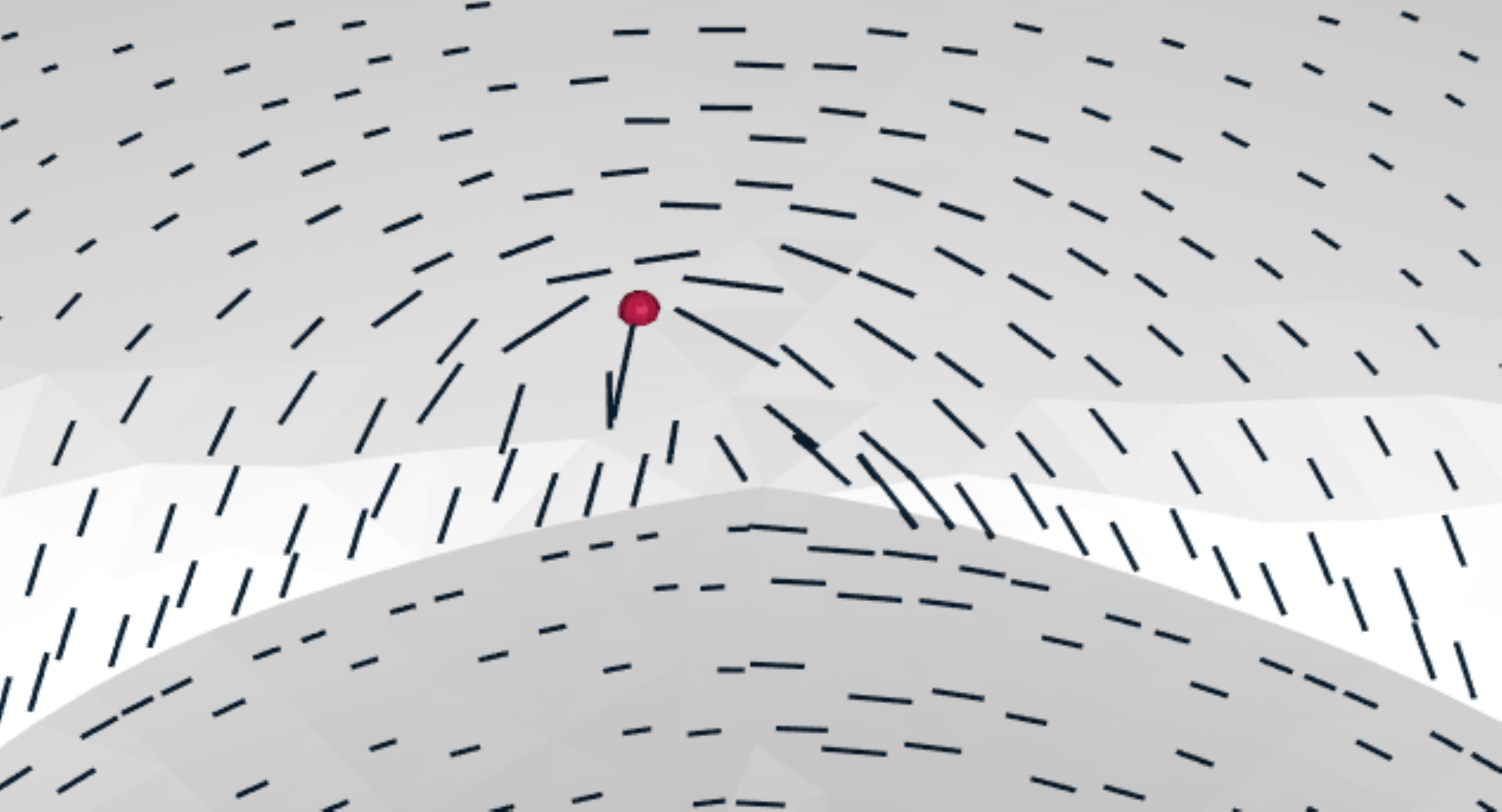}\\   
		\includegraphics[width=0.46\columnwidth]{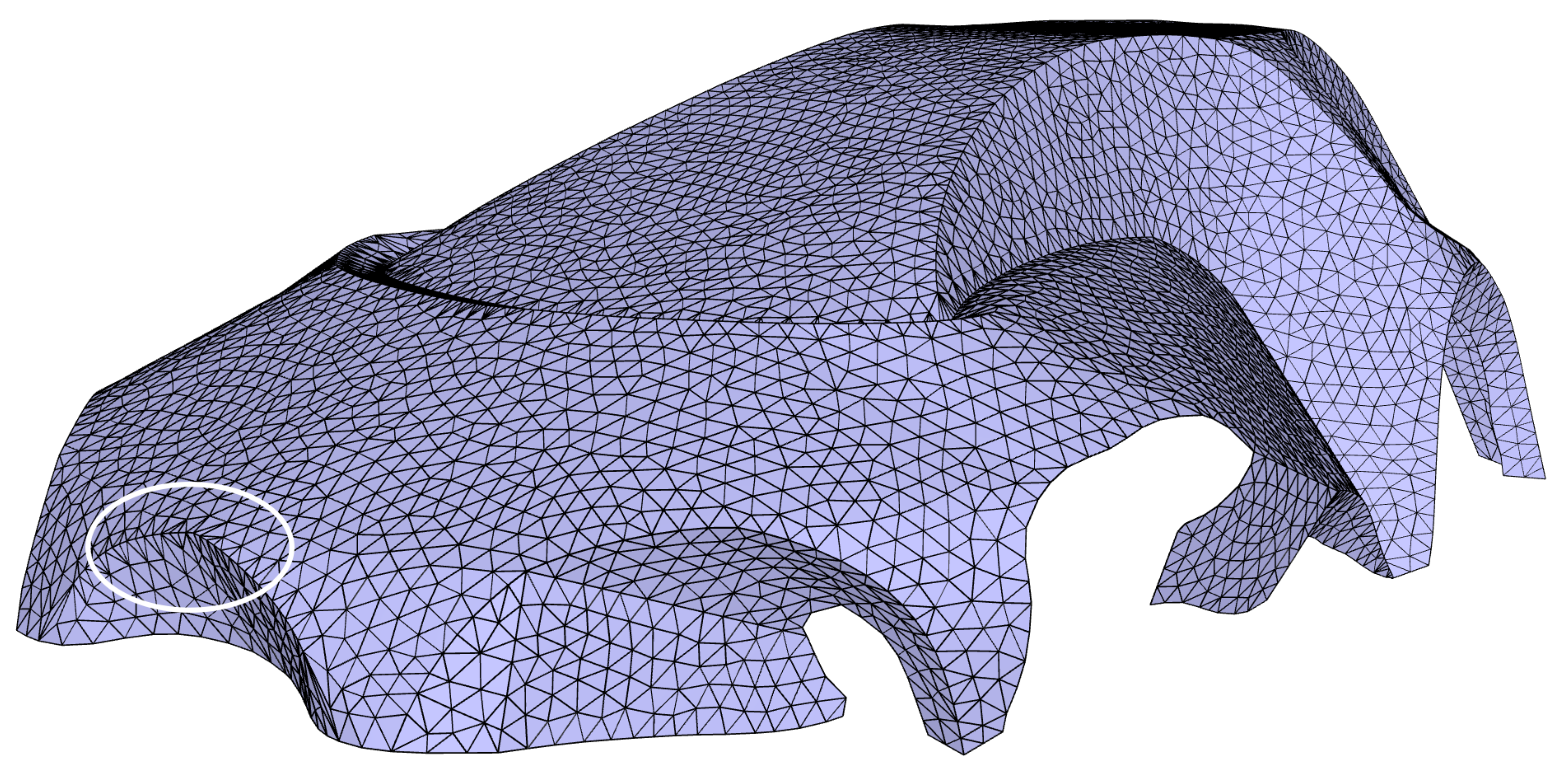} &
		\includegraphics[width=0.46\columnwidth]{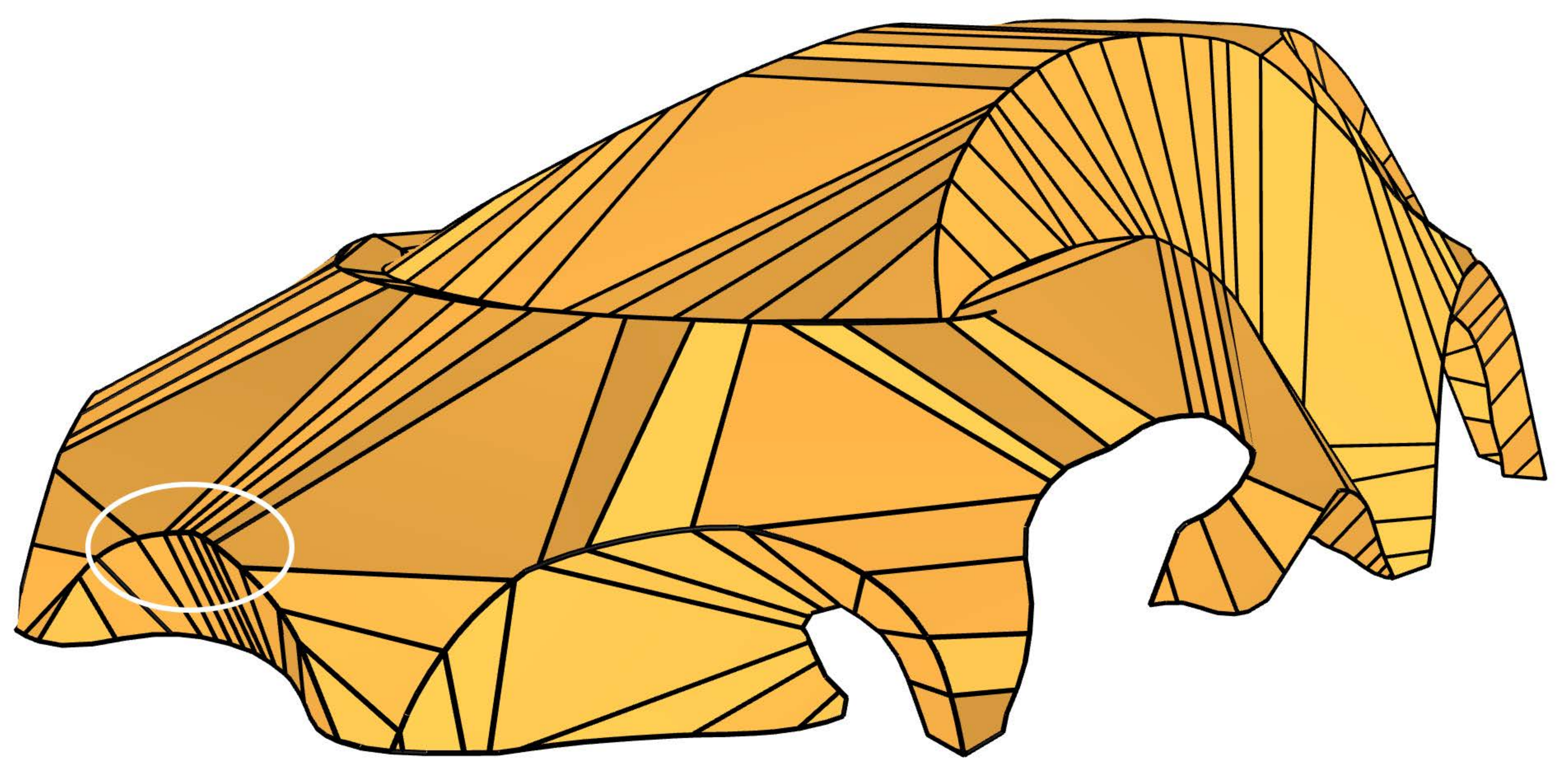} \\
		\multicolumn{2}{c}{\includegraphics[width=0.46\columnwidth]{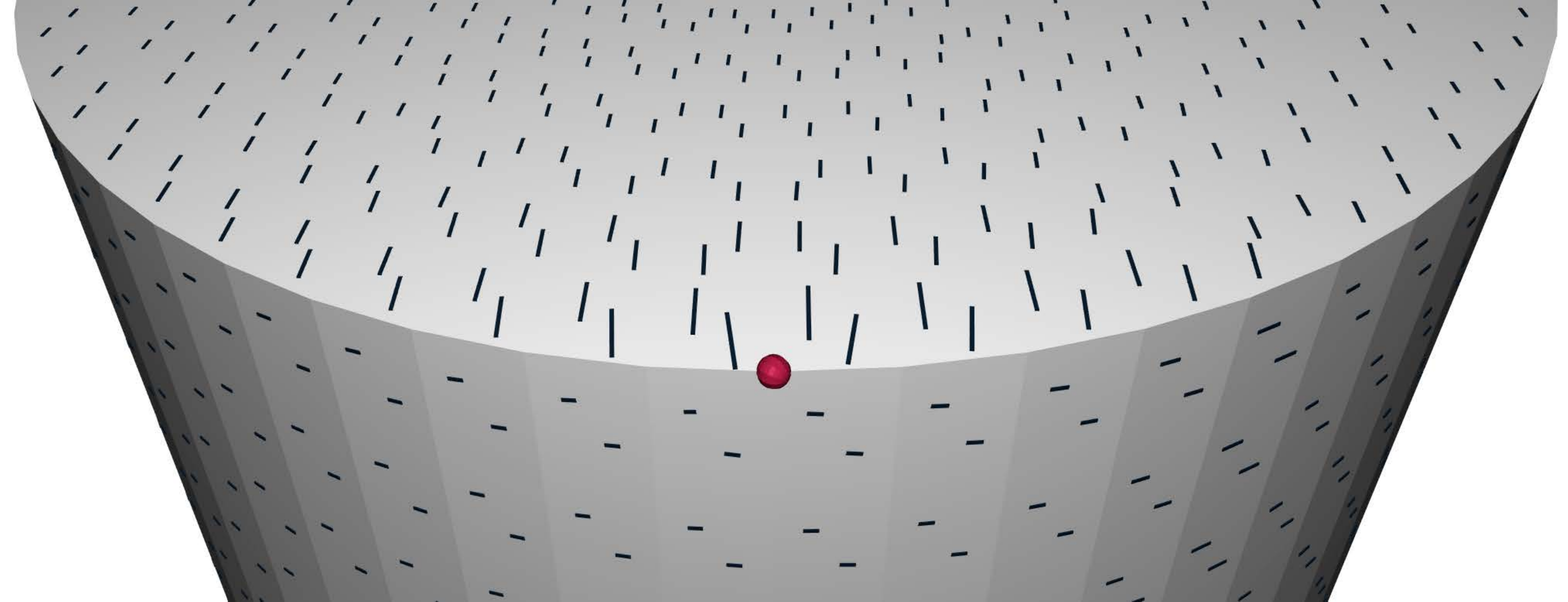}}
	\end{tabular}
	\caption{When a cone apex of non-trivial curvature is present in the input mesh, our method might get stuck oscillating between solutions with different singularity configurations. The top row shows our method struggling to put the singularity exactly on the crease and oscillating between two solutions for the circled part of the car model (courtesy of \cite{curved_folding_kilian}, designed by Gregory Epps). The middle row shows the input mesh and our obtained result. In comparison, the bottom row shows a successfully placed singularity on the crease of the cylinder model from \figref{fig:piecewisedev}.}
	\label{fig:oscillation}
\end{figure}

\paragraph{Limitations}
Our method is not entirely triangulation independent, as shown in \figref{fig:triangulation-direction}. If the input meshing is at odds with the principal curvature directions, this leads to poor ruling estimation and diminished performance of our algorithm in terms of planarity error. This is most noticeable near the corners of the given input, where there is relatively little data for our algorithm to align to. In order to minimize bias introduced by the triangulation, it is advisable to triangulate polygonal input meshes with higher valences by inserting a new vertex at the face center and connecting it to the vertices of the original polygon in a triangle fan.

Furthermore, as we treat curved folds in identical manner to creases (namely we assume they delineate a developable surface piece), there is no guarantee that the curved folded surface as a whole remains developable in one piece. An interesting direction for future work would be to incorporate known geometrical constraints at curved folds into our method. 

{If a cone apex is present in the mesh (see \secref{sec:developable_background}), the natural behaviour encouraged by our algorithm is to place a singularity on the crease to compensate for the curvature of the seam (see the cylinder example in~\figref{fig:piecewisedev} and \figref{fig:oscillation}, bottom). Nevertheless, our algorithm may fail to put the singularity exactly on the seam, depending on discretization. As a result, our optimization might get stuck, oscillating between nearby solutions with different singularity configurations (\figref{fig:oscillation}). Even though our method nominally fails to converge for the car example in \figref{fig:oscillation}, a reasonable result close to the expected one is still obtained.}

Our method may struggle with thin features, e.g., as part of a piecewise developable, as these can often provide no alignment information at all. In the future it would be interesting to see how the vector field on surrounding developable pieces can be used to add constraints to these thin features, since in the final meshing we wish to guarantee continuity throughout the pieces.
As shown in \figref{fig:analytical-comparison}, our output quality with respect to Hausdorff distance, as well as mean and maximal planarity error increases as the input resolution increases. Our method is therefore dependent on the input resolution but still performs well on low resolution inputs. 

Finally, we have no theoretical guarantees that the ruling-aligned edges in our output mesh do not intersect, although we never see this happen in our experiments. In torsal regions, the guiding field discourages overlapping behavior, as rulings on a developable are ordered. For planar regions, the guiding field contains more noise, but here the smoothness requirement for the vector field (and thus the corresponding parameterization) strongly discourages crossovers.

\begin{figure}
	\setlength{\tabcolsep}{2pt}   
	\begin{tabular}{cc}
		\includegraphics[width=0.43\columnwidth]{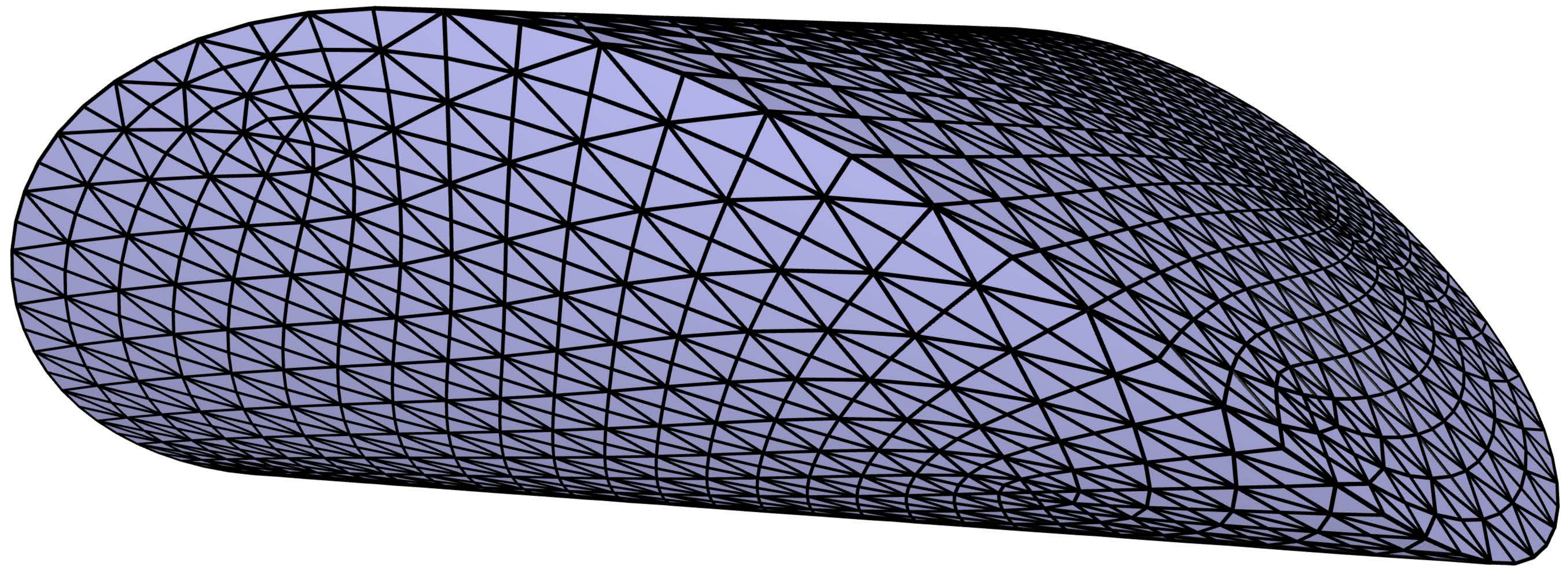} &
		\includegraphics[width=0.43\columnwidth]{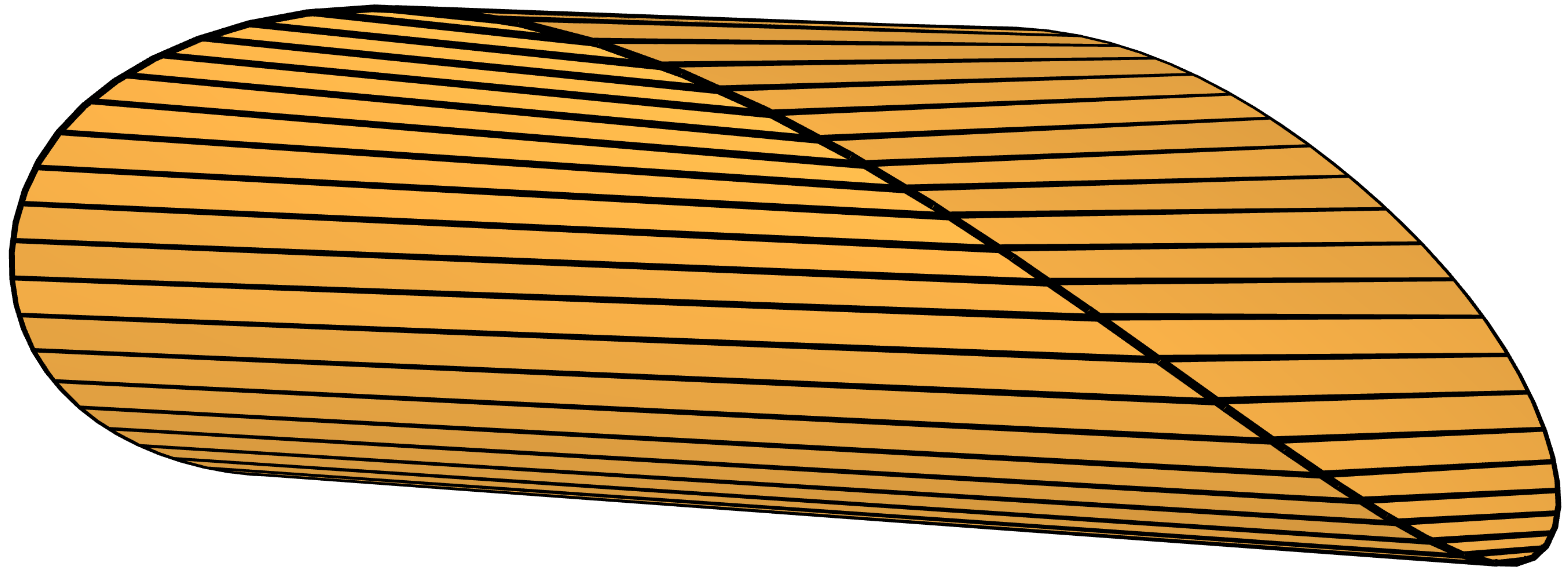}\\
		\includegraphics[width=0.43\columnwidth]{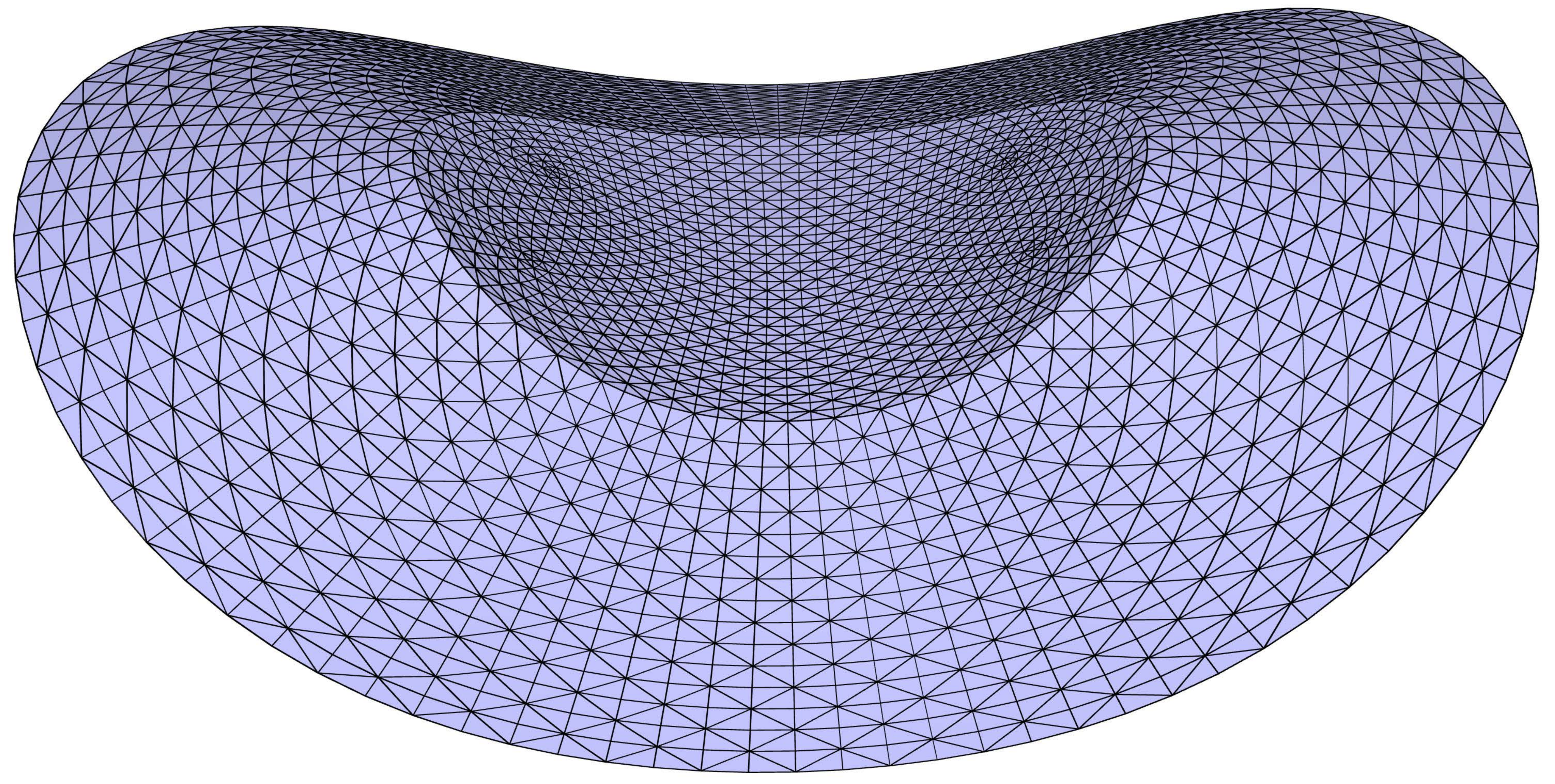}&
		\includegraphics[width=0.43\columnwidth]{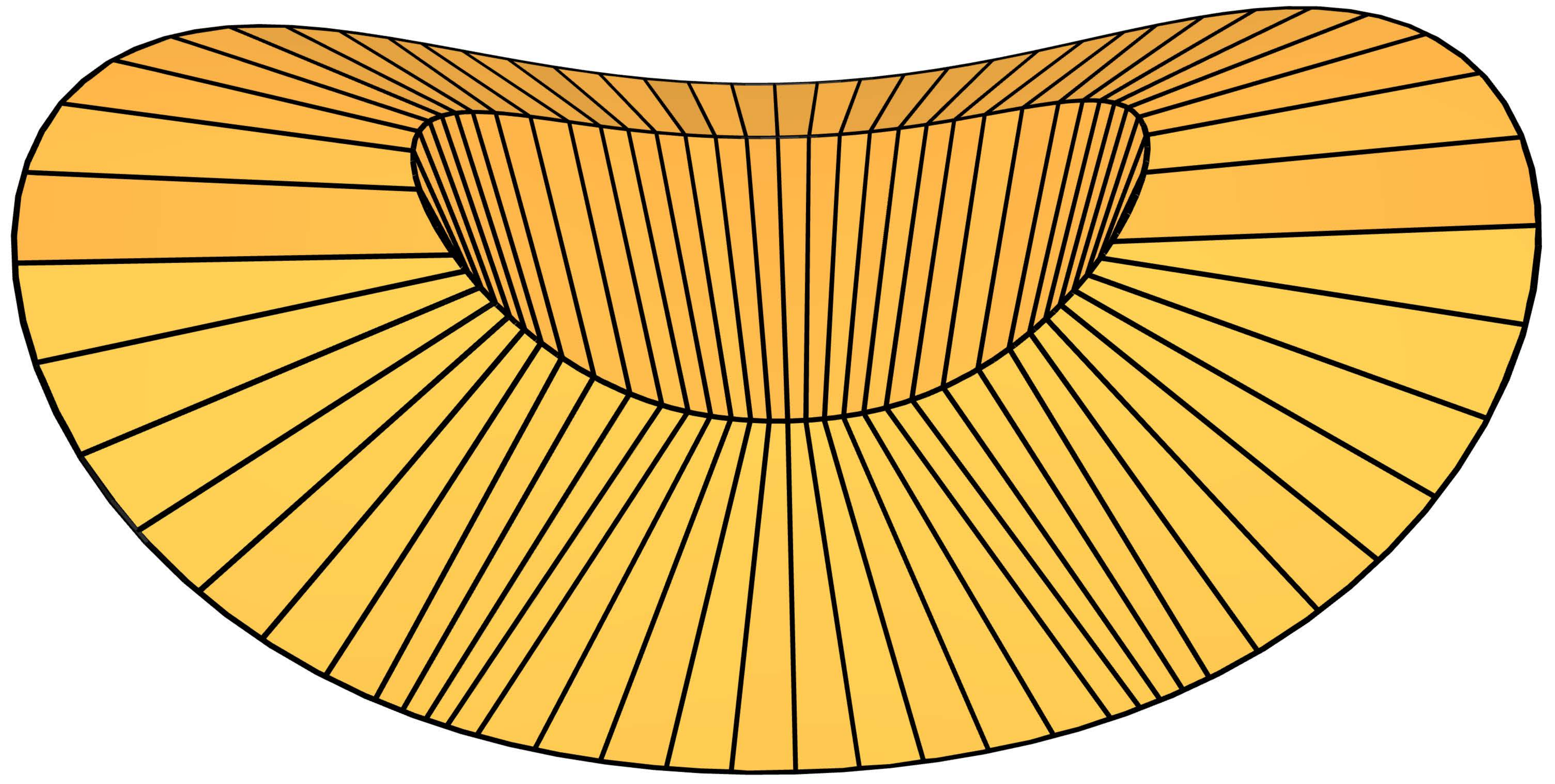}\\
		\includegraphics[width=0.35\columnwidth]{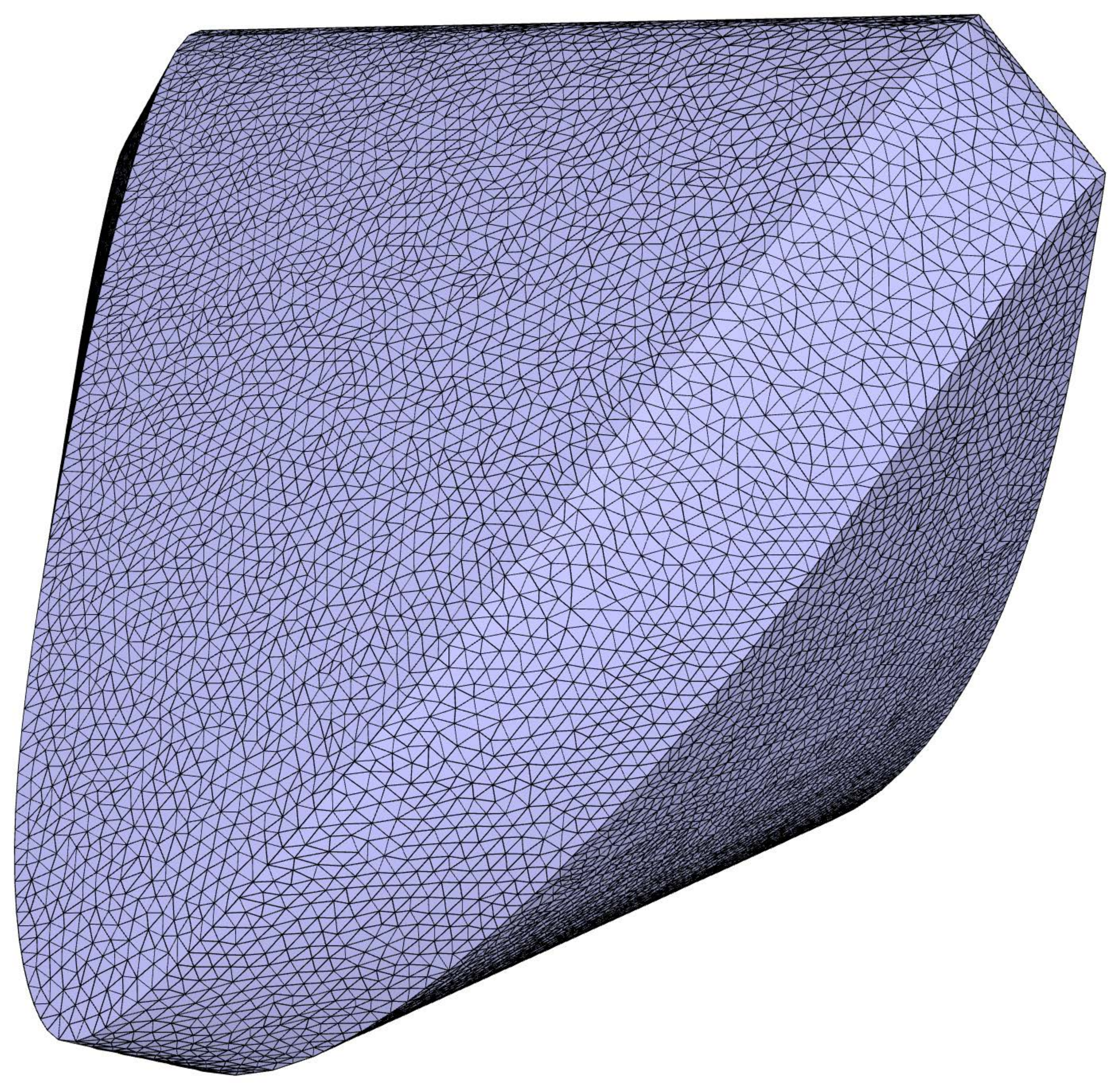}&
		\includegraphics[width=0.35\columnwidth]{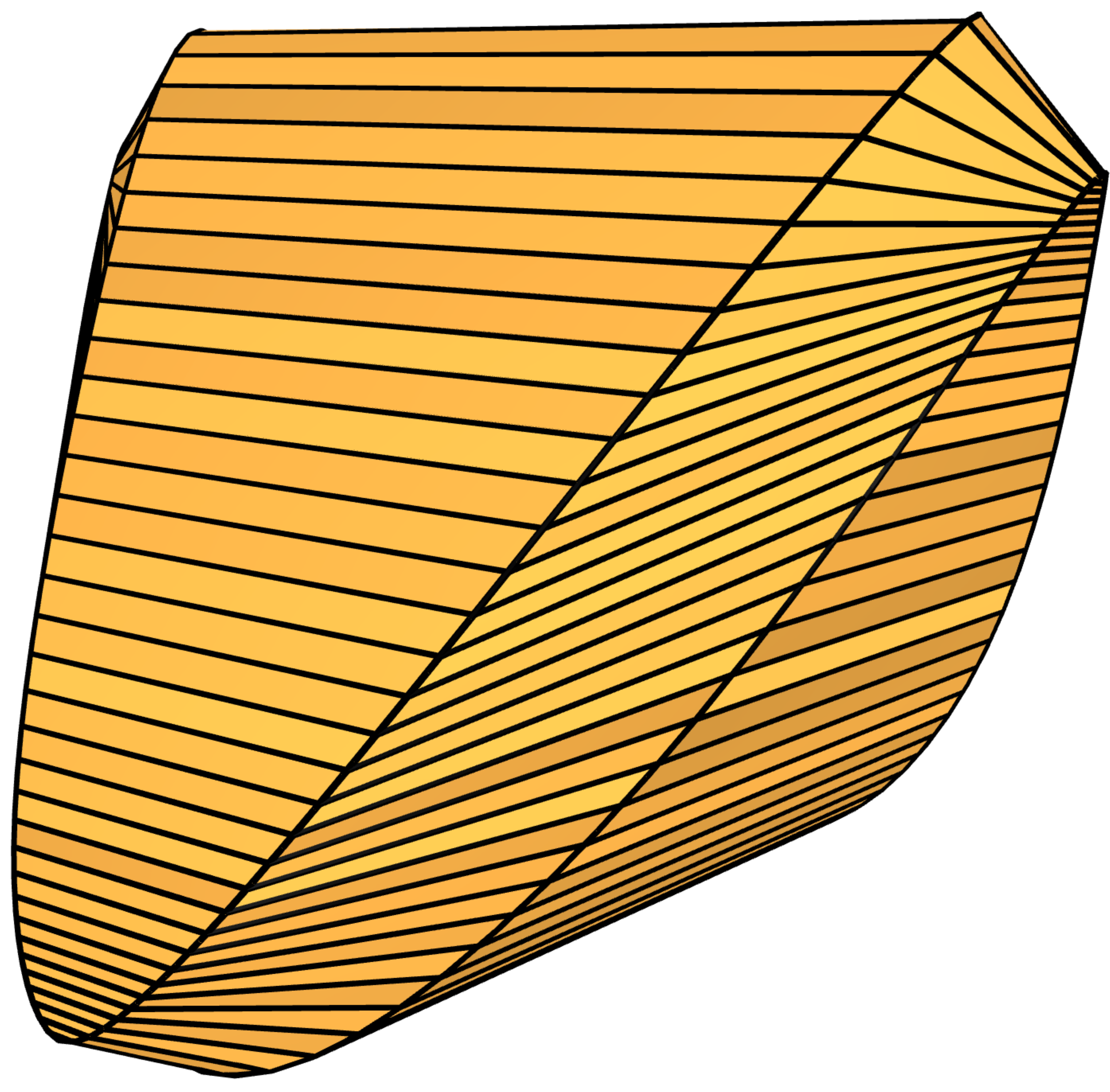}\\
		\includegraphics[width=0.37\columnwidth]{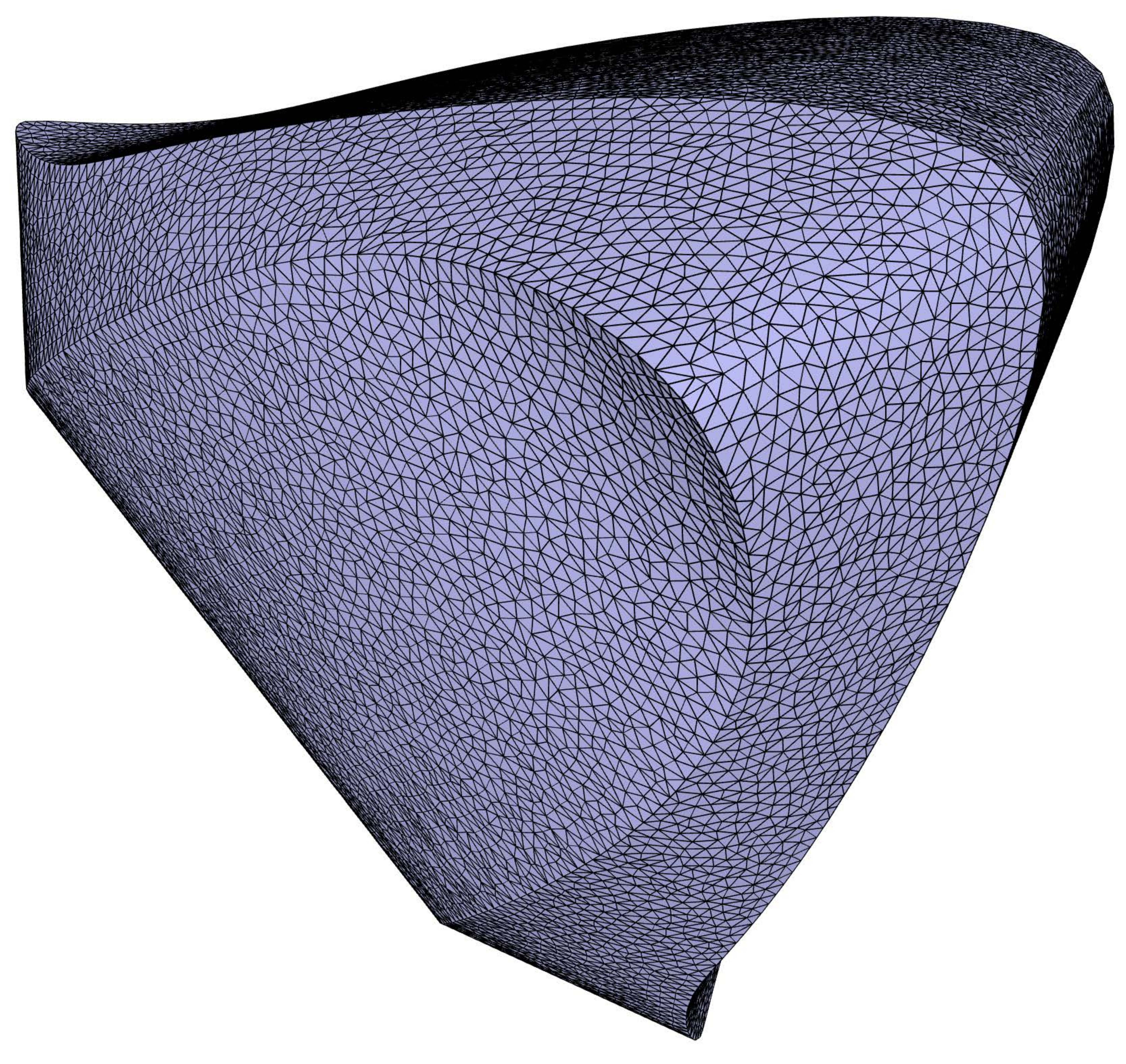}&
		\includegraphics[width=0.37\columnwidth]{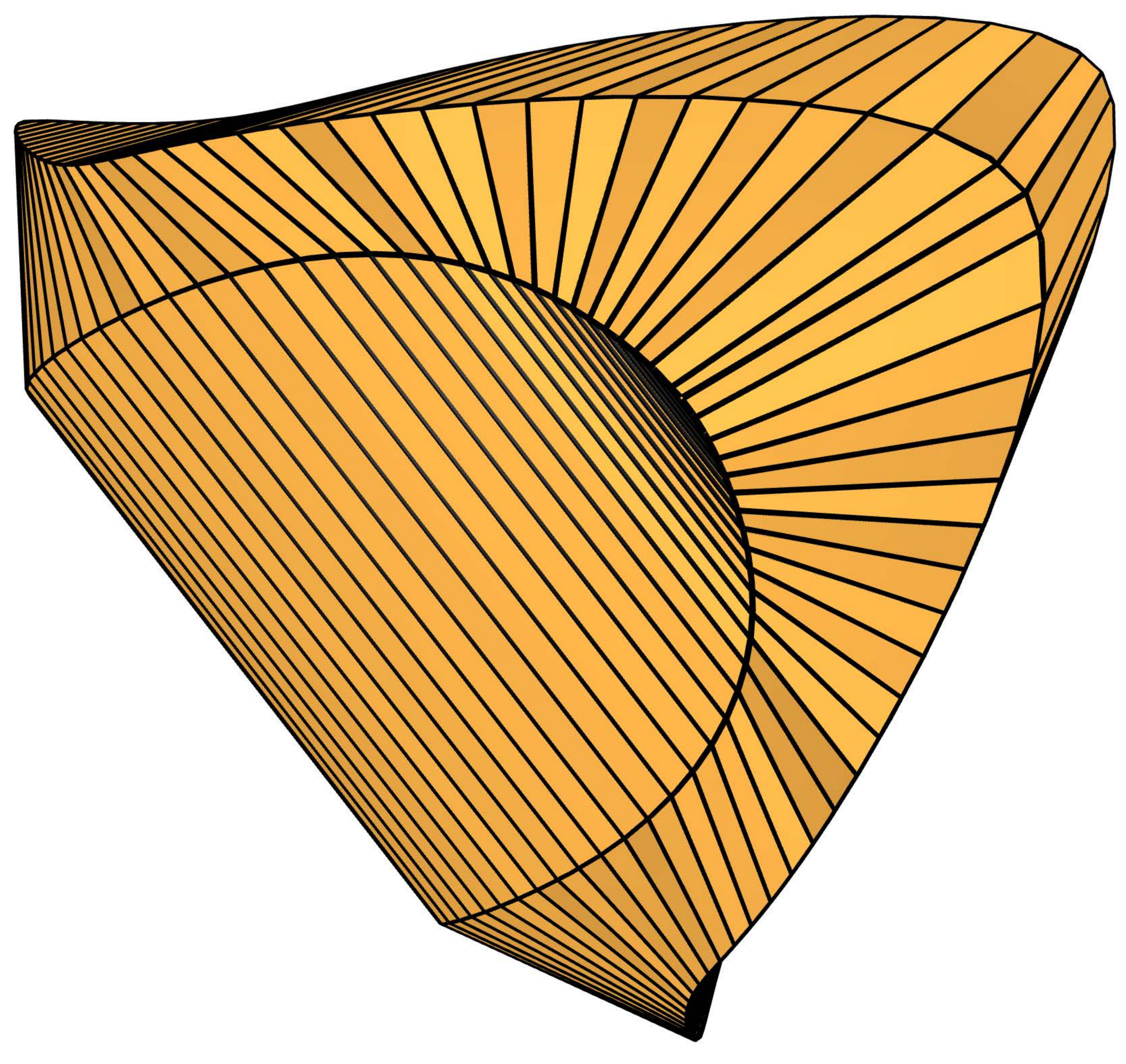}
	\end{tabular}
	\caption{Sphericons and D-forms are piecewise developable surfaces with creases connecting the individual pieces, and therefore can be remeshed with our method. The top two models are courtesy of \cite{CBP_Pottmann_Isometry:2020}, the bottom two are courtesy of \cite{pottmann_new}.}
	\label{fig:sphericonsdforms}
\end{figure}

\begin{figure}
	\setlength{\tabcolsep}{2pt}   
	\begin{tabular}{cc}
		\includegraphics[width=0.38\columnwidth]{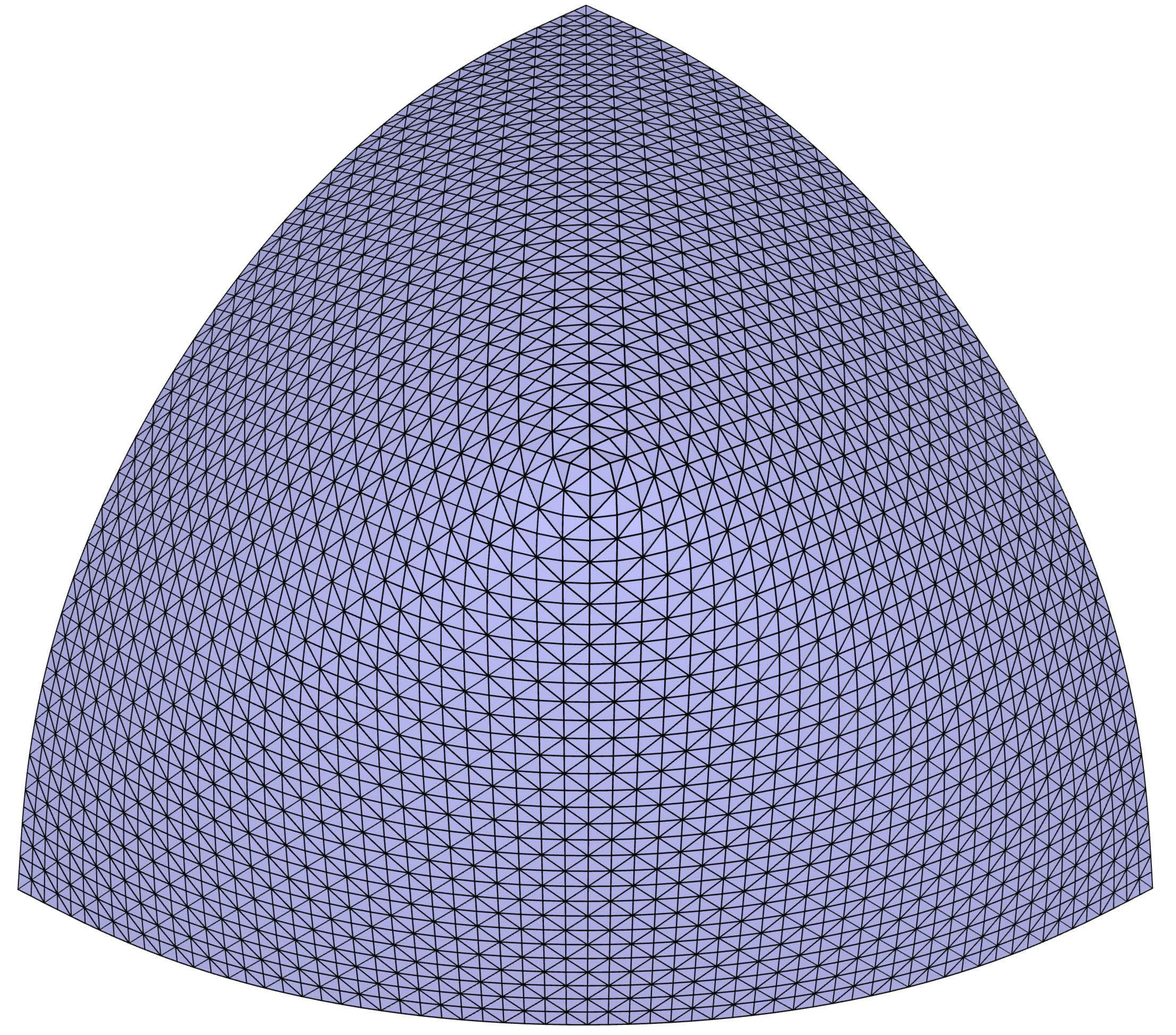} &
		\includegraphics[width=0.38\columnwidth]{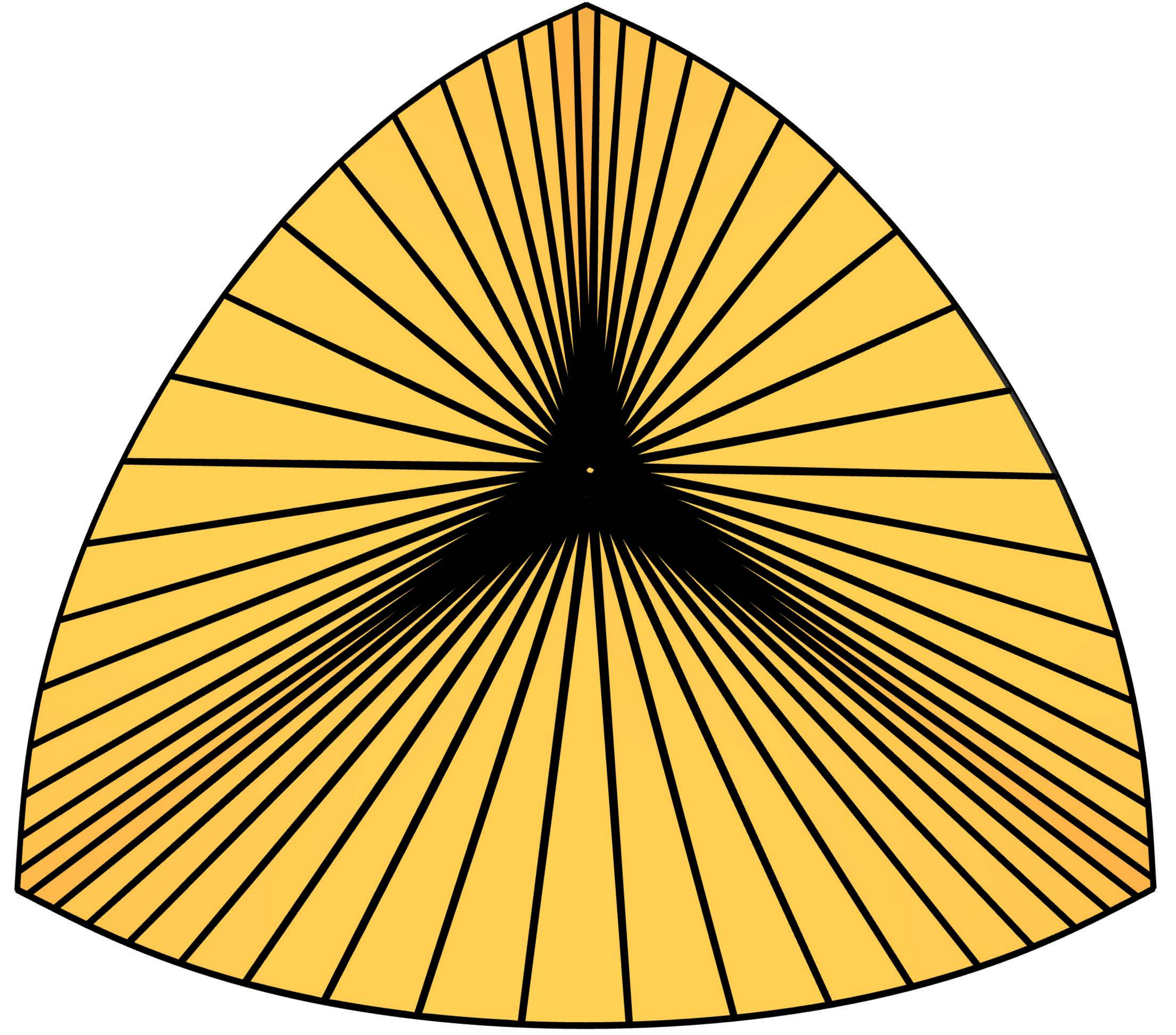}\\
		\includegraphics[width=0.48\columnwidth]{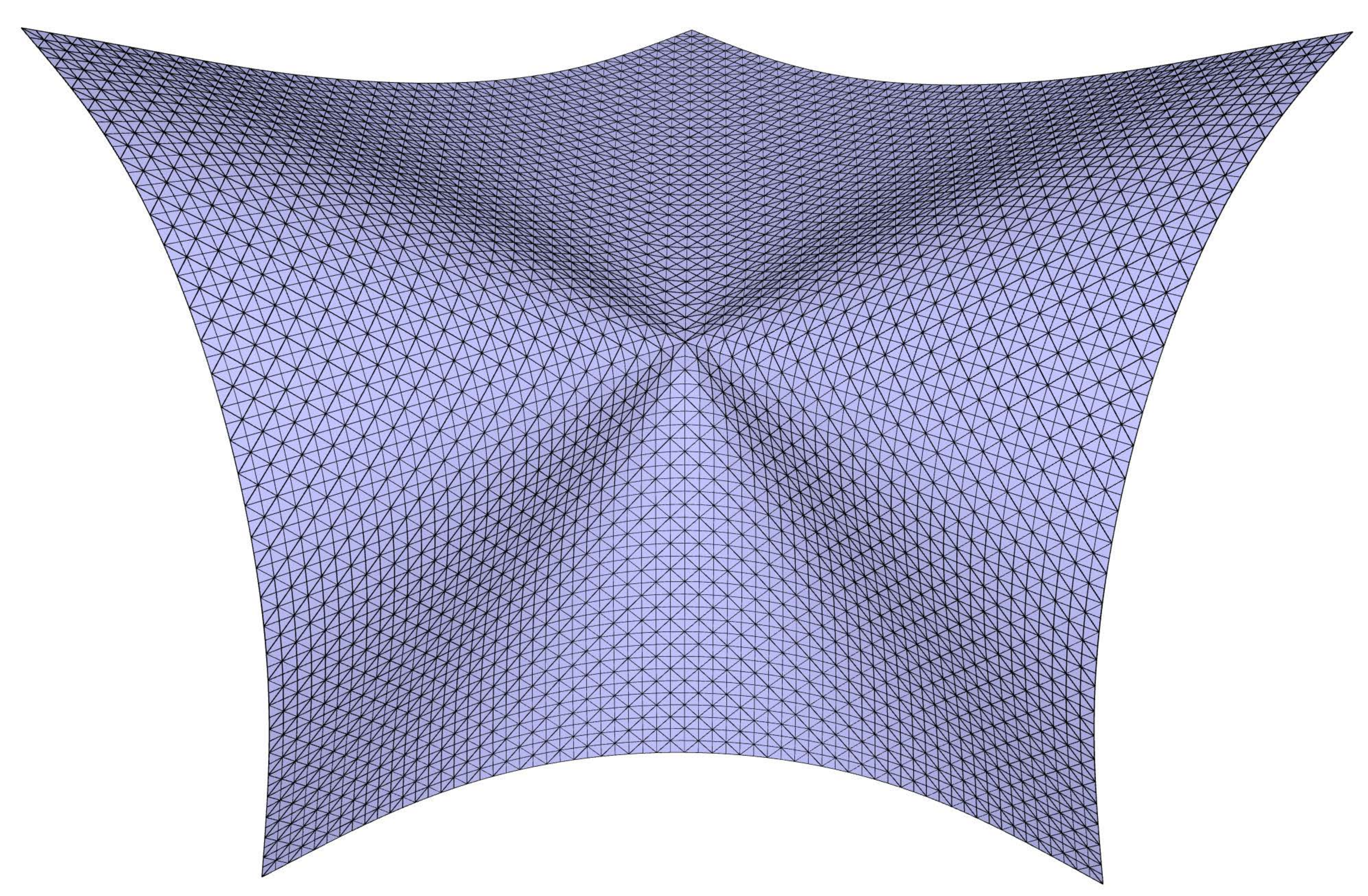}&
		\includegraphics[width=0.48\columnwidth]{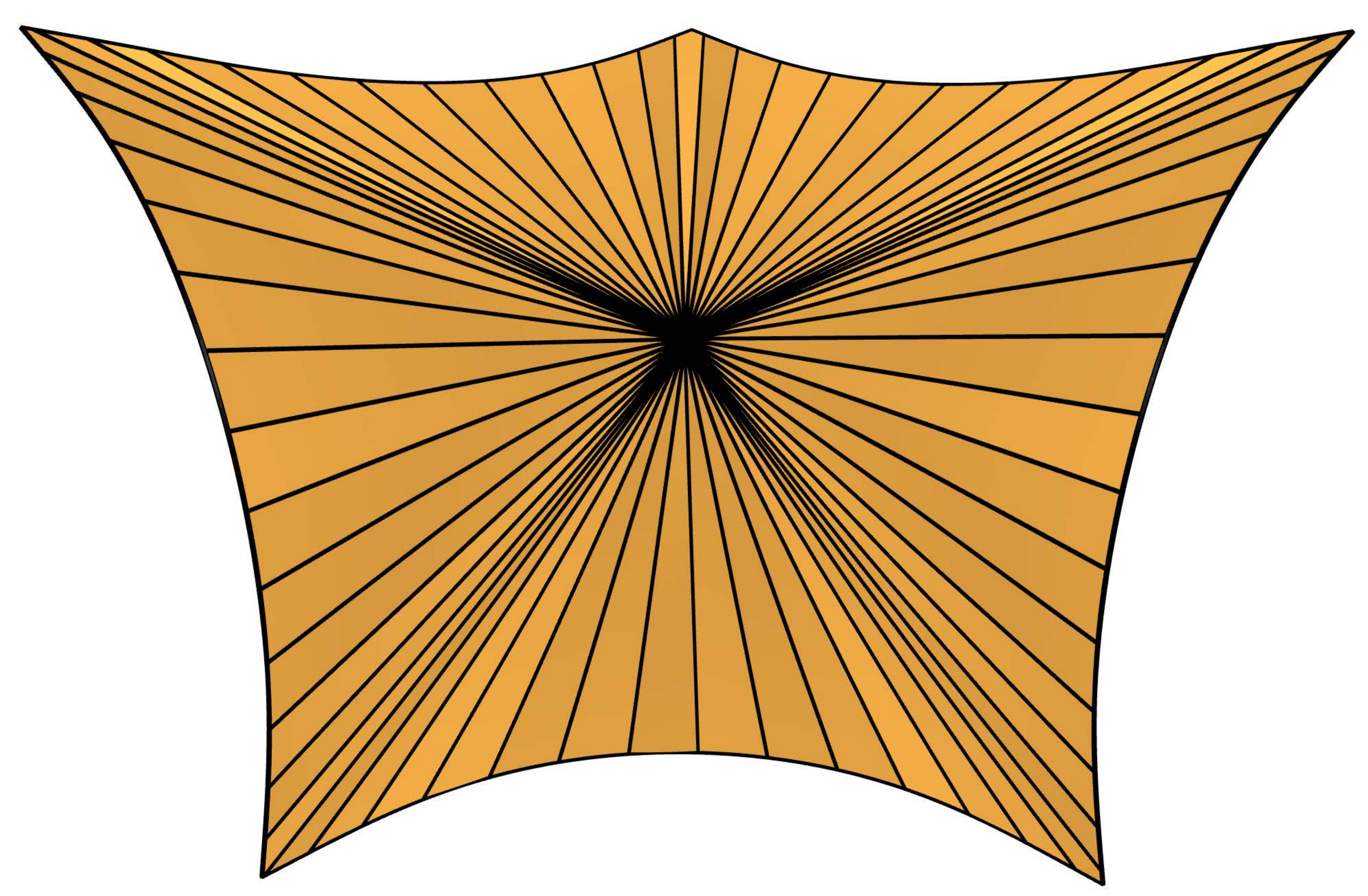}\\
		\includegraphics[width=0.45\columnwidth]{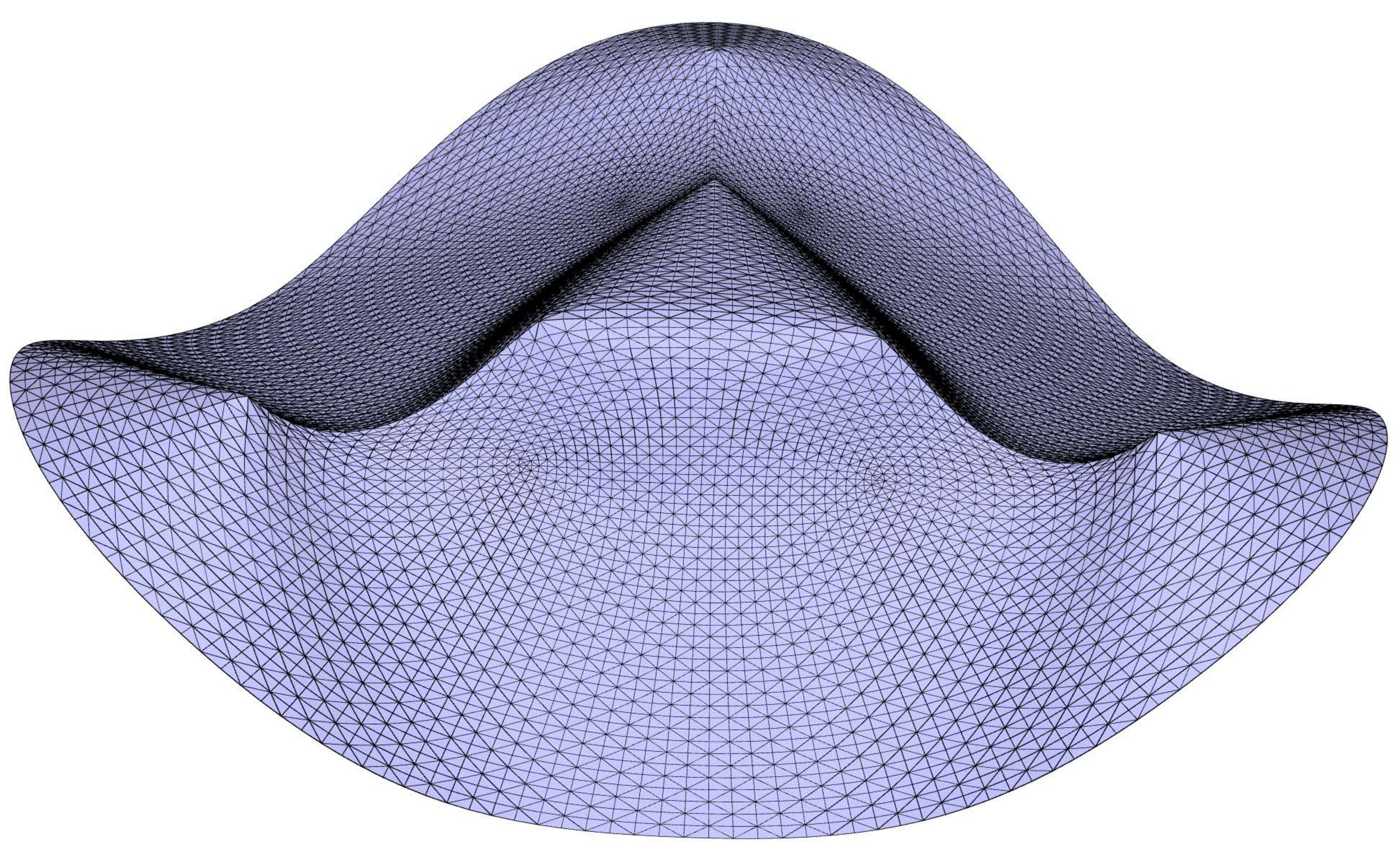}&
		\includegraphics[width=0.45\columnwidth]{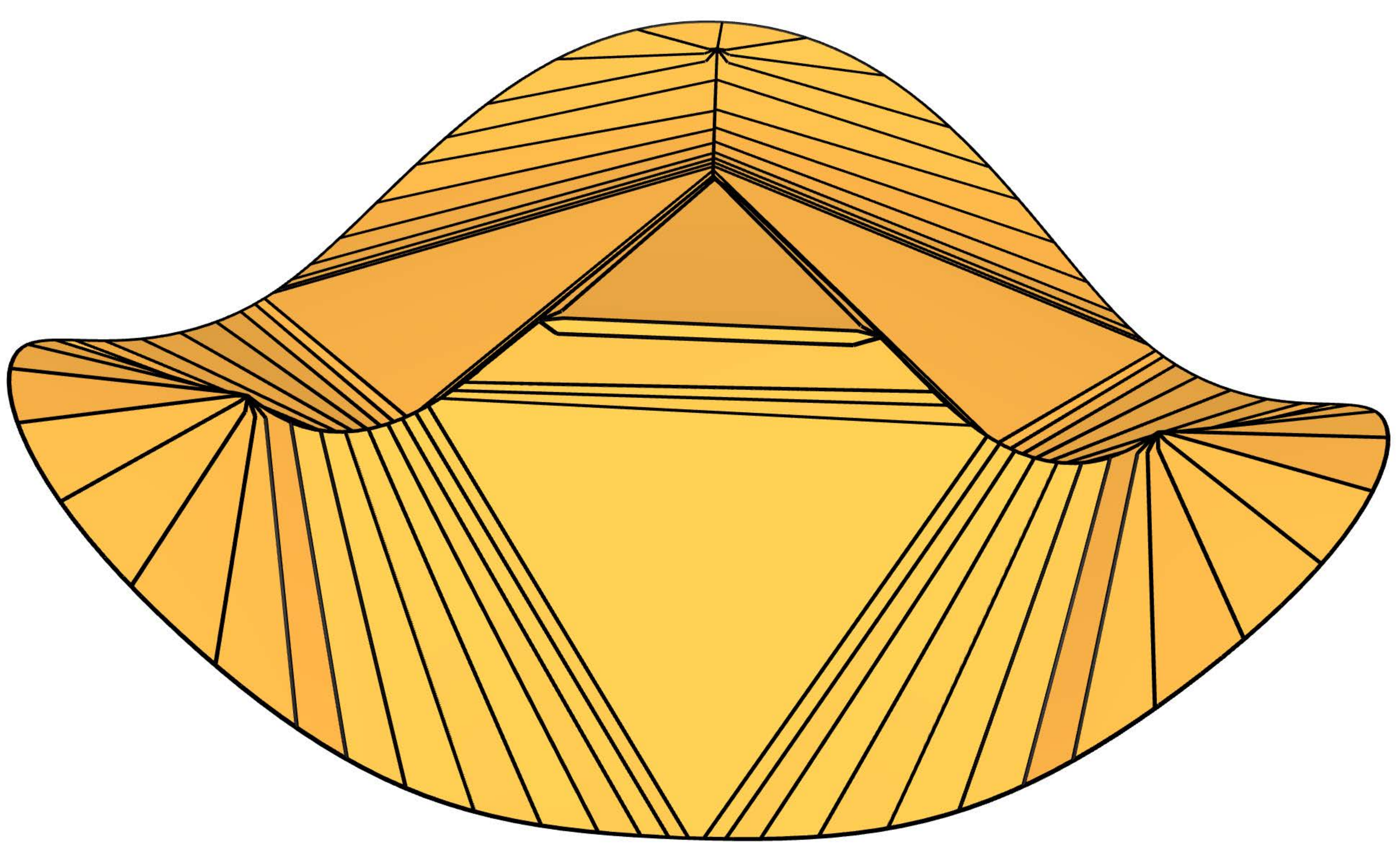}
	\end{tabular}
	\caption{Our method applied to glued developable surfaces that include cone apexes. The bottom model contains open creases that end in cone apexes. These models are courtesy of \cite{CBP_Pottmann_Isometry:2020}.}
	\label{fig:glued}
\end{figure}

\begin{figure}
	\setlength{\tabcolsep}{2pt}   
	\begin{tabular}{cc}
		\includegraphics[width=0.38\columnwidth]{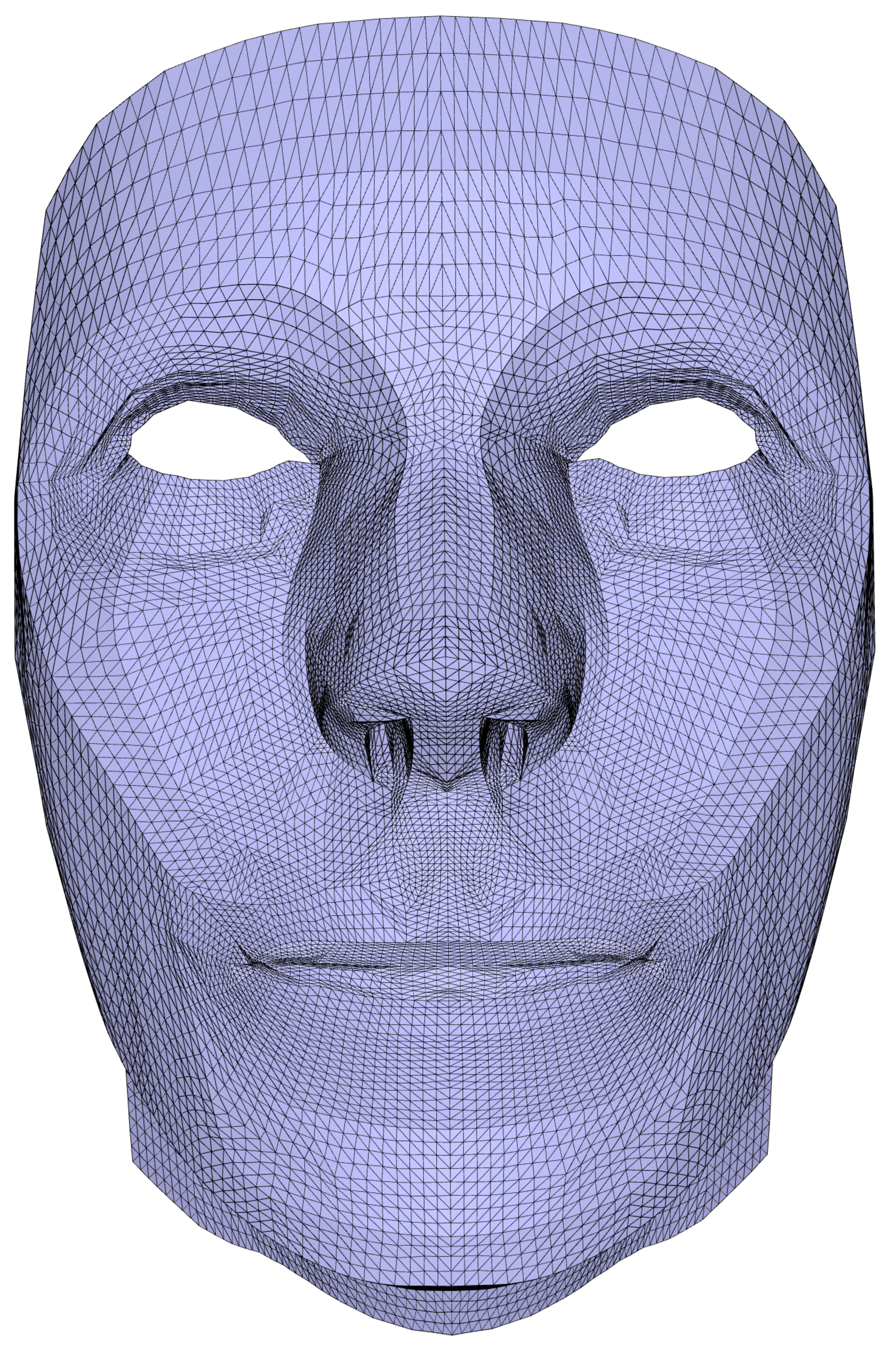}&
		\includegraphics[width=0.38\columnwidth]{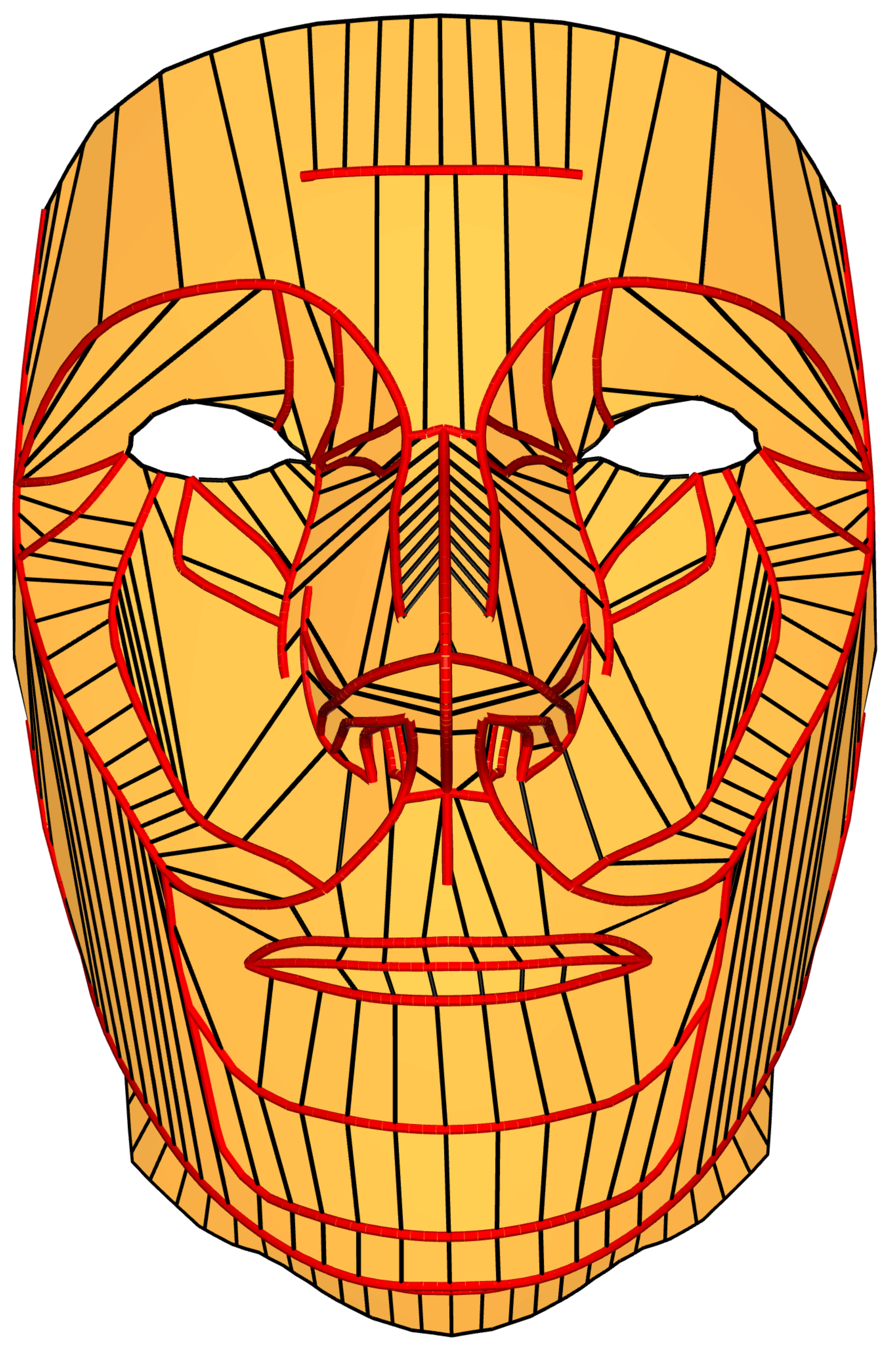}\\
		\includegraphics[width=0.43\columnwidth]{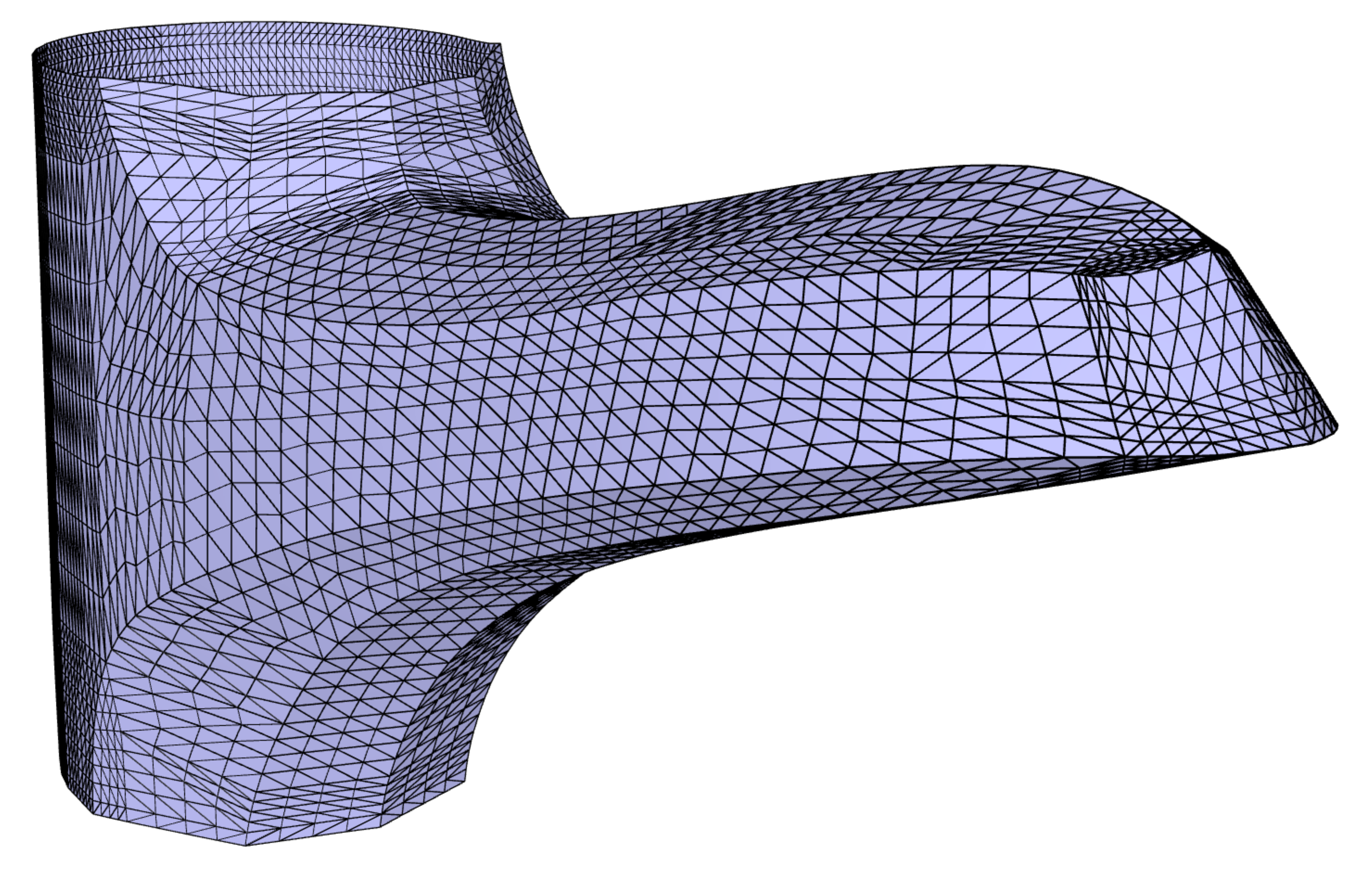} &
		\includegraphics[width=0.43\columnwidth]{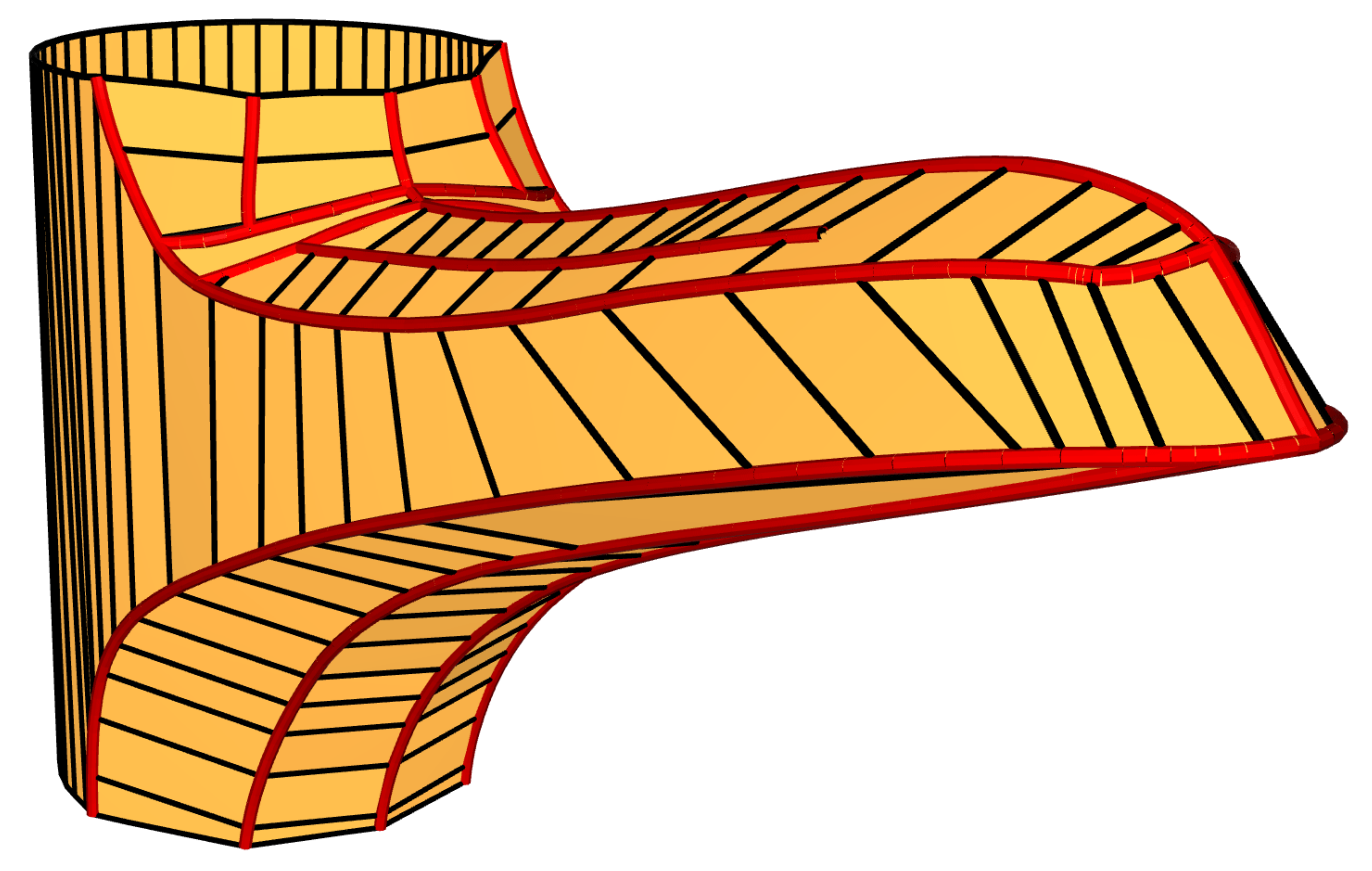}\\
	\end{tabular}
	\caption{Our method can handle open creases when they are defined as mesh boundaries. Examples of open creases can be seen on the forehead of the mask or on the top of the faucet. Since these creases are defined as mesh boundaries, seamless parameterization and meshing across them is no longer guaranteed. The red lines highlight the set of crease edges $\set{E}_c$. The models are courtesy of \cite{stein_dev}.}
	\label{fig:Keenan}
\end{figure} 

\begin{figure}
	\setlength{\tabcolsep}{2pt}   
	\resizebox*{!}{0.9\textheight}{%
		\begin{tabular}{cc}
			\includegraphics[width=0.4\columnwidth]{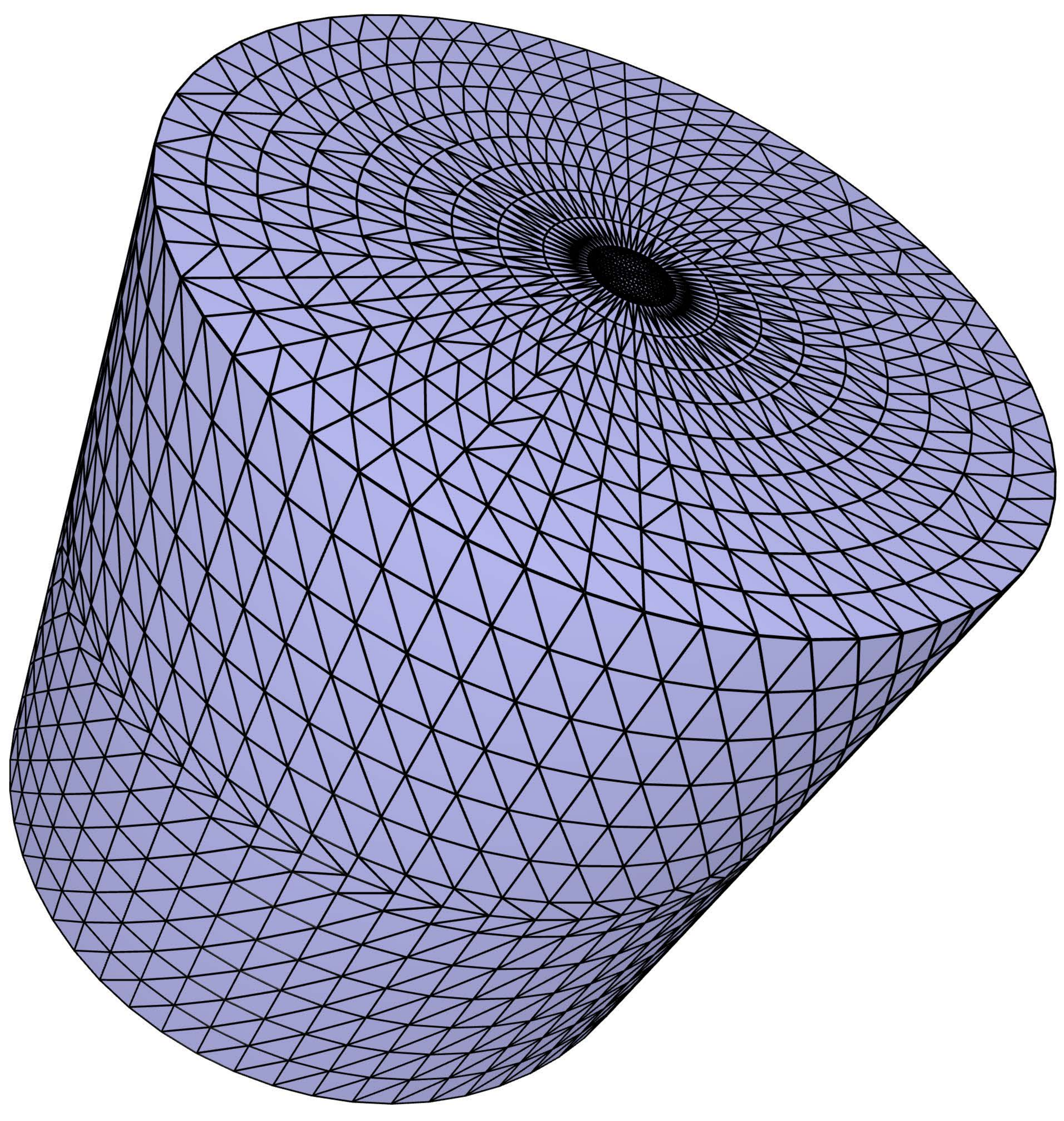} &
			\includegraphics[width=0.4\columnwidth]{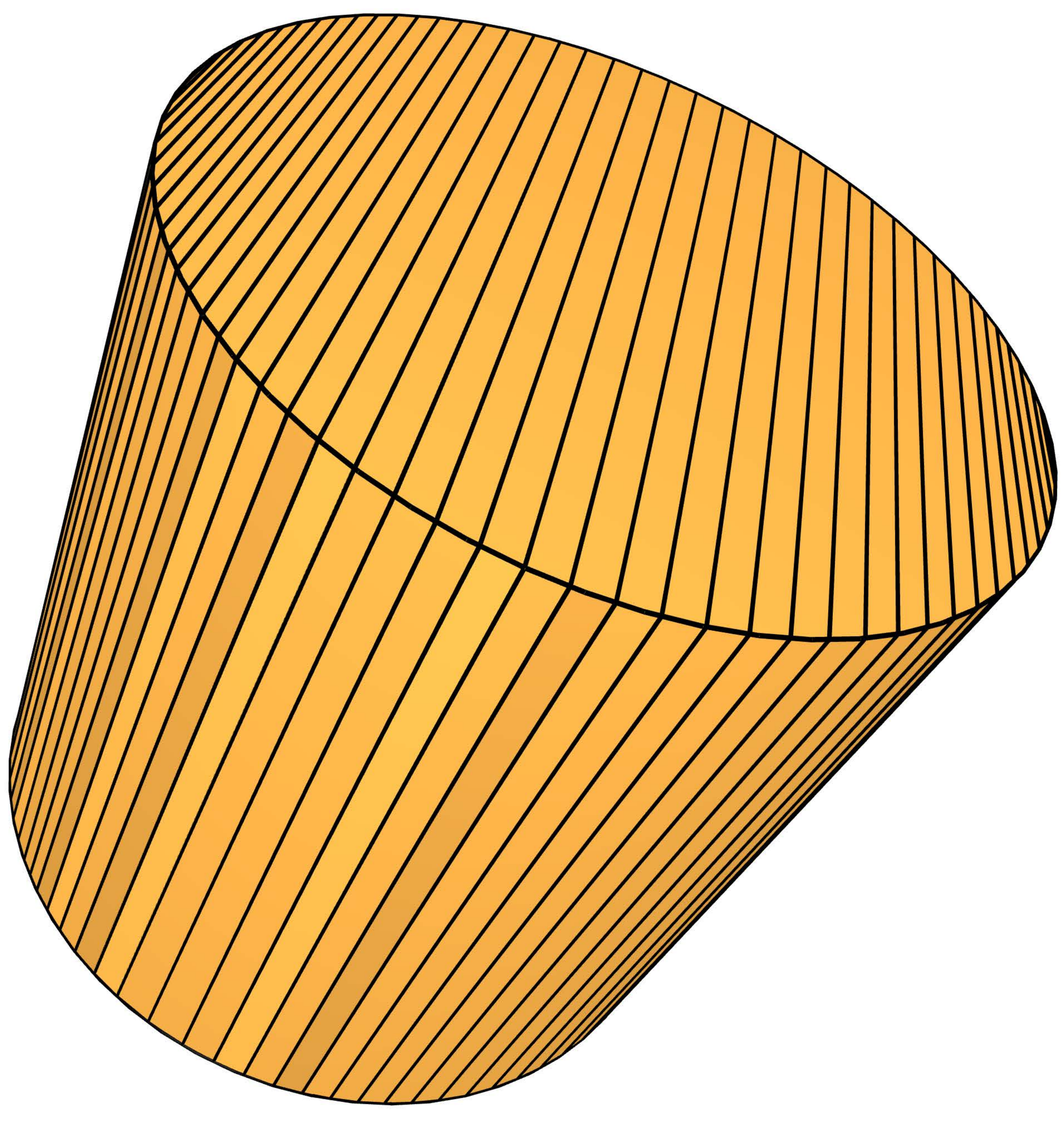}\\ \midrule
			\includegraphics[width=0.4\columnwidth]{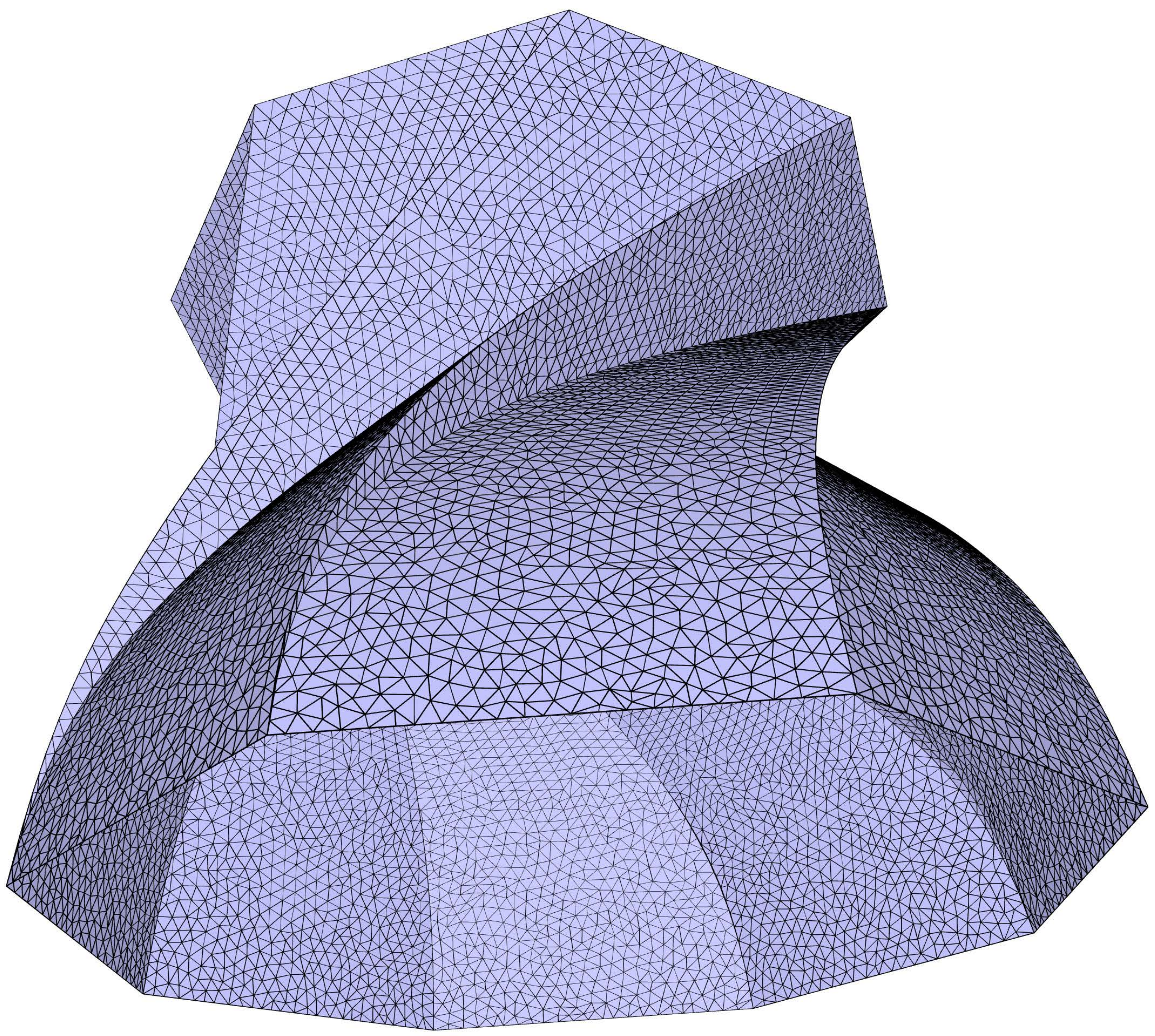}&
			\includegraphics[width=0.4\columnwidth]{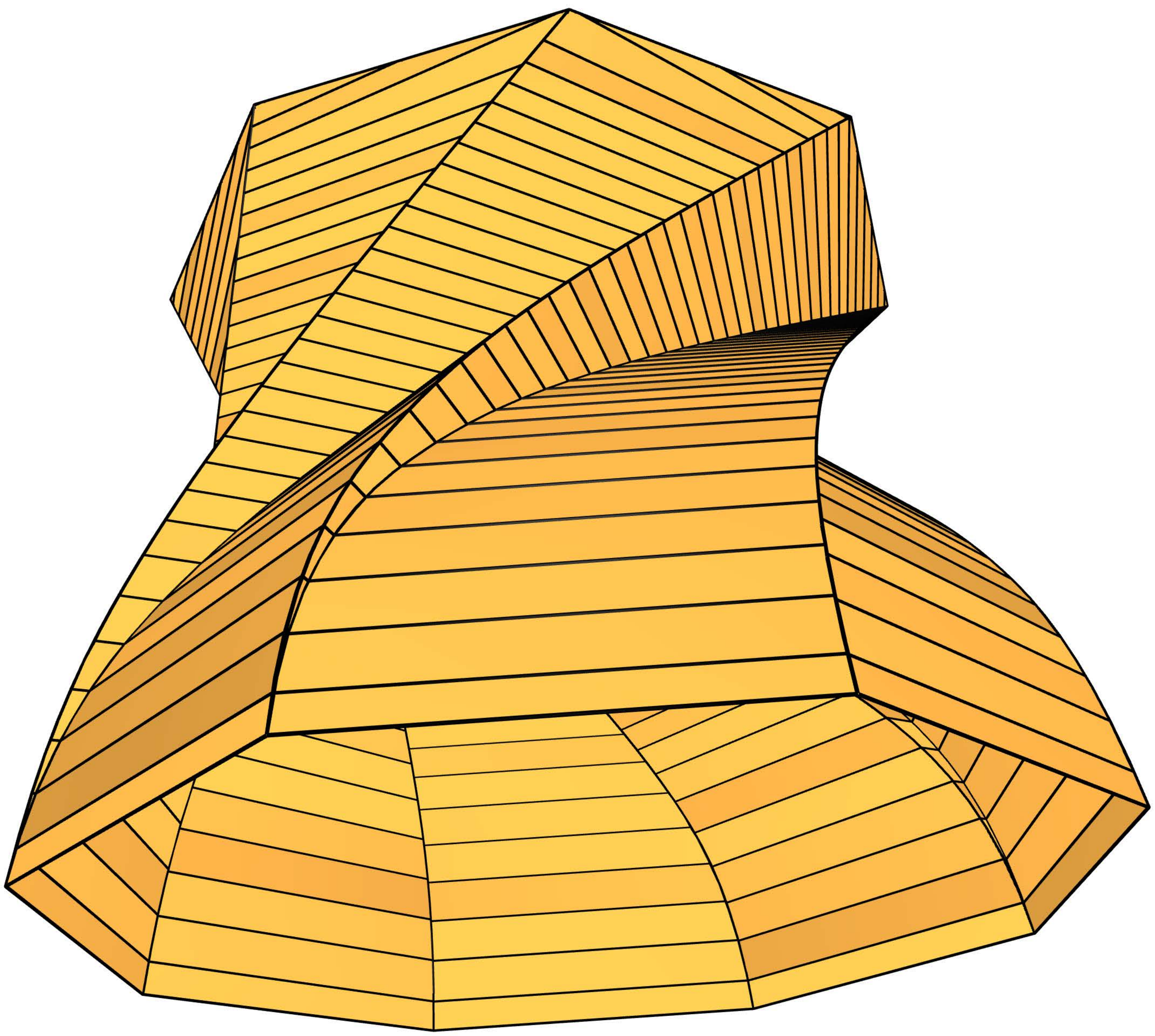}\\
			\includegraphics[width=0.35\columnwidth]{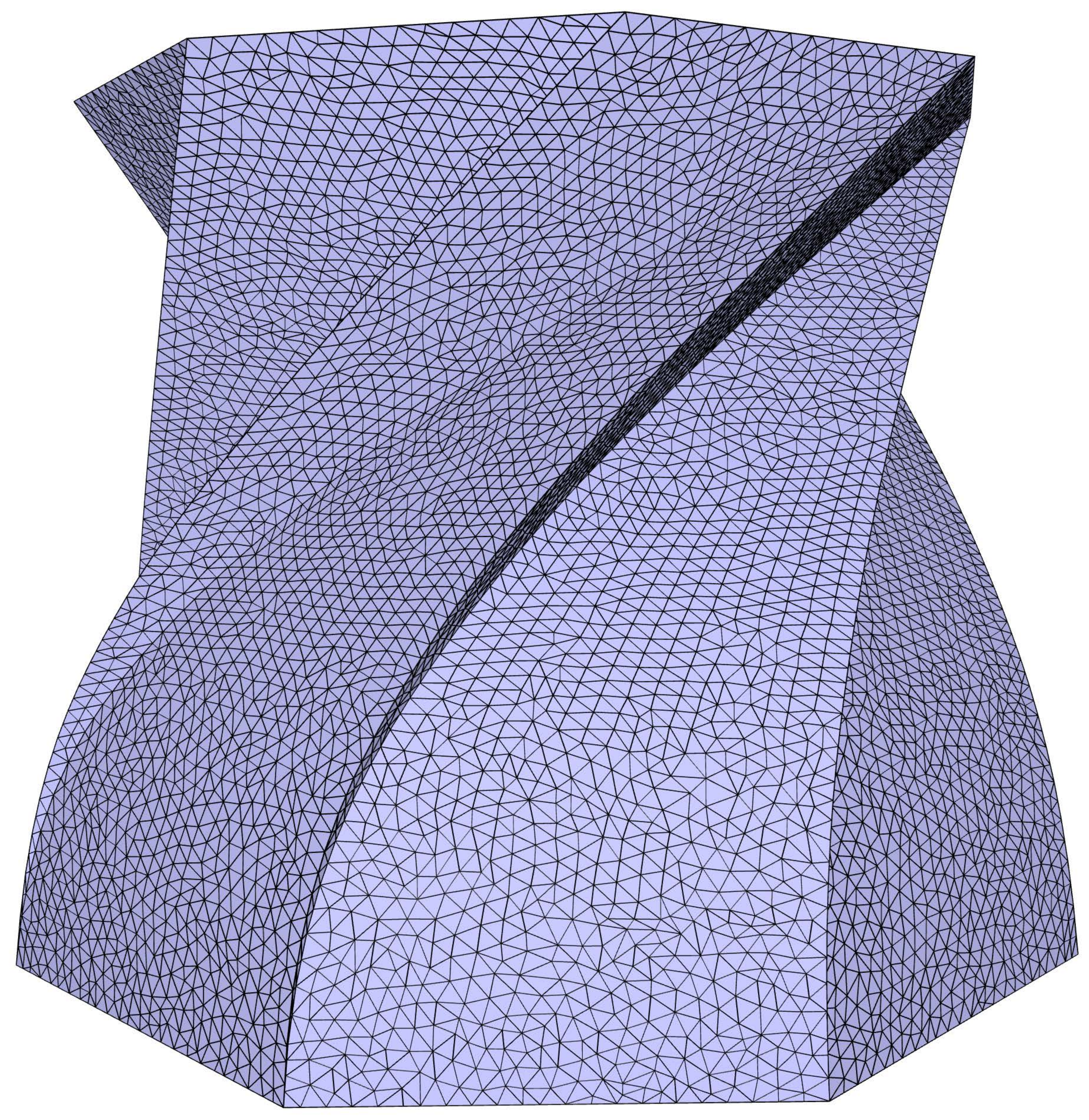}&
			\includegraphics[width=0.35\columnwidth]{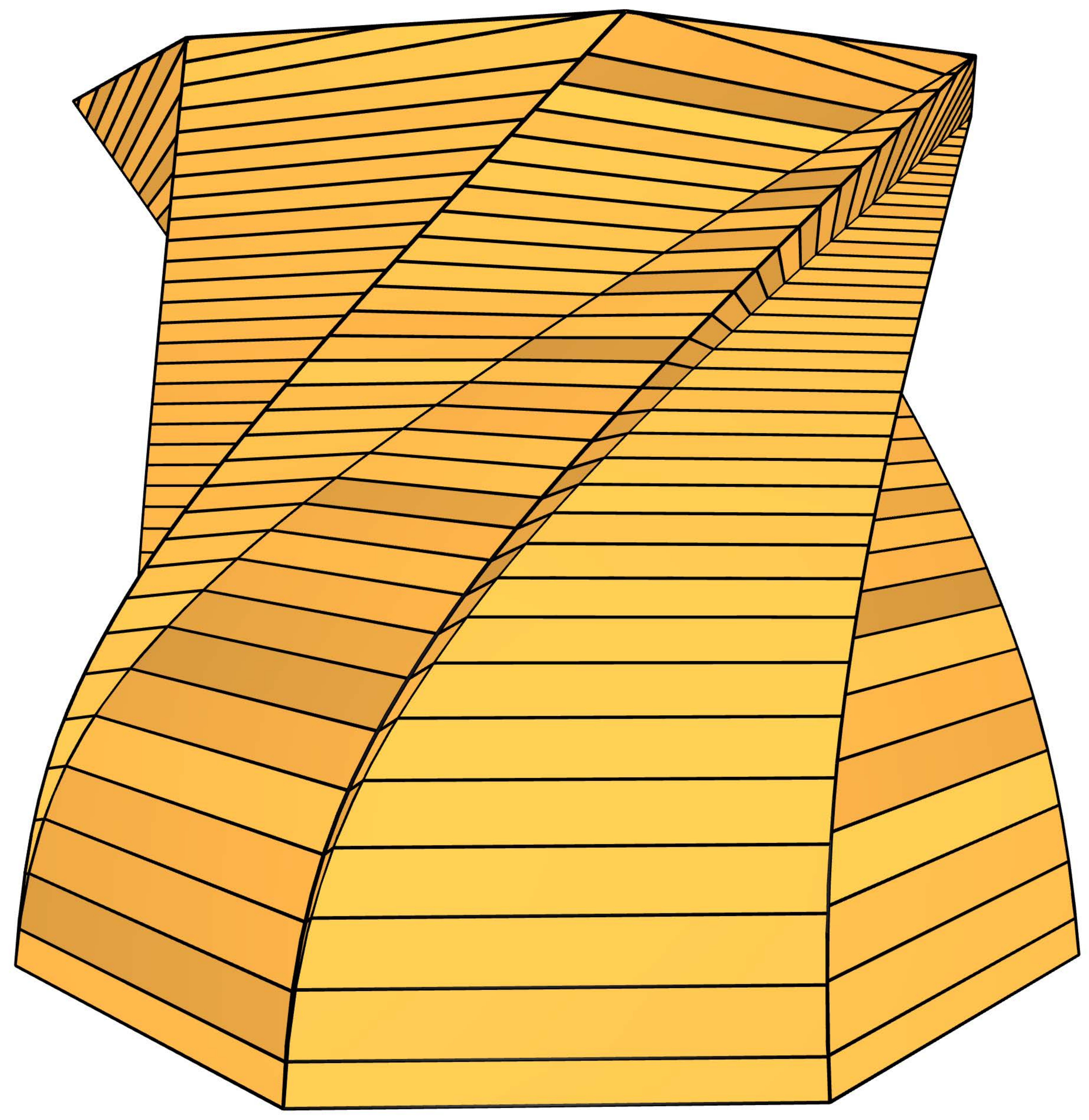}\\
			\includegraphics[width=0.35\columnwidth]{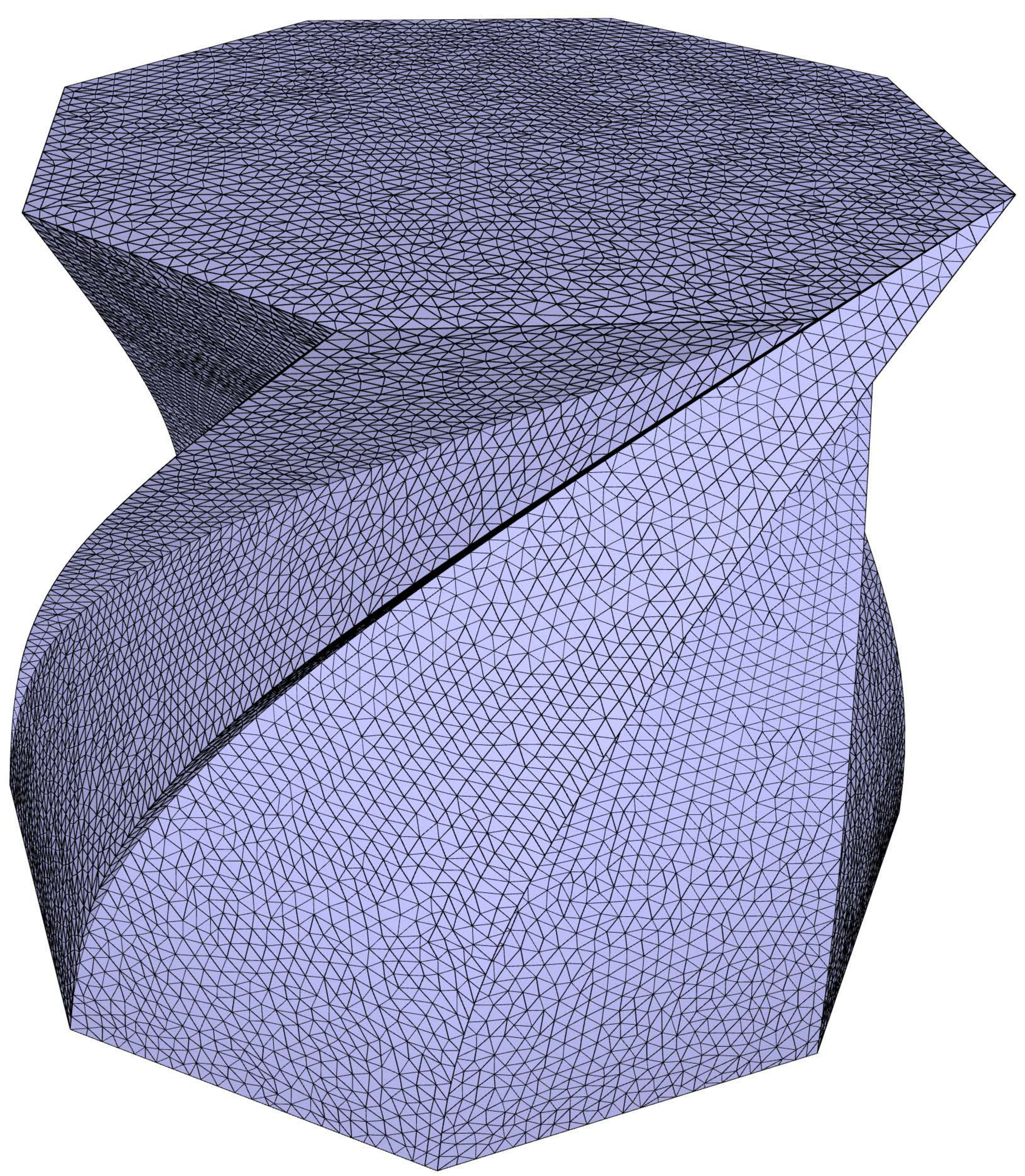}&
			\includegraphics[width=0.35\columnwidth]{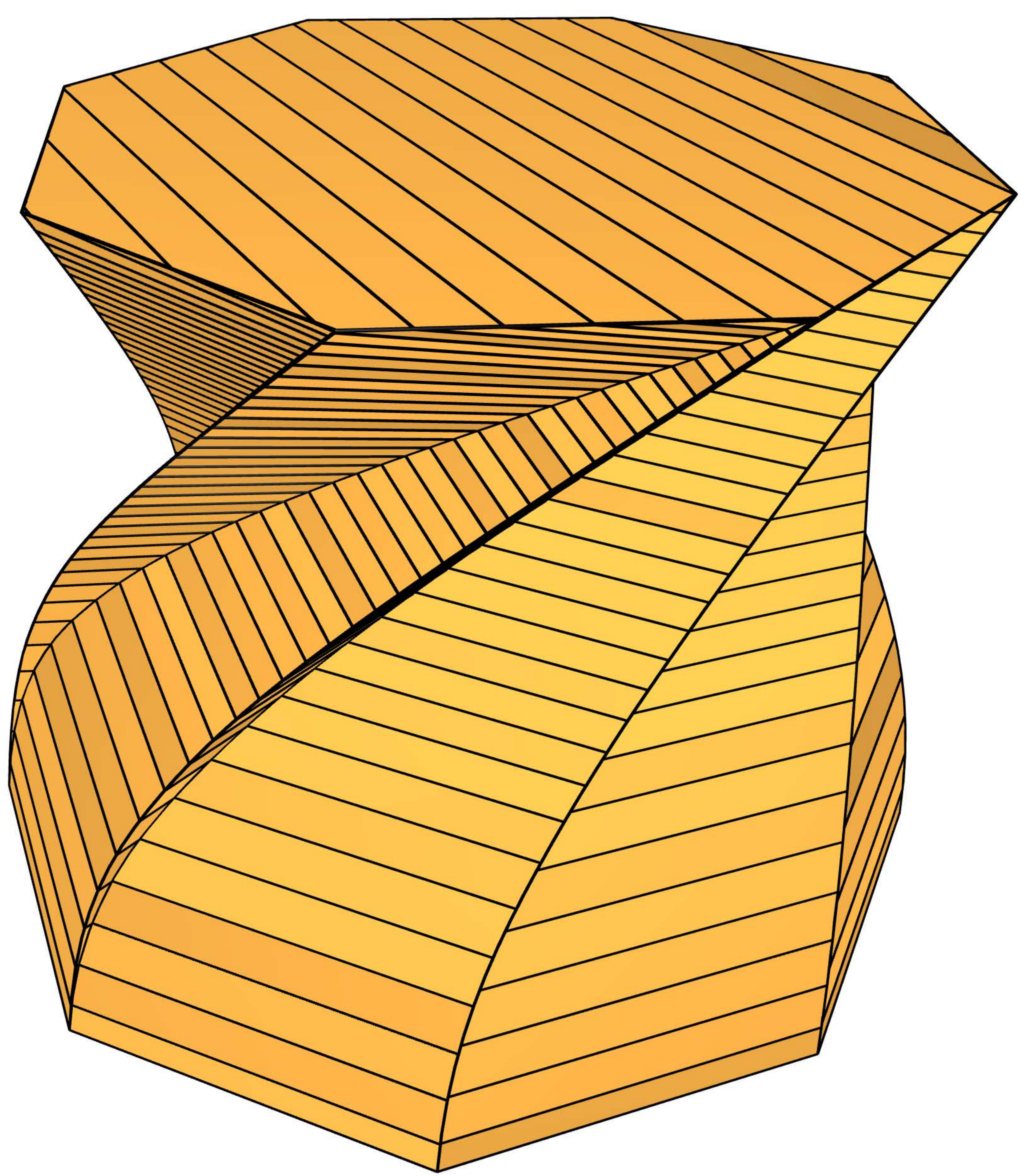}\\ \midrule
			\includegraphics[width=0.43\columnwidth]{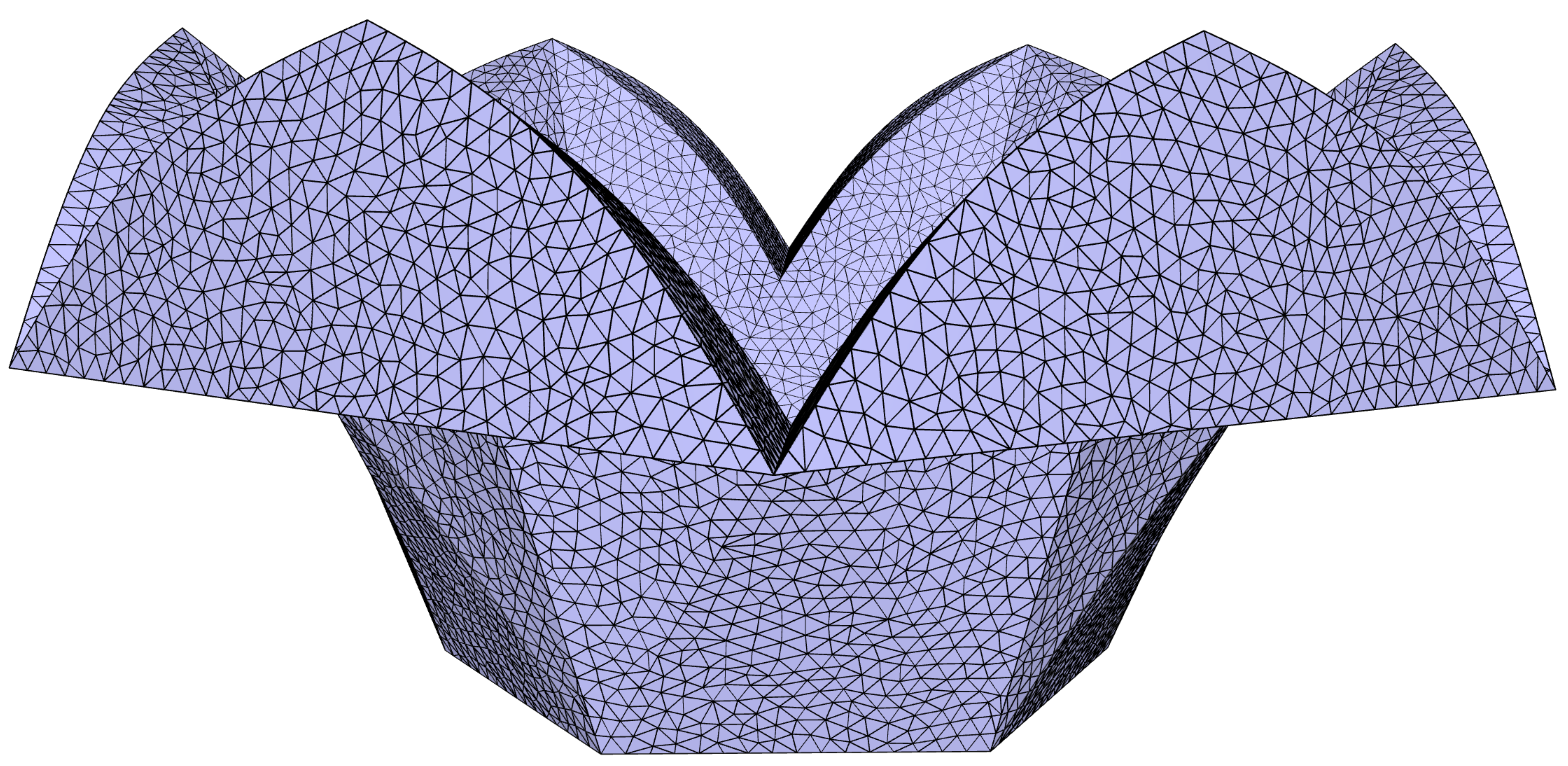} &
			\includegraphics[width=0.43\columnwidth]{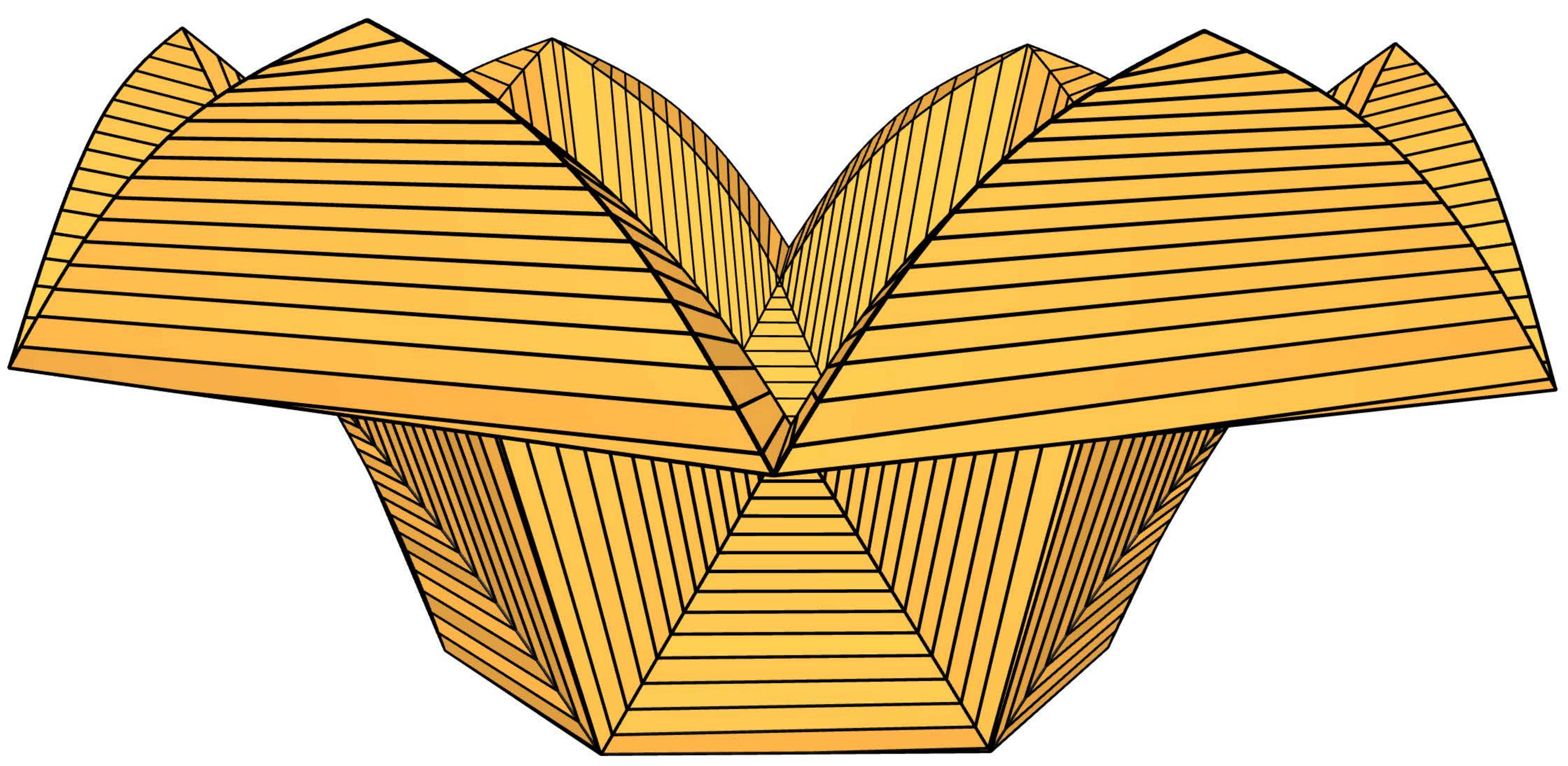}\\
			\includegraphics[width=0.4\columnwidth]{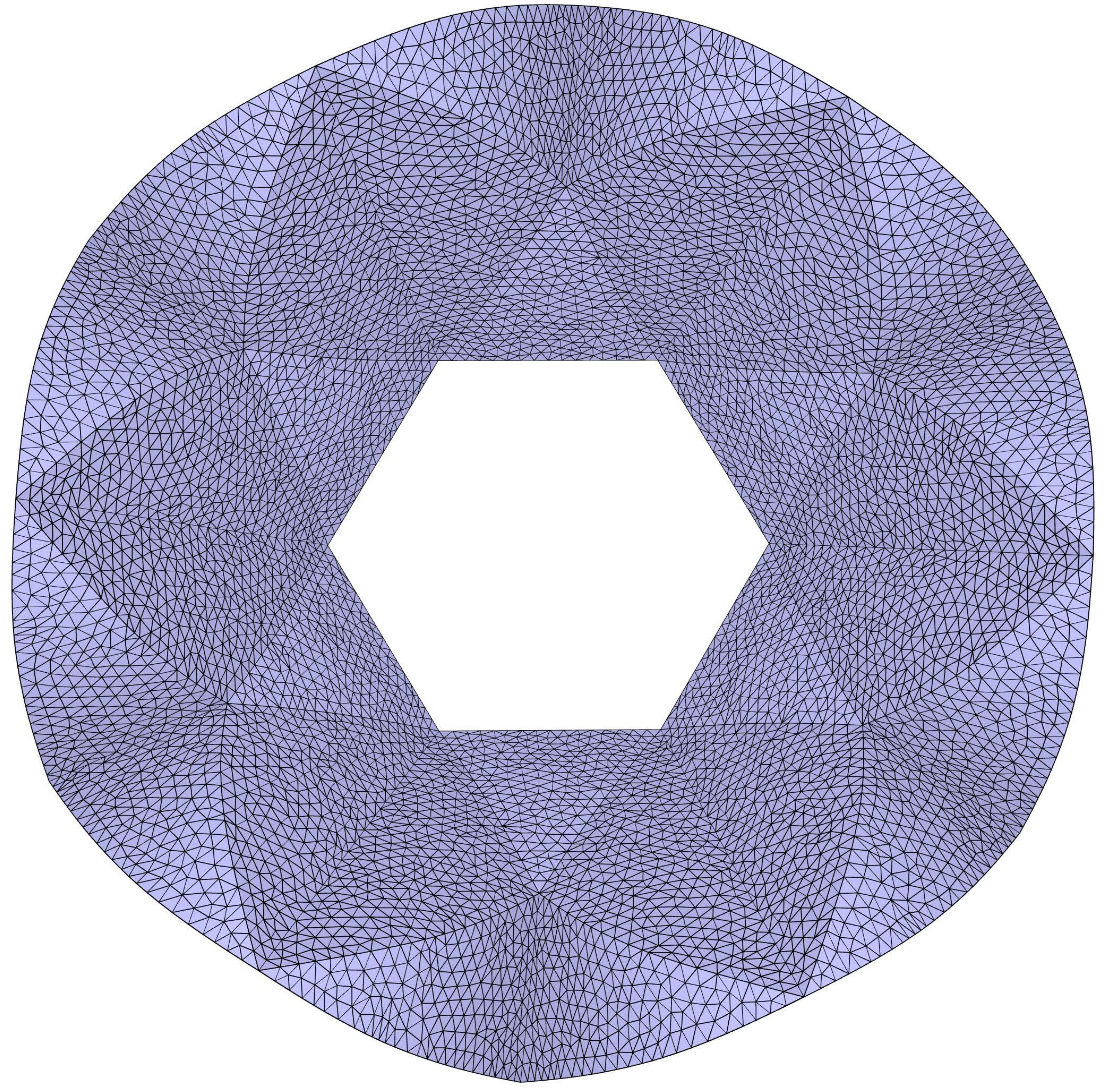} &
			\includegraphics[width=0.4\columnwidth]{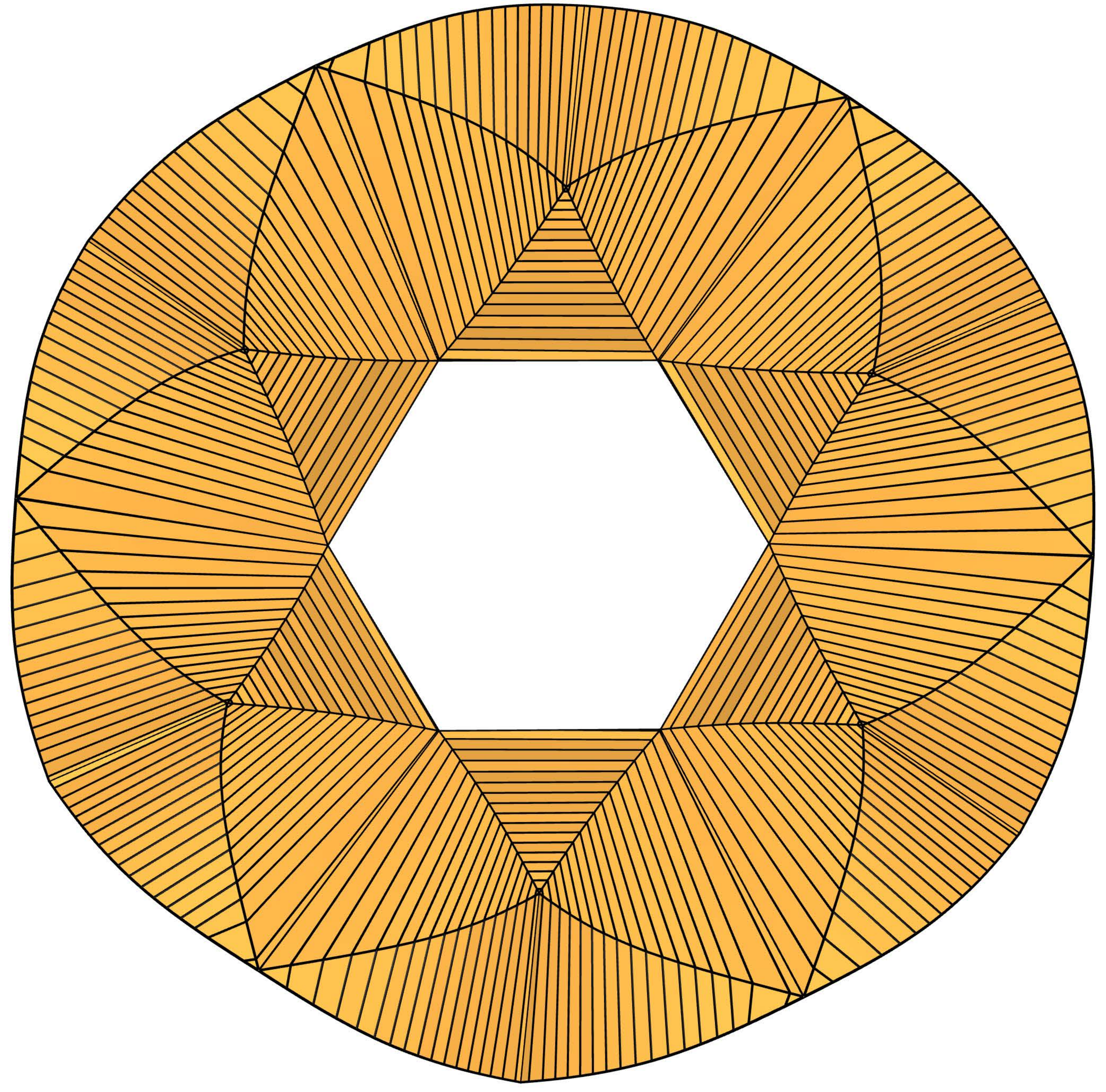}\\ \midrule
			\includegraphics[width=0.5\columnwidth]{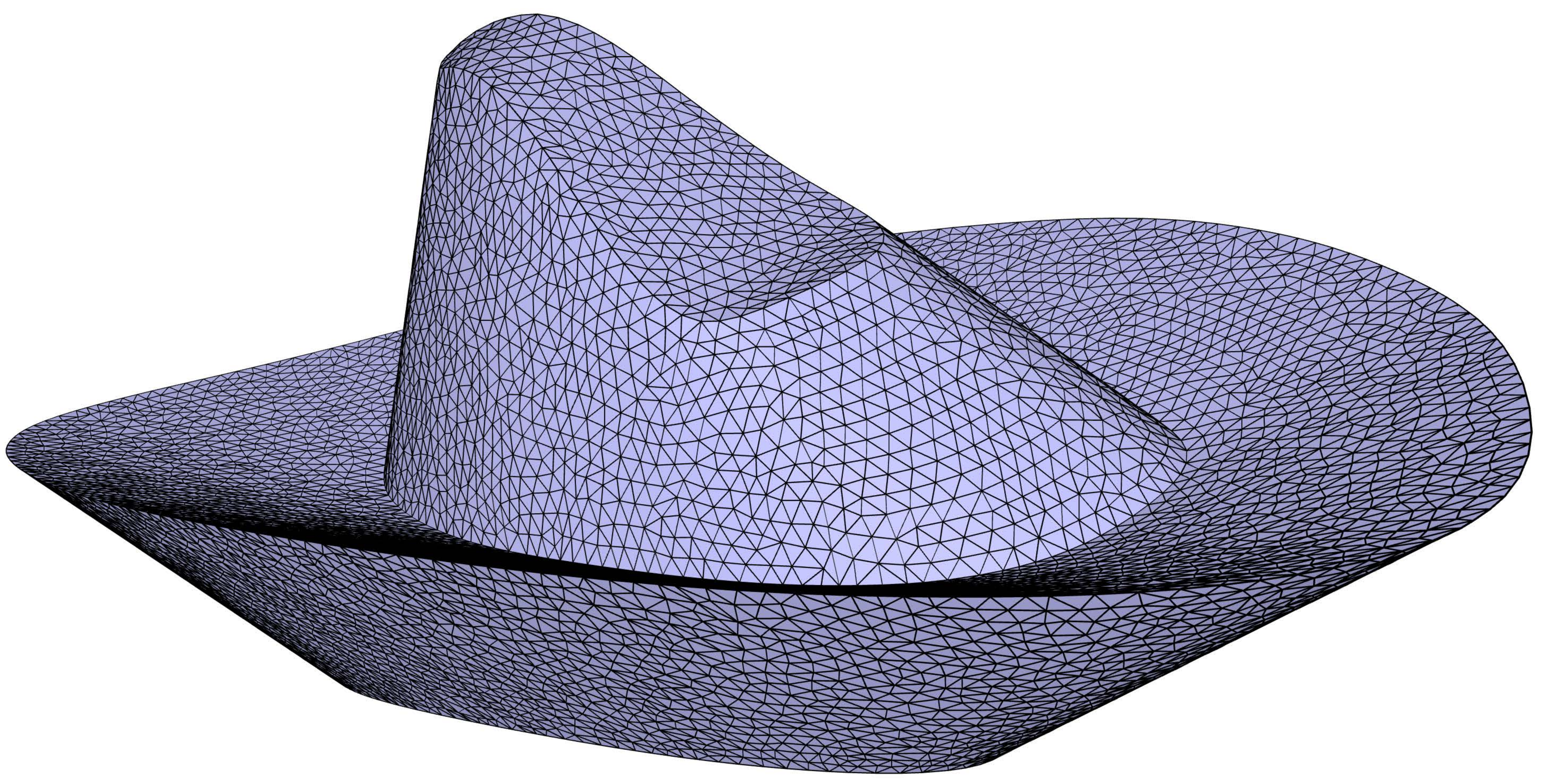}&
			\includegraphics[width=0.5\columnwidth]{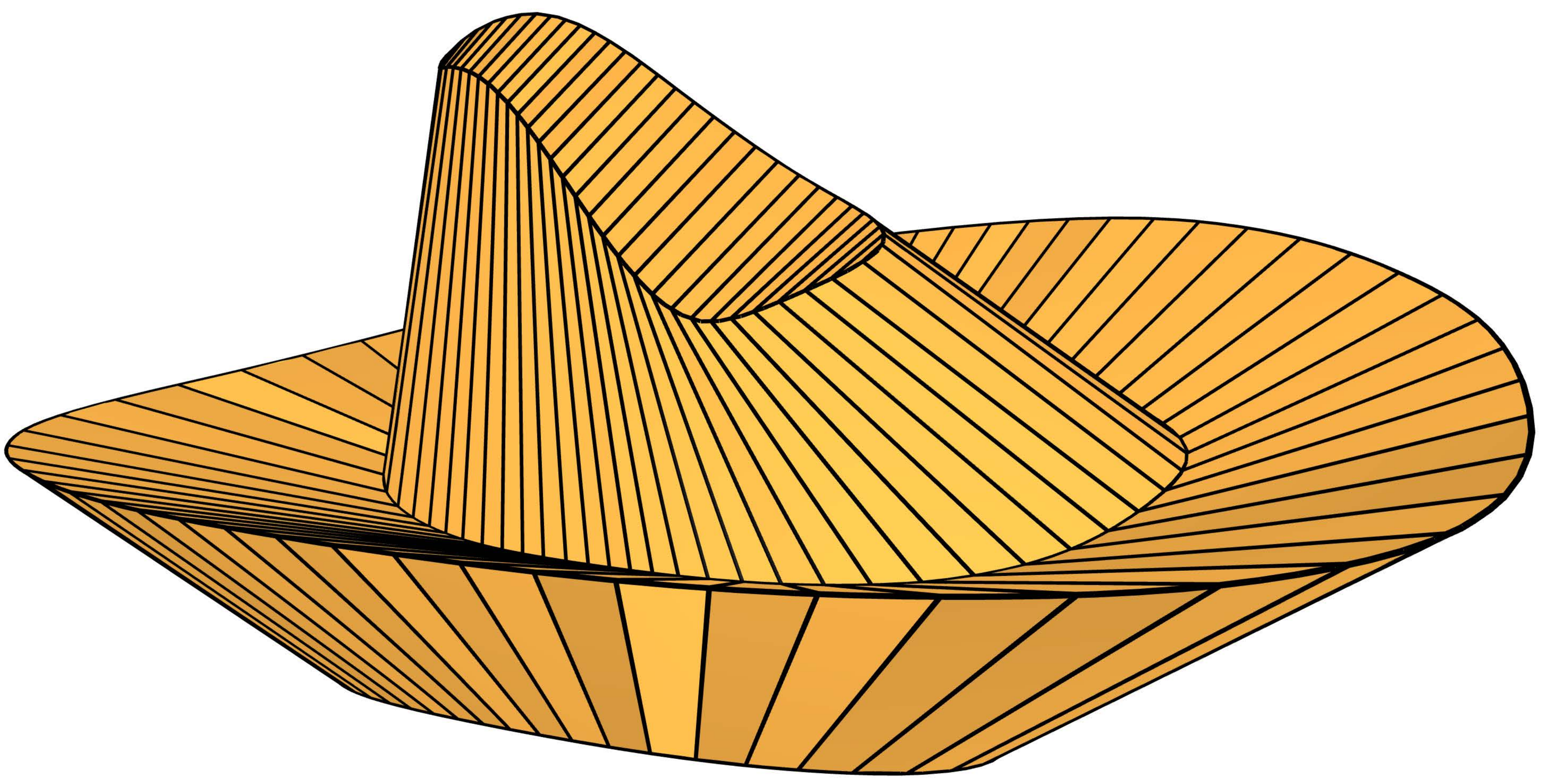}\\
		\end{tabular}%
	}
	\caption{Our method can handle piecewise developable surfaces with or without boundary and of different genera. These models are courtesy of \cite{pottmann_new}.}
	\label{fig:piecewisedev}
\end{figure}

\begin{figure*}[t]
	\centering
	\setlength{\tabcolsep}{2pt}
	\begin{tabular}{ccccc}
		\includegraphics[height=0.12\textheight]{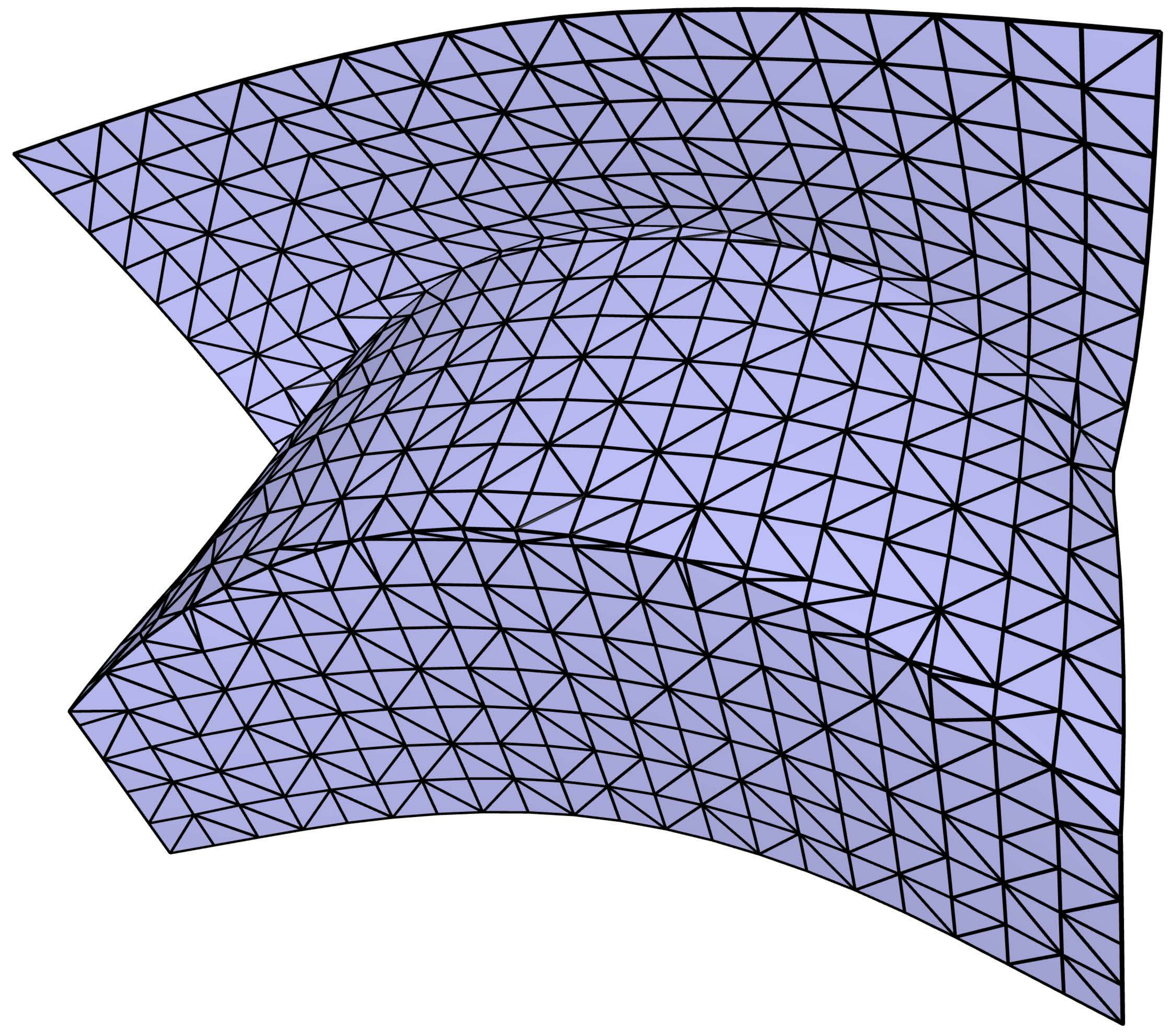}& 
		\includegraphics[height=0.12\textheight]{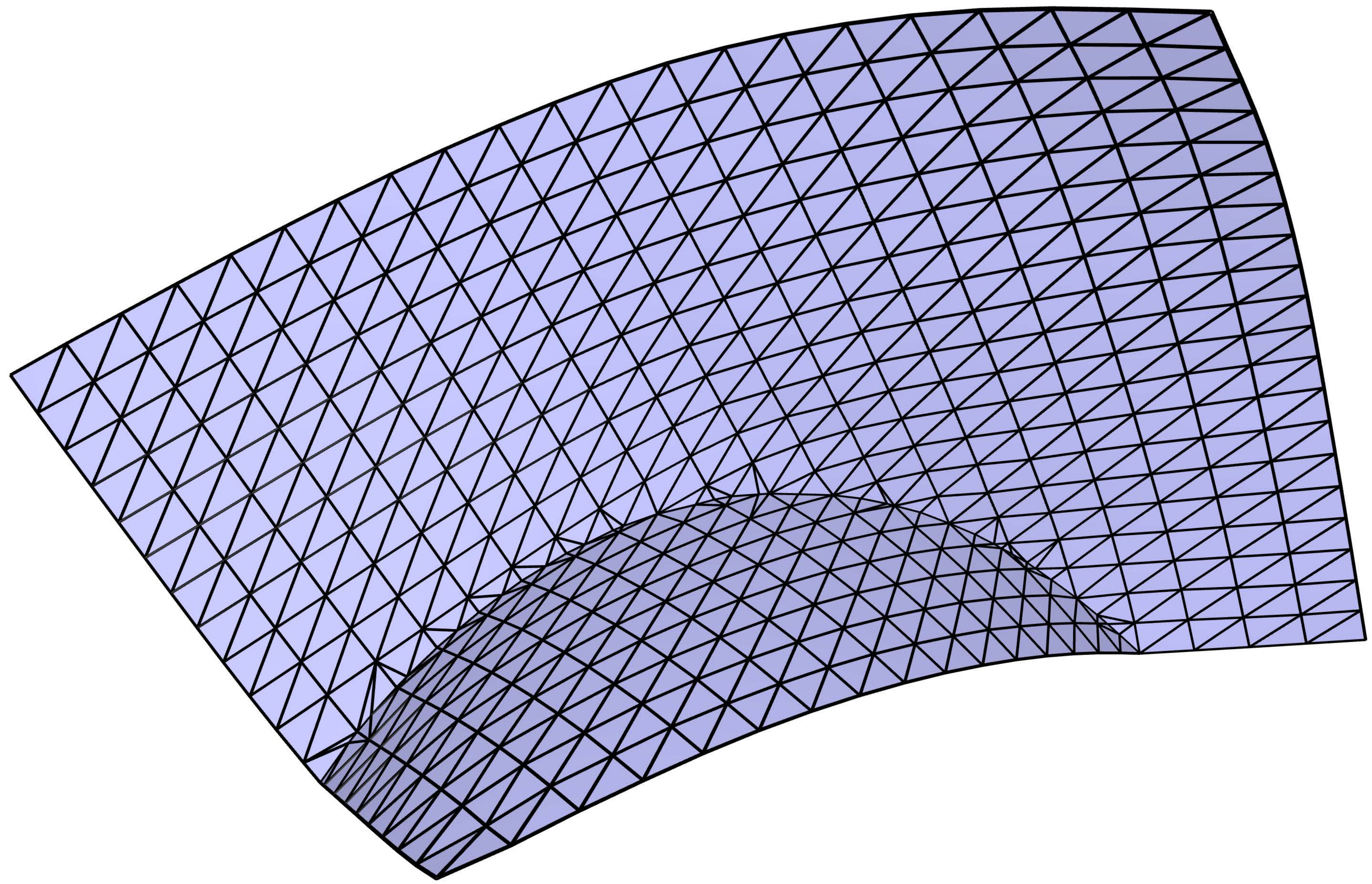}&
		\includegraphics[height=0.12\textheight]{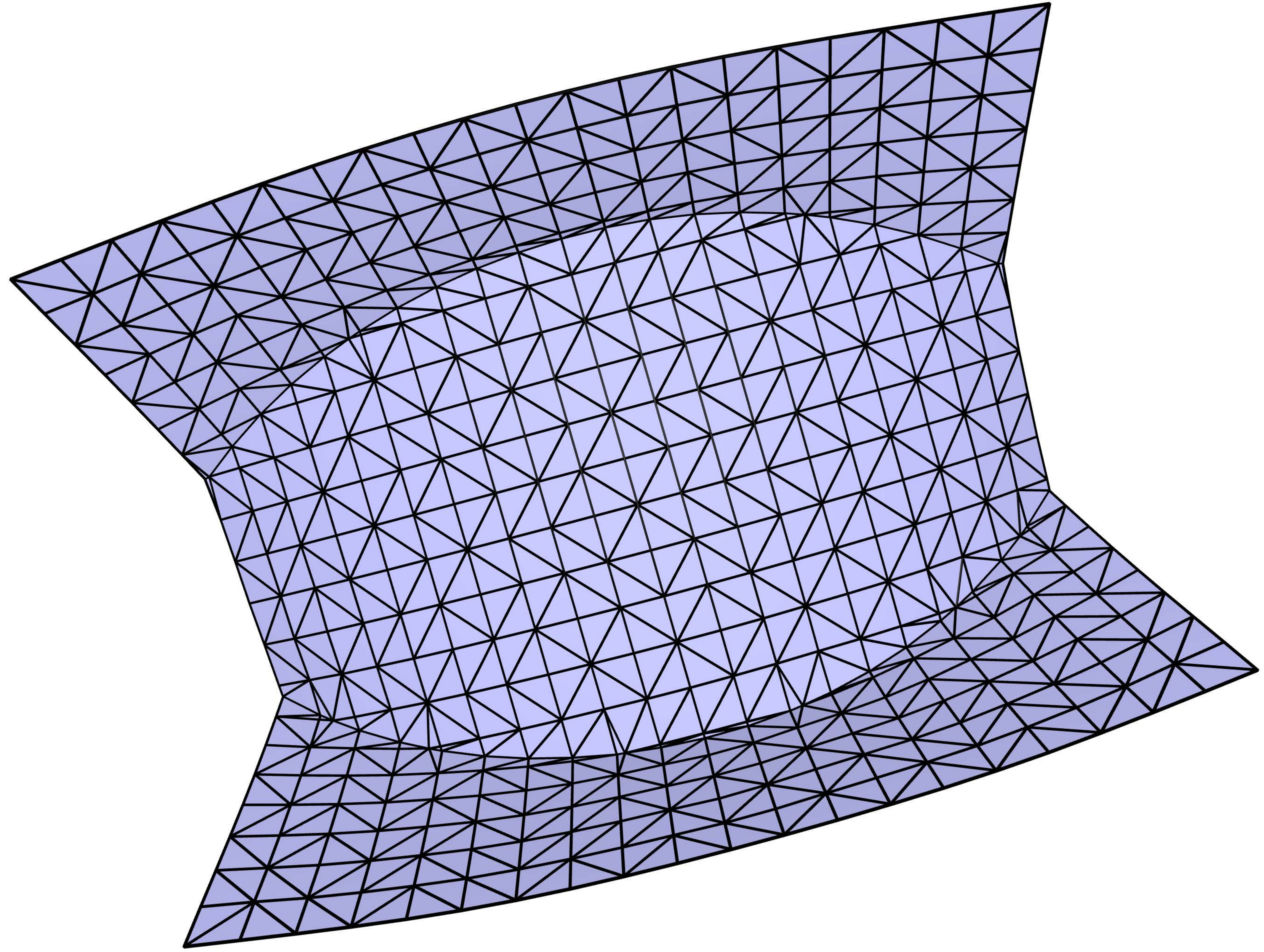}&
		\includegraphics[height=0.12\textheight]{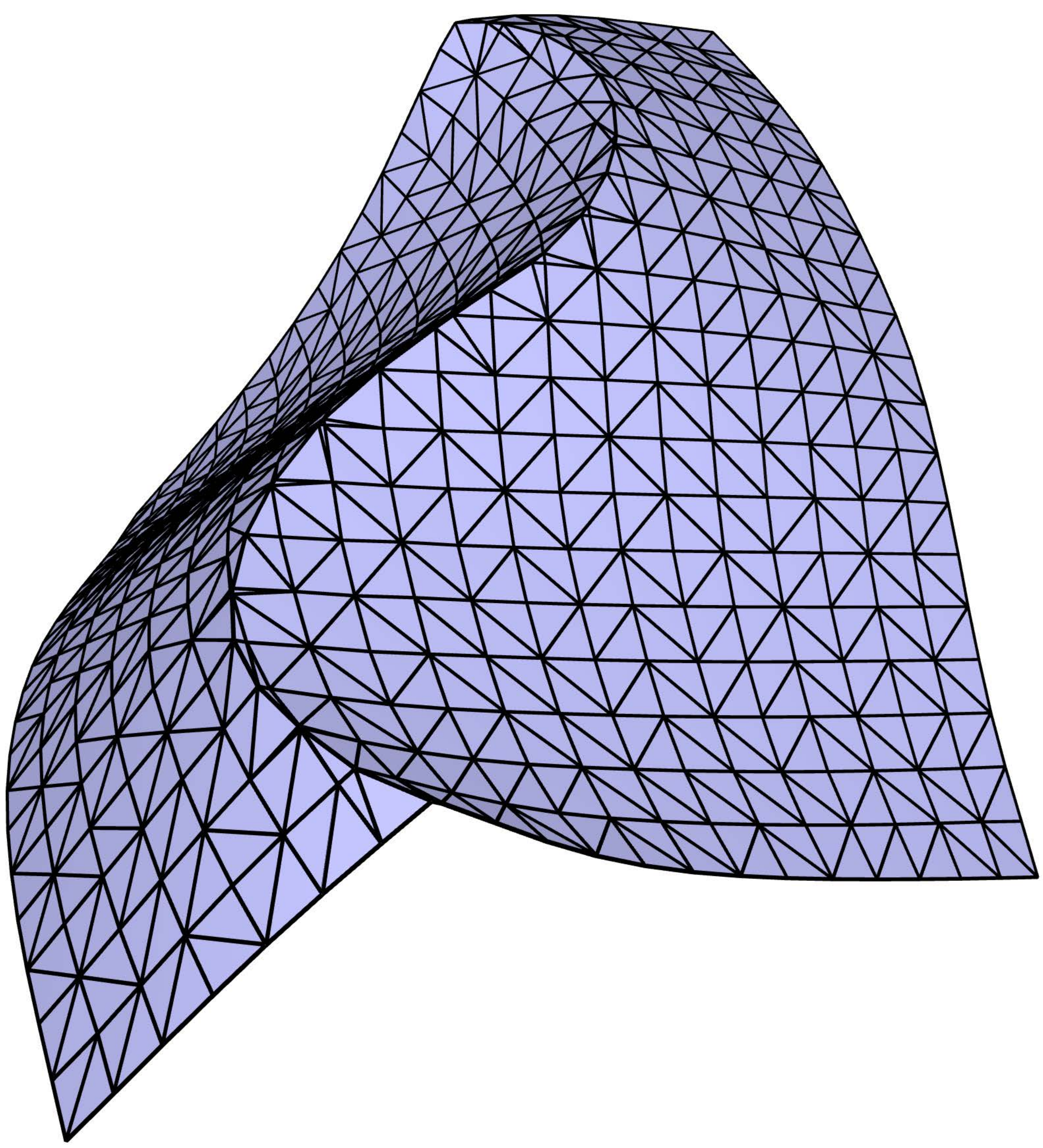}&
		\includegraphics[height=0.12\textheight]{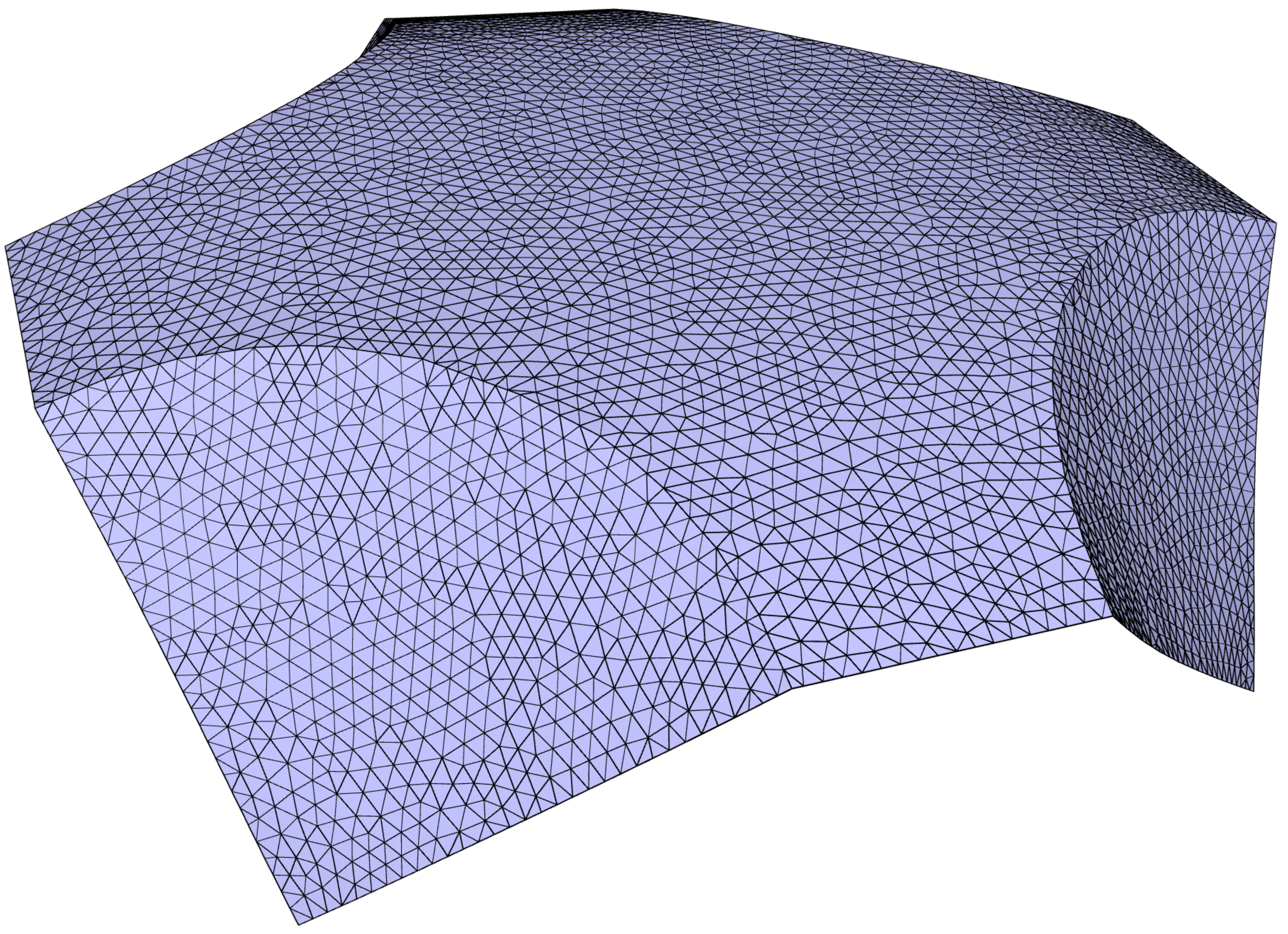}
		\\
		\includegraphics[height=0.12\textheight]{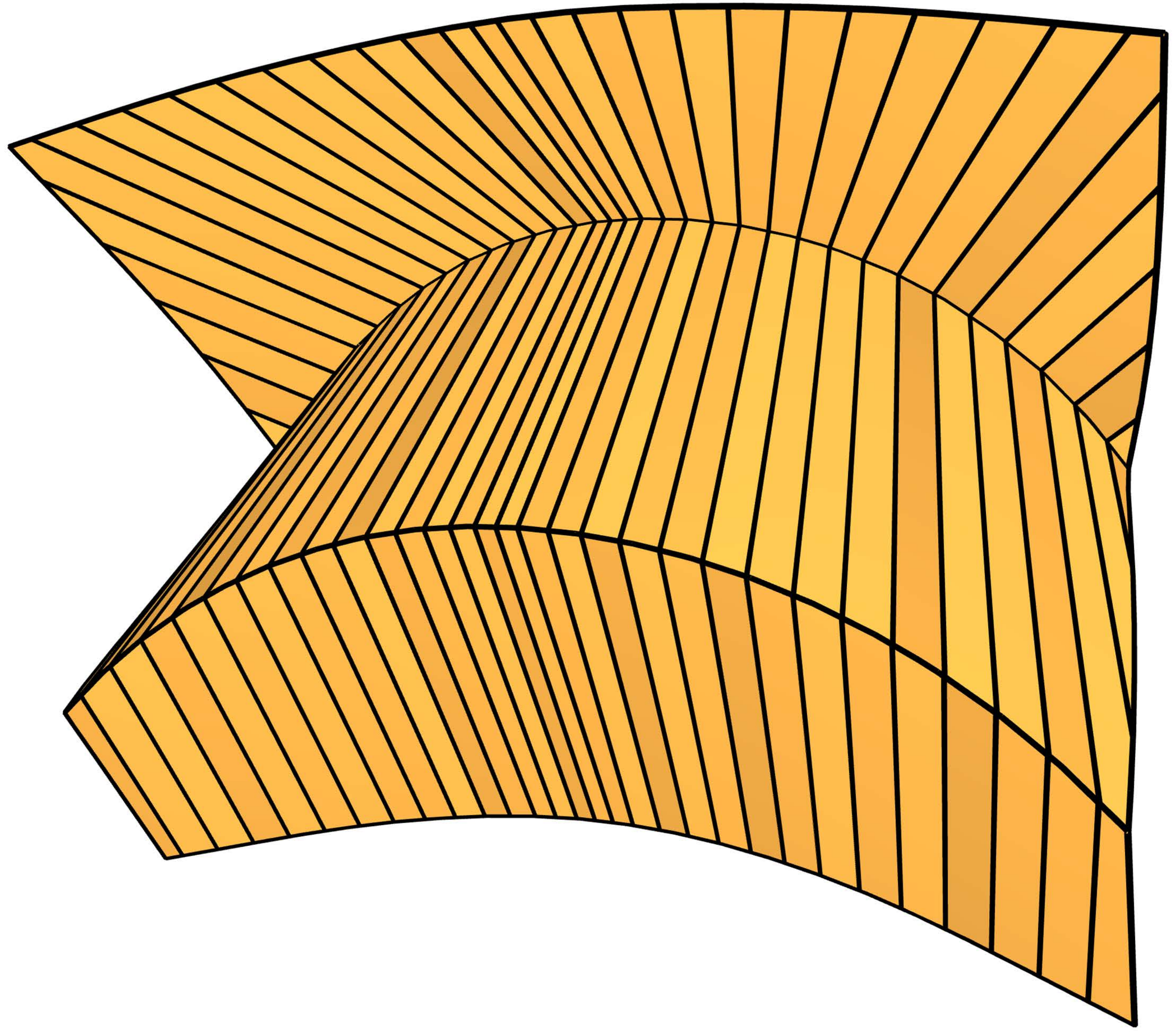}& 
		\includegraphics[height=0.12\textheight]{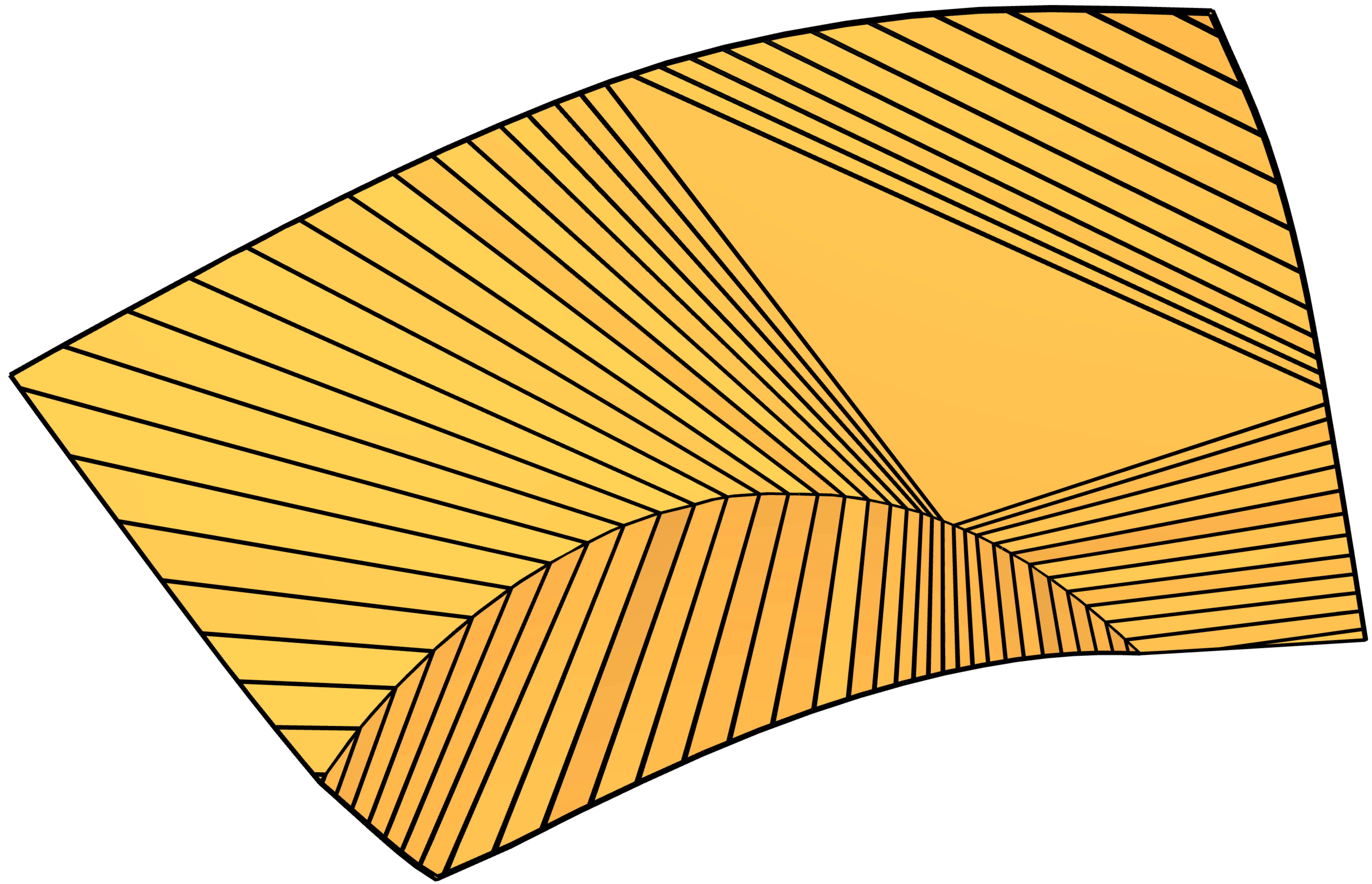}&
		\includegraphics[height=0.12\textheight]{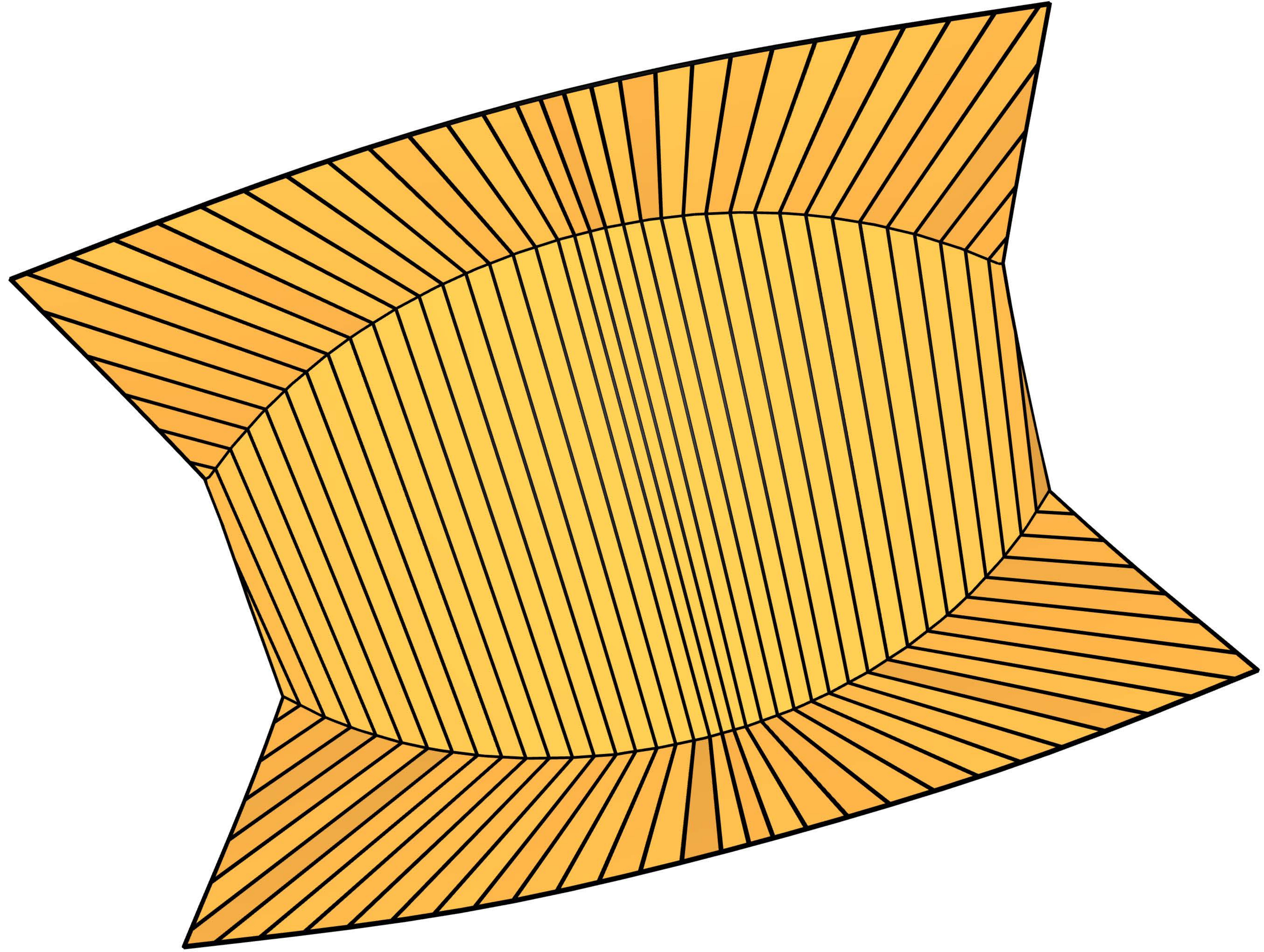}&
		\includegraphics[height=0.12\textheight]{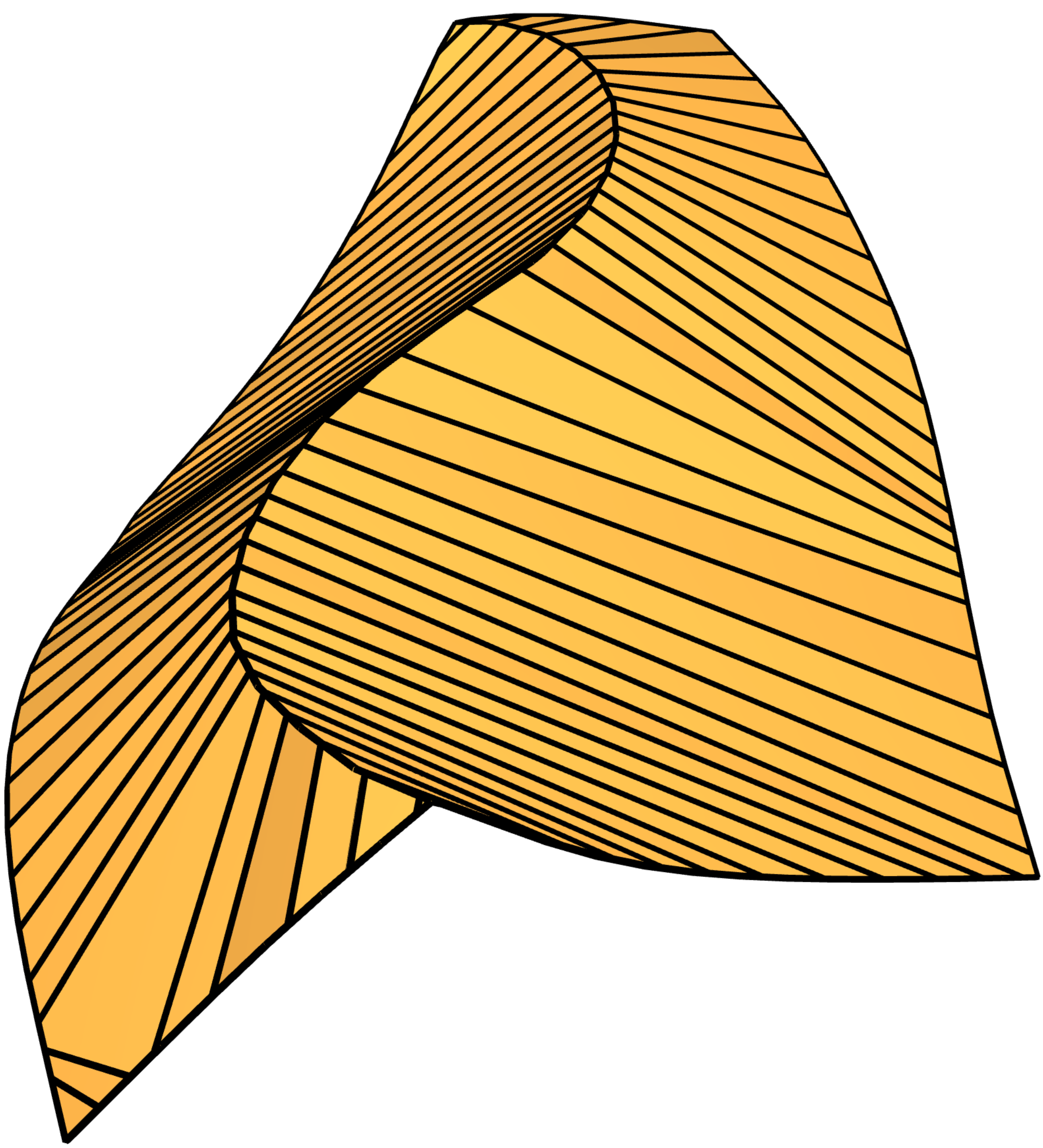}&
		\includegraphics[height=0.12\textheight]{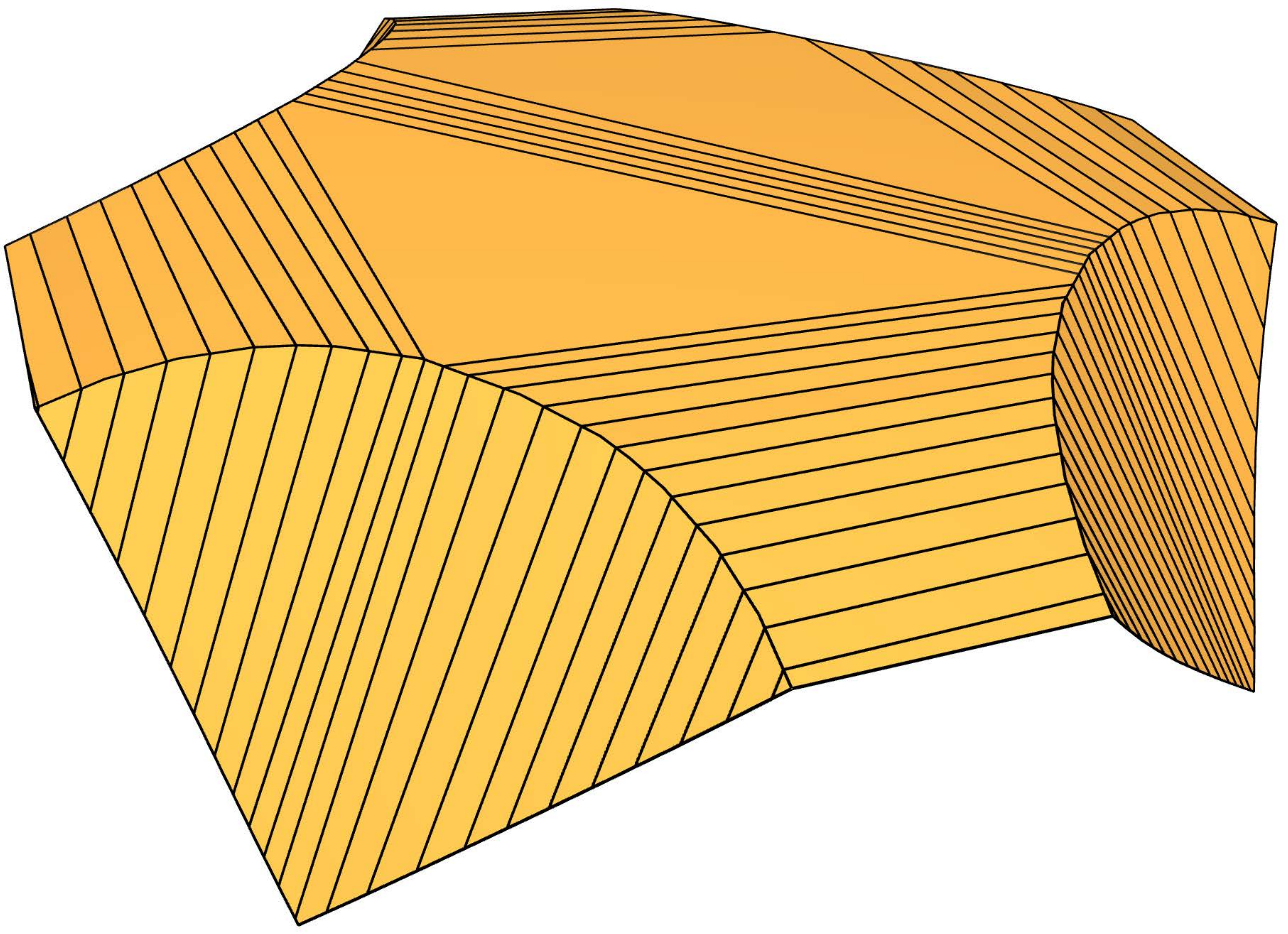}\\
	\end{tabular}
	\caption{For surfaces with curved folds our method produces meshes with faces that align well along the folds. Models courtesy of \cite{Rabinovich:CurvedFolds:2019}.}
	\label{fig:curvedfolds}
\end{figure*}

\begin{figure*}[t]
	\centering
	\setlength{\tabcolsep}{4pt}
	\begin{tabular}{cccc}
		\includegraphics[height=0.12\textheight]{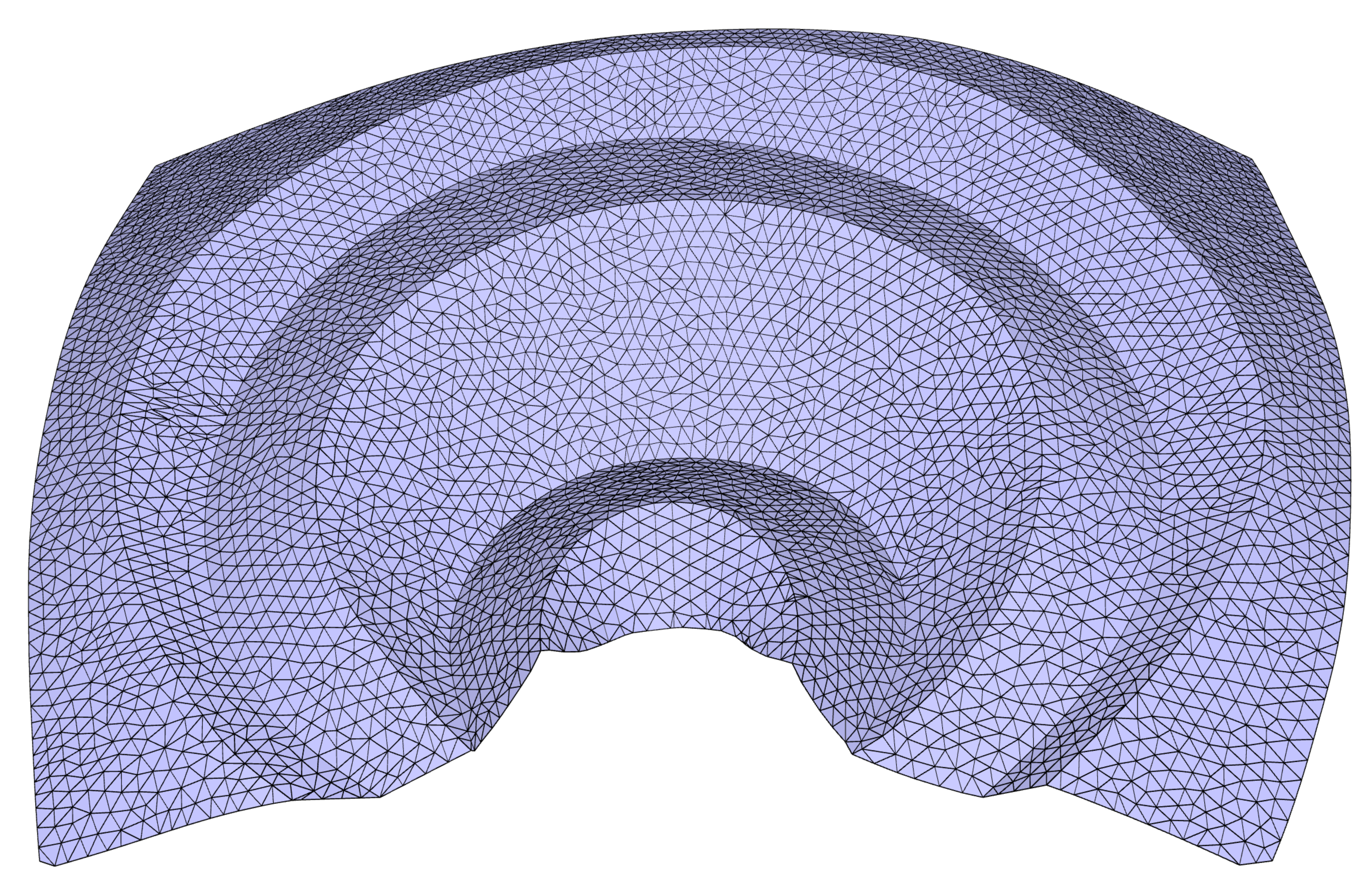}&
		\includegraphics[height=0.12\textheight]{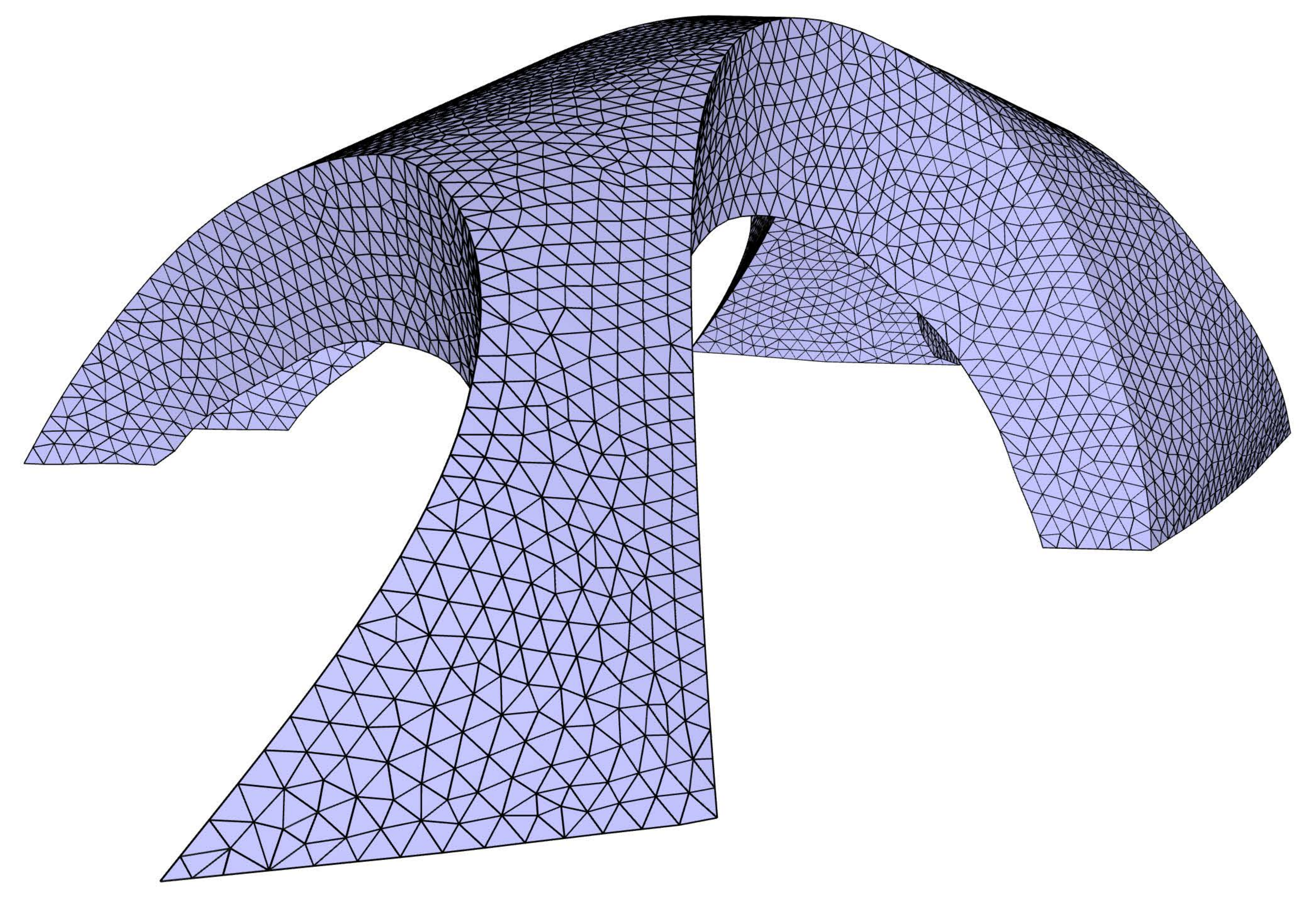}&
		\includegraphics[height=0.12\textheight]{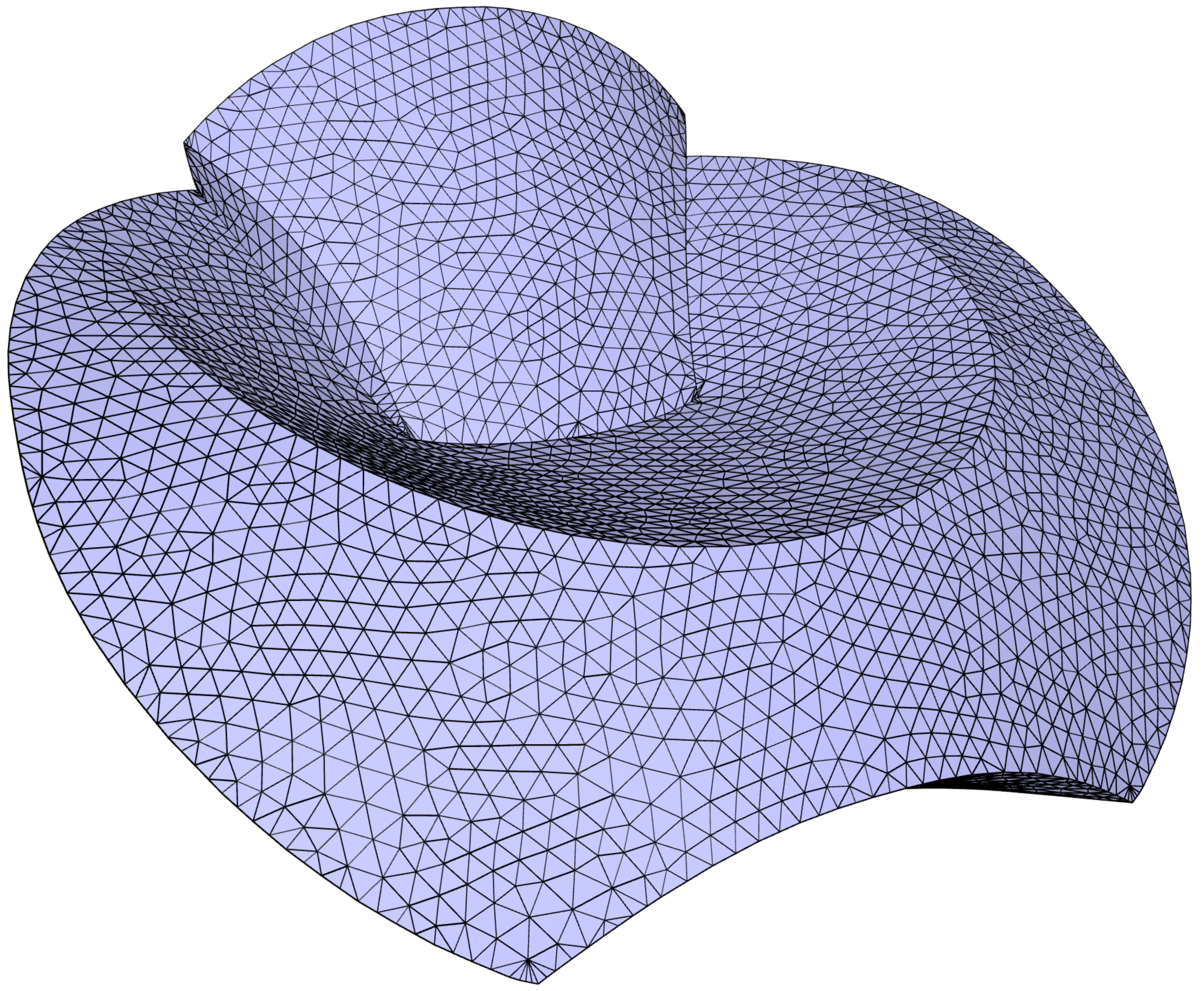}&		
		\includegraphics[height=0.12\textheight]{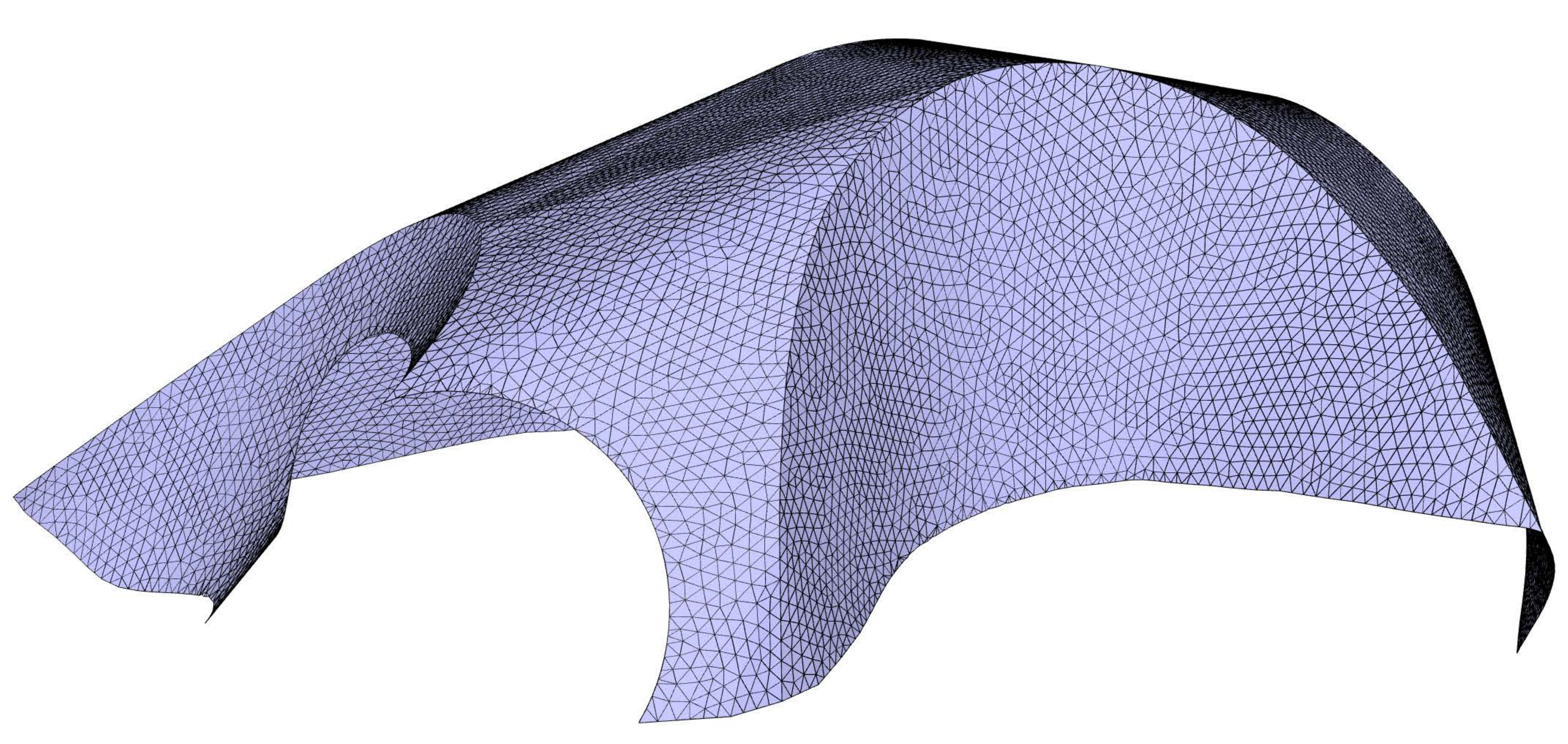}
		\\
		\includegraphics[height=0.12\textheight]{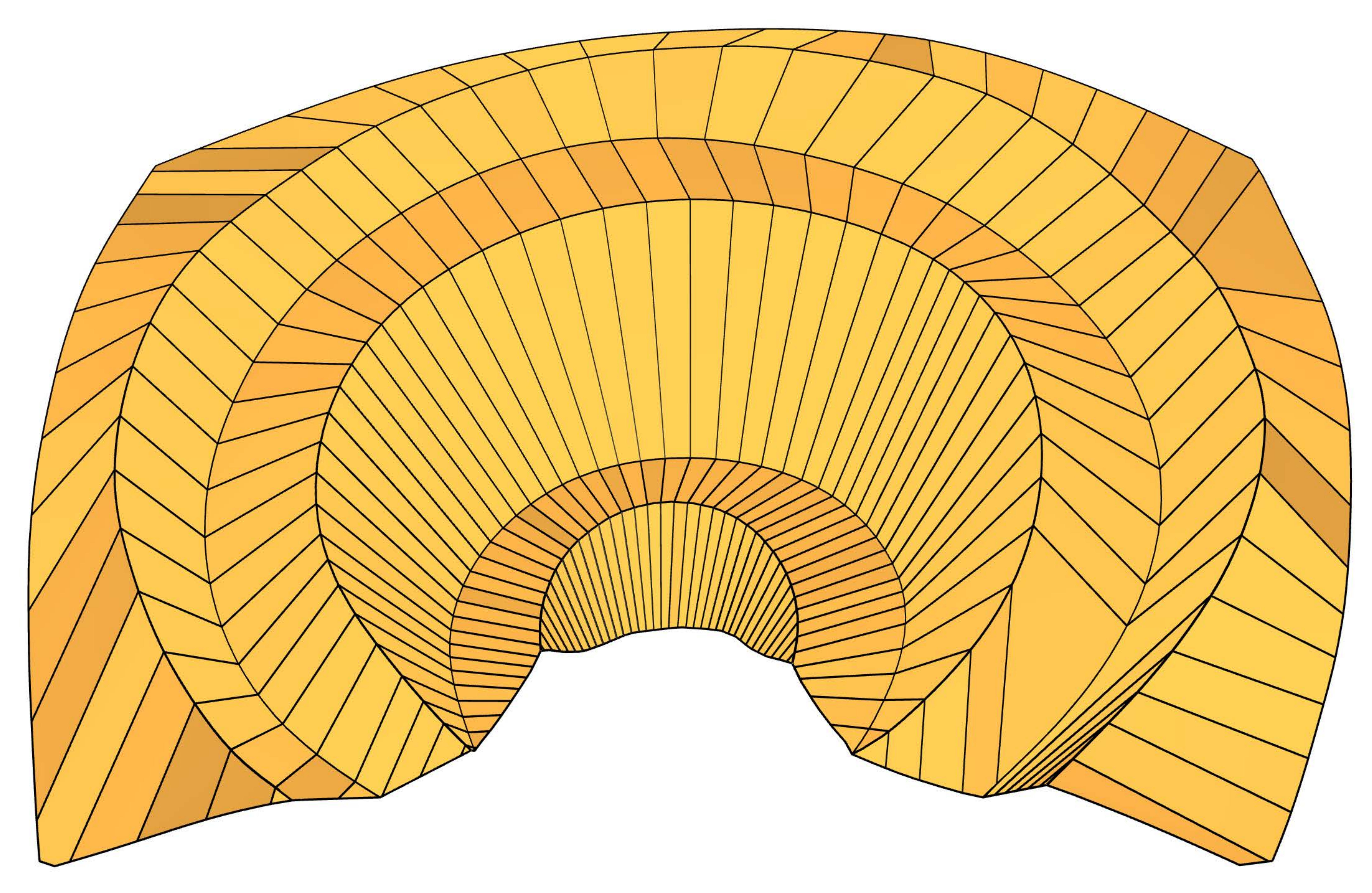}&
		\includegraphics[height=0.12\textheight]{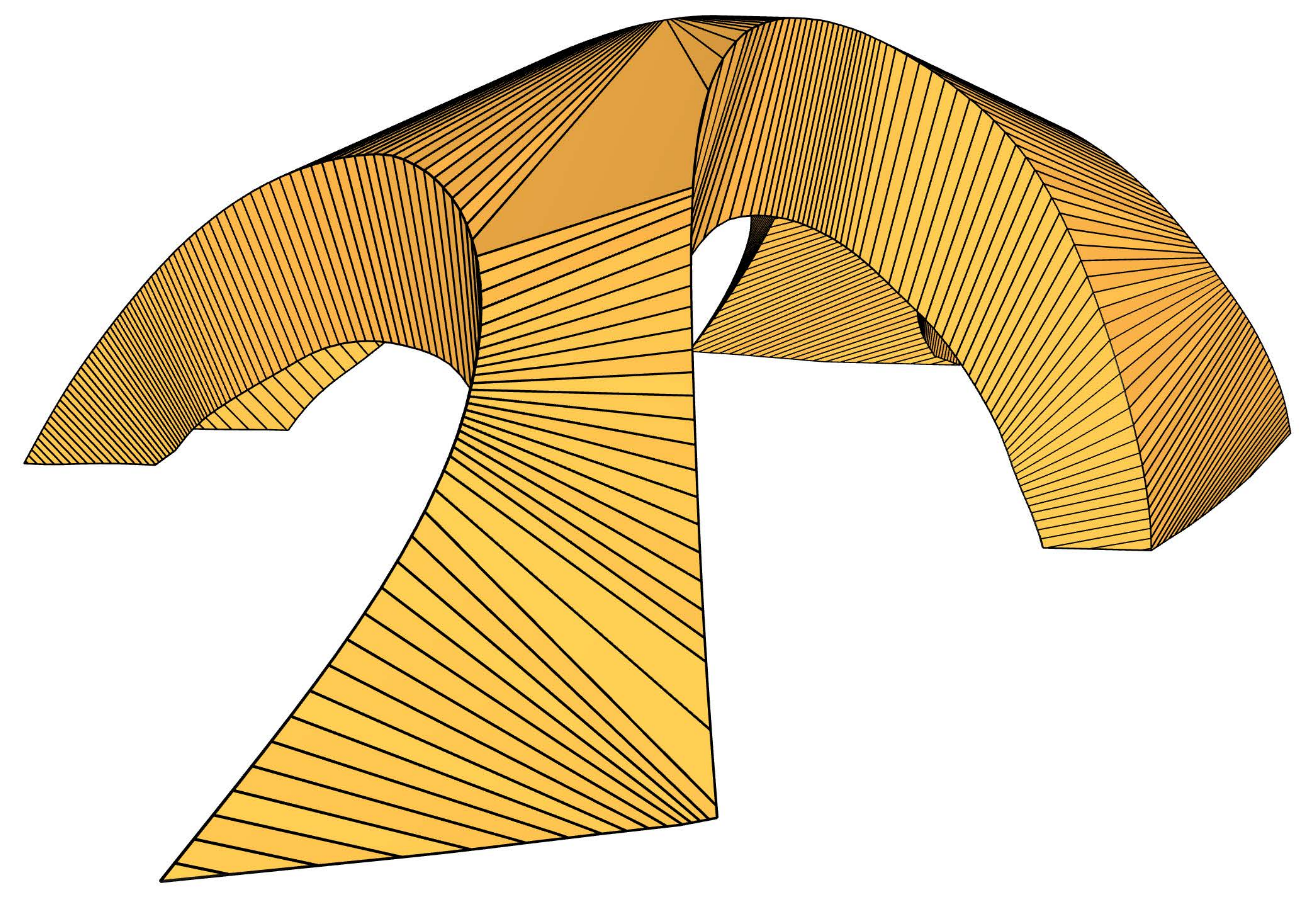}& 
		\includegraphics[height=0.12\textheight]{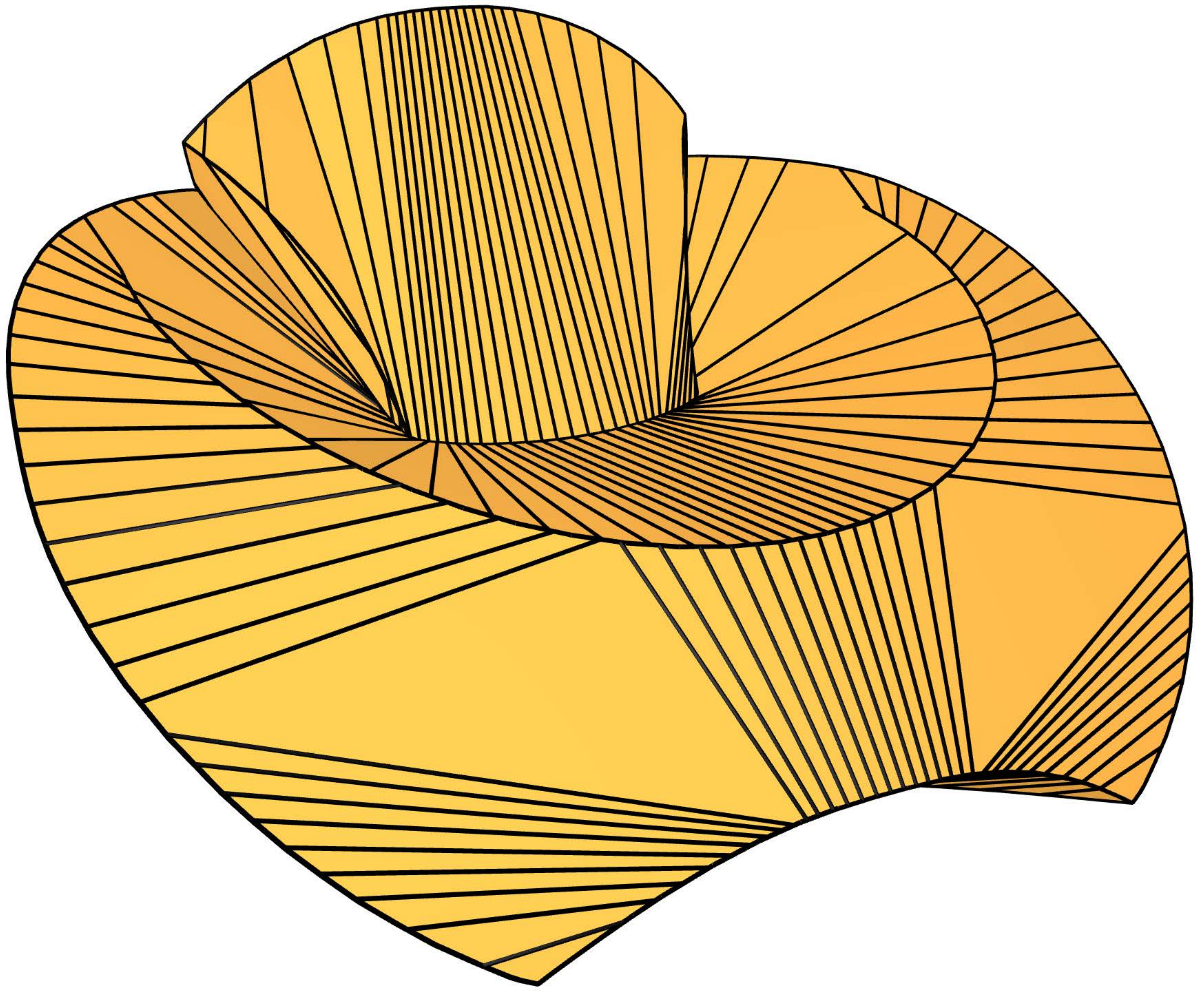}&
		\includegraphics[height=0.12\textheight]{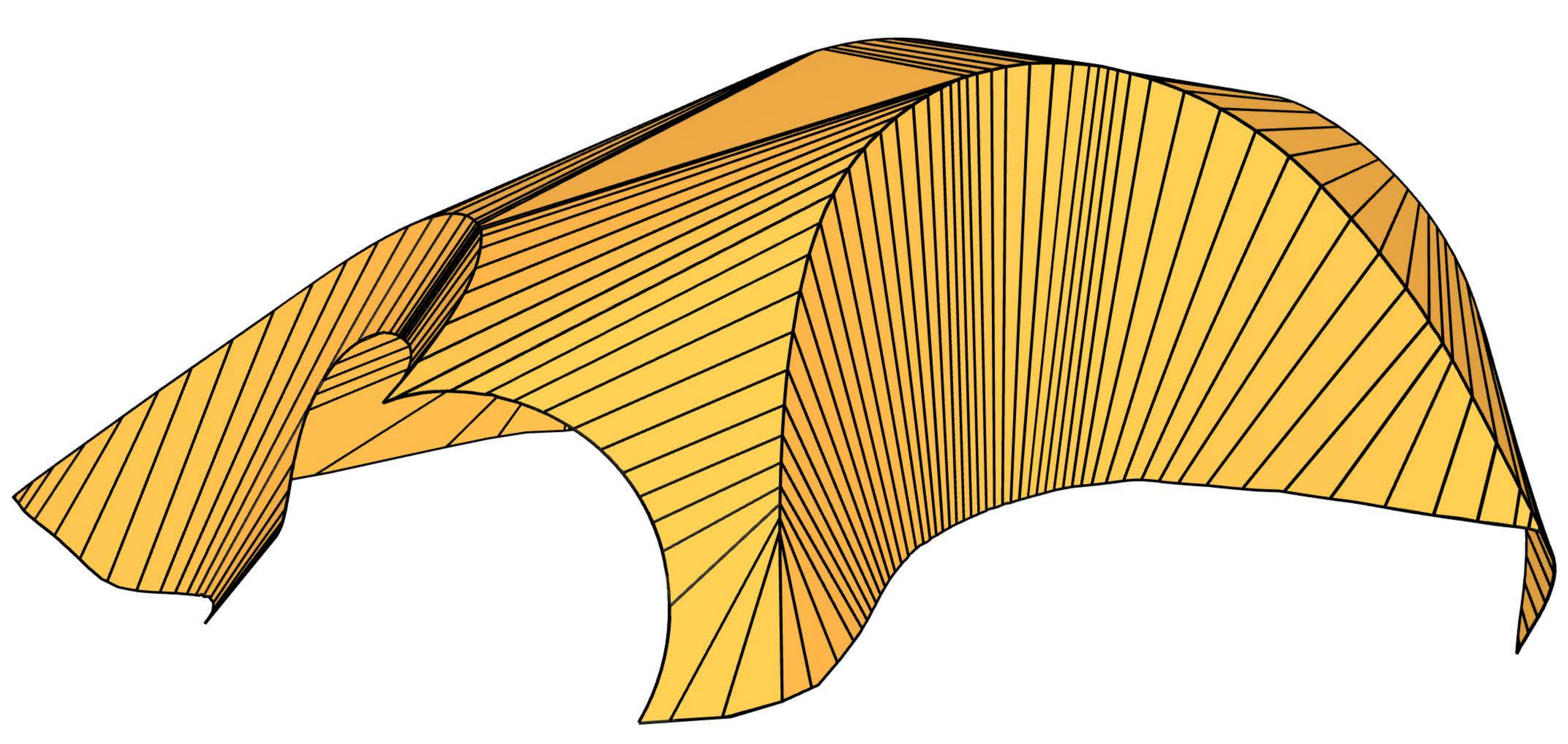}
		\\
	\end{tabular}
	\caption{Surfaces made with complicated curved folding patterns can also be handled by our method. Models courtesy of \cite{curved_folding_kilian}. Chair model courtesy of \cite{Kilianstringactuated}, based on a crease pattern designed by Benjamin Sp\"{o}th, owned by Joris Laarman Lab.}
	\label{fig:curvedfoldingKilian}
\end{figure*}


\section{Conclusion}
We presented an algorithm that converts developable surfaces represented by triangle meshes to a discrete curvature line parameterization, i.e., polygonal meshes with planar faces, where all interior edges correspond to rulings. This conversion from an unstructured triangulation of a developable surface to a curvature-aligned PQ mesh is an important step in the developable modeling and fabrication pipeline, for which thus far a robust practical solution was missing.

\setlength{\intextsep}{0pt}%
\setlength{\columnsep}{3pt}%
\begin{wrapfigure}{r}{0.35\linewidth}
	\centering
	\includegraphics[width=\linewidth]{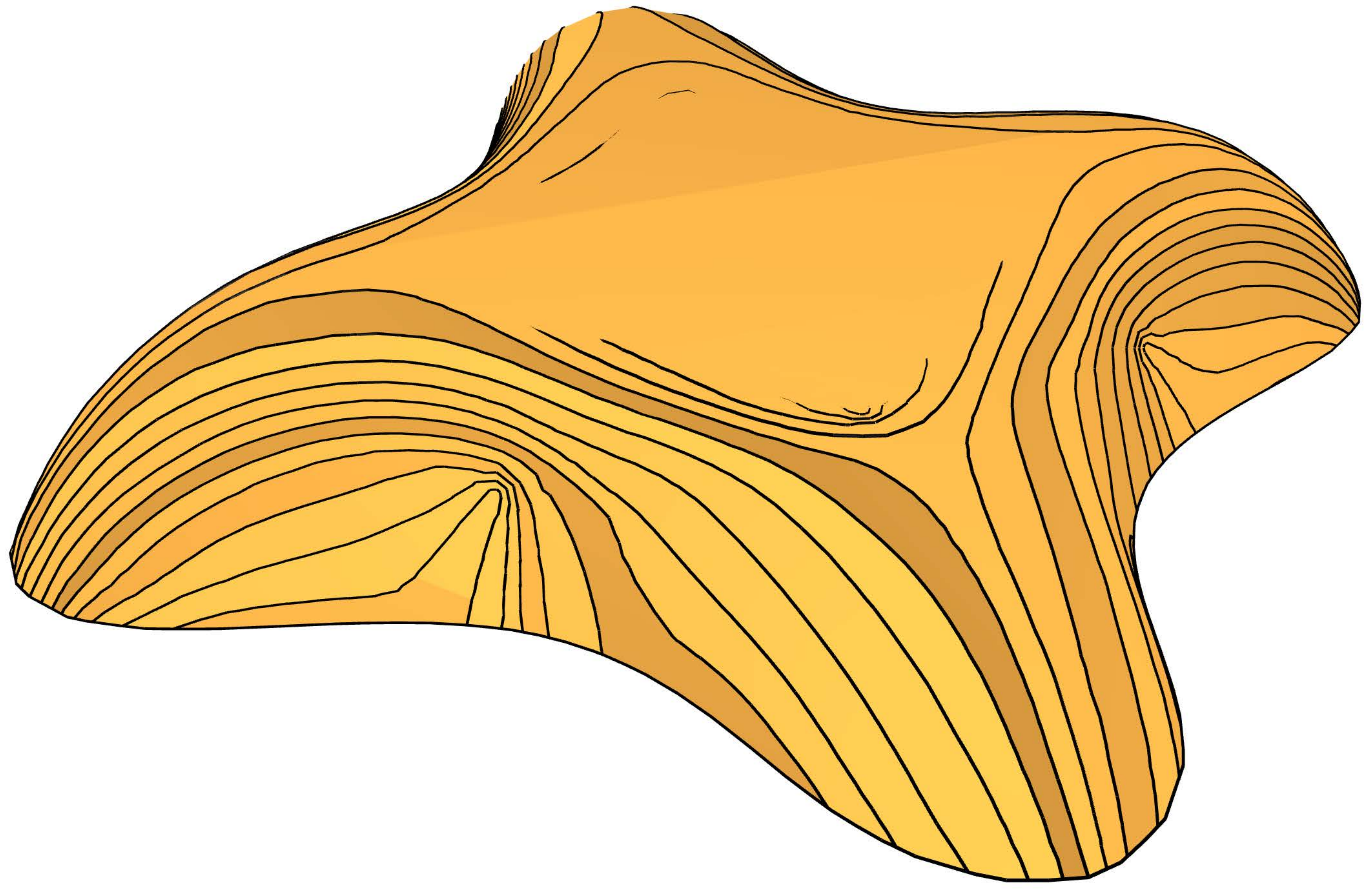}
\end{wrapfigure}
Our method can only be expected to work well for nearly developable input shapes. We show a non-developable example in the inset, where the edge simplification step is omitted as it would degenerate the mesh. 
It would be interesting to see how we can adapt our remeshing algorithm to be usable for the approximation of non-developable surfaces by developable patches. Another venue for further study would be the automatic tuning of the values $\omega_s$ and $\omega_a$ based on the noise levels of the estimated input rulings. 
Finally, it is conceivable to adapt the output mesh resolution based on the local curvature, allowing a denser representation in more curved areas.

\begin{figure*}[p]
	\centering
	\setlength{\tabcolsep}{0pt}   
	\includegraphics[width=\linewidth]{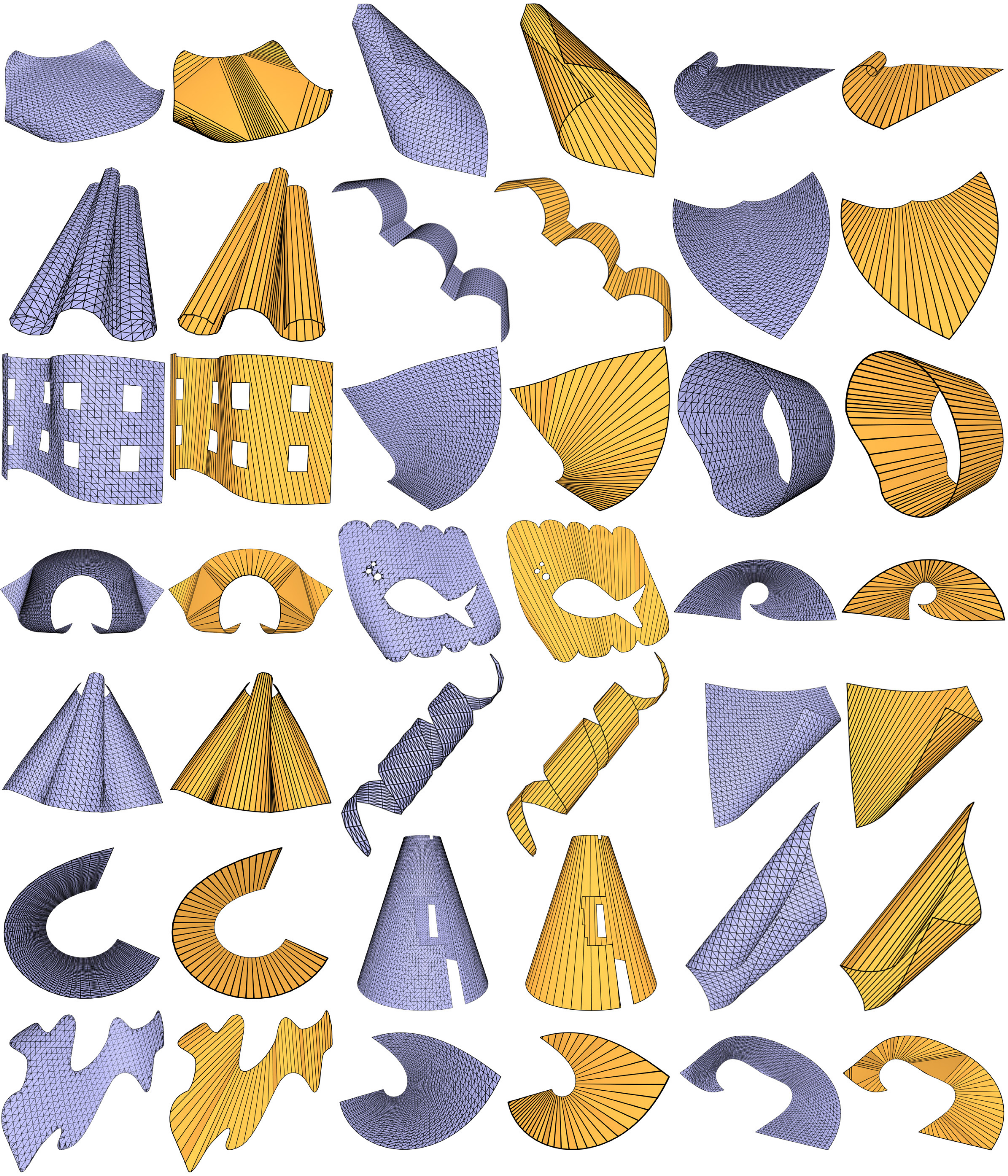}
	\caption{Various remeshing results obtained with our method. Note that our method can handle a wide variety of developable geometry and topology, including cylindrical topology, multiple holes and sophisticated boundary shapes.
	}
	\label{fig:galleryNewwide}
\end{figure*}

\begin{acks}
	The authors would like to thank Michael Rabinovich and Helmut Pottmann for illuminating discussions. This project has received funding from the European Research Council (ERC) under the European Union’s Horizon 2020 research and innovation programme (Grant agreement No. 101003104).
\end{acks}
\balance
\bibliographystyle{ACM-Reference-Format}
\bibliography{references}

\appendix

\captionsetup[subfloat]{labelformat=empty}
\begin{table*}
	\caption{{Statistics of our results, reported by figure number (in scanline order within figures that show multiple results). 
			We record the number of output mesh vertices $|\set{V}'|$ and faces $|\set{F}'|$, the number of optimization iterations needed to reach convergence, as well as the maximum and mean face planarity error $p$ (in percentage of the average diagonal length of the face). We also report the Hausdorff distance between the input and output mesh.
			Many of our meshes meet the common planarity tolerance of $\leq 1\%$ without any planarization optimization.}}
	\label{tab:planarity}
	\centering
	\setlength{\tabcolsep}{4.6pt}
	\begin{threeparttable}
		\subfloat[][]{
			\begin{tabular}{lcccccc}
			\toprule
			Fig. & 
			$|\set{V}'|$ &
			$|\set{F}'|$ &
			\small Iter. &
			\small $p_\text{max}$ [\%]&
			\small $p_\text{mean}$ [\%]&
			\small $h$[\%]\\
			\midrule
			\ref{fig:teaser} & 
			693 &
			137 &
			47 &
			0.38 &
			0.06 &
			0.46 \\ \midrule
			\ref{fig:general_planarization_fail}& 
			270 &
			74 &
			5 &
			1.80 & 
			0.41 &
			0.67\\  \midrule
			\ref{fig:quad_remesher_fail} & 
			180 &
			51 &
			8 &
			1.53& 
			0.28&
			0.45 \\
			\ref{fig:quad_remesher_fail} & 
			206 &
			44 &
			5 &
			0.59 &
			0.15 &
			0.85 \\
			\ref{fig:quad_remesher_fail} & 
			218 &
			50 &
			9 &
			0.88 &
			0.18 &
			1.07 \\
			\ref{fig:quad_remesher_fail} & 
			190 &
			48 &
			5 &
			0.95 &
			0.19 &
			2.55 \\  \midrule
			\ref{fig:pipeline} & 
			190 &
			48 &
			5 &
			0.95 &
			0.19 &
			2.55 \\  \midrule
			\ref{fig:editing-sequence} & 
			208 &
			45 &
			4 &
			0.41 &
			0.11 &
			0.66 \\
			\ref{fig:editing-sequence} & 
			206 &
			44 &
			5 &
			0.59 &
			0.15 &
			0.85 \\
			\ref{fig:editing-sequence} & 
			210 &
			46 &
			3 & 
			0.54 &
			0.16 &
			0.41 \\
			\ref{fig:editing-sequence} & 
			208 &
			45 &
			4 &
			0.45 &
			0.17 &
			1.39 \\
			\ref{fig:editing-sequence} & 
			218 &
			50 &
			9 &
			0.88 &
			0.18 &
			1.07 \\  \midrule
			\ref{fig:planarization} & 
			288 &
			107 &
			52 &
			11.84&
			1.98 &
			0.48 \\
			\ref{fig:planarization} & 
			288 &
			107 &
			- &
			0.00&
			0.00 &
			0.95 \\  \midrule
			\ref{fig:schwartztri} & 
			520 &
			110 &
			6 &
			2.58 &
			0.71 &
			0.09 \\ \midrule
			\ref{fig:different-resolutions} & 
			164 &
			23 &
			3 &
			0.52 &
			0.24&
			0.47 \\
			\ref{fig:different-resolutions} & 
			210 &
			46 &
			3 &
			0.54 &
			0.16&
			0.41 \\
			\ref{fig:different-resolutions} & 
			298 &
			90 &
			3 &
			0.32 &
			0.10&
			0.41 \\  \midrule
			\ref{fig:analytical-comparison} & 
			384 &
			68 &
			7 &
			1.08 &
			0.34 &
			0.31 \\
			\ref{fig:analytical-comparison} & 
			632 &
			67 &
			7 &
			1.22 &
			0.10 &
			0.26 \\
			\ref{fig:analytical-comparison} & 
			1130 &
			66 &
			36 &
			0.25 &
			0.02 &
			0.24 \\  \midrule
			\ref{fig:noisyinput} & 
			176 &
			29 & 
			18 &
			3.64 &
			1.04 &
			0.56 \\
			\ref{fig:noisyinput} & 
			156 &
			19 & 
			22 &
			5.42 &
			2.51 &
			1.34 \\  \midrule
			\ref{fig:triangulation-direction} & 
			254 &
			68 & 
			19 &
			3.83 &
			0.42 &
			0.63 \\
			\ref{fig:triangulation-direction} & 
			254 &
			68 & 
			23 &
			3.82 &
			0.42 &
			0.66 \\
			\ref{fig:triangulation-direction} & 
			258 &
			70 & 
			23 &
			3.34 &
			0.48 &
			0.21 \\  \midrule
			\ref{fig:oscillation} & 
			2194 &
			447 &
			299\tnote{***} &
			4.78 &
			0.34 &
			2.90 \\ \midrule
			\ref{fig:sphericonsdforms} & 
			202 &
			100 &
			12 &
			0.14 &
			0.05 &
			0.20 \\
			\ref{fig:sphericonsdforms} & 
			324 &
			100 &
			10 &
			0.46 &
			0.14 &
			0.22 \\
			\ref{fig:sphericonsdforms} & 
			858 &
			274 &
			9 &
			1.66 &
			0.09 &
			1.15 \\
			\ref{fig:sphericonsdforms} & 
			966 &
			272 &
			6 &
			1.14 &
			0.08 &
			0.13 \\ \midrule
			\ref{fig:glued} & 
			306 &
			54 &
			2 &
			0.15 &
			0.04 &
			0.15 \\
			\ref{fig:glued} & 
			438 &
			54 &
			6 &
			0.14 &
			0.04 &
			1.91 \\
			\ref{fig:glued} & 
			578 &
			93 &
			41 &
			2.55 &
			0.30  &
			1.09 \\ 
			\bottomrule
		\end{tabular}
	}
\hspace{0.05\linewidth}
\subfloat[][]{
		\begin{tabular}[t]{lcccccc}
			\toprule
			Fig. & 
			$|\set{V}'|$ &
			$|\set{F}'|$ &
			\small Iter. &
			\small $p_\text{max}$ [\%]&
			\small $p_\text{mean}$ [\%]&
			\small $h$[\%]\\
			 \midrule
			\ref{fig:Keenan} & 
			2455 &
			440 &
			46 &
			2.65 &
			0.25 &
			3.29 \\
			\ref{fig:Keenan} & 
			972 &
			195 &
			17 &
			1.71 &
			0.26 &
			0.53 \\
			\midrule
			\ref{fig:piecewisedev} & 
			288&
			170 &
			18 &
			0.19 &
			0.05 &
			0.09 \\
			\ref{fig:piecewisedev}\tnote{*} & 
			1970 &
			567 &
			49 &
			0.82 &
			0.05 &
			0.04 \\
			\ref{fig:piecewisedev}\tnote{**} & 
			1947 &
			636 &
			18 &
			0.26 &
			0.05 &
			0.01 \\
			\ref{fig:piecewisedev} & 
			738 &
			211 &
			15 &
			0.84 &
			0.10 &
			0.14 \\  \midrule
			\ref{fig:curvedfolds} & 
			258 &
			98 &
			4 &
			1.16 &
			0.24 &
			0.23 \\
			\ref{fig:curvedfolds} &
			243 &
			80 &
			44 &
			0.67 &
			0.20 &
			0.62 \\
			\ref{fig:curvedfolds} &
			280 &
			115 &
			6 &
			1.98 &
			0.16 &
			0.35 \\
			\ref{fig:curvedfolds} &
			232 &
			82 &
			21 &
			2.24 &
			0.32 &
			0.42 \\
			\ref{fig:curvedfolds} & 
			693 &
			137 &
			47 &
			0.38 &
			0.06 &
			0.46\\  \midrule
			\ref{fig:fabrication} & 
			196 &
			52 &
			4 &
			2.39 &
			0.37 &
			0.26 \\ 
			\ref{fig:fabrication} & 
			210 &
			28 &
			9 &
			3.49 &
			0.43 &
			1.47 \\\midrule
			\ref{fig:curvedfoldingKilian} & 
			1115 &
			275 &
			16 &
			1.62 &
			0.19 &
			0.28 \\
			\ref{fig:curvedfoldingKilian} & 
			1603 &
			747 &
			14 &
			0.31 &
			0.06 &
			1.90 \\
			\ref{fig:curvedfoldingKilian} & 
			873 &
			188 &
			57 &
			0.75 &
			0.11 &
			3.89 \\
			\ref{fig:curvedfoldingKilian} & 
			1473 &
			294 &
			10 &
			1.84 &
			0.12 &
			0.18 \\	
	\midrule
			\ref{fig:galleryNewwide} & 
			254 &
			68 &
			15 &
			3.63 &
			0.41 &
			0.65 \\
			\ref{fig:galleryNewwide} & 
			210 &
			46 &
			4 &
			1.25 &
			0.19 &
			0.23 \\
			\ref{fig:galleryNewwide} & 
			204 &
			43 &
			20 &
			7.18 &
			0.50 &
			0.34 \\
			\ref{fig:galleryNewwide} & 
			212 &
			47 &
			3 &
			0.56 &
			0.33 &
			0.60 \\
			\ref{fig:galleryNewwide} & 
			390 &
			70 &
			7 &
			3.93 &
			0.31 &
			0.67 \\
			\ref{fig:galleryNewwide} & 
			204 &
			43 &
			3 &
			0.69 &
			0.14 &
			0.10 \\
			\ref{fig:galleryNewwide} & 
			422 &
			88 &
			5 &
			3.35 &
			0.56 &
			0.85 \\
			\ref{fig:galleryNewwide} & 
			208 &
			45 &
			7 &
			0.43 &
			0.10 &
			0.46 \\
			\ref{fig:galleryNewwide} & 
			226 &
			63 &
			8 &
			0.40 &
			0.12 &
			0.18 \\
			\ref{fig:galleryNewwide} & 
			328 &
			75 &
			40 &
			0.78 &
			0.18 &
			0.69 \\
			\ref{fig:galleryNewwide} & 
			1033 &
			71 &
			7 &
			1.72 &
			0.34 &
			0.38 \\
			\ref{fig:galleryNewwide} & 
			218 &
			50 &
			4 &
			2.35 &
			0.20 &
			0.32 \\
			\ref{fig:galleryNewwide} & 
			242 &
			58 &
			4 &
			0.80 &
			0.36 &
			1.29 \\
			\ref{fig:galleryNewwide} & 
			288 &
			107 &
			52 &
			11.84&
			1.98 &
			0.48 \\
			 \ref{fig:galleryNewwide} & 
			 208 &
			 45 &
			 4 &
			 0.38 &
			 0.11 &
			 0.20 \\
			\ref{fig:galleryNewwide} & 
			214 &
			48 &
			9 &
			0.15 &
			0.05 &
			0.08 \\
			\ref{fig:galleryNewwide} & 
			502 &
			53 &
			16 &
			0.32 &
			0.06 &
			0.09 \\	
			\ref{fig:galleryNewwide} & 
			194 &
			45 &
			3 &
			1.19 &
			0.24 &
			0.40 \\
			\ref{fig:galleryNewwide} & 
			609 &
			47 &
			7 &
			2.89 &
			0.60 &
			0.21 \\ 
			\ref{fig:galleryNewwide} & 
			204 &
			43 &
			7 &
			0.49 &
			0.13 &
			0.39 \\
			\ref{fig:galleryNewwide} & 
			288 &
			67 &
			16 &
			1.31 &
			0.20 &
			0.82 \\
			\bottomrule
		\end{tabular}
	}
		\begin{tablenotes}
			\item[*] Same model displayed from 3 viewing angles. \item[**] Same model displayed from 2 viewing angles. \item[***] Maximal number of iterations, not converged. 
		\end{tablenotes}
	\end{threeparttable}
\end{table*}

\end{document}